%% file: disertace.tex
\pdfoutput=1
\documentclass[12pt, twoside]{article}
 \usepackage{graphics,fancyheadings,sectsty, color, a4wide}
\usepackage{natbib}
\begin{document}
\long\def\symbolfootnote[#1]#2{\begingroup%
\def\thefootnote{\fnsymbol{footnote}}\footnote[#1]{#2}\endgroup}

\long\def\symbolfootnotetext[#1]#2{\begingroup%
\def\thefootnote{\fnsymbol{footnote}}\footnotetext[#1]{#2}\endgroup}

\def\figurename{\it\small Figure}
\def\tablename{\it\small Table}
\newcommand{\mps}{m\,s$^{-1}$}
\newcommand{\arcsec}{$^{\prime\prime}$}
\newcommand{\degr}{$^\circ$}

\allsectionsfont{\sf\bfseries}
\include{casopisy}
\begin{titlepage}
\begin{center}
{\sf 
{\large Charles University in Prague\\[1mm]
Faculty of Mathematics and Physics}

\vspace{50pt}
{\Huge\bfseries DOCTORAL THESIS}

\vspace{50pt}
\resizebox{5cm}{!}{\includegraphics{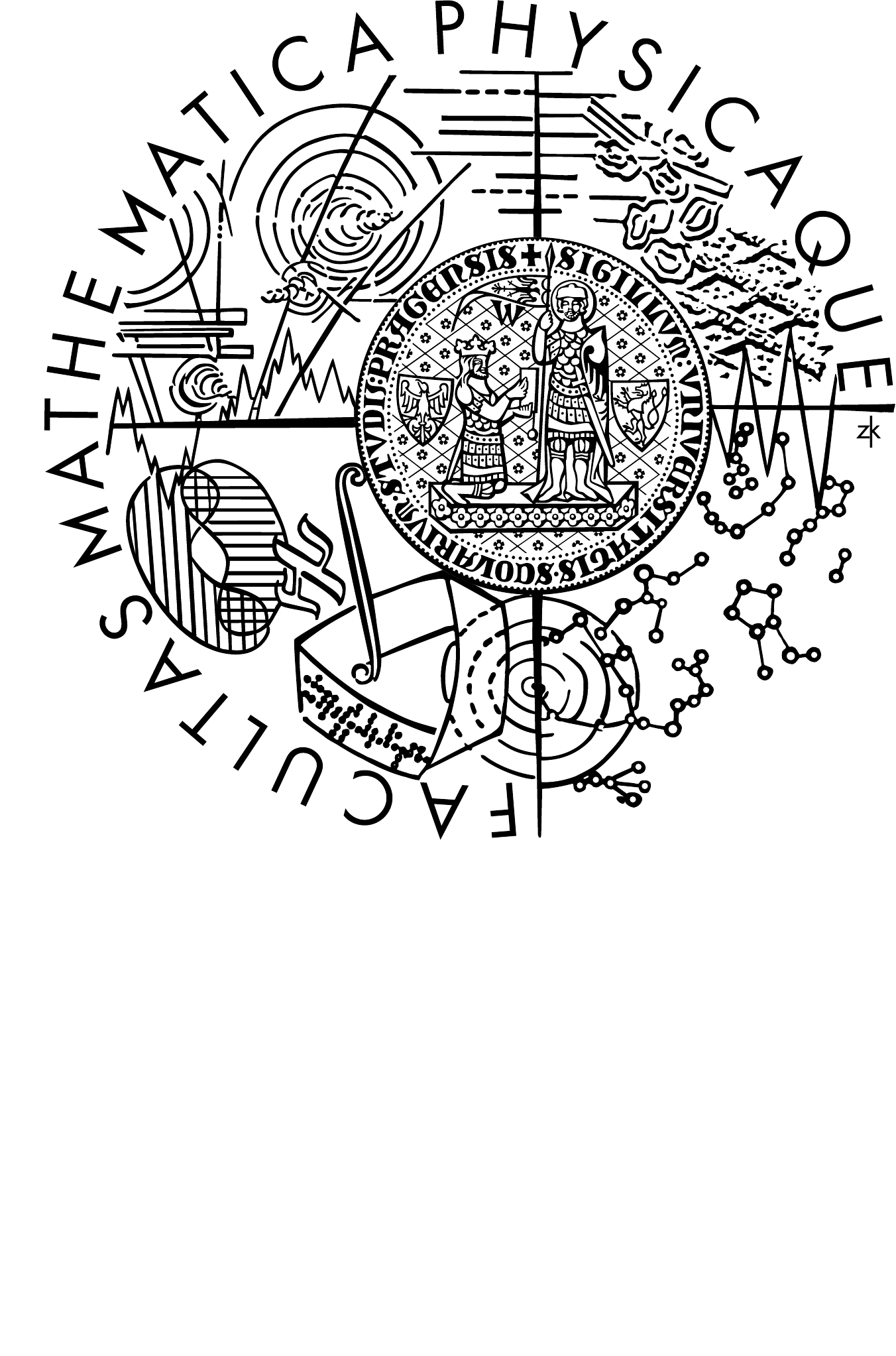}}

{\Large Michal \v{S}vanda}

\vspace{20pt}
{\Large\bfseries Velocity Fields \\
\vspace{5pt}
in the Solar Photosphere}

\vspace{40pt}
{\large Astronomical Institute (v. v. i.)\\[1mm]
 Academy of Sciences of the Czech Republic\\[1mm]
 Observatory Ond\v{r}ejov}\\
\vspace{10pt}
Supervisor: {\bfseries RNDr. Michal Sobotka, CSc.}\\
Advisers: {\bfseries Ing. Mirek Klva\v{n}a, CSc., RNDr. Pavel Ambro\v{z}, CSc.}\\[3mm]
{\sl A dissertation submitted to Committee for Doctoral Study Programme F1 of MFF UK in partial fulfillment of the requirements for the degree of Doctor of Philosophy.}
}
\end{center}
\end{titlepage}

\newpage
\thispagestyle{empty}
\vspace*{5cm}
\newpage
\thispagestyle{empty}
\vspace*{5cm}
\newpage

\pagestyle{empty}
\noindent Some parts of this thesis were made under an effective support of Czech Science Foundation under the grants No. 205/04/2129 and 205/03/H144, of Grant Agency of Academy of Sciences of the Czech Republic under grant No. IAA 3003404, and ESA-PECS under grant No. 98030. The data used in this thesis were kindly provided by the SoHO/MDI consortium (SoHO is the project of international cooperation between ESA and NASA) and NSO/Kitt Peak, which are produced cooperatively by NSF/NOAO, NASA/GSFC, and NOAA/SEL. \\[-1mm]

\noindent Most of the parts of this thesis were done using the resources of Solar Oscillation Investigation group of Stanford University, Palo Alto, USA. I thank the staff of this institution for their hospitality during my stays there.\\[-1mm]

\noindent I would like to thank to my supervisor and advisers Michal Sobotka, Mirek Klva\v{n}a, and Pavel Ambro\v{z} for a patient leadership and sharing many practical aspects of scientific work with me, to my collaborators Sasha Kosovichev, Junwei Zhao, and Thierry Roudier for a fruitful collaboration, which I hope will continue in the future. \\[-1mm]

\noindent At last but not the least I would like to thank to my beloved Jana for her patience and support during the whole period.

\vspace{7.5cm}
\noindent I declare that this thesis was written by my own with the use of cited resources. I agree with the borrowing
of this work.

\vspace{2cm}
\begin{center}\begin{tabular}{ccc}
\rule{5cm}{1pt} & \rule{2cm}{0pt} &  \rule{5cm}{1pt} \\
In~Ond\v{r}ejov  & &  Michal \v{S}vanda \\
\end{tabular}\end{center}

\newpage
\tableofcontents

\include{anotace}
\newpage
\vspace*{5cm}
{\it \noindent Mottoes:\\
\rule[1mm]{5cm}{1pt} \\[5mm]

\noindent If at first you don't succeed, try, try, try, try, try, try, try, try again.

\begin{flushright} Jack O'Neill, Stargate SG-1, Window of Opportunity \end{flushright}
\vspace{3mm}

\noindent If you don't follow your dreams, you might as well be a vegetable. 

\begin{flushright} Burt Munro, The World's Fastest Indian \end{flushright}


} 

\newpage
\pagestyle{fancy}
\lhead[\sf\slshape \thepage]{\sf \fancyplain{}%
   {\leftmark}}
\chead{}
\rhead[\sf \fancyplain{}{\leftmark}]{\sf\slshape \thepage}
\lfoot{}
\cfoot{}
\rfoot{}

\newpage \setcounter{page}{1}\include{preface}
\newpage \include{introduction}
\newpage \include{method_synthetic}

\newpage \include{timedistance}
\newpage \include{real}

\newpage \include{long}
\newpage \include{filament}
\newpage \include{meridional}
\newpage \include{conclusions}

\newpage \include{appendix_a}

\pagestyle{fancy}
\lhead[\sf\slshape \thepage]{\sf \fancyplain{}%
   {\leftmark}}
\chead{}
\rhead[\sf \fancyplain{}{\leftmark}]{\sf\slshape \thepage}
\lfoot{}
\cfoot{}
\rfoot{}

\newcommand{\SortNoop}[1]{}

\newpage \include{mypubs}
\end{document}

%% file: casopisy.tex
\newcommand\aj{{AJ}}%
\newcommand\araa{{ARA\&A}}%
\newcommand\apj{{ApJ}}%
\newcommand\apjl{{ApJ}}%
\newcommand\apjs{{ApJS}}%
\newcommand\ao{{Appl.~Opt.}}%
\newcommand\apss{{Ap\&SS}}%
\newcommand\aap{{A\&A}}%
\newcommand\aapr{{A\&A~Rev.}}%
\newcommand\aaps{{A\&AS}}%
\newcommand\azh{{AZh}}%
\newcommand\baas{{BAAS}}%
\newcommand\jrasc{{JRASC}}%
\newcommand\memras{{MmRAS}}%
\newcommand\mnras{{MNRAS}}%
\newcommand\pra{{Phys.~Rev.~A}}%
\newcommand\prb{{Phys.~Rev.~B}}%
\newcommand\prc{{Phys.~Rev.~C}}%
\newcommand\prd{{Phys.~Rev.~D}}%
\newcommand\pre{{Phys.~Rev.~E}}%
\newcommand\prl{{Phys.~Rev.~Lett.}}%
\newcommand\pasp{{PASP}}%
\newcommand\pasj{{PASJ}}%
\newcommand\qjras{{QJRAS}}%
\newcommand\skytel{{S\&T}}%
\newcommand\solphys{{Sol.~Phys.}}%
\newcommand\sovast{{Soviet~Ast.}}%
\newcommand\ssr{{Space~Sci.~Rev.}}%
\newcommand\zap{{ZAp}}%
\newcommand\nat{{Nature}}%
\newcommand\iaucirc{{IAU~Circ.}}%
\newcommand\aplett{{Astrophys.~Lett.}}%
\newcommand\apspr{{Astrophys.~Space~Phys.~Res.}}%
\newcommand\bain{{Bull.~Astron.~Inst.~Netherlands}}%
\newcommand\fcp{{Fund.~Cosmic~Phys.}}%
\newcommand\gca{{Geochim.~Cosmochim.~Acta}}%
\newcommand\grl{{Geophys.~Res.~Lett.}}%
\newcommand\jcp{{J.~Chem.~Phys.}}%
\newcommand\jgr{{J.~Geophys.~Res.}}%
\newcommand\jqsrt{{J.~Quant.~Spec.~Radiat.~Transf.}}%
\newcommand\memsai{{Mem.~Soc.~Astron.~Italiana}}%
\newcommand\nphysa{{Nucl.~Phys.~A}}%
\newcommand\physrep{{Phys.~Rep.}}%
\newcommand\physscr{{Phys.~Scr}}%
\newcommand\planss{{Planet.~Space~Sci.}}%
\newcommand\procspie{{Proc.~SPIE}}%

%% file: anotace.tex
\newpage
\enlargethispage{1cm}\vspace*{-1cm}

{\small
\thispagestyle{empty}
\noindent Title: \emph{Velocity Fields in the Solar Photosphere}\\
Author: \emph{Michal \v{S}vanda}, michal@astronomie.cz\\
Institute: \emph{Astronomical Institute (v. v. i.) of the Academy of Sciences of the Czech Republic}\\
Supervisor: \emph{RNDr. Michal Sobotka, CSc., msobotka@asu.cas.cz}\\
Advisers: \emph{Ing.~Mirek~Klva\v{n}a,~CSc., mklvana@asu.cas.cz,} \\ 
\color{white}{Advisors:}\color{black}{ }\emph{RNDr.~Pavel~Ambro\v{z},~CSc., pambroz@asu.cas.cz}\\
Study branch: \emph{f1 -- Theoretical physics, astronomy, and astrophysics}
\begin{center}\bf Abstract\end{center}
Large-scale velocity fields in the solar photosphere remain a mystery in spite of many years of intensive studies. In this thesis, the new method of the measurements of the solar photospheric flow fields is proposed. It is based on local correlation tracking algorithm applied to full-disc dopplergrams obtained by Michelson Doppler Images (MDI) on-board the Solar and Heliospheric Observatory (SoHO). The method is tuned and tested on synthetic data, it is shown that the method is capable of measuring of horizontal velocity fields with an accuracy of 15~\mps. It is also shown that the method provides the measurements comparable with time-distance local helioseismology. The method is applied to real data sets. It reproduces well known properties of solar photospheric velocity fields. Moreover, the case studies show an evidence about the influence of the changes in the flow field topology on the stability of the eruptive filament and support the theory of the dynamical disconnection of bipolar sunspots from their magnetic roots. The method has a great perspective in the future use. The meridional flux transportation speed is also studied and it is shown that the direct measurement may differ from time-distance local helioseimology in the areas occupied by the strong magnetic field. This result has an impact to the flux transport dynamo models, which use the meridional speed as the essential observational input parameter. \\

}

%% file: preface.tex
\section{Preface}

This thesis presents the collection of results obtained during my Ph.D. study program on the topic of surface velocity fields in the photosphere of the Sun and connected topics. Rather than the topic as a whole, some subtopics were studied. The main result of this work is a data set, which can and will be used for other analyses during next years. 

In Section~\ref{sect:introduction} I present a short overview about the physics and known properties of the solar surface flow fields. In Section~\ref{sect:lct} the principle of the local correlation tracking method, which was used in the most of the work done, is explained. The method was applied to full-disc dopplergrams. It was tuned using synthetic data. For this purpose, a code generating synthetic solar dopplergrams with known properties was developed. It is described in Section~\ref{sect:method} as well as the optimal values of free parameters involved in the method, and the noise and accuracy estimation. 

The performance of the method is verified comparing its results with results of the modern-most method of measurements of surface and sub-surface flow fields. In Section~\ref{sect:comparizon} it is shown that both methods applied to the same set of data provide comparable results. The processed dataset and the visualization of the results used during the analysis is described in Section~\ref{sect:real}. 

In the next Sections, the particular results obtained with the proposed method are described: The long-term properties and periodicities in Section~\ref{sect:long} and the flows under the eruptive filament and their evolution during the eruptive phase in Section~\ref{sect:filament}.

In the Section~\ref{sect:meridional}, a different but very similar topic is studied. The meridional flux transport process is an essential property of the solar global dynamo performance. Using a different dataset (magnetic butterfly diagram) and a different technique, the meridional flux transportation speed is measured and compared with the results of the time-distance helioseismology. It is shown that the surface measurements are biased by motions around local magnetic regions. This result is important for the flux transport dynamo models. 

The flow fields in the previous Sections concerned only the surface of the Sun. In Section~\ref{sect:stellar} it is shown that the local correlation tracking method could work also on stellar data coming from the Doppler imaging. So far, the study of this problem is in the very early stage, therefore it is attached as an appendix. However, it has a large perspective into the future because of lots of data coming from robotic telescopes.

%% file: introduction.tex
\section{Introduction}
\label{sect:introduction}

The Sun is the closest star -- this fact allows us to resolve individual features on its surface and in its atmosphere. Using many types of observations, we can collect a large amount of data describing the behaviour of the solar plasma in various phenomena. The Sun is a variable star -- the magnetic activity undergoes the main cycle with a length of 22~years (reversal of a global magnetic field) which is composed of two consecutive 11~years `spot' cycles. The most visible evidence of the solar cyclicity is the change of the number, size, and shape of sunspots. However, evidences of such cyclicity may be found also in the total solar irradiance, number of solar flares, or the shape of the solar corona. The Sun exhibits also cycles with different lengths (from few minutes to many centuries) and different properties. Better knowledge of the physics lying under solar magnetic variability and active phenomena will improve our attempts to predict solar activity. 

\subsection{Structure and dynamics of the photospheric velocity field, especially in large-scale}

The first evidence about the velocity fields in the solar photosphere comes from Christoph Scheiner, who in 1630 noticed that sunspots near the equator traverse the solar disc faster than sunspots in higher latitudes. \cite{1859MNRAS..19...81C} used series of sunspot drawings to infer the differential rotation rate and the inclination of the solar rotation axis. Since Carrington's measurement of the differential rotation this phenomenon was confirmed many times using many techniques. For details see an older review by \cite{1985SoPh..100..141S} or a recent review by \cite{2000SoPh..191...47B}. The existence of the differential rotation was the first evidence about the movements of objects in the solar photosphere. Since then many other types of motions were detected. 

The energy coming from a thermonuclear fusion in the solar core is from approx. 0.7~$R_\odot$ carried by the convection. The dynamic behaviour in the solar photosphere is therefore mainly driven by the plasma motions in the underlying convection zone. The most evident manifestation of the convective behaviour is the solar granulation. It is observed in white-light as a cellular pattern with a characteristic size of 1,000~km and a lifetime of 3--17~minutes \citep{1986AA...168..330D}. Motions of granules are studied mainly using local correlation tracking technique; see e.\,g. recent study using the TRACE white light data by \cite{2002AA...387..672K}. From such studies it is known that granules are carried by the flow field of the larger scale and form the cellular-like pattern of supergranulation. On smaller scales, exploding granules seem to form cells smaller than supergranules -- the mesogranulation -- that are also advected within the supergranular cells \citep{2005AA...444..245L}. Nevertheless, the convection power spectrum in the photosphere obtained from Doppler measurements peaks at granulation scales ($l \ge 1000$, $l$ is a spherical harmonic degree), with a secondary peak at $l\sim 120$ corresponding to supergranulation. There is no evidence \citep{2000SPD....31.0504H} about the peak corresponding to mesogranules or other cellular-like convection mode. From this reason, the existence of mesogranulation as a separated convection mode is still in doubt. 

\subsubsection{Supergranules}

The supergranulation pattern is clearly seen as a peak in a convection power spectrum with $l\sim 120$ \citep{2000SPD....31.0504H}. This corresponds to size of roughly 30~Mm. Supergranules were discovered by \cite{1956MNRAS.116...38H} when studying variations of the equatorial rotation velocity using the autocorrelation method. He found a pattern with a characteristic size of 26~Mm and a characteristic velocity amplitude of 170~\mps. A more detailed study done using full-disc dopplergrams was performed by \cite{1962ApJ...135..474L}, who basically confirmed \citeauthor{1956MNRAS.116...38H}'s results. In the continuing studies by \cite{1963ApJ...138..631N} and \cite{1964ApJ...140.1120S} also for the first time appeared the name of the new convection mode -- ``super-granulation''. After a decades of the studies, the true origin of supergranulation remains a mystery. There exist many realistic simulations of the solar convection zone, however in none of them it was reported the convection mode with the properties of supergranules. Convective motions at supergranular scales have been reported in global simulations by \cite{2002ApJ...581.1356D}, who focused on the upper regions of the solar convection zone. Higher spatial resolution than in other global simulations was achieved by limiting the simulation domain to radii between 0.92--0.98~$R_\odot$ and by imposing a four-fold periodicity in longitude. The resulting pattern exhibits a hierarchy of scales, from supergranular-scale mottling to a network of larger cells and extended north-south downflow lanes more comparable to the deep-shell simulations. Although provocative, it is premature to identify this small-scale convection pattern too closely with supergranulation on the Sun, because other statistical parametres of the results of simulation do not agree with the parametres of real supergranulation. Solar supergranulation may involve dynamics, which is not captured in these global simulations such as ionization effects or self-organization processes involving smaller-scale granules \citep{2003ApJ...597.1200R}. On the other hand, there is no evidence of supergranulation in the detailed simulations focused on granulation. The origin of supergranulation remains unclear, \cite{2005AA...430L..57R} proposed the collective interaction of the granules, as its cause. In particular, the author showed that the formation of the long-lived large scale pattern can be obtained by computing the advection of many small-scale short-lived granular downflows.

The determined size of the supergranular cells is sensitive to a method used for the measurement of this quantity. Most often it is used the value of 30~Mm. \cite{1956MNRAS.116...38H}, using the autocorrelation method, got the size of 26~Mm. \cite{1989SoPh..120....1W} got in a detailed study using the autocorrelation method of full-disc dopplergrams the size of 31.2$\pm$2.3~Mm. \cite{2000ApJ...534.1008S} used a tesselation method to get a typical size of the supergranule. They applied the method to full-disc MDI dopplergrams (2\arcsec\,px$^{-1}$) and also to Ca~II K~filtergrams obtained at the South Pole Observatory (3.2\arcsec\,px$^{-1}$). They found 10.5~Mm in the photosphere and 14--26~Mm in the chromosphere as a typical size of the supergranule. However, they argued that the results obtained by the tesselation method are strongly influenced by the resolution of images. Using statistical tests they concluded that the most probable size of the supergranule in the photosphere and the chromosphere is 25.9~Mm. Also \cite{1996AAS...188.0201H} noted that the autocorrelation method has a tendency to overestimate (1.5--2~times) the size of real structures. The depth of the supergranulation, inferred using local helioseismology and using various studies, is supposed to have a value of 8~Mm \citep{1998soho...418..581D} to 15~Mm \citep{2003soho...12..417Z}. 

The lifetime of supergranulation is one of the crucial parameters. One of the early studies \citep{1989SoPh..120....1W} used the dopplergrams measured at NSO Kitt Peak. They segmented the series of images, labeled the supergranules and tracked each of them in the whole series. Using exponential fit they obtained the characteristic lifetime of 50--80 hours. They also mentioned some of cells that survived over one week. In the recent study by \cite{2000SPD....31.0106D}, which is a part of DeRosa's thesis, the authors made a carefull study of supergranular lifetimes, made a histogram of their lifetimes, and reported that the long-duration datasets contain several instances, where individual supergranules are
recognizable for time scales as long as 50 hours, though most cells persist for about 25 hours that is often quoted as a supergranular lifetime. \cite{2004SoPh..221...23D} found the mean lifetime of 22.5~hours. They also studied the relation between the lifetime and the size of the supergranules and found that supergranules probably exist in two regimes. For small supergranules (under 27~Mm) there can be observed almost linear increase of the mean lifetime with size, for larger (above 27~Mm) is the mean lifetime nearly constant (33~hours). This behaviour can be related to the loss of structural coherence by the largest supergranules, leading to a fragmentation event in smaller parts.

The internal velocity field in supergranules is nearly horizontal and is not easy to be measured. E.\,g. \cite{2002SoPh..205...25H} used full-disc MDI dopplergrams with solar rotation, meridional circulation and $p$-modes of solar oscillations removed. They decomposed the dopplergram in the series of annuli of different radius $r$, which is related to the heliocentric angle $\rho=\arcsin{r/R}$, where $R$ is the solar radius. Authors assumed that the mean line-of-sight velocity $v_d$ in the annulus may be calculated using
\begin{equation}
\left< v_d^2(\rho)\right>=\left< v_v^2\right>+\left[ \left< v_h^2\right>-\left< v_v^2\right>\right]\sin^2{\rho},
\end{equation}
where $v_v$ is the vertical and $v_h$ the horizontal component of the internal supergranular velocity field. Using the least-square fit they obtained the typical values of the internal horizontal velocity of $\left< v_h\right>=(258\pm 1)$~\mps{} and of the internal vertical velocity of $\left< v_v\right>=(29\pm 2)$~\mps. It is coherent with many older studies \citep[e.\,g.][]{1962ApJ...135..474L}

\cite{1964ApJ...140.1120S} noted that supergranule boundaries correspond with the chromospherical Ca~II~K network. This behaviour is coherent with the idea of the advected magnetic flux tubes, which are advected towards the boundaries of supergranular cells and concentrate there. Similar network can be observed also in other `metal' lines.  

\subsubsection{Giant cells?}
There is no doubt that large-scale velocity structures with $l \le 100$ exist in the convection power spectra obtained from surface measurements or local helioseismology. However, there has not been found a clear evidence about the characteristic pattern of such structures or obtained the qualitative diagnostics of large-scale convective motions. In the velocity power spectra, there are two peaks denoting existence of granulation and supergranulation. At low $l$ the power of velocity mode at $l$ drops almost linearly with $l$ \citep{2000SPD....31.0504H}. Of course this fact could also mean that there is no characteristic pattern in large-scale. 

The giant cells were noted is some older studies (\citeauthor{1970SoPh...14...80B}, \citeyear{1970SoPh...14...80B} or  \citeauthor{1987BAICz..38...92B}, \citeyear{1987BAICz..38...92B}) connected with the distribution of the magnetic field in the solar photosphere. Also \cite{1997HvaOB..21....9A} used low-resolution magnetic maps as tracers and found some evidence about the existence of giant cells. \cite{2000SPD....31.0403M} found pattern with characteristic size 3--10-times larger than the size of supergranules with lifetime greater than 10~days. When analysing power spectra of full-disc MDI dopplergrams, they concluded that the physical origin of such giant cells and the supergranulation is the same -- `their' giant cells are just larger and more stable. They also found the correspondence between `large-scale supergranules' and the length of large and stable bipolar sunspot groups.

Recently, several groups have reported long-lived features in dopplergrams which are highly correlated in longitude, corresponding to azimuthal wavenumbers of $m = 0$--8 (angular extent $>45\,^\circ$) but with a narrow latitudinal extent of not more than about 6\,$^\circ$. Although \cite{1998Natur.394..653B} interpret these features as giant convection cells, \cite{2001ApJ...560..466U} argues that they more likely comprise a spectrum of inertial oscillations, possibly related to Rossby wave modes and perhaps also to torsional oscillations. Evidence for a different giant cell structure has been presented by \cite{2004ApJ...608.1167L}. They studied the supergranulation pattern using correlation tracking and found a tendency for north-south alignment of supergranular cells. Such an alignment would be expected if the supergranulation were advected by larger-scale, latitudinally-elongated lanes of horizontal convergence such as those commonly seen in numerical simulations of solar convection \citep{2004ApJ...614.1073B}. Advection by such structures may also help to explain why the supergranulation pattern appears to rotate faster than the surrounding plasma measured by Doppler shift.

We can conclude that there is still not any clear evidence about the existence of the giant cells. It is assumed that the size of such cells should be between 200 and 400~Mm, lifetime greater than one week and the internal velocity field is expected mostly horizontal with the magnitude of few meters per second. Early numerical simulations \citep{1991MNRAS.252P...1S} showed that the existence of giant cell convection is not neccessary for the heat transport within the whole convection zone. They concluded that the existence of convection cells with size of 10--20~Mm forming at the base of the convection zone with a distance of 30--40~Mm between neighbouring cells is sufficient for heat transport to the surface. From such travelling cells, the supergranulation is formed in the shallow subsurface layer. Recent global simulations by \citeauthor{2004ApJ...614.1073B} (e.\,g. \citeyear{2004ApJ...614.1073B}) show global convection of many kinds, parameters of whose are sensitive to the state parameters of plasma in simulations. 

Global scale convection is observed using interferometric techniques in the close supergiant star Betelgeuse ($\alpha$~Orionis) -- see \cite{1990MNRAS.245P...7B}.

\subsubsection{Differential rotation}

Differential rotation is an obvious, yet poorly understood, solar phenomenon. Systematic measurements of its parametres were started by \citeauthor{1859MNRAS..19...81C} (e.\,g. \citeyear{1859MNRAS..19...81C}), who derived the regression formula describing the differentiality of the spot rotation in form
\begin{equation}
\omega=A+B\sin^{7/4}{b},
\end{equation}
where $b$ is the heliographic latitude and $\omega$ is the angular rotation rate. 

Since then, there exist many methods of measuring the solar rotation. They have one common denominator -- the results differ with the dataset, with the method, and with the resolution. 

Basically, the differential rotation is described as an integral of the zonal component $v_\varphi$ of the studied flow field. The integrated flow field may be obtained using spectroscopic method, using tracer-type measurements, or using helioseismic inversions. The latest case allows to measure the solar rotation not only as a function of heliographic latitude, but also as a function of depth. From the helioseismic inversion we know that throughout the convective envelope, the rotation rate decreases monotonically toward the poles by about 30~\%. Angular velocity contours at mid-latitudes are nearly radial. Near the base of the convection zone, there is a sharp transition between differential rotation in the convective envelope and nearly uniform rotation in the radiative interior. This transition region has become known as the solar tachocline. The rotation rate of the radiative interior is intermediate between the equatorial and polar regions of the convection zone. Thus, the radial angular velocity gradient across the tachocline is positive at low latitudes and negative at high latitudes, crossing zero at a latitude of about 35\,$^\circ$ \citep[e.\,g.][]{2003ARAA..41..599T}. In addition to the tachocline, there is another layer of comparatively large radial shear in the angular velocity near the top of the convection zone. At low and mid-latitudes there is an increase in the rotation rate immediately below the photosphere, which persists down to $r \sim 0.95~R_\odot$. The angular velocity variation across this layer is roughly 3~\% of the mean rotation rate and according to the helioseismic analysis of \cite{2002SoPh..205..211C} $\omega$ decreases within this layer approximately as $r^{-1}$. At higher latitudes the situation is less clear. The radial angular velocity gradient in the subsurface shear layer appears to be smaller and may switch sign. 

Currently, the surface measurements of the solar rotation are expressed in the form 
\begin{equation}
\omega=A+B\sin^2{b}+C\sin^4{b}\ .
\label{eq:difrot}
\end{equation}

\noindent In this formula, $A$ is equivalent to the equatorial rotation rate, $B$ and $C$ describe the differentiality. The difficulty of such expression lies in the coupling of $B$ and $C$ by an inverse correlation. It makes some issues when comparing different measurements made using different methods and different datasets. One remedy to this problem is to fix the ratio $B/C$. However, this technique is not satisfactory enough for comparison of very different types of measurements. Another approach, which completely solves the problem, is to use as a base function of the fit not even powers of $\sin{b}$, but to use an orthogonal base functions, such as Gegenbauer polynomials \citep{1984SoPh...94...13S}. 

Generally, the measured solar rotation is more rigid when measured using larger-scale objects, such as coronal holes or large-scale background magnetic field. There are known many relations of the differential rotation profile to the phase of the progressing solar cycle -- see e.\,g. \cite{2003SoPh..212...23J}. For overview of the solar differential rotation measurements see \cite{1985SoPh..100..141S} or a more recent review by \cite{2000SoPh..191...47B}. 

Current (magneto)hydrodynamic simulation basically reveal the structure of the solar rotation. They use many hydrodynamic quantities to reproduce the measured rotation profile -- angular momentum transport by convection, Reynolds stresses, and also angular momentum transport by meridional circulation, which seems to be essential for many types of simulations. See e.\,g. \cite{1993AA...279L...1K} or \cite{2004ApJ...614.1073B}. However, the results are not so satisfying. On the positive side, the angular velocity exhibits a realistic latitudinal variation, with little radial variation above mid-latitudes. On the negative side, the low-latitude angular velocity contours are somewhat more cylindrical than suggested by helioseismology, with more radial shear. Furthermore, at present there is little tendency for simulations such as these to form rotational shear layers near the top and bottom of the convection zone. Although these simulations do exhibit non-periodic angular velocity fluctuations of about the right amplitude relative to helioseismic inversions \citep[a few percent; see][]{2000SoPh..192...59M}, there is currently little evidence for systematic behavior such as torsional oscillations. Since the radiative interior possesses much more mechanical and thermal inertia than the convective envelope, the differential rotation in the convection zone may be sensitive to the complex dynamics occuring in the tachocline. In other words, we may not fully understand the rotation profile in the convection zone until we get the tachocline right. A realistic tachocline is probably also a prerequisite to achieving the solar-like dynamo cycles and wave-mean flow interactions which appear to be responsible for torsional and tachocline oscillations. 

\subsubsection{Oscillations in the rotation pattern}

The ``torsional oscillations'', in which narrow bands of faster than average rotation, interpreted as zonal flows, migrate towards the solar equator during the sunspot cycle, were discovered by \cite{1980ApJ...239L..33H}. At latitudes below about 40\,$^\circ$, the bands propagate equatorward, but at higher latitudes they propagate poleward. The low-latitude bands are about 15\,$^\circ$ wide in latitude. The flows were studied in surface Doppler measurements \citep{2001ApJ...560..466U}, and also using local helioseismology \citep{1997ApJ...482L.207K}. The surface pattern of torsional oscillations penetrate deep in the convection zone, possibly to its base, as studied by \cite{2002Sci...296..101V}. The magnitude of the angular velocity variation is about 2--5~nHz, which is roughly 1~\% (5--10~\mps) of the mean rotation rate. The direct comparison between different techniques inferring the surface zonal flow pattern \citep{2006SoPh..235....1H} showed that the results are pretty coherent.
 
The surface magnetic activity corresponds well with the oscillation pattern -- the magnetic activity belt tend to lie on the poleward side of the faster-rotating low-latitude bands. The magnetic activity migrate towards the equator with the low-latitude bands of the torsional oscillations as the sunspot cycle progresses \citep{2004ApJ...603..776Z}. Some studies \citep[e.\,g.][]{2002ApJ...575L..47B} suggest that meridional flows may diverge out from the activity belts, with the equatorward and poleward flows well correlated with the faster and slower bands of torsional oscillations.
  
The other type of oscillations detected in the rotation pattern were first reported by \cite{2000Sci...287.2456H}. It is completely different from the torsional oscillations. It has a period of 1.3~years and the origin of them is localized around the tachocline at the base of the convection zone. This type of oscillations depict as a periodic change of the rotation rate, the amplitude is about 3~nHz at the equator and slightly larger at a latitude of 60\,$^\circ$. So far, there is no evidence of latitudinal propagation. The study of oscillations near the tachocline is currently at the limits of the sensitivity of helioseismic inversions. However, \cite{2003soho...12..409T} found evidence about the 1.3~yr oscillations with variations of the amplitude as the solar cycle progresses.
 
\subsubsection{Meridional circulation}

The differential rotation is the axisymmetric component of the mean longitudinal flow, $<v_\varphi>$. The axisymmetric flow in the meridional plane, $<v_\theta>$ and $<v_r>$, is generally known as the meridional circulation. The meridional circulation in the solar envelope is much weaker than the differential rotation, making it relatively difficult to measure. Although it can in principle be probed using global helioseismology, the effect of meridional circulation on global acoustic oscillations is small and may be difficult to distinguish from rotational or other effects. 
 
Two principal methods are used to measure the meridional flow: feature tracking and direct Doppler measurement. There are several difficulties complicating the measurements of the meridional flow using tracers. Sunspots and filaments do not provide sufficient temporal and spatial resolution for such studies. Sunspots also cover just low latitudinal belts and do not provide any informations about the flow in higher latitudes. Doppler measurements do not suffer from the problem associated to the tracer-type measurements, however they introduce another type of noisy phenomena into account. It is difficult to separate the meridional flow signal from the variation of the Doppler velocity from the disc center to the limb. Using different techniques the parametres of the meridional flow show large discrepancies. It is generaly assumed that the solar meridional flow is in the close subphotospherical layers poleward with one cell. Such flow is also produced by early global hydrodynamical simulation such as \cite{1982ApJ...256..316G}. As reviewed by \cite{1996ApJ...460.1027H}, the surface or near sub-surface velocities of the meridional flow are generally in range 1--100~\mps, the most often measured values lie in the range of 10--20~\mps. The flow has often a complex latitudinal structure with both poleward and equatorward flows, multiple cells, and large asymmetries about the equator. \cite{2004ApJ...603..776Z} used the time-distance helioseismology to infer the properties of the meridional flow in years 1996--2002. They found the meridional flows of an order of 20~\mps, which remained poleward during the whole period of observations. In addition to the poleward
meridional flows observed at the solar minimum, extra meridional circulation cells of flow converging toward the
activity belts are found in both hemispheres, which may imply plasma downdrafts in the activity belts. These
converging flow cells migrate toward the solar equator together with the activity belts as the solar cycle evolves. \cite{2002ApJ...575L..47B} measured the meridional flow (and torsional oscillations) using the time-distance helioseismology and found the residual meridional flow showing divergent flow patterns around the solar activity belts below a depth of 18~Mm or so. It needs to be noted that the reverse flow (equatorward), which is assumed to flow at the base of the convection zone with velocities of $\sim 1$~\mps{} was not observed yet, although e.~g. \cite{2003ApJ...589..665H} interpret the migration of the sunspot groups during the sunspot cycle as evidence of the deep recurrent meridional flow with the average magnitude of 1.2~\mps. 

Recent global simulations \citep[e.\,g.][]{2004ApJ...614.1073B} also reveal the pattern of the meridional flow. Using the input parametres variation, the agreement between the model and measured velocities can be obtained. The modern dynamo flux-transport models use the meridional flow and the differential rotation as the observational input. In the models by Dikpati et al. (\citeauthor{2006ApJ...638..564D}, \citeyear{2006ApJ...638..564D}, \citeauthor{2006ApJ...649..498D}, \citeyear{2006ApJ...649..498D}, or \citeauthor{2006AGUFMSH22A..06D}, \citeyear{2006AGUFMSH22A..06D}) the `retrograde` meridional flow at the base of the convection zone is calculated from the continuity equation. They found the turnover time of the single meridional cell of 17--21~years. The meridional flow is assumed to be essential for the dynamo action, global magnetic field reversal and forecast of the ongoing solar cycles. Using the measurements of the sub-surface meridional flow they are able to reconstruct the magnetic activity in the past cycles and also predict the activity in the ongoing cycles. 

\subsubsection{Solar subsurface weather}

The most substantial recent advance in the search for large-scale non-axisymmetric motions in the solar envelope has been the mapping of horizontal flows by local helioseismology. After subtracting the contributions from differential rotation and meridional circulation, the residual flow maps reveal intricate, evolving flows on a range of spatial scales \citep[e.\,g.][]{2004ApJ...603..776Z}. Such flow patterns have become known as solar subsurface weather \citep[SSW; ][]{2002Sci...296...64T}. Such maps derived using different local helioseismic inversions (ring diagram or time-distance) are within the resolution quite stable \citep{2004ApJ...613.1253H}. 

The inferred SSW patterns show a high correlation with magnetic activity, becoming more complex at solar maximum. Near the surface, strong horizontal flows converge into active regions and swirl around them, generally in a cyclonic sense (counter-clockwise in the northern hemisphere and clockwise in the southern hemisphere). Deeper down, roughly 10~Mm below the photosphere, the pattern reverses; here flows tend to diverge away from active regions. The topology of the measured SSW is poorly reproduced in recent global simulations. The knowledge of the long-term behaviour of the SSW is essential in the field of investigation of the coupling between velocity and magnetic field and contribute in the theory of the solar dynamo. 

The inferrence of the `surface weather` using surface measurements is in principle the main aim of this thesis.

\subsection{Methods of measurements of the large-scale velocity fields}

Basically, there are three methods of measuring the photospheric velocity fields:

\begin{enumerate}
\item \emph{Direct Doppler measurement} -- provides only one component (line-of-sight) of the velocity vector. These velocities are generated by local photospheric structures, amplitudes of which are significantly greater than amplitudes of the large-scale velocities. The complex topology of such structures complicates an utilisation for our purpose. Analysing this component in different parts of the solar disc led to very important discoveries (e.\,g. supergranulation -- \citeauthor{1956MNRAS.116...38H}, \citeyear{1956MNRAS.116...38H} and \citeauthor{1962ApJ...135..474L}, \citeyear{1962ApJ...135..474L}).

\item \emph{Tracer-type measurement} -- provides two components of the velocity vector. When tracing some photospheric tracers, we can compute the local horizontal velocity vectors in the solar photosphere. Tracking motions of sunspots across the solar disc led to the discovery of the differential rotation \citep{1859MNRAS..19...81C}.

\item \emph{Local helioseismology} -- provides a full velocity vector. The local helioseismology (see \citeauthor{1996ApJ...461L..55K}, \citeyear{1996ApJ...461L..55K}, \citeauthor{2001ApJ...557..384Z}, \citeyear{2001ApJ...557..384Z}, or \citeauthor{2004ApJ...603..776Z}, \citeyear{2004ApJ...603..776Z}) is a very promising method using the information about the solar oscillations to infer the structure and also the dynamics in the convection zone. 
\end{enumerate}

\noindent In this work we used mostly the tracer type method, local correlation tracking in particular, applied to the surface structures.

The method needs a tracer -- a significant structure recorded in different frames, the lifetime of which is much longer than the time lag between the correlated frames. We decided to use the supergranulation pattern in the full-disc dopplergrams, acquired by the Michelson Doppler Imager \citep[MDI; ][]{1995SoPh..162..129S} onboard Solar and Heliospheric Observatory (SoHO). We assume that supergranules are carried as objects by the large-scale velocity field. This velocity field is probably located beneath the photosphere, so that the resulting velocities will describe the dynamics in both the photospheric and subphotospheric layer. The existence of the supergranulation on almost the whole solar disc (in contrast to magnetic structures) and its large temporal stability make the supergranulation an excellent tracer.

\subsubsection{Local correlation tracking}
\label{sect:lct}
Since the photosphere is a very thin layer (0.04~\% of the solar radius), the large-scale photospheric velocity fields have to be almost horizontal. Then, the tracer-type measurement should be sufficient for mapping the behaviour of such velocities. In this field the local correlation tracking (LCT) method is very useful.

This method was originally designed for the removal of the seeing-induced distortions in image sequences \citep{1986ApOpt..25..392N} and later used for mapping the motions of granules in the series of white-light images \citep{1988ApJ...333..427N} under the name \emph{local cross-correlation}. The method works on the principle of the best match of two frames that record the tracked structures at two different instants. 

The algorithm is applicable to two frames $I_1$ and $I_2$ having the same dimension that were captured in different instants. There is a time lag $\tau$ between both frames, which has to be smaller than the lifetime of used tracer. For every pixel in the first frame a subframe (correlation window) is chosen, described by the coordinates of its centre $(x_0,y_0)$ and the size $p$. Since in this study we used the Gaussian-weighted window, $p$ is equal to the $FWHM$ of the Gaussian profile. The parameter $p$ is selected according to used tracer, so that $p$ is larger than a characteristic size of the tracer. Let the subframe in image $I_1$ is $S_1(x_0, y_0)$.

\begin{figure}
\resizebox{\textwidth}{!}{\includegraphics{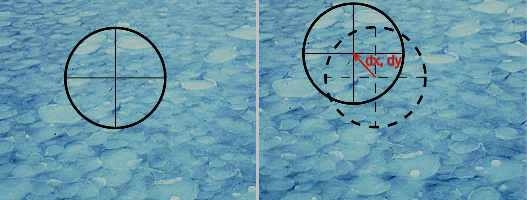}}
\caption{Simple scheme explaining the performance of a LCT technique.}
\label{fig:lct}
\end{figure}

$S_1(x_0, y_0)$ is compared with a subframe in image $I_2$ of the same size, which has its centre in the coordinates \mbox{$(x_0+\delta x,y_0+\delta y)$}. Let the subframe in image $I_2$ is \mbox{$S_2(x_0+\delta x,y_0+\delta y)$}. The proper motion of tracers in the point $(x_0, y_0)$ is defined by a displacement $(\delta x,\delta y)$, which maximise the correlation of $S_1$ and $S_2$. See Fig.~\ref{fig:lct}.

In this work we defined the correlation function $C(x_0+\delta x, y_0+\delta y)$ as weighted absolute difference of $S_1$ and $S_2$:
\begin{equation}
C(x_0,y_0,\delta x,\delta y)=W(p)\cdot\left| S_1(x_0,y_0)-S_2(x_0+\delta x,y_0+\delta y) \right|,
\end{equation}
where $W(p)$ is the weighting function, which depends on the size of the correlation window. We chosed $W(p)$ with a Gaussian shape given by
\begin{equation}
W(x,y)=e^{-\frac{x^2+y^2}{\sigma^2}},
\end{equation}
where $\sigma=0.424661p$ and the formula is valid for $x \in \left\langle -2p,2p\right\rangle$ and $y \in \left\langle -2p,2p\right\rangle$.

The correlation of each pair of subframes is performed for various values of $\delta x$, $\delta y$. In this work we use three integer values $\delta x$ and $\delta y$ given by
\begin{equation}
\delta x=0,\,\pm shift;\ \delta y=0,\,\pm shift,
\end{equation}
where $shift$ is a free parameter of LCT program. 

In our case for every pixel of the image with coordinates $(x_0,y_0)$, where $x_0 \in \left\langle 0,N_x\right\rangle$ and $y_0 \in \left\langle 0,N_y \right\rangle$ ($N_x$ and $N_y$ describe dimensions of input image) we obtain a 3$\times$3-matrix $M(x_0,y_0)$ containing the values of correlation of subframes $S_1$ and $S_2$ when displaced by $\delta x$, $\delta y$.
\begin{equation}
M(x_0,y_0)=\left( \begin{array}{lll} C(x_0,y_0,-s,+s) & C(x_0,y_0,0,+s) & C(x_0,y_0,+s,+s)\\ C(x_0,y_0,-s,0) & C(x_0,y_0,0,0) & C(x_0,y_0,+s,0) \\ C(x_0,y_0,-s,-s) & C(x_0,y_0,0,-s) & C(x_0,y_0,+s,-s) \end{array}\right),
\label{eq:M}
\end{equation}
where $s=shift$.

Extremal value of $M$ (maximum in case of the correlation is used, minimum if the absolute difference of subframes is used) is equivalent to the best match between $S_1$ and $S_2$. Corresponding values of $\delta x$ and $\delta y$ mean the displacement of the tracer in point $(x_0,y_0)$ of the input frames. To obtain the subpixel precision, a surface is fitted to the values of $M$ and the extremum is found on the fitted surface. Some numerical tests \citep{1988ApJ...333..427N} showed that the best results are obtained by a biquadratic surface given by
\begin{equation}
f(x,y)=a_1+a_2x+a_3y+a_4x^2+a_5y^2.
\label{eq:minimum}
\end{equation}
According to \cite{1988ApJ...333..427N} the polynomial surface of order less than 2 overestimate the displacements, while polynomial surface of order larger that 2 underestimates the displacements significantly. After some algebra \citep{1991PhDT.......137D} we obtain for coefficients $a$ of (\ref{eq:minimum}) fitted to the matrix (\ref{eq:M}) formulae
\begin{eqnarray}
a_1 & = & \frac{1}{2}M[0,0]+\frac{1}{2}M[2,0]+\frac{1}{2}M[0,2]+\frac{1}{2}M[2,2]+
\frac{1}{2}M[0,1]+\frac{1}{2}M[2,1]-\nonumber \\
& & -M[1,0]-M[1,1]-M[1,2], \nonumber \\
a_2 & = & \frac{1}{2}M[0,0]+\frac{1}{2}M[2,0]+\frac{1}{2}M[0,2]+\frac{1}{2}M[2,2]+
\frac{1}{2}M[1,0]+\frac{1}{2}M[1,2]-\nonumber \\
& & -M[0,1]-M[1,1]-M[2,1], \nonumber \\
a_3 & = & M[0,0]-M[2,0]-M[0,2]+M[2,2], \nonumber \\
a_4 & = & -M[0,0+M[2,2]+M[2,0]-M[0,1]+M[2,1]-M[0,2], \nonumber \\
a_5 & = & -M[0,0+M[2,2]+M[1,0]-M[2,0]+M[0,2]-M[1,2].
\end{eqnarray}

Extremal shift of the best fit can be calculated using
\begin{eqnarray}
min_x&=&\frac{1}{D}\left( a_3a_5-\frac{8}{3}a_2a_4\right), \nonumber\\
min_y&=&\frac{1}{D}\left( a_3a_4-\frac{8}{3}a_1a_5\right),
\end{eqnarray}
where $D=\frac{3}{2}\left(\frac{64}{9}a_1a_2-a_3a_3 \right)$. The formulae assume that the absolute value of difference of subframes is used as the measure of correlation. The displacement of the tracers in point $(x_0, y_0)$ is then given by
\begin{eqnarray}
d_x&=&min_x\cdot shift \nonumber \\
d_y&=&min_y\cdot shift.
\end{eqnarray}

The extrapolation in matrix $M$ makes the resulting displacements unreliable, so that the displacements have to satisfy the conditions $d_x \in \left\langle -shift,+shift\right\rangle$, $d_y \in \left\langle -shift,+shift\right\rangle$. The desired behaviour can be accomplished by a suitable choice of free parameters of the LCT program. 

The result of LCT application to two frames is a two-dimensional map containing in each pixel two components of the displacement vector ${\mathbf d}=(d_x,d_y)$. The vector velocity field is obtained using
\begin{equation}
{\mathbf v} = \frac{\mathbf d}{\tau}.
\end{equation}

For the evaluation of LCT in a time series of dopplergrams we used the modified program {\tt flowmaker.pro} implemented in IDL by \cite{molowny}.

LCT was recently used for tracking many features in various types of observations, especially for tracking the granules in high-resolution white-light images (e.\,g. \citeauthor{1999ApJ...511..436S}, \citeyear{1999ApJ...511..436S}, \citeauthor{2000ApJ...544.1155S}, \citeyear{2000ApJ...544.1155S}). The same method was recently used for calculation of large-scale flow fields using low resolution magnetograms (\citeauthor{2001SoPh..198..253A}, \citeyear{2001SoPh..198..253A} and \citeauthor{2001SoPh..199..251A}, \citeyear{2001SoPh..199..251A}). \cite{2004SoPh..223...39C} measured the helicity injected to a flaring active region using LCT applied to MDI full-disc magnetograms. \cite{2004ApJ...610.1148W} modified the LCT algorithm introducing the induction equation, which allows to determine all three components of surface flow field when tracking photospheric magnetograms. \cite{2003SPD....34.0708S} used LCT method for measurement of the chromospheric rotation profile when tracking full-disc H$\alpha$ observations obtained at the Big Bear Solar Observatory.

\label{label:lct_under}In some studies it has been shown that LCT technique underestimates the real velocities
due to the smoothing of processed data by a correlation window. For
example in \citeauthor{2006astro.ph..8204G}
(\citeyear{2006astro.ph..8204G}) the underestimation by 33~\% is found. Georgobiani's study was
done using different LCT code applied to simulated data and the
results were compared with time-distance method applied to $f$-mode.
This means that the correction factor is different for different
settings of LCT method, and, therefore, should be determined
(calibrated) empirically for each particular study. Application of LCT to MDI Dopplergrams 
by \citeauthor{2004ApJ...616.1242D} (\citeyear{2004ApJ...616.1242D})  showed the underestimation by 
more than 30~\%. Some other studies also denote the 
underrepresenting of magnitudes by LCT (\citeauthor{1999ApJ...511..436S}, \citeyear{1999ApJ...511..436S} -- 20~\%, 
\citeauthor{1999AA...349..301R}, \citeyear{1999AA...349..301R} -- 25~\%).

%% file: method_synthetic.tex
\section{Method and tests on synthetic data}
\label{sect:method}
\symbolfootnotetext[0]{\hspace*{-7mm} $\star$ This chapter was in a condensed form published in {\v S}vanda, M., Klva{\v n}a, M., and Sobotka, M., 2006, \emph{Large-scale horizontal flows in the solar photosphere. I. Method and tests on synthetic data}, Astronomy and Astrophysics, 458, 301--306.}

A recent experience with applying this method to observed data \citep[e.\,g.][]{svanda05} has shown that for the proper setting of the parameters and for the tuning of the method, synthetic (model) data with known properties are needed. The synthetic data for the analysis come from a simple numerical simulation (SISOID code = \emph{SI}mulated \emph{S}upergranulation as \emph{O}bserved \emph{I}n \emph{D}opplergrams), with the help of which we can reproduce the supergra\-nulation pattern in full-disc dopplergrams. The design of the synthetic dopplergrams with known parameters is very important for the calculation of the vector velocity fields, so that we carried out the simulation carefully to get the correct and valuable results about the abilities of the method. 

\subsection{The SISOID code}
\begin{figure}[!htb]
\resizebox{0.49\textwidth}{!}{\includegraphics{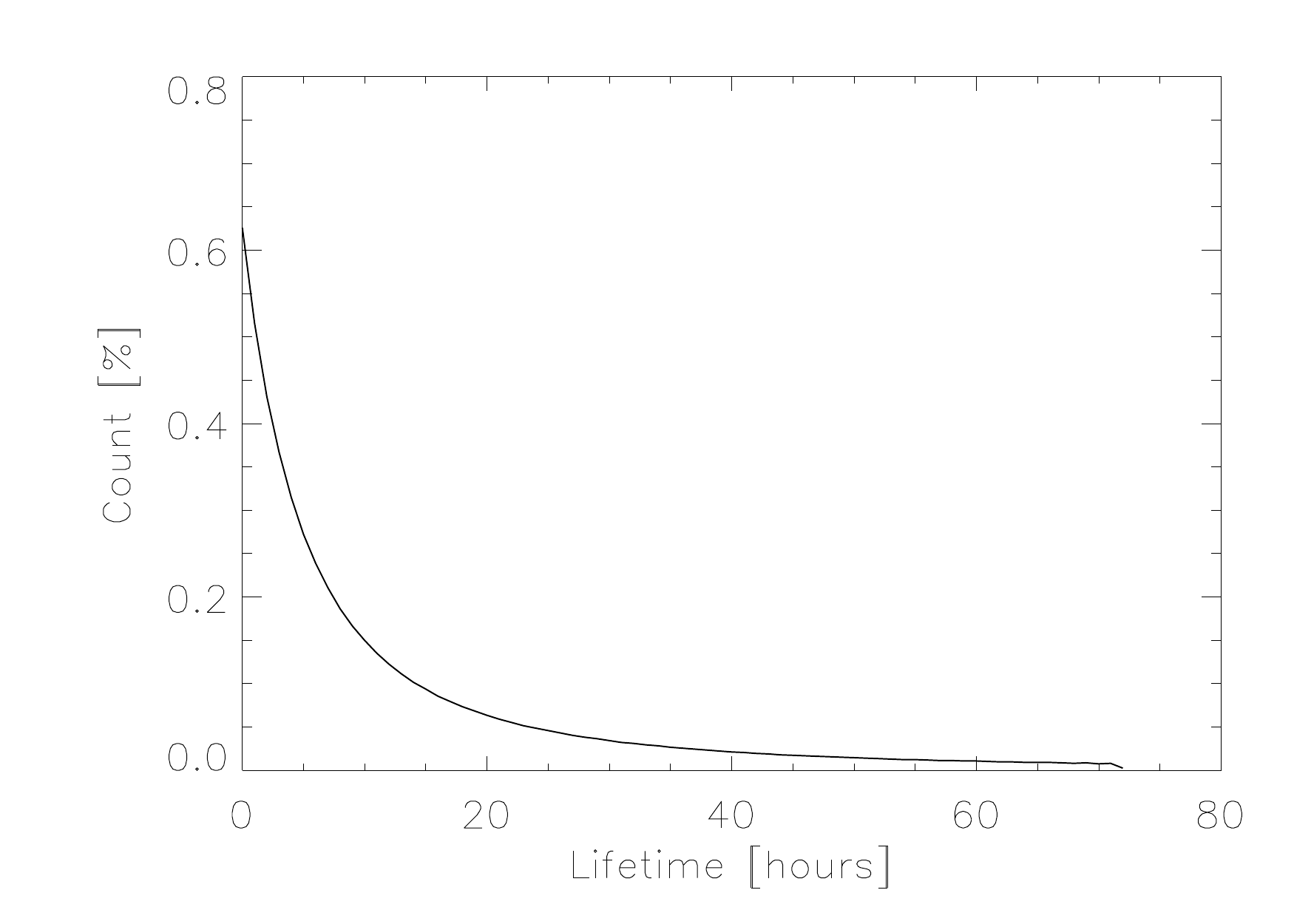}}
\resizebox{0.49\textwidth}{!}{\includegraphics{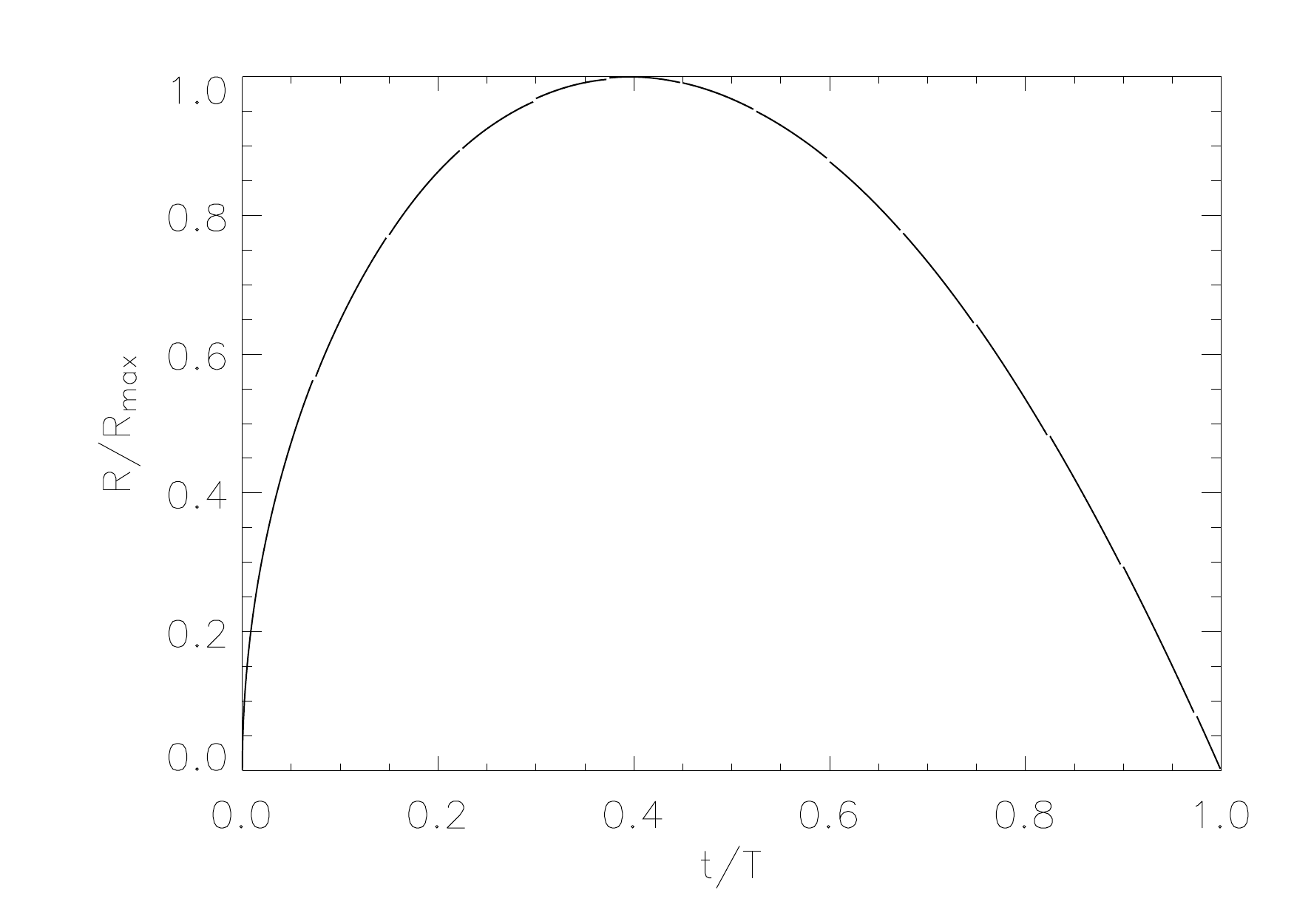}}
\caption{\emph{Left} -- The histogram of supergranule lifetime used in the SISOID code. Adopted from \cite{2000SPD....31.0106D}. \emph{Right} -- Approximated evolution of the diameter of synthetic supergranule during its lifetime as used in the SISOID code.}
\label{fig:sg_lifetime}
\end{figure}

\begin{figure}[!bt]
\resizebox{0.49\textwidth}{!}{\includegraphics{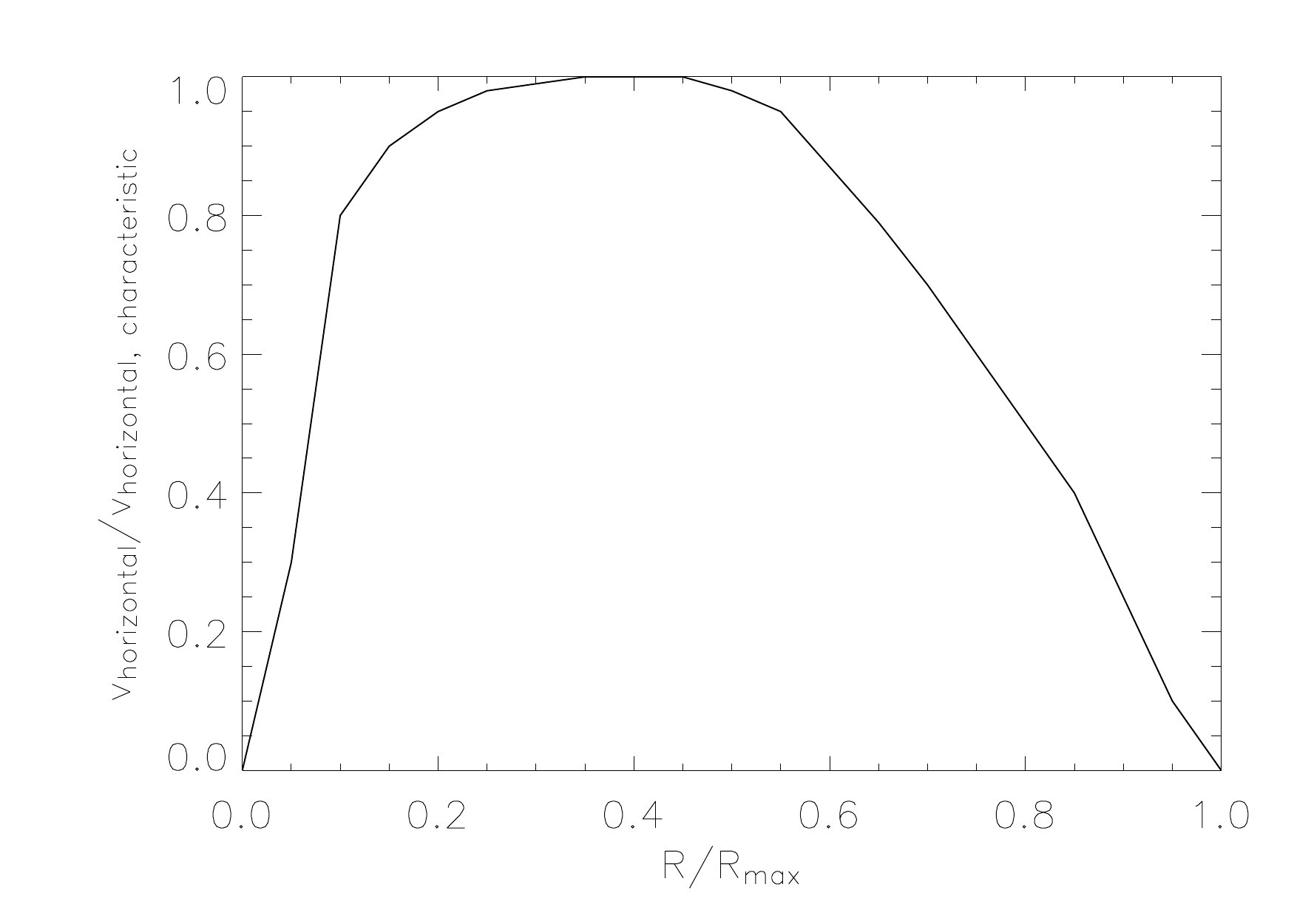}}
\resizebox{0.49\textwidth}{!}{\includegraphics{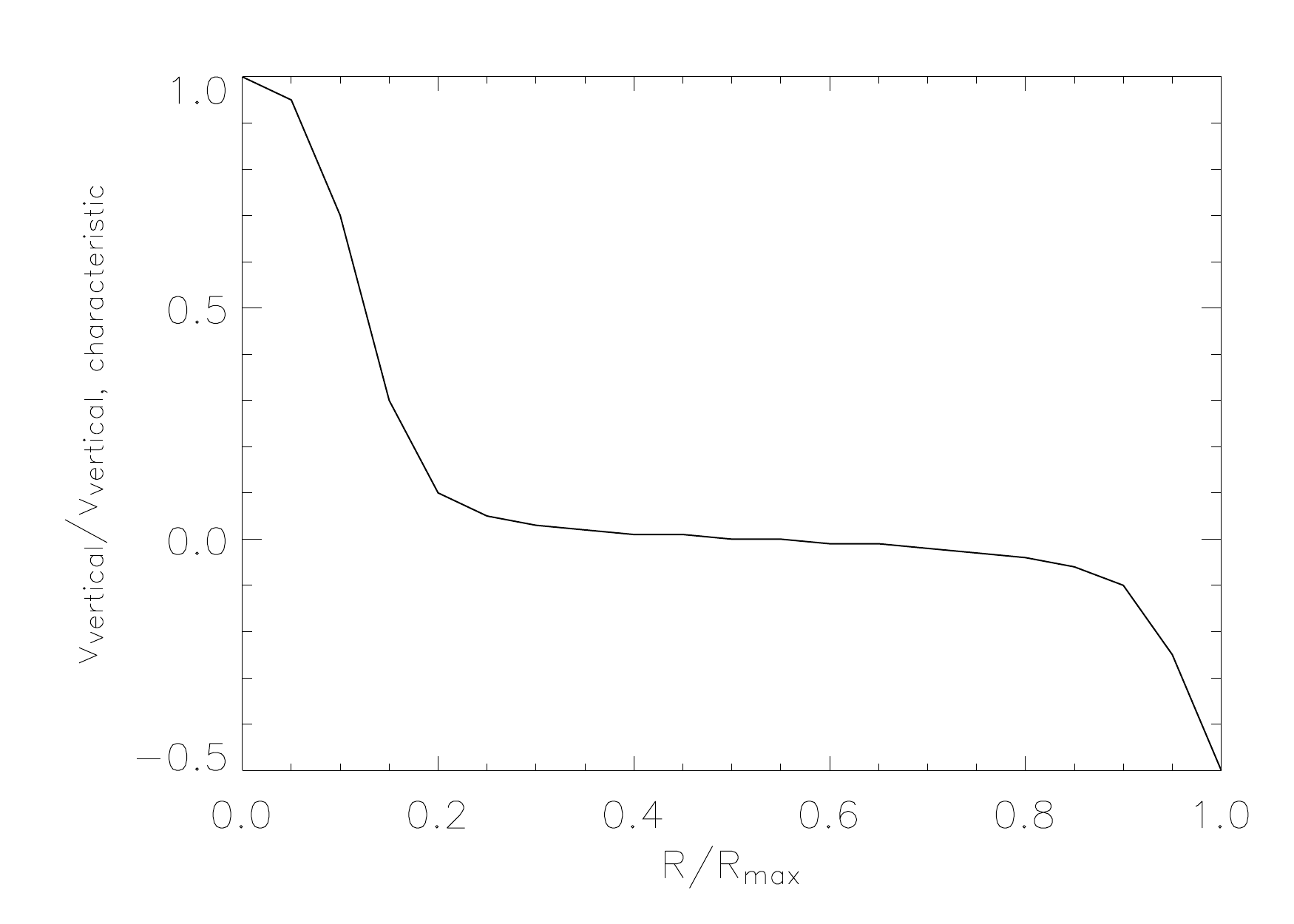}}
\caption{The dependence of the values of horizontal (\emph{left}) and vertical (\emph{right}) components of the internal supergranular velocity field on the radial distance from the centre of the synthetic supergranule, used in the SISOID code.}
\label{fig:sg_internal_flows}
\end{figure}

The SISOID code is not based on physical principles taking place in the origin and evolution of supergranulation, but instead on a reproduction of known parameters that describe the supergranulation. Individual synthetic supergranules are characterised as centrally symmetric features described by their position, lifetime (randomly selected according to the measured distribution function of the supergranular lifetime -- \citeauthor{2000SPD....31.0106D}, \citeyear{2000SPD....31.0106D}, see Fig.~\ref{fig:sg_lifetime} left), maximal diameter (randomly according to its distribution function that is basically normal with the mean of 31\,200~km and the variation of 2\,300~km -- \citeauthor{1989SoPh..120....1W}, \citeyear{1989SoPh..120....1W}) and characteristic values of their internal horizontal and vertical velocity components (randomly according to their distribution function with a normal shape described by parametres $<v_h>=258\pm 1$~\mps, $<v_v>=29\pm2$~\mps{} -- \citeauthor{2002SoPh..205...25H}, \citeyear{2002SoPh..205...25H}). The magnitude of the horizontal and vertical component of the internal velocity field depends only on the relative distance from the centre of the cell and is aproximated by curves shown in Fig.~\ref{fig:sg_internal_flows}.

The most important simplification in the SISOID code is that individual supergranules do not influence each other, but simply overlap. The final line-of-sight velocity at a certain point is given by the sum of line-of-sight velocities of individual synthetic supergranules at the same position.

New supergranules can arise inside the triangle of neighbouring supergranules (identification of such triangles is done by the Delaunay triangulation algorithmed by \citeauthor{barry}, \citeyear{barry}) only when the triangle is not fully covered by other supergranules and when any of the supergranules located at the vertices of the triangle is not too young, so that in the future it could fully cover the triangle. The position of the origin of the new supergranule is the centroid of the triangle; each vertex is weighted by the size of its supergranule. The diameter evolves according to Fig.~\ref{fig:sg_lifetime} right -- in the lifetime the supergranule grows from zero (during first 40 per cent of its lifetime) to its maximum diameter and shrinks to zero again (during last 60 per cent of its lifetime). All the supergranules behave in this way, which roughly approximates the real behaviour of the convection cells. 

The SISOID simulation is done in the pseudocylindrical Sanson-Flamsteed coordinate grid \citep{2002A&A...395.1077C}; the transformation from the heliographic coordinates is given by the formulae:
\begin{equation}
x=\vartheta \cos \varphi,\ y=\varphi,
\label{eq:sanson}
\end{equation}
where $x$ and $y$ are coordinates in the Sanson-Flamsteed coordinate system, and $\vartheta$ and $\varphi$ are heliographic coordinates originating at the centre of the disc. At each step an appropriate part of the simulated supergranular field is transformed into heliographic coordinates. The output of the program is a synthetic dopplergram of the solar hemisphere in the orthographical projection to the disc. We assume in our simulation that the Sun lies in an infinite distance from the observer and that the $P$ (position angle of the solar rotation axis) and $b_0$ (heliograhic latitude of the centre of solar disc) angles are known.

\begin{figure}[!b]
\resizebox{\textwidth}{!}{\includegraphics{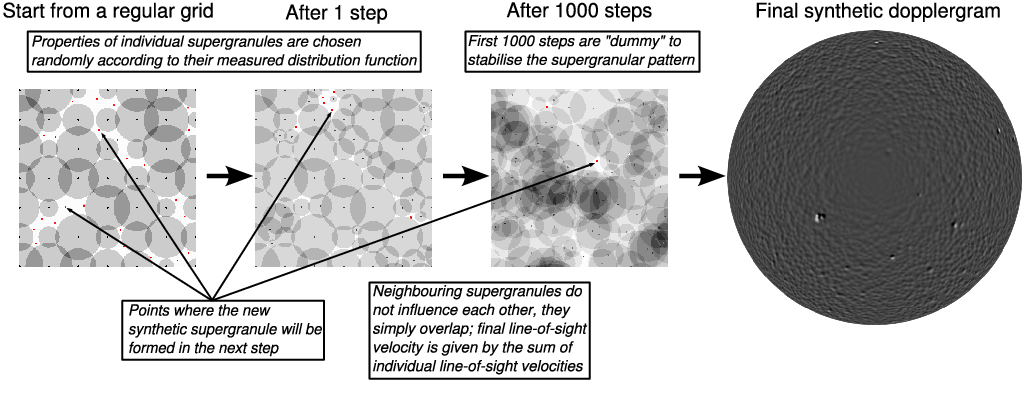}}
\caption{The cartoon describing how the SISOID code works and forms the final synthetic dopplergram}
\label{fig:cartoon1}
\end{figure}

Each step in calculation includes the evaluation of the parameters of individual supergranules, then small old supergranules under the threshold (2~Mm in size) are removed from the simulation and all the triangles are checked, whether a new supergranule can arise inside them (see Fig.~\ref{fig:cartoon1}). This step in the SISOID code corresponds to 5~minutes in real solar time. The computation is always started from the regular grid. Properties of ``supergranules'' are chosen randomly according to their real distribution functions. The first 1000 steps are ``dummy'', i.~e., no vector velocity field is included and no synthetic dopplergram is calculated. This starting interval is taken for the stabilisation of the supergranular pattern. In the next steps, the model vector velocity field is already introduced. This field influences only the positions of individual cells. The dopplergram is calculated every third step. For one day in real solar time, 96 dopplergrams are calculated.

\begin{figure}[!t]
\resizebox{0.49\textwidth}{!}{\includegraphics{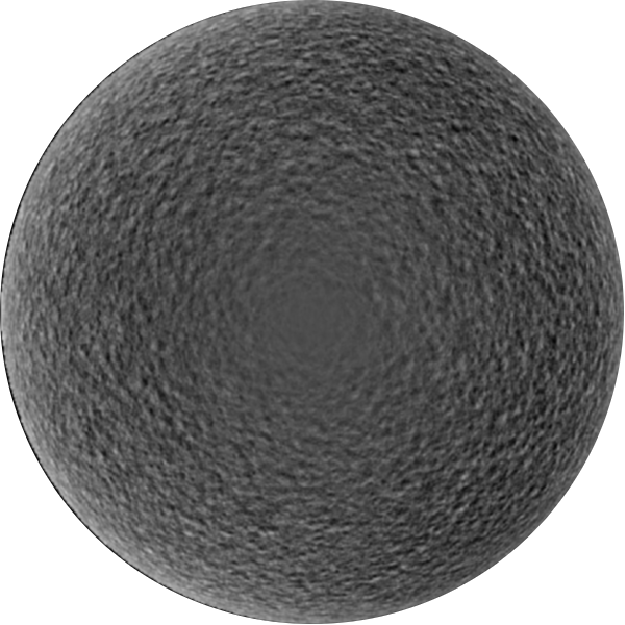}}
\resizebox{0.49\textwidth}{!}{\includegraphics{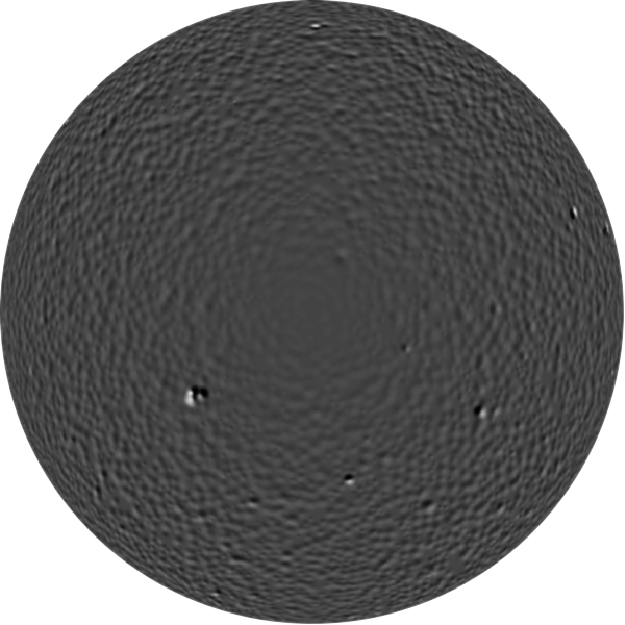}}
\caption{Comparison of the real dopplergram (\emph{left}, observed by MDI/SoHO) and
the simulated one (\emph{right}). Both images are visually very similar. The black colour means line-of-sight velocity $-700$~m\,s$^{-1}$ (towards an observer) while white represents $+700$~m\,s$^{-1}$.}
\label{fig:synthdopplergrams}
\end{figure}

The model velocity field with Carrington rotation added is applied according to the assumption of the velocity analysis that the supergranules are carried by a velocity field on a larger scale. In the simulation, the position of individual supergranules is influenced, and no other phenomena are taken into account. These synthetic dopplergrams are visually similar to the real observed dopplergrams (see Fig.~\ref{fig:synthdopplergrams}).

\subsection{Method of data processing}
The MDI onboard SoHO acquired the full-disc dopplergrams at a high cadence in certain periods of its operation  -- one observation per minute. These campaigns were originally designed for studying the high-frequency oscillations. The primary data contain lots of disturbing effects that have to be removed before ongoing processing: the rotation line-of-sight profile, $p$-modes of solar oscillations. We detected some instrumental effects connected to the data-tranfer errors. It is also known that the calibration of the MDI dopplergrams is not optimal \citep[e.\,g.][]{2002SoPh..205...25H} and has to be corrected to avoid systematic errors. While examining long-term series of MDI dopplergrams, we have met systematic errors connected to the retuning of the interferometer. We should also take those geometrical effects into account (finite observing distance of the Sun, etc.) causing bias in velocity determination. According to \cite{2000SoPh..195..219S} for example, the bias coming out of a perspective is about 2~m\,s$^{-1}$, and it depends on the position on the disc. It has been proven \citep[e.\,g.][]{liu} that MDI provides reliable velocity measurements when the magnetic field is lower than 2000~Gauss. The velocity observation by MDI will induce up to 100~\% error if the magnetic field is higher than 3000~Gauss due to the magnetic sensitivity of the used Ni\,I line and the limitations of computational algorithm, which cause crosstalks between measured MDI dopplergrams and magnetograms. The removal of these effects will be described in detail in Section~\ref{sect:real}, while the synthetic data used in this study do not suffer from these phenomena.

As input to the data processing we take a one-day observation that contains 96 full-disc dopplergrams in 15-minute sampling. Structures in these dopplergrams are shifted with respect to each other by the rotation of the Sun and by the velocity field under study. The primary data must be pre-processed by removal of the manifestation of the Carrington rotation and by the suppression of the $p$-modes. 

First, the shift caused by the rotation has to be removed. For this reason, the whole data series (96 frames) is ``derotated'' using Carrington rotation rate, so that the heliographic longitude of the central meridian is equal in all frames and also equals the heliographic longitude of the central meridian of the central frame of the series. This data-processing step causes the central disc area (``blind spot'' caused by prevailing horizontal velocity component in supergranules) in the derotated series to move with the Carrington rate. During the ``derotation'' the seasonal tilt of the rotation axis towards the observer (given by $b_0$ -- heliographic latitude of the centre of the disc) is also removed, so that $b_0=0$ in all frames. 

Then the data series is transformed to the Sanson-Flamsteed coordinate system to remove the geometrical distortions caused by the projection of the sphere to the disc. Parallels in the Sanson-Flamsteed pseudocylindrical coordinate system are equispaced and projected at their true length, which makes it an equal area projection. Formulae of the transformation from heliographic coordinates are given by Eq.~(\ref{eq:sanson}).

\begin{figure}[!]
\resizebox{\textwidth}{!}{\includegraphics{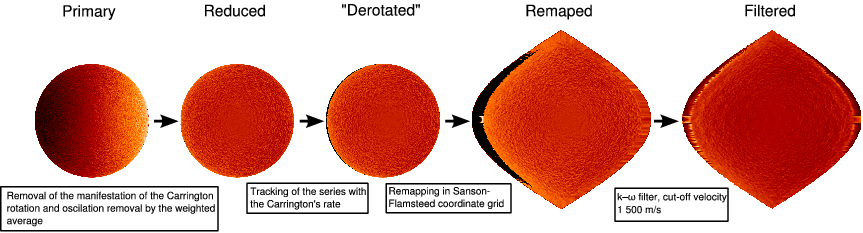}}
\caption{The scheme showing individual steps during the preparation of the dopplergrams for the LCT technique.}
\label{fig:cartoon2}
\end{figure}

The noise coming from the evolutionary changes in the shape of individual supergranules and the motion of the ``blind spot'' in the data series with the Carrington rotational rate are suppressed by the $k$-$\omega$ filter in the Fourier domain (\citeauthor{1989ApJ...336..475T}, \citeyear{1989ApJ...336..475T}, \citeauthor{1997ApJ...480..406H}, \citeyear{1997ApJ...480..406H}). The cut-off velocity is set to 1\,500~m\,s$^{-1}$ and has been chosen on the basis of empirical experience. The procedure can be seen in Fig.~\ref{fig:cartoon2}.

The existence of the differential rotation complicates the tracking of the large-scale velocity field, because the amplitudes and directions of velocities of the processed velocity field have a significant dispersion. We have found that, when the scatter of magnitudes is too large, velocities of several hundred m\,s$^{-1}$ cannot be measured precisely by the LCT algorithm where the displacement limit for correlation was set to detect velocities of several tens of m\,s$^{-1}$. Therefore the final velocities are computed in two steps. The first step provides a rough information about the average zonal flows using the differential rotation curve
\begin{equation}
\omega = A+B\sin^2 b+C\sin^4 b\ ,
\label{eq:fay}
\end{equation}
and calculating its coefficients.

In the second step this average zonal flow is removed from the data series, so that during the ``derotation'' of the whole series the differential rotation inferred in the first step and expressed by~(\ref{eq:fay}) is used instead of the Carrington rotation. The scatter of the magnitudes of the motions of supergranules in the data transformed this way is much smaller, and a more sensitive and precise tracking procedure can be used.

The LCT method is used in both steps. In the first step, the checked range of velocity magnitudes is set to 200~m\,s$^{-1}$, but the accuracy of the calculated velocities is roughly 40~m\,s$^{-1}$. In the second step the range is only 100~m\,s$^{-1}$ with much better accuracy. The lag between correlated frames equals in both cases 16 frame intervals (i.~e. 4 hours in solar time), and the correlation window with FWHM 30~pixels equals 60\arcsec{} on the solar disc in the linear scale. In one observational day, 80 pairs of velocity maps are calculated and averaged.

For the calculation we use the adapted program {\tt flowmaker.pro} originally written in IDL by \cite{molowny}. The algorithm has a limitation in the range of displacements that are checked for each pixel. The quality of correspondence (in our case the sum of absolute differences of both correlation windows) is computed in nine discrete points, then the biquadratic surface is fitted through these nine points, and an extremum position \citep{1991PhDT.......137D} is calculated (see Section~\ref{sect:lct}). The final displacement vector is equivalent to the position of the extremum.

\subsection{Results of the synthetic data experiment}
\label{sect:synthetic_results}
In our tests we have used lots of variations of simple axisymmetric model flows (with a wide range of values of parameters describing the differential rotation and meridional circulation) with good success in reproducing the models. When comparing the resulting vectors of motions with the model ones, we found a systematic offset in the zonal component equal to $v_{\rm offset,\ zonal}=-15\ \rm m\,s^{-1}$. This constant offset appeared in all the tested model velocity fields and comes from the numerical errors during the  ``derotation'' of the whole time series. For the final testing, we used one of the velocity fields obtained in our previous work \citep{svanda05}. This field approximates the velocity distribution that we may expect to observe on the Sun. The model flows have structures with a typical size of 60\arcsec{}, since they were obtained with the correlation window of this size.

\begin{figure}
\resizebox{0.50\textwidth}{!}{\includegraphics{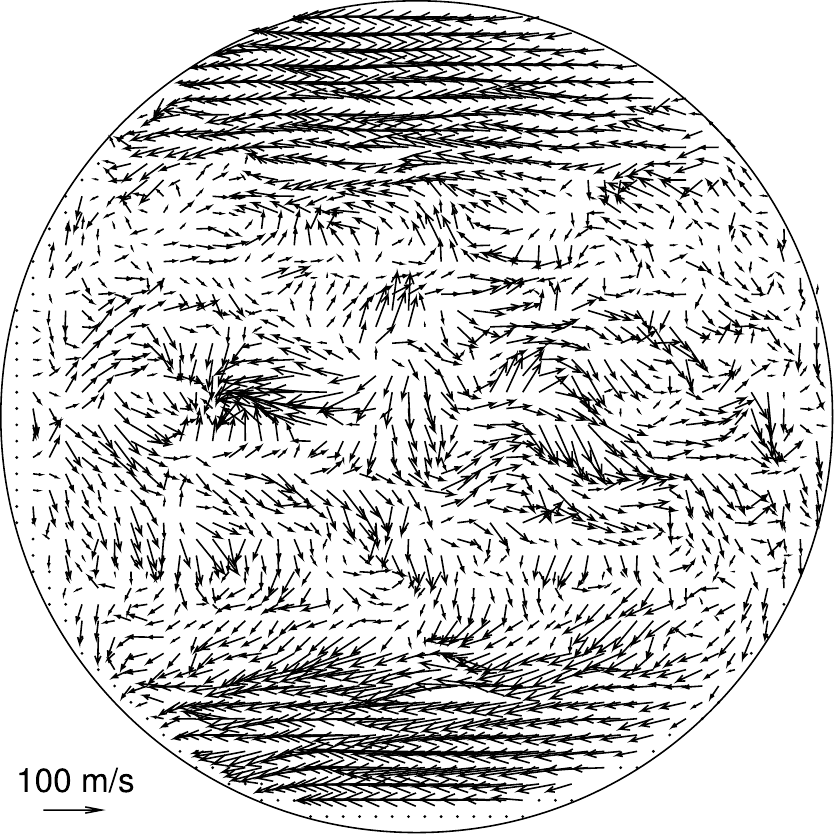}}
\resizebox{0.50\textwidth}{!}{\includegraphics{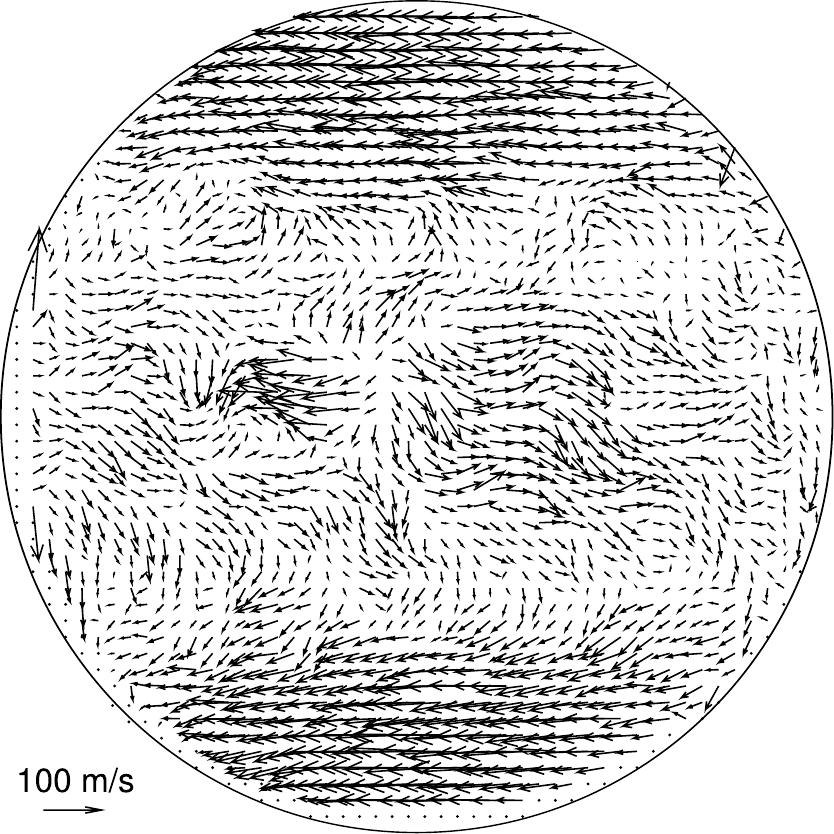}}
\caption{\emph{Left} -- model vector velocity field. \emph{Right} -- a velocity field that was computed by applying the LCT method to the synthetic supergranulation pattern with the imposed model field. The arrow lengths, representing the velocity magnitudes, have the same scale. The images are visually very similar, however the magnitudes of calculated velocities are underestimated.}
\label{fig:sipky}
\end{figure}

\begin{figure}[!t]
\resizebox{0.49\textwidth}{!}{\includegraphics{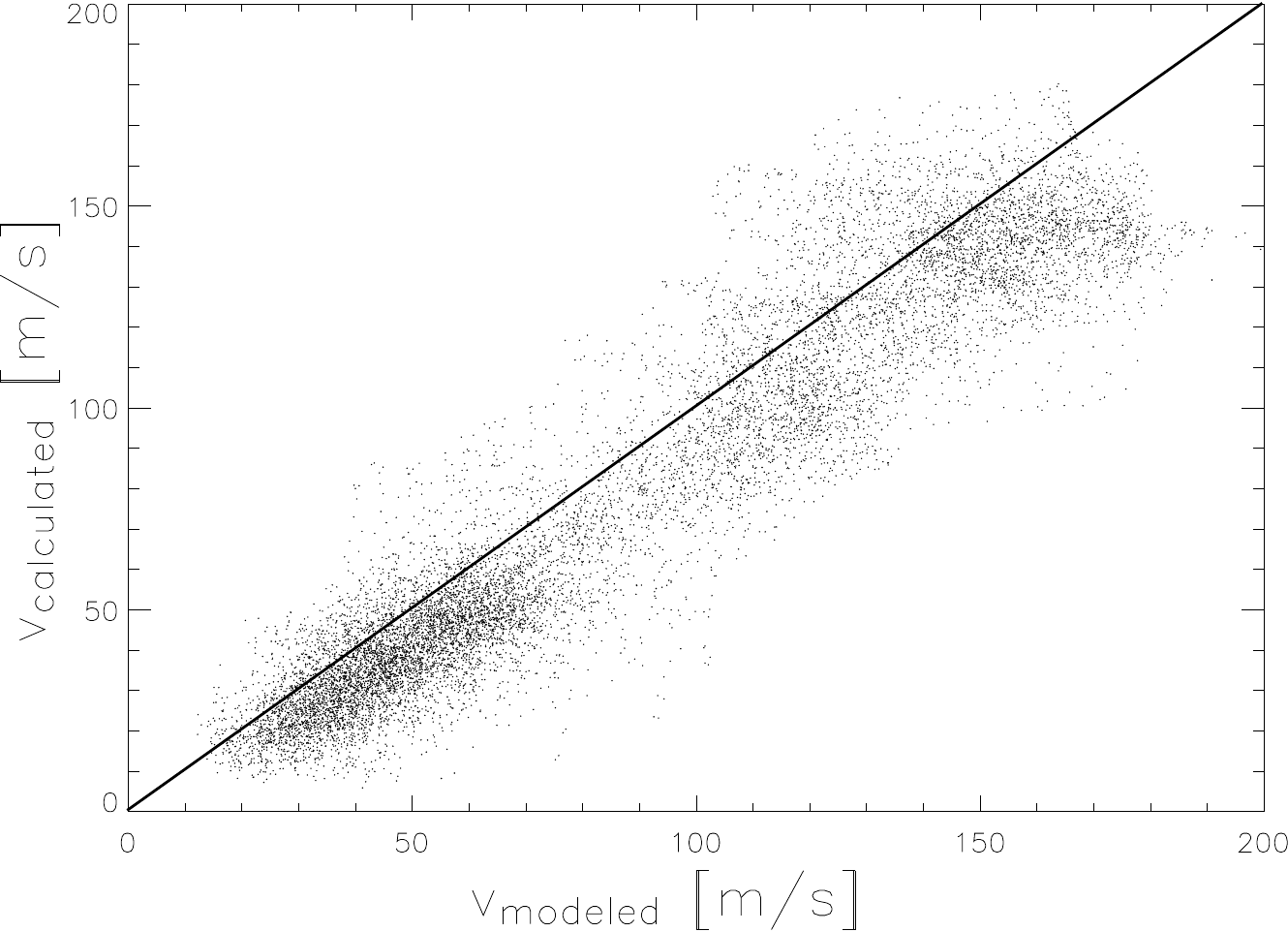}}
\resizebox{0.483\textwidth}{!}{\includegraphics{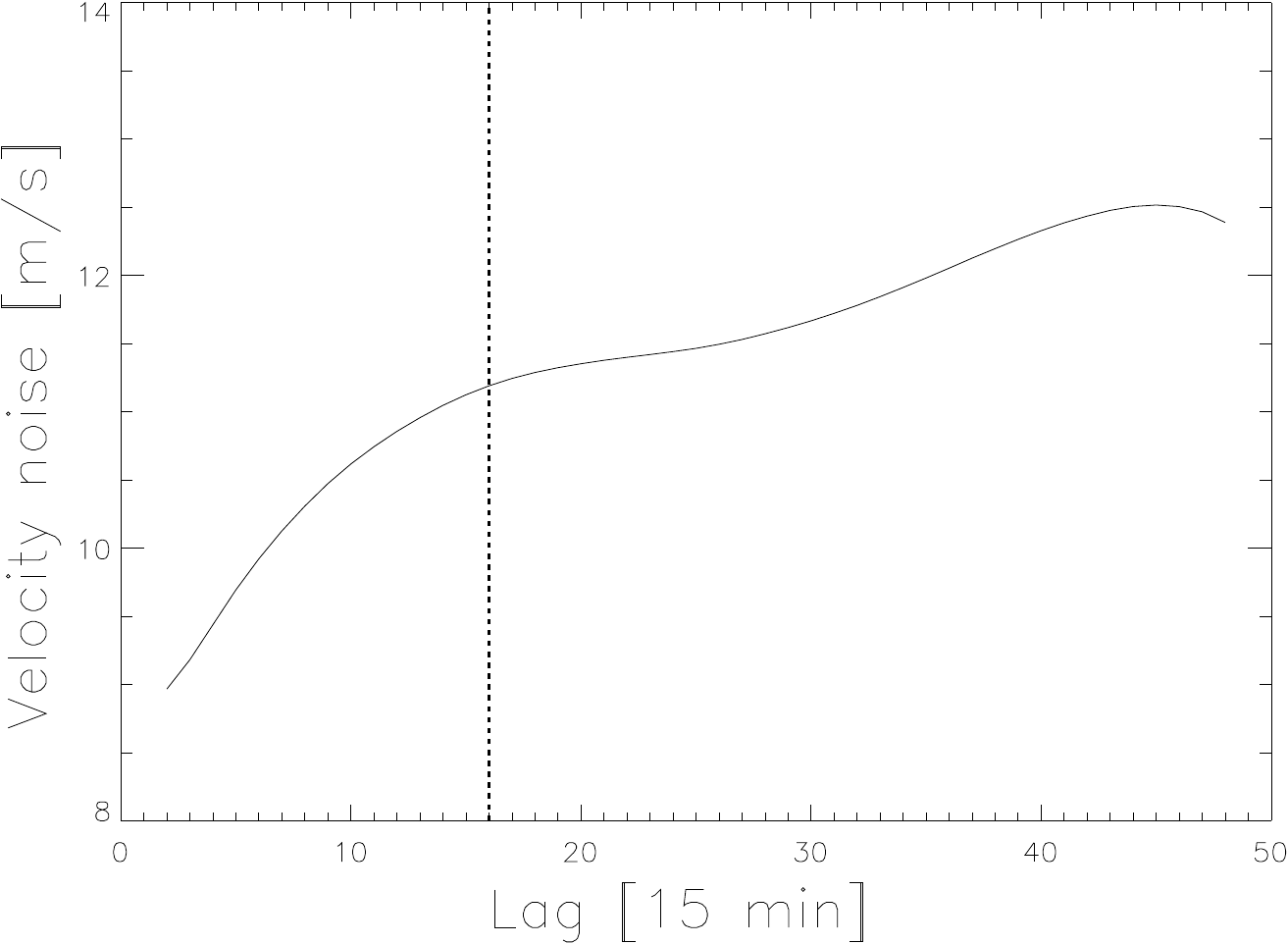}}
\caption{\emph{Left} -- Scatter plot for inference of the calibration curve. Magnitudes of calculated velocities are slightly underestimated by LCT, but the linear behaviour is clearly visible. A line representing the 1:1 ratio is displayed. The calibration affects only the magnitudes of the flows, while the directions do not need any correction. \emph{Right} -- Dependence of the 1-$\sigma$-error of the calculated velocity on the time lag when no model velocity field was introduced and only the evolution of supergranules have been taken into account. The dashed line represents lag 16 (4 hours) that is usually used in our method.}
\label{fig:kalibrace}
\end{figure}

The calculated velocities (with $v_{\rm offset,\ zonal}=-15\ \rm m\,s^{-1}$ corrected) were compared with the model velocities (Fig.~\ref{fig:sipky}). Already from the visual impression it becomes clear that most of vectors are reproduced very well in the direction, but the magnitudes of the vectors are not reproduced so well. Moreover, it seems that the magnitudes of vectors are underestimated. This observation is confirmed when plotting the magnitudes of the model vectors versus the magnitudes of the calculated vectors (Fig.~\ref{fig:kalibrace} left). The scatter plot contains more than 1 million points, and most of the points concentrate along a strong linear dependence, which is clearly visible. This dependence can be fitted by a straight line that can be used to derive the calibration curve of the magnitude of calculated velocity vectors. The calibration curve is given by the formula
\begin{equation}
v_{\rm cor}=1.13\,v_{\rm calc},
\label{eq:calibration}
\end{equation}
where $v_{\rm calc}$ is the magnitude of velocities coming from the LCT, and $v_{\rm cor}$ the corrected magnitude. The directions of the vectors before and after the correction are the same. The uncertainty of the fit can be described by 1-$\sigma$-error 15~m\,s$^{-1}$ for the velocity magnitudes under 100~m\,s$^{-1}$ and 25~m\,s$^{-1}$ for velocity magnitudes greater than 100~m\,s$^{-1}$. The uncertainties of approx. $15\ \rm m\,s^{-1}$ have their main origin in the evolution of supergranules. We studied the dependence of the error of velocity determination on the time lag used when no model velocity field was introduced. We found that this dependence is slowly increasing with the time lag (Fig.~\ref{fig:kalibrace} right) due to the evolution of individual supergranules. Evolution of supergranules is only one part of story, but it gives the lower limit of accuracy that can be obtained by this method. We also ran a test of the LCT sensitivity on the evolution of supergranules when a known underlying velocity field is introduced and came to similar results.

\begin{figure}[!t]
\resizebox{0.50\textwidth}{!}{\includegraphics{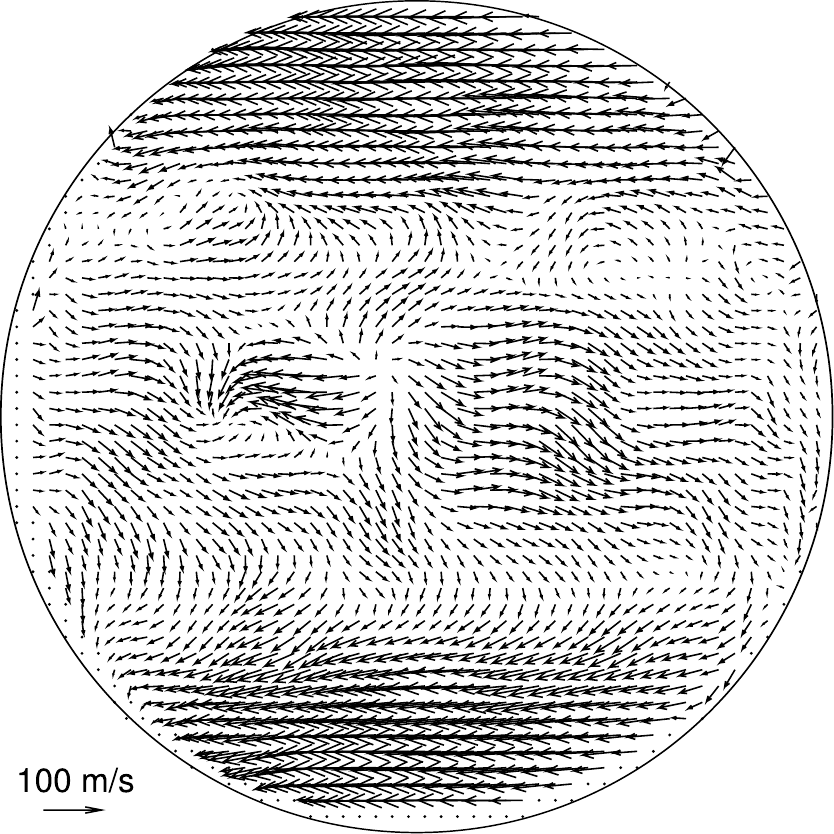}}
\resizebox{0.50\textwidth}{!}{\includegraphics{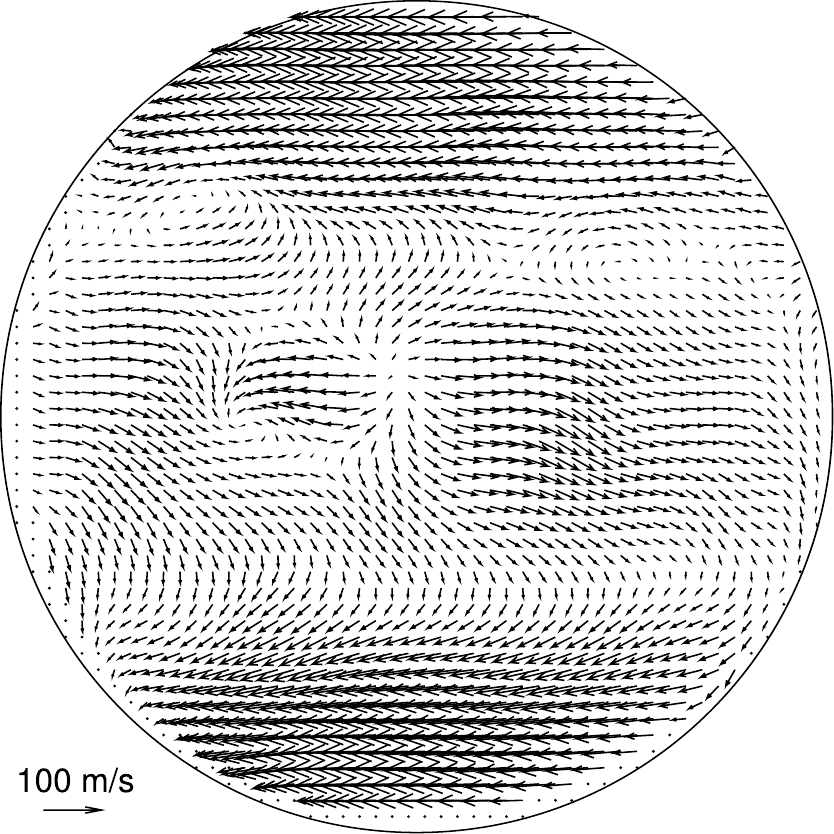}}
\caption{Influence of the calculated velocity field on the choice of FWHM of the correlation window. \emph{Left} -- 120\arcsec, \emph{right} -- 200\arcsec. The model velocity field is the same as in Fig.~\ref{fig:sipky}.}
\label{fig:fwhm}
\end{figure}

\begin{figure}[!t]
\resizebox{0.50\textwidth}{!}{\includegraphics{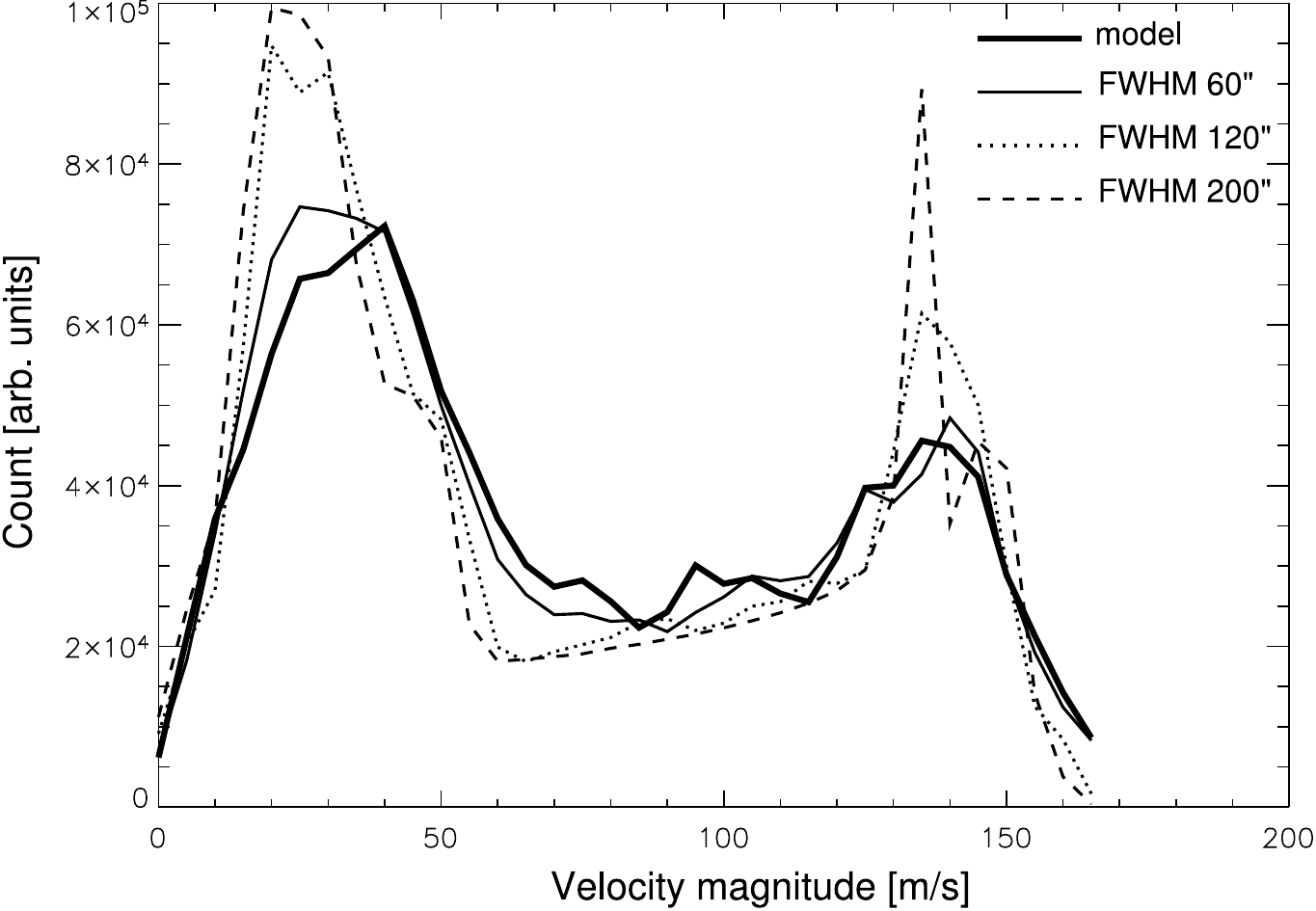}}
\caption{Histograms of velocity magnitudes for various FWHM of the LCT algorithm.}
\label{fig:histograms}
\end{figure}
We tested the sensitivity of the method to the choice of values of FWHM of the correlation window and of the lag between correlated frames in the LCT method. We found that our method is practically insensitive to the choice of the time lag between correlated frames when the lag is in the interval of 10--24 (2.5--6 hours). The larger the lag we choose, the lower the velocities we are able to detect. On the other hand, we have to take into account that a larger lag between correlated frames causes more noise in calculated results coming from evolutionary changes of supergranules and probably also from evolutionary changes in the velocity field under study. According to our tests, for a time lag greater than 30 (7.5~hours), the numerical noise raises very fast. The lag 16 (4 hours) seems to be a good tradeoff between sensitivity and noise. 

The choice of different FWHMs of the correlation window changes the spatial resolution according to FWHM. The general character of the vector field is preserved within the limits of resolution (cf.~Fig.~\ref{fig:fwhm}). Larger FWHM causes a smoothing of results and an underestimation of vector magnitudes (cf. Fig.~\ref{fig:histograms}). We found the used parameters (FWHM 60\arcsec, lag 4~hours) to be the best compromise, however these values can be changed during the work on real data.

Using the synthetic data generated by the SISOID code we have verified that the proposed method is reliable when measuring the large-scale velocity fields in the solar photosphere using the LCT method applied to full-disc dopplergrams.

%% file: timedistance.tex
\section{Direct comparison to the time-distance helioseismology}
\label{sect:comparizon}

\symbolfootnotetext[0]{\hspace*{-7mm} $\star$ This chapter was in the condensed form published as \v{S}vanda, M., Zhao, J., and Kosovichev, A. G., 2007, \emph{Comparison of Large-Scale Flows on the Sun Measured by Time-Distance Helioseismology
and Local Correlation Tracking}, Solar Physics, 241(1), 27--37.}

Velocity fields in one instant may be calculated using different techniques. However, the results obtained by different methods may have discrepancies. These can be caused by the nature of the methods, e.~g. due to different types of averaging, and also because of the use of different datasets from different instruments. In addition, various disturbing effects  can be important. Therefore, we decided to compare the results obtained by two different methods, time-distance helioseismology and local correlation tracking (LCT), using the same set of data: high-cadence dopplergrams covering almost one Carrington rotation obtained from Michelson Doppler Imager (MDI, \citeauthor{1995SoPh..162..129S}, \citeyear{1995SoPh..162..129S}).

Both methods provide surface or near-surface velocity vector fields. However, the results of these methods can be interpreted
differently. While local helioseismology measures intrinsic plasma
motions (through advection of acoustic waves), LCT measures apparent
motions of structures (granules or magnetic elements). It is
known that some structures do not necessarily follow the flows of
the plasma on the surface. For example, supergranulation appears to rotate
faster than the plasma \citep{2000SoPh..193..333B}, which may be
caused by travelling waves \citep{2003Natur.421...43G} or may 
be explained also as projection effect \citep{2006ApJ...644..598H}. 
Some older studies \citep[see e.~g.][]{1991BAAS...23.1033R} also suggest 
that the difference in flow properties measured on the basis of structures motions and 
plasma motions is caused by deeper anchor depth of these structures. An evolution of pattern 
may also play significant role (e.~g. due to emergence of
magnetic elements). Another possibility is
that surface structures are not coherent features, but patterns 
traveling with a different group velocity than the surface plasma 
velocity, such as occurs for the features present in simulations of 
travelling-wave convection \citep[e.~g.][]{1996ApJ...457..933H}.

Some attempts to compare the results of local helioseismology and the
LCT method for large scales, with characteristic size 100~Mm and
more, have been carried out by \cite{2005ASPC..346....3A}, but his results were inconclusive. The
correlation coefficient describing the match of the velocity maps
obtained by local helioseismology and the LCT method was close to
zero. Nevertheless, there were compact and continuous regions of the
characteristic size from 30 to 60 heliographic degrees with a good
agreement between the two methods, so that one could not conclude
that the results were completely different. In his study many
factors could play a significant role: the techniques were applied to
different types of datasets (LCT was applied to low resolution
magnetograms acquired at Wilcox Solar Observatory, and the
time-distance method used MDI dopplergrams). Both techniques had
very different spatial resolution, and also the accuracy of the
measurements was not well known.

We decided to avoid these problems and analyze the same data set
from the MDI instrument on SoHO. The MDI provides approximately two
months of continuous high-cadence (1 minute cadence) full-disc
dopplergrams each year. This \emph{Dynamics Program} provides data
suitable for helioseismic studies, and also for the local correlation
tracking of supergranules. Thus, this is a perfect opportunity to
compare performance and results of two different techniques using
the same set of data, and avoid effects of observations with
different instruments or in different conditions.

\subsection{Data preparation}

The selected dataset consists of 27 data-cubes from March 12th,
2001, 0:00~UT to April 6th, 2001, 0:31~UT, where each third day was
used, and in these days three 8.5-hour long data-cubes were
processed (so that every third day in the described interval was
fully covered by measurements). Each data-cube is composed of 512
dopplergrams (with spatial resolution of 1.98$^{\prime\prime}\,\rm px^{-1}$) 
at a one-minute cadence (so that covering 8 hours and 32
minutes). All the frames of each data-cube were tracked with a rigid
rate of 2.871~$\mu$rad\,s$^{-1}$, remapped to the Postel
projection with a resolution of 0.12\,$^\circ$\,px$^{-1}$ 
(1,500~km\,px$^{-1}$ at the center of the disc), 
and only a central meridian region was selected for the ongoing processing
(with size of 256$\times$924~px covering 30 heliographic degrees in
longitude and running from $-54\,^\circ$ to $+54\,^\circ$ in
latitude), so that effects of distortions due to the projection do
not play a significant role.

Tracked data-cubes were used to perform the time-distance analysis.
From all the frames in each data-cube the mean dopplergram (like
Fig.~\ref{fig:examples} left) was subtracted to suppress the influence
of velocity structures like supergranulation and to highlight the
signals of $p$-modes of solar oscillations. The surface gravity wave
($f$-mode) has different dispersion characteristics than $p$-modes
used in this study, and, therefore, it is filtered out from the
$k$--$\omega$ diagram before computing the travel times. 
The $f$-mode, if not filtered out, will disturb $p$-modes measurements, and it is also not
straightforward to perfrom inversions if not separating two different modes. The $p$-mode 
inversions are less sensitive to the surface flows than $f$-mode data, but still recover the 
large-scale flows well.

$P$-modes of solar oscillations have their origin in the 
solar convection zone and travel through the solar interior
to the surface. The time of the excursion of the wave bulk depends 
on the speed of sound and on the velocity field describing the mass flow
in the layers of the solar interior, through which the bulk is travelling.
In the time-distance technique, the travel times of oscillations bulk from 
the point in the photosphere (central point) to surrounding annuli around 
this point are measured. The radius of the studied diameter of the annuli is 
related to the depth in the interior, where the studied oscillation mode is reflected back
to the surface. Travel times are measured by the cross-correlation between Doppler velocities 
in the central point and velocities in the selected annuli around this point. 

The mass flow velocities in the interior are calculated from the differences
of travel times from the central point to the surrounding annuli and the travel 
times from the surrounding annuli to the central point when the state properties 
in the affected layers of the solar interior are known. In this study the theoretical
travel ray approximation is derived from the solar model S \citep{1996Sci...272.1286C}. 
Dividing the annuli into sectors the underlying flow field of selected orientations can be inferred.
For details see \cite{1996ApJ...461L..55K}, \cite{2001ApJ...557..384Z}, or \cite{2004ApJ...603..776Z}.

The  time-distance inversion results were smoothed by a Gaussian
with FWHM of 30~px to match the resolution to the LCT method, and
only the horizontal components ($v_x$, $v_y$) of the full velocity
vector were used.

While for the time-distance method the $p$-modes of solar
oscillations play a crucial role, they significantly influence the
performance of the LCT method in a negative way. The oscillations
are clearly visible in the dopplergrams, and, thus, cause random
errors in the calculation of displacements. Therefore, before
applying the LCT method the oscillation signals must be suppressed.
For our high-cadence data it is possible to do this using  temporal
averaging. According to \cite{1988SoPh..117....1H} it is better to use a Gaussian type of
temporal averaging than the boxcar one. We average the dopplergrams
over 31~minute periods with weights given by a formula
\begin{equation}
w(\Delta t)=  \exp\left[{\frac{(\Delta
t)^2}{2a^2}}\right]-\exp\left[{\frac{b^2}{2a^2}}\right] \left(1+\frac{b^2-(\Delta
t)^2}{2a^2}\right),
\label{eq:average}
\end{equation}
where $\Delta t$ is the time between a given frame and the central one
(in minutes), $b=16$~min and $a=8$~min. We verified that this filter
suppresses the solar oscillations in the 2--4 mHz frequency band by
a factor of more than five hundred.

The other issue significantly influencing the performance of the LCT
 method is the change of  contrast and background intensity caused by
solar rotation. Due to tracking the Doppler images with rotation,
the magnitude of the line-of-sight component of the solar rotational
velocity changes from frame to frame and affects the LCT results.
The method interprets these changes as motion towards east,
mainly in the central part of the solar disc, where the contrast in
the structures of dopplergrams is very low (see Fig.~\ref{fig:examples}
left). We suppress the influence of the moving background by
subtraction of a polynomial surface fit of the third order. We have
tested that this provides almost the same results as the other
possible procedures: local removal of the mean values and unsharp
masking. Subtraction of the polynomial fit is not so sensitive
to anomalies in the dopplergrams, caused by regions with strong
magnetic field.

The LCT method used in this study is described by the following
parameters: the time-lag between correlated frames is 120 frames (2
hours), the correlation window has a Gaussian shape with FWHM of 30
px, the correlation is measured by the sum of absolute differences
of subframes (it is faster than calculation of the correlation
coefficient and provides the same results), the extremum position
is calculated using the nine-point method of \cite{1991PhDT.......137D}. For each
data-cube, the results of all the correlated pairs are averaged, so
that the method provides an averaged flow field in 8.5~hours in the
same sense as the time-distance analysis.

\subsection{Results}

\begin{figure}[!t]
\resizebox{0.49\textwidth}{!}{\includegraphics{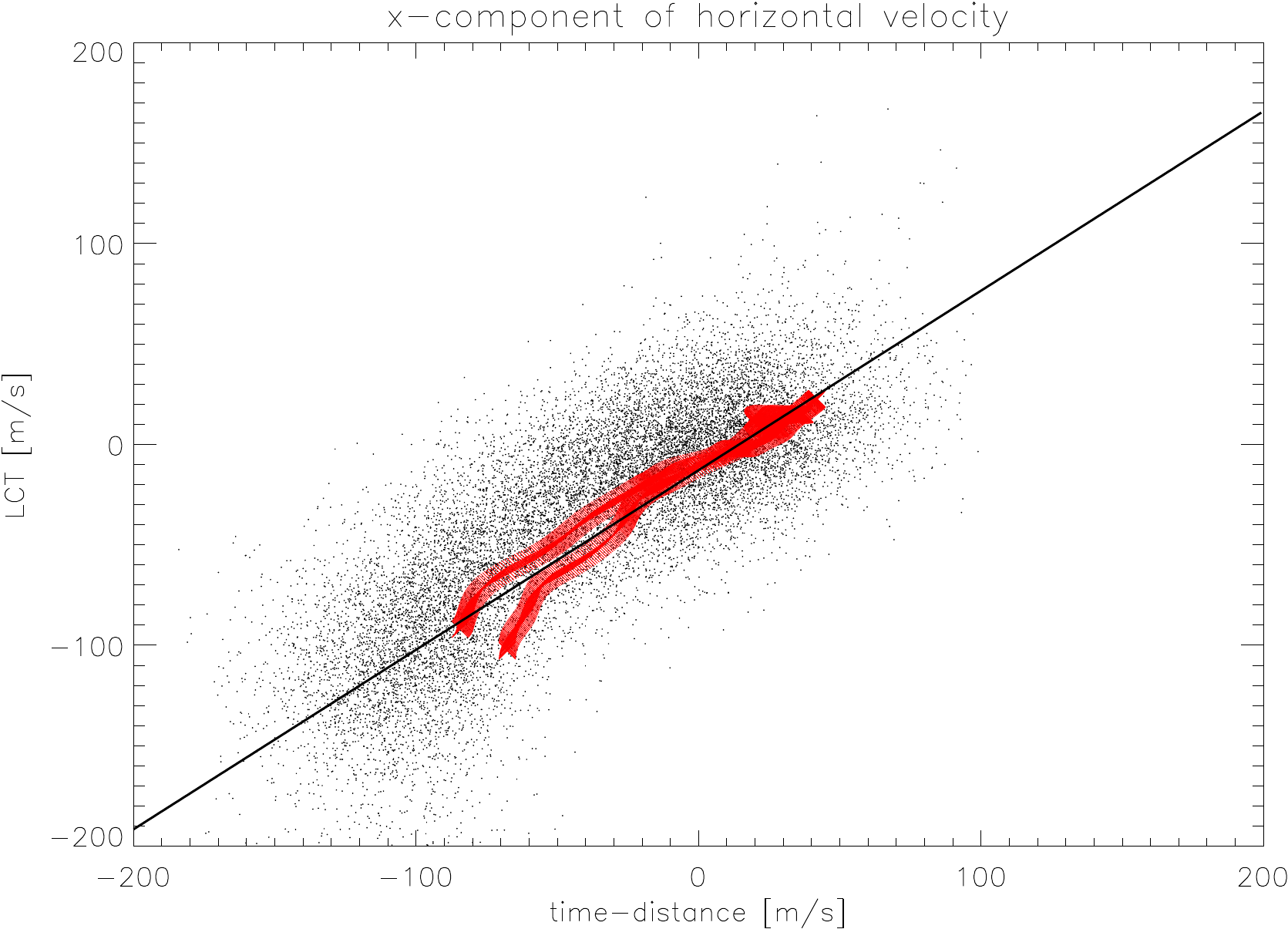}}
\resizebox{0.49\textwidth}{!}{\includegraphics{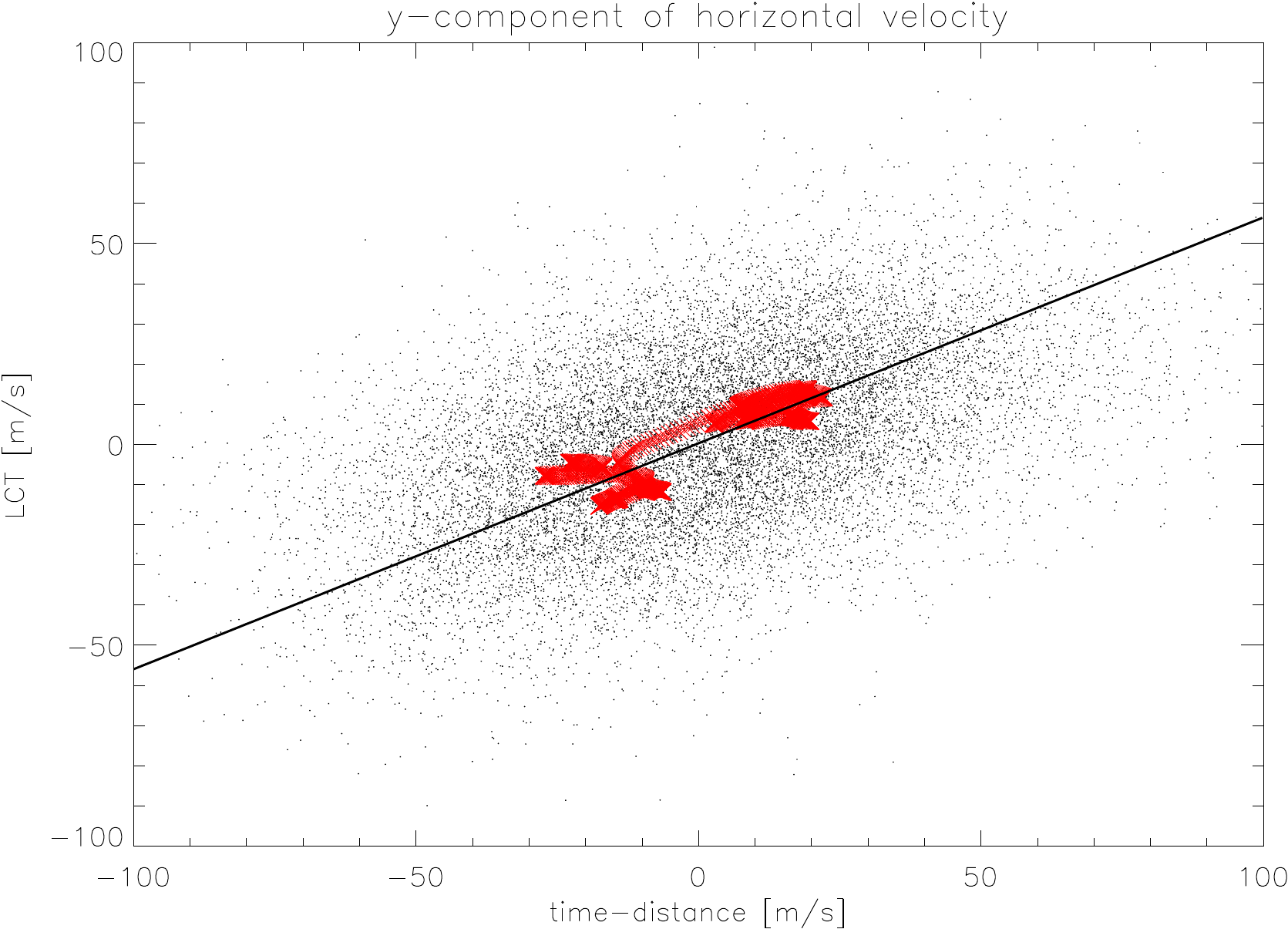}}
\caption{\emph{Left} -- $v_{x,{~\rm LCT}}$ versus $v_{x,{~\rm t-d}}$
plot. Red crosses denote mean zonal velocities (differential
rotation) which have been used for  fitting of the calibration
formula (\ref{vx_fit}). \emph{Right} -- the same for $v_y$
component, the regression fit is described by Eq.~(\ref{vy_fit}).}
\label{fig:vx,vy}
\end{figure}

\subsubsection{Statistical processing}
\label{statistics}

The results containing 27 horizontal flow fields were statistically
processed to obtain the cross-calibration curves for these methods.
It is generally known (see discussion on page~\pageref{label:lct_under})
that the LCT method slightly underestimates the
velocities; thus, the results should be corrected by a certain
factor. From the comparison of the $x$-component of velocity (cf.
Fig~\ref{fig:vx,vy} left) we obtained parameters of a linear fit given
by (numbers in parentheses denote a 1$\sigma$-error of the regression
coefficient)
\begin{equation}
v_{x,{~\rm LCT}}=0.895(0.008) v_{x,{~\rm t-d}}-12.6(0.3)~{\rm m\,s^{-1}} .
\label{vx_fit}
\end{equation}
The correlation coefficient between $v_{x,{~\rm LCT}}$ and
$v_{x,{~\rm t-d}}$ is $\rho=0.80$. We assume that the time-distance
measurements for $v_{x,{~\rm t-d}}$ are correct, and the magnitude
of the LCT measurements, $v_{x,{~\rm LCT}}$, must be corrected
according to the slope of Eq.~(\ref{vx_fit}). This correction factor
has a value of 1.12, which is in perfect agreement with the
correction factor of 1.13 found in the tests of the same LCT code
using synthetic dopplergrams with the same resolution and similar
LCT parameters (see Section~\ref{sect:method}). We assume that both velocity
components obtained with the LCT method should be corrected by this
factor. 

The regression line of $v_y$ component (Fig.~\ref{fig:vx,vy} right)
is
\begin{equation}
v_{y,{~\rm LCT}}=0.56(0.01) v_{y,{~\rm t-d}}+0.4(0.2)~{\rm m\,s^{-1}} .
\label{vy_fit}
\end{equation}
After the slope correction using the $v_x$ fits, the regression
curve is slightly different:
\begin{equation}
v_{y,{~\rm LCT}}=0.63(0.01) v_{y,{~\rm t-d}}+0.4(0.2)~{\rm m\,s^{-1}},
\end{equation}
with the correlation coefficient between $v_{y,{~\rm LCT}}$ and
$v_{y,{~\rm t-d}}$ close to 0.47. The slope of the linear fit
differs significantly from the expected value 1.0. For this study we
decided to correct the $y$-component of the time-distance results.

We have tested that this asymmetry is not related to the LCT
technique. The tests did not show any preference in direction of flows
measured by LCT or any dependence of the results on the size
of the field of view (which does not have a square shape in our case). Also in the study performed in Section~\ref{sect:method} based on synthetic data the asymmetry between the zonal and the meridional
component was not encountered. 

We found two possibilities that could explain this behaviour. The
first explanation is a drift of the supergranular pattern towards
the equator. This case does not explain why the meridional
velocities from both techniques seem to be proportional to each
other. A systematic drift would rather be depicted as a systematic
constant shift, or a shift depending on the latitude. However, the
meridional components of velocities are generally rather small, so
that the errors of the measurements can play a significant role and
the proportional behavior can be only apparent. 

The second
explanation is based on unspecified asymmetries influencing travel-time
measurements, for instance, due to different sensitivity of the MDI
instrument to $p$-modes propagating in the east-west and north-south
directions. As it has been studied
recently \citep{2006astro.ph..8204G}, a 
comparison between $f$-mode time-distance and LCT applied to
realistic numerical simulation did not show such asymmetry. The asymmetry in the east-west and
north-south directions observed by time-distance helioseismology
was on the contrary noticed in a recent study based on numerical simulated
data \citep{2006astro.ph.12551Z}. So that the asymmetry takes
place only in the $p$-modes inversions and should be further 
investigated in more details.

\begin{figure}[!t]
\begin{center}
\resizebox{0.75\textwidth}{!}{\includegraphics{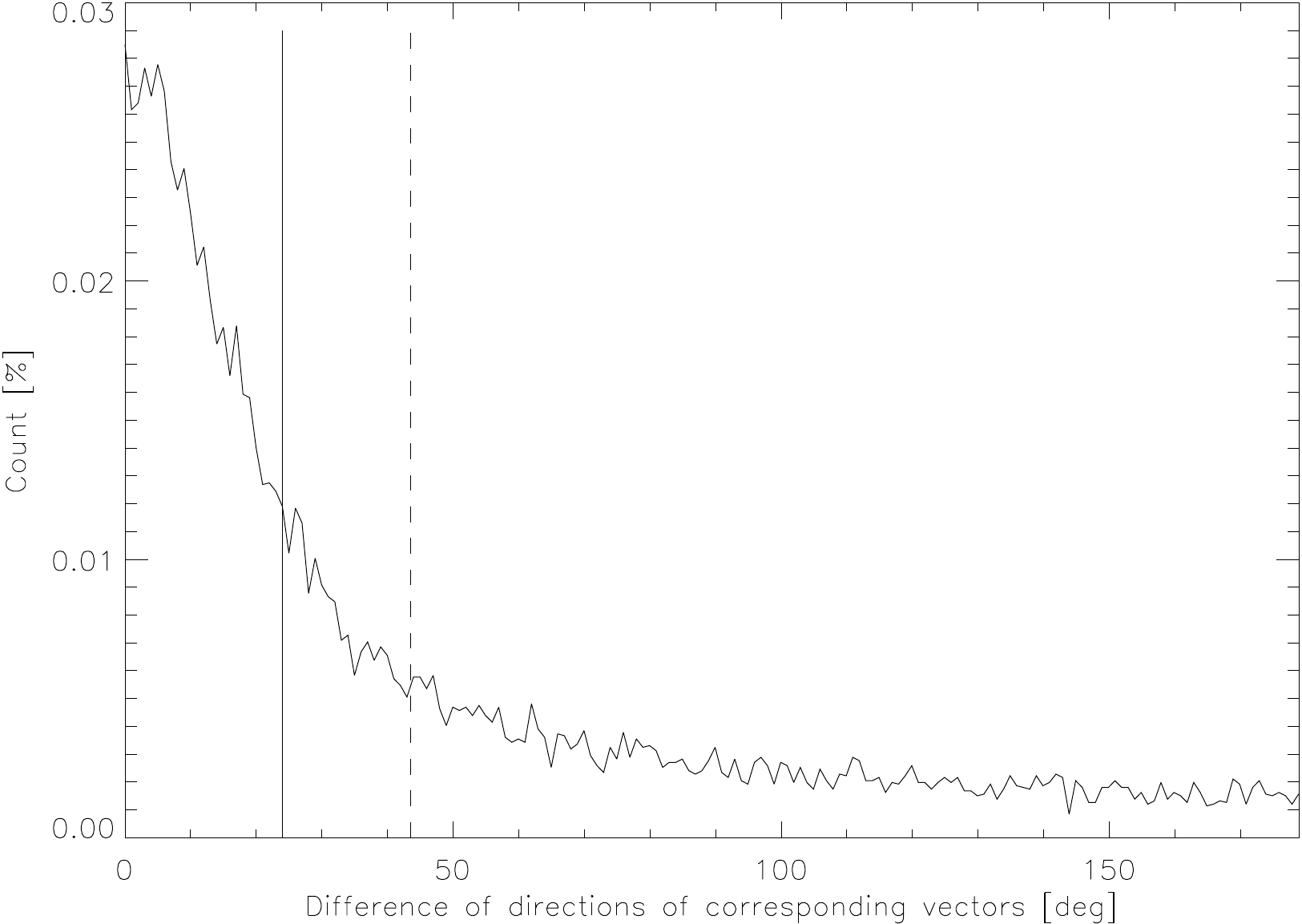}}
\end{center}
\caption{Histogram
of the angular differences ($\Delta \varphi$) between directions of
the velocity vectors obtained by the time-distance and LCT
techniques. Dashed vertical line denotes the mean value and solid
vertical line represents the median of $\Delta \varphi$.} \label{fig:histy}
\end{figure}
The final calibration formulae providing the best statistical
agreement between the velocities calculated using both methods are:

\begin{eqnarray}
v_{x, \rm LCT, corr} & = & 1.12 v_{x, \rm LCT, calc} \label{calbeg}\\
v_{y, \rm LCT, corr} & = & 1.12 v_{y, \rm LCT, calc} \\
v_{x, \rm t-d, corr} & = & v_{x, \rm t-d, calc} \\
v_{y, \rm t-d, corr} & = & 0.63 v_{y, \rm t-d, calc}\label{calend},
\end{eqnarray}
where the index \emph{corr} denotes the corrected value, and the index
\emph{calc} denotes the original calculated value.

After the corrections, as presented in the histogram of
Fig.~\ref{fig:histy}, the differences between the directions of the
velocity vectors ($\Delta \varphi$) calculated by these techniques are quite
reasonable. The mean value of the distribution is 43.56\,$^\circ$,
however, the mean value is not a good indicator in this case because
the distribution function is not normal. The median value is
24.02\,$^\circ$; and 66.6~\% points have the difference in the
corresponding vector directions under 45\,$^\circ$. 

Instead of computation of the correlation coefficient of the arguments 
of both vector fields we decided to compute a magnitude-weighted cosine 
of $\Delta \varphi$. This quantity is given by 
\begin{equation}
\rho_{\rm W} = \frac{\sum |{\mathbf v}_{t-d}| \frac{|{\mathbf v}_{t-d} \cdot {\mathbf v}_{LCT}|}
{|{\mathbf v}_{t-d}||{\mathbf v}_{LCT}|}}{\sum |{\mathbf v}_{t-d}|}, 
\label{eq:weightedcosine}
\end{equation}
where ${\mathbf v}_{t-d}$ is the time-distance vector field, ${\mathbf v}_{LCT}$ is the LCT vector field
and the summation is performed over all vectors in the field. The closer this quantity is to 1, the 
better is the agreement between both vector fields. Larger vectors are weighted more than smaller ones. 

We have found that in our case $\rho_{\rm W}=0.86$, which means an almost perfect match.

\subsubsection{Mean velocities}
\begin{figure}[!t]
\resizebox{0.49\textwidth}{!}{\includegraphics{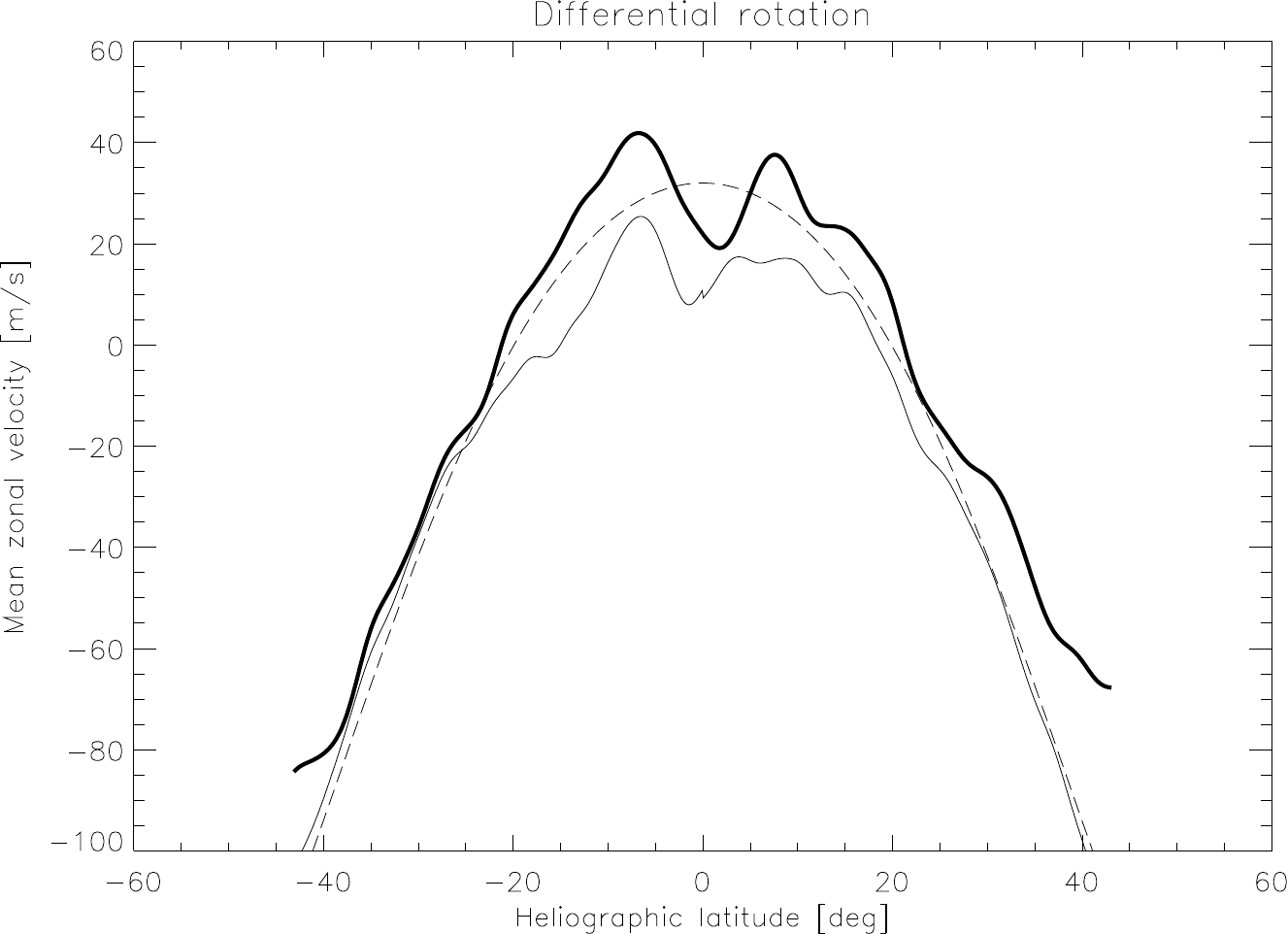}}
\resizebox{0.49\textwidth}{!}{\includegraphics{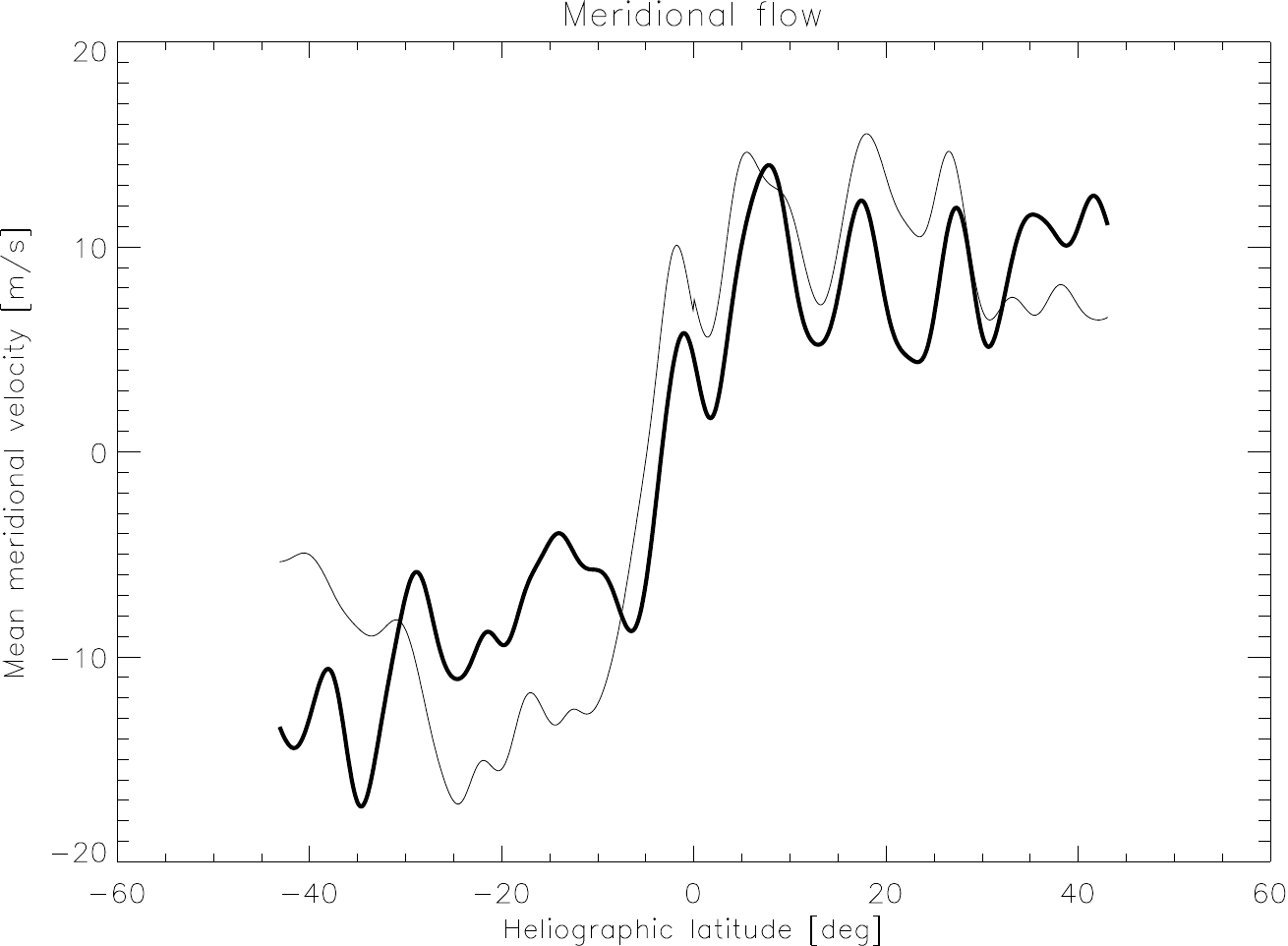}}
\caption{\emph{Left:} Thick curve shows the mean zonal velocity as a function of
latitude (differential rotaton) obtained from the time-distance
data, thin curve shows the mean zonal velocity obtained from the LCT
data. Dashed line represents a standard rotation profile
(``Snodgrass rate''). The mean zonal velocities are plotted in the
coordinate system rotating rigidly with 2.871~$\mu$rad\,s$^{-1}$.
\emph{Right:} Velocities of the mean meridional flow as a function of
latitude obtained by the time-distance technique (thick curve) and
by the LCT method (thin curve). The correlation coefficient for the
mean zonal flow is $\rho=0.98$ and for the mean meridional flow
$\rho=0.88$. } \label{fig:mean}
\end{figure}

\begin{figure}[!t]
\resizebox{0.49\textwidth}{!}{\includegraphics{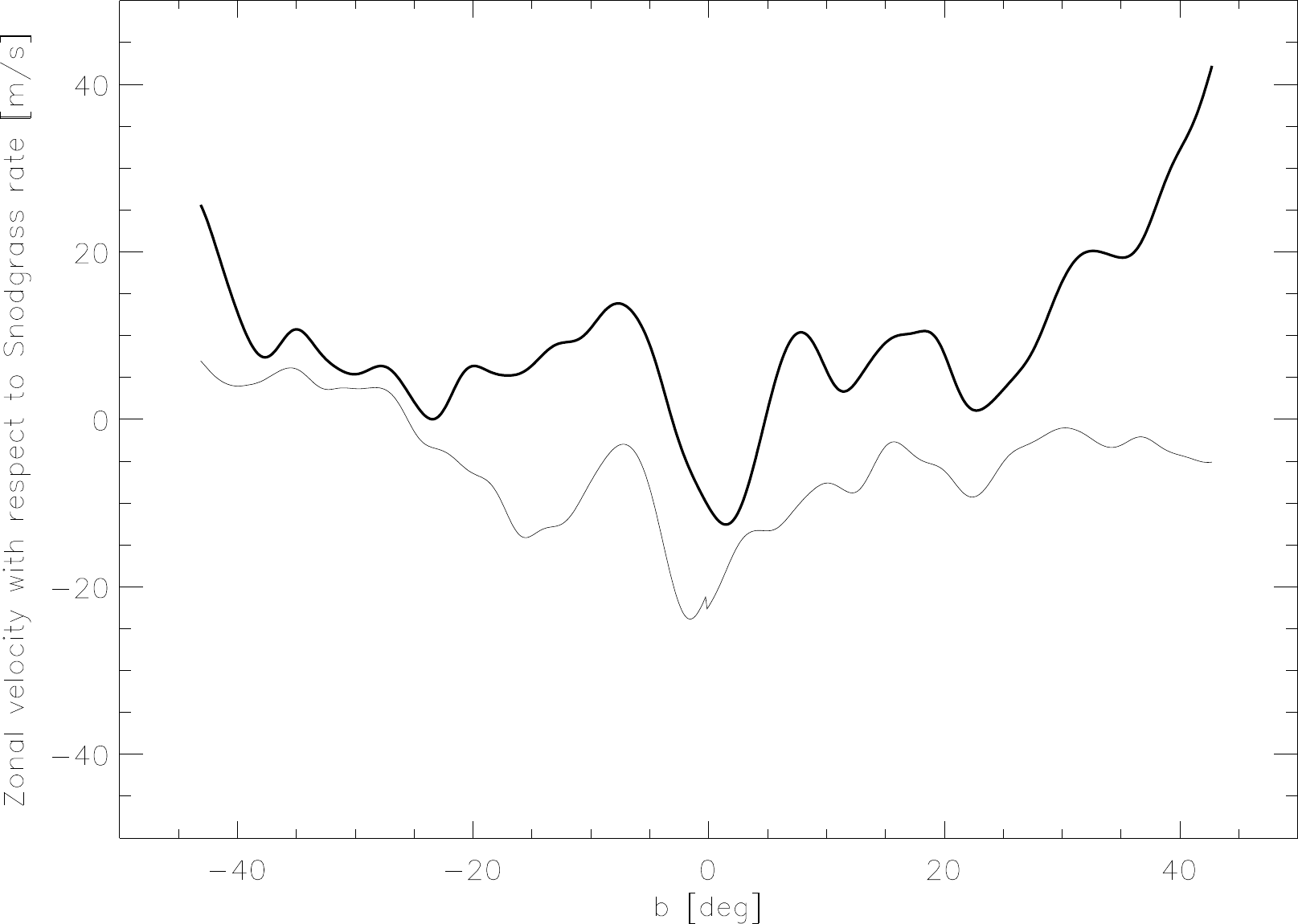}}
\caption{Mean zonal flows (like in Fig.~\ref{fig:mean} left) with the standard rotation profile
(``Snodgrass rate'') removed. The systematic shift between the curves obtained from time-distance
results (thick line) and LCT results (thin line) is evident and is in average 14.1~\mps.} \label{fig:mean1}
\end{figure}
In addition to the detailed comparison of the vector fields, we
compare the mean flows, the differential rotation and the meridional
circulation. These flows can be quite simply calculated from the
results of both techniques. In both cases, they provide the mean
zonal and mean meridional flows for the Carrington rotation
No.~1974. The results are displayed in Fig.~\ref{fig:mean}, where the
differential rotation curves are compared with a standard profile
from \cite{1990ApJ...351..309S} in the left panel. It can be clearly seen
that the latitudinal profiles for both techniques are very similar
and also that the mean velocities do not differ so much in
magnitude. The correlation coefficients are $\rho=0.98$ for the
zonal flow and $\rho=0.88$ for the meridional flow. In the
differential rotation curves, the LCT results give a little slower
rotation, which is also seen from Eq.~(\ref{vx_fit}). The mean difference
of average zonal velocities obtained by both techniques is 14.1~m\,s$^{-1}$ (see 
Fig.~\ref{fig:mean1}). 

It is coherent with the systematic shift found in Section~\ref{sect:method}, i.~e. a systematic offset of $-$15~\mps{} caused by the change of the
projection effect due to the ``derotation'' of the studied data series. We believe
that the offset found between the LCT and time-distance results is caused by the same reason.
  
We conclude that for the mean flows the results obtained by two such
different techniques agree very well.

\subsubsection{Detailed comparison}

 \begin{figure}[!t]
 \begin{center}
 \resizebox{!}{10cm}{\includegraphics{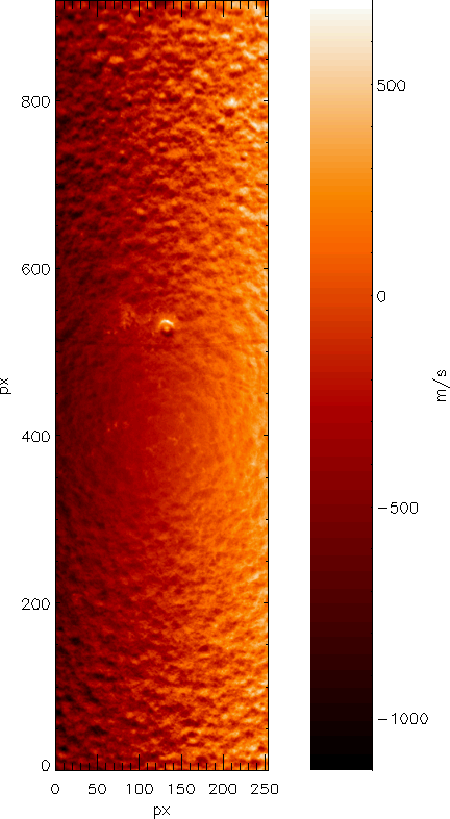}}
 \resizebox{!}{10cm}{\includegraphics{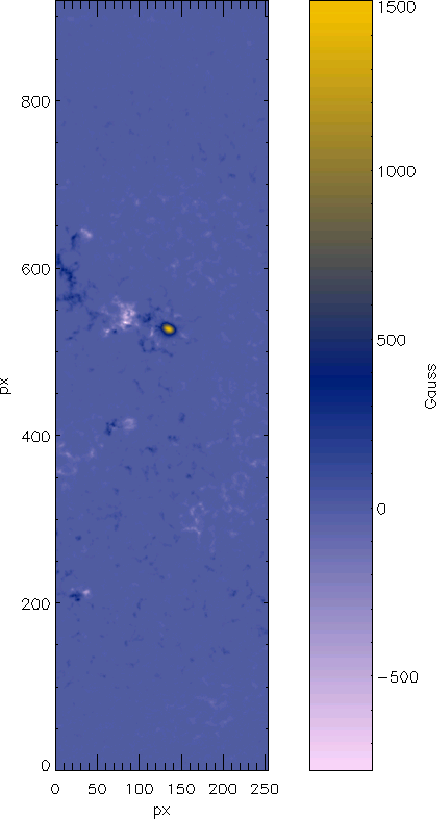}}
 \end{center}
 \caption{Part of the solar
 disc chosen for the detailed comparison of the results of
 time-distance and LCT method. The 8.5-hour measurements are centered at
 March 24th, 2001, 4.16~UT. One pixel represents 0.12\,$^\circ$ in
 heliographic coordinates. \emph{Left} -- Averaged MDI dopplergram.
 Note the low contrast of the supergranular cells in the center of
 the images. This is the ``blind spot'' caused by the prevailing
 horizontal motions in the photosphere with this resolution.
 \emph{Right} -- Averaged MDI magnetogram.} \label{fig:examples}
 \end{figure}

\begin{figure}[!t]
\begin{center}
 \resizebox{!}{12cm}{\includegraphics{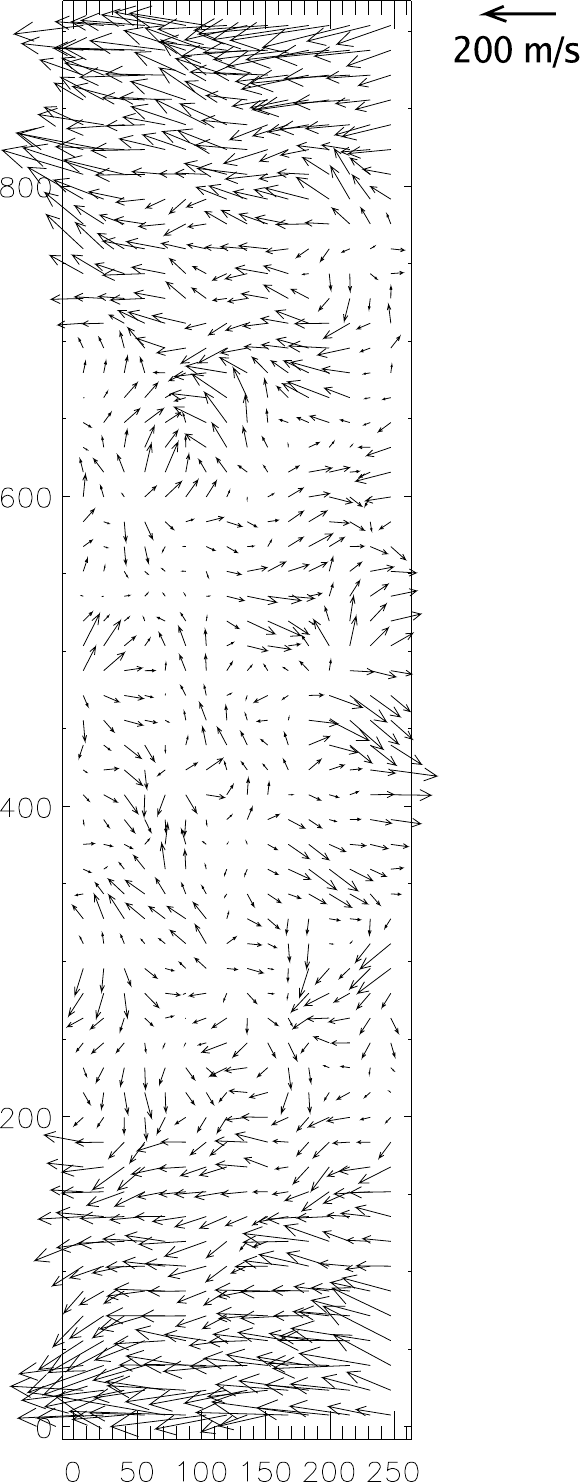}}
 \resizebox{!}{12cm}{\includegraphics{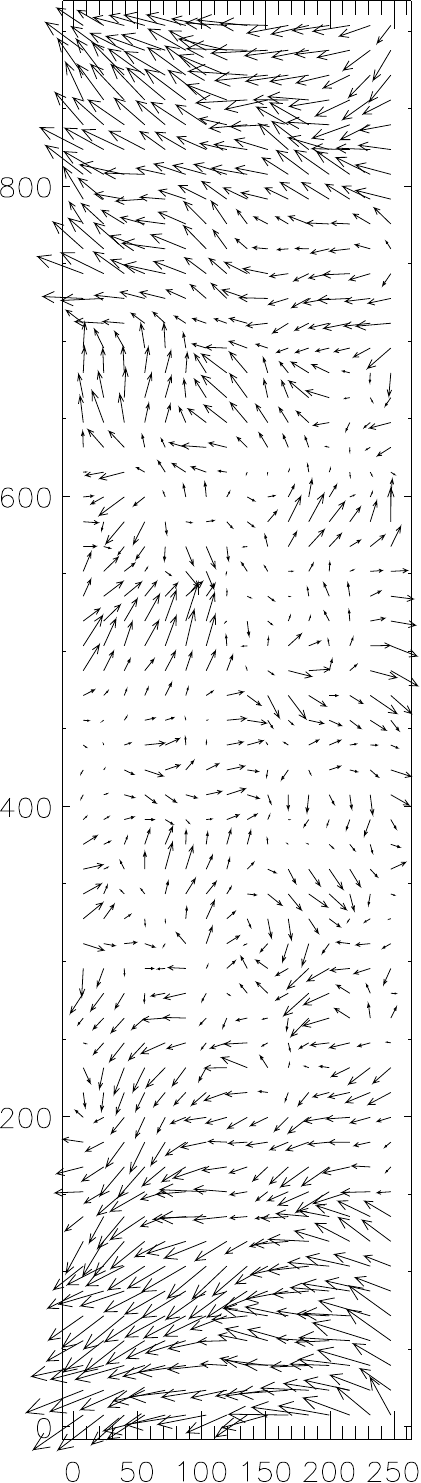}}
\end{center}
\caption{\emph{Left} --
velocity field obtained by the LCT method. \emph{Right} -- vector
field obtained by the time-distance technique. Both plots are
centered at heliographic coordinates $b_0=0.0\,^\circ$,
$l_0=214.3\,^\circ$, units on both axes are pixels in the data frame
with resolution 0.12\,$^\circ$\,px$^{-1}$ in the Postel's
projection.} \label{fig:arrows}
\end{figure}
\begin{figure}[!t]
\begin{center}
 \resizebox{!}{10cm}{\includegraphics{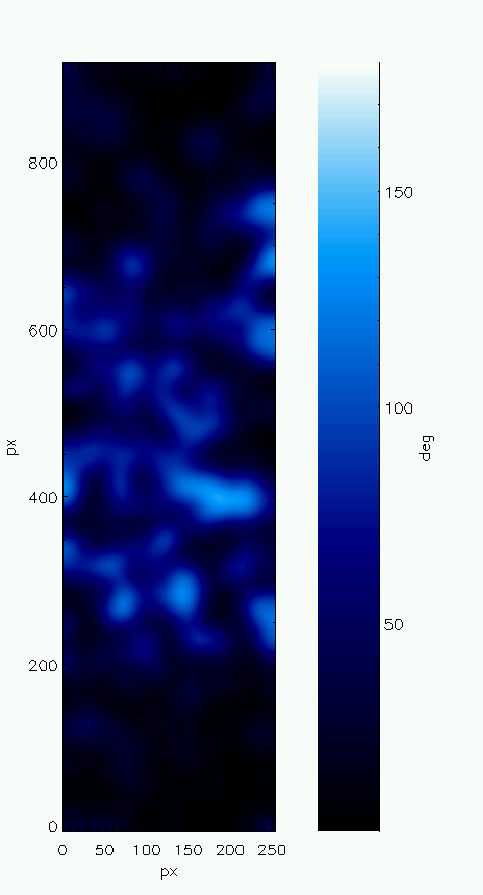}}
 \resizebox{!}{10cm}{\includegraphics{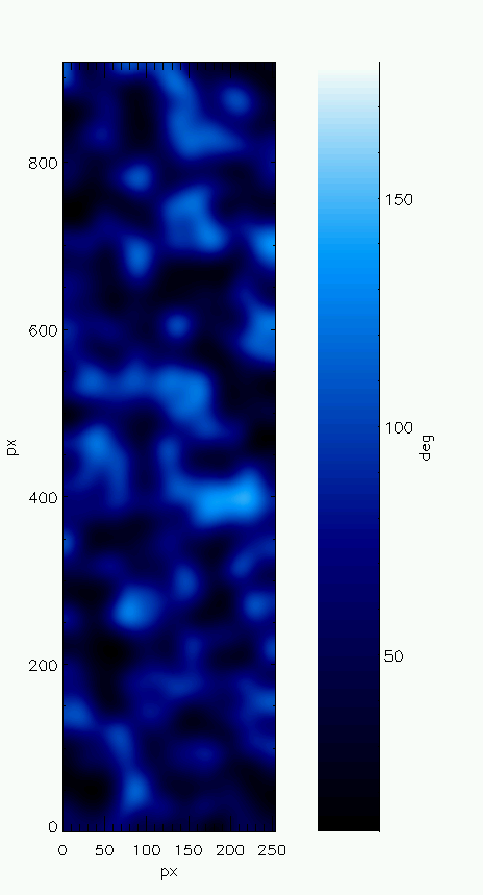}}
\caption{Difference of
phases of corresponding vectors ($\Delta\varphi$). \emph{Left} -- for the full vector data, \emph{right} -- for the data with mean zonal flow subtracted. In the second case the magnitudes of vectors everywhere in the studied area are comparable and the inaccuracies of individual measurements equally important. With the mean zonal flow removed, the distribution of $\Delta\varphi$ becomes more uniform in the studied area.} \label{fig:difphases}
\end{center}
\end{figure}

For a detailed comparison of the flow fields, we selected one data
cube, representing 8.5-hour measurements centered at 4:16UT of March
24th, 2001 ($l_0$=214.3\,$^\circ$; see the averaged MDI Dopplergram
and magnetogram in Fig.~\ref{fig:examples}). In this map, the
correlation coefficient for the $x$-component of the velocity is
$\rho=0.82$ and for the $y$-component: $\rho=0.58$, and for the
vector magnitude: $\rho=0.73$. The vector plots of the flow fields
obtained by both techniques, shown in Fig.~\ref{fig:arrows}, seem to be
quite similar, in general; however many differences can be seen. The
regions  where the differences are most significant correspond to
relatively small (under 50~m\,s$^{-1}$) velocities. This is clear
from the map of the differences between the vector directions
displayed in Fig.~\ref{fig:difphases}.

The statistics of the differences between the directions of the
corresponding vectors ($\Delta \varphi$) is presented in
Fig.~\ref{fig:histy}. Values of $\Delta\varphi$ slightly anti-correlate
with the averaged magnitude of the corresponding vectors
($\rho=-0.58$). We think that this is due to uncertainties of both
techniques. From our tests using synthetic data, it became clear
that the inaccuracy of the local correlation tracking code is
15~m\,s$^{-1}$ for velocities smaller than 100~m\,s$^{-1}$ and
25~m\,s$^{-1}$ for velocities larger than 100~m\,s$^{-1}$, for both
components (see Section~\ref{sect:method}). We think that the 10\% accuracy for the
time-distance velocity vectors is a reasonable estimate. 
As \cite{2001ApJ...557..384Z} stated, cross-talk 
effects between horizontal flows and vertical flow components of flow velocities
affect the time-distance inversion results. The cross-talk 
effect prevents us from inverting the vertical velocity 
correctly, but does not block the determination of horizontal velocities
satisfactorily \citep{2003soho...12..417Z}. 
However, the vertical  velocities are not discussed in this Section because they are not 
measured by the LCT technique.
Obviously, the inaccuracy in one component may cause a significant change of the
direction of the horizontal vector for small velocities; and, hence,
the agreement of both techniques in such areas is not as good as in
the areas of high velocities.

The vector velocity field may be also influenced by the temporal evolution of 
the traced pattern. We have tested using the full-disc MDI dopplergrams that temporal
changes at mesogranular and smaller scales are effectively filtered out by a $k$--$\omega$
filter. The temporal evolution of the supergranular pattern may, in the worst case, significantly
influence the calculated vector velocity field in the close (roughly equal to the FWHM of the
chosen correlation window) vicinity of a rapidly changing (e.~g. disappearing) supergranule.

\subsection{Conclusions}

\begin{figure}[!t]
\begin{center}
 \resizebox{!}{12cm}{\includegraphics{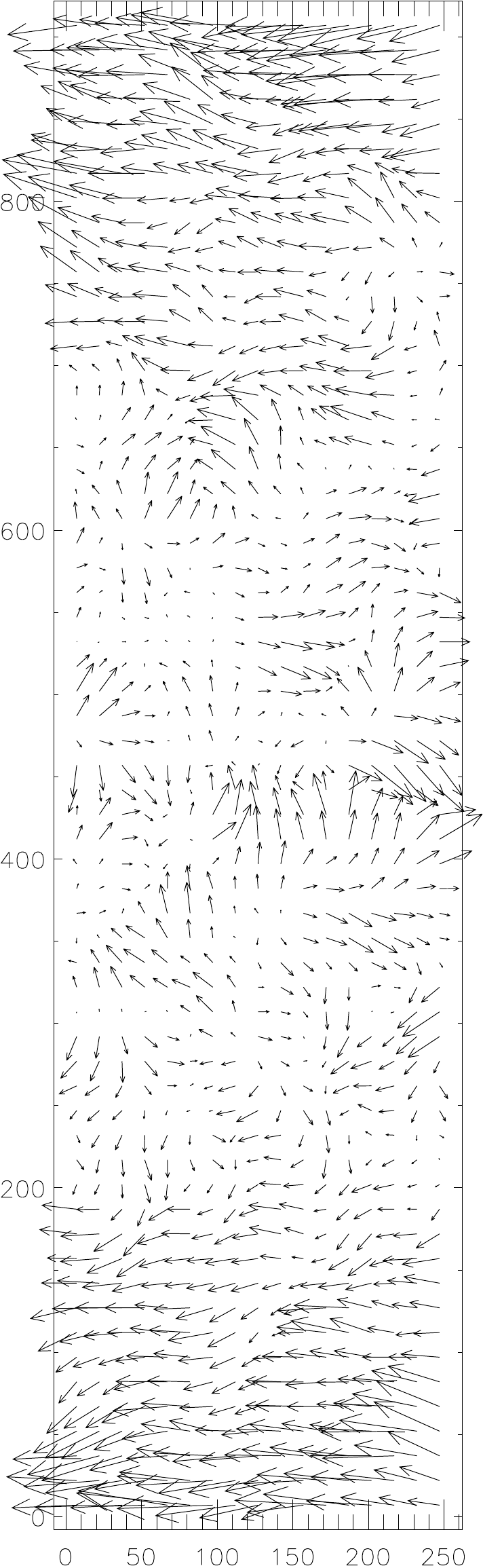}}
 \resizebox{!}{12cm}{\includegraphics{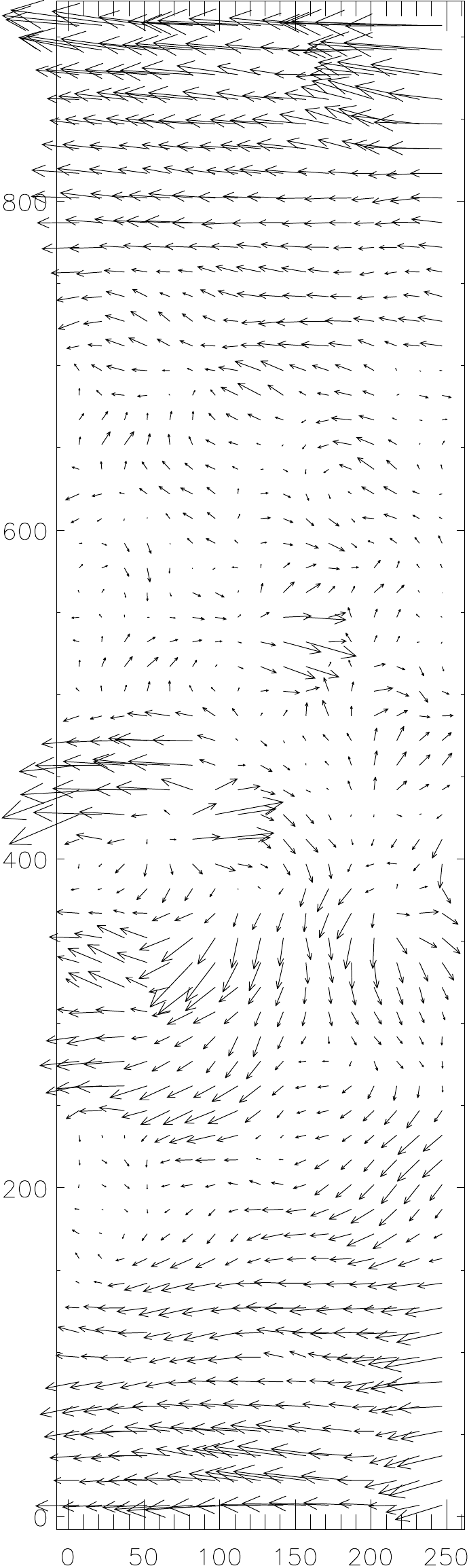}}
\end{center}
\caption{\emph{Left} --
Velocity field obtained by the LCT method applied to the data cube used in the helioseismic study. \emph{Right} -- Velocity field
obtained by the `standard' LCT method described in Section~\ref{sect:method}. One can see that velocity field obtained using the
same LCT code and different data processing may differ. The magnitude weighted cosine has a value of $\rho_W=0.71$, so both 
fields are still similar. There are two factor in the data processing that cause the difference: Velocity field obtained
from the helioseismic data cube are in the Postel projection and the velocity field using the `standard` method was calculated using the lag of 4~hours 
and averaged over 24~hours, while velocity field obtained from the helioseismic data cube was calculated using the lag of 2~hours and averaged over 8.5~hours.} 
\label{fig:cube_vs_classic}
\end{figure}

Flow velocity fields on the solar surface obtained by two different
techniques, time-distance helioseismology and local correlation
tracking (LCT), were compared. Despite the fact that the first
technique uses $p$-modes of solar oscillations to compute the
velocity field (in the data, the large-scale structures like
supergranulation were suppressed), while the other one uses
large-scale supergranulation pattern from averaged Doppler images as
tracers for the velocity vectors determination (and requires
$p$-modes removal), we found that both results match reasonably
well. We have confirmed some recent studies that the LCT method
slightly underestimates the actual velocities (as the consequence of
a smoothing procedure), and we determined empirical correction factors.
After the corrections, the match in global velocity structures, mean
zonal and meridional flows, is very good. The results of a detailed
comparison of the vector velocity fields are not so satisfactory.
However, the correlation coefficient for individual components of
the flows is positive and significant, so we conclude that a
meaningful match is found. It is shown that the largest disagreement
is caused by very small velocities in some regions, where the errors
of both methods become quite significant.

%% file: real.tex
\section{Application to real full-disc data}
\label{sect:real}

\subsection{Processed data}

In agreement with the aim of processing of the full-disc MDI dopplergrams we have processed all the suitable dopplergrams measured in the \emph{Dynamic campaigns} approximately two months each year since 1996. MDI provides in this way the most homogeneous material, which can be used for such studies. Similar data series are measured by the GONG network, but it is based on ground-based instruments and the measurements are strongly influenced by the turbulence in the Earth's atmosphere.

Among all, the MDI data suffer of some issues. The gaps in the data series are the most important problem. In some years, the gaps were present almost every day and had a length of several hours. They are mostly caused by the misoperation of the instrument. Other important issues are caused by the errors during the transmission of the data from SoHO to the central data storage. They usually depict as missing parts of the frame. Some frames also have a wrong orientation. 

Since the homogeneous non-interrupted series are needed for the current study, we have to deal with the described errors. The missing parts of the frames and the issue of the wrong orientation are non-correctable, therefore we delete all the frames with missing values and inconsistent telemetry in headers of the frames. The gaps are filled with linearly interpolated frames to run the reduction routines smoothly. 

\begin{figure}[!h]
\centering
\resizebox{0.40\textwidth}{!}{\includegraphics{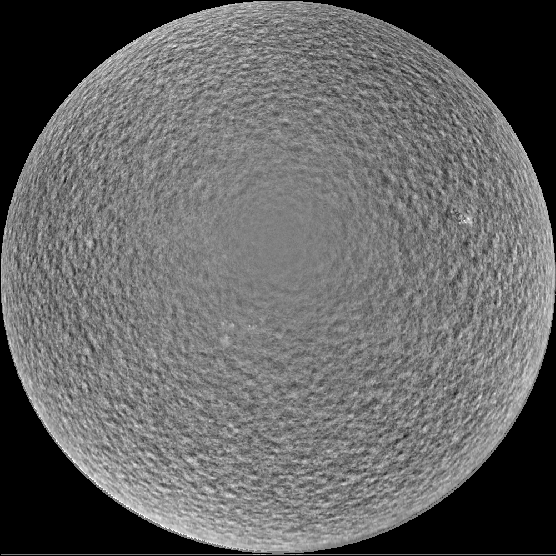}}
\resizebox{0.40\textwidth}{!}{\includegraphics{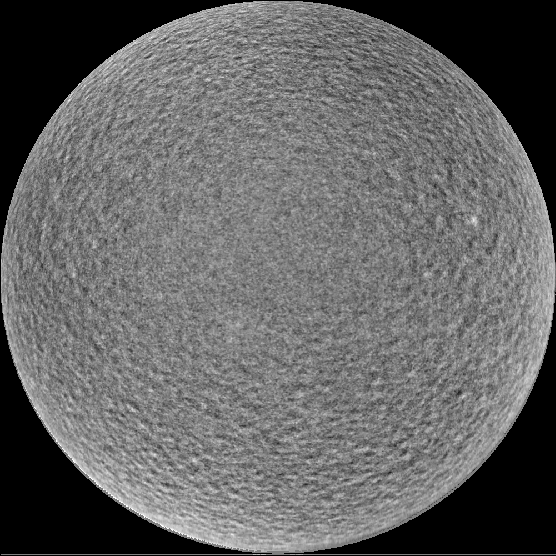}}
\caption{Influence of the linear interpolation of missing frames in the series undergoing the temporal average oscillations removal. \emph{Left} -- dopplergram not influenced by the linear interpolation, \emph{right} -- $p$-modes of solar oscillation are clearly visible in the averaged dopplergram, which was calculated from the data series containing a significant amount of linearly interpolated frames.}
\label{fig:dopplerinterp}
\end{figure}

The linear interpolation of the missing frames influence the data processing in a negative way. During the temporal averaging in order to suppress the $p$-modes of the solar oscillations, the series containing interpolated frames do not fulfill the assumptions of the processing. This in fact leads to the enhancement of the solar oscillations signature in dopplergrams -- see Fig.~\ref{fig:dopplerinterp}. 

\begin{figure}[!t]
\centering
\resizebox{0.49\textwidth}{!}{\includegraphics{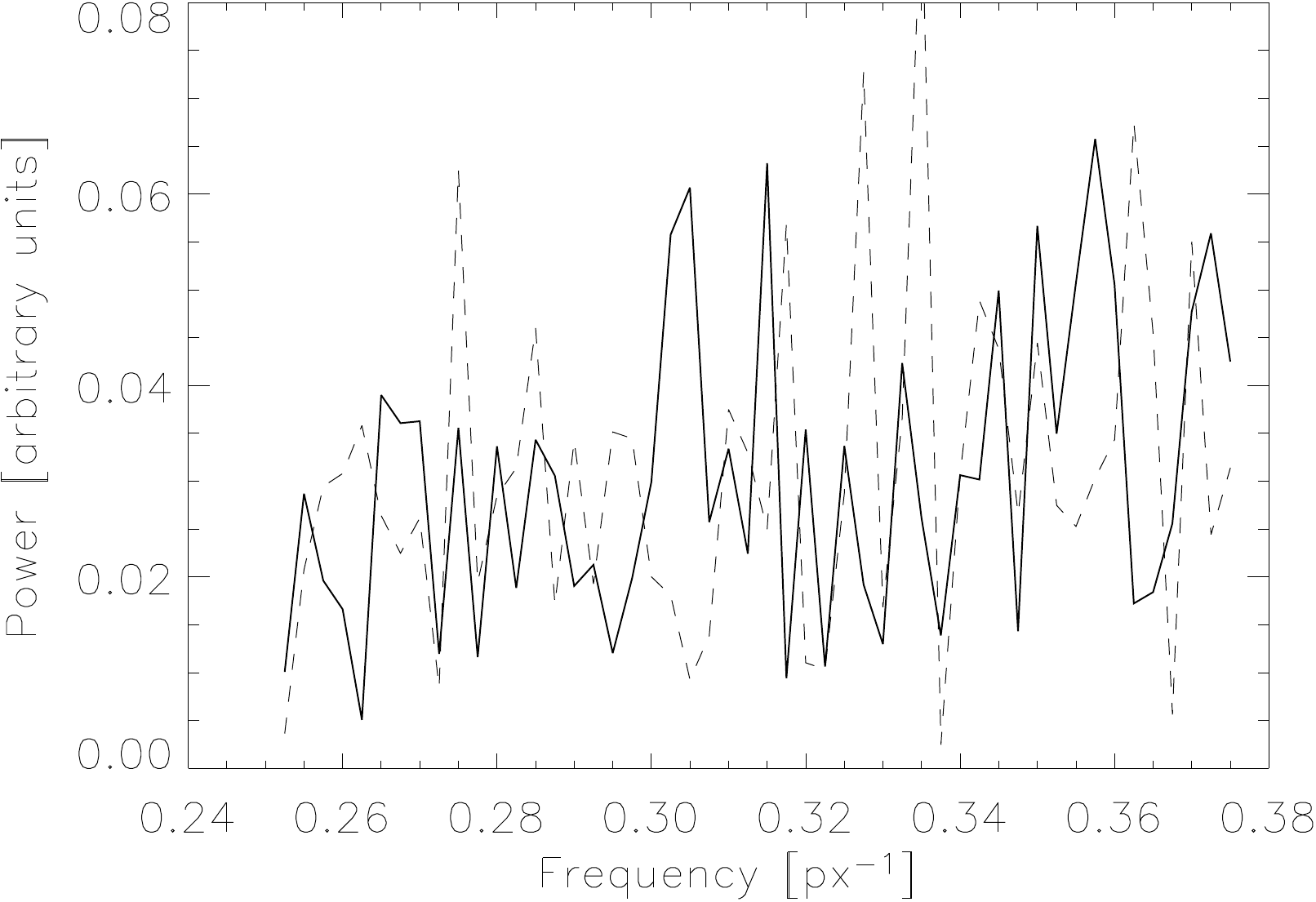}}
\caption{The part of the power spectra of the dopplergram central area. The excess of power in the band of 0.32--0.34~px$^{-1}$ in ``bad'' dopplergrams (dashed line) with respect to the ``good'' one (solid line) is clearly visible.}
\label{fig:powerspectra}
\end{figure}

Dopplergrams influenced by the linear interpolation are not used in the ongoing data processing routine, because the signal of $p$-modes confuses the LCT algorithm. The difference between a ``bad'' dopplergram and a ``good'' one is seen mainly in the centre of the disc, where in the ``good'' dopplergram the signal of supergranulation is weak, while in the ``bad'' one  the signal of $p$-modes is clearly visible on the centre of the disc. The ``bad'' dopplergram can be therefore easily detected on the basis of excess of power in the power spectra in the band of 0.32--0.34~px$^{-1}$, which corresponds to the size of 3~px (4500~km) -- see Fig.~\ref{fig:powerspectra}. 

The processed data belonging to the \emph{Dynamics campaign} are summarized in Tab.~\ref{tab:series}. Each day, two 24-hours series separated by 12 hours were processed. If the series contains some ``bad'' dopplergrams with a manifestation of the linear interpolation in the primary data, the whole series was not processed. We have tested that the interpolation in the 24-hours series strongly influences the behaviour of the $k$--$\omega$ filter and the LCT program.

\begin{table}[!t]
\begin{tabular}{lcc}
\bfseries From date -- To date & \bfseries No. of day in series & \bfseries No. of processed series\\
\hline
\rule{0pt}{18pt}23 May 1996 -- 24 Jul 1996 & 62 & 118\\
13 Apr 1997 -- 14 Jul 1997 & 92 & 181\\
9 Jan 1998 -- 10 Apr 1998 & 91 & 174 \\
13 Mar 1999 -- 28 May 1999 & 76 & 98\\
27 May 2000 -- 13 Jul 2000 & 47 & 65\\
28 Feb 2001 -- 28 May 2001 & 88 & 105\\
18 Mar 2002 -- 3 Jun 2002 & 77 & 83\\
8 Aug 2003 -- 23 Sep 2003 & 46 & 2 \\
18 Oct 2003 -- 24 Nov 2003 & 37 & 22 \\
4 Jul 2004 -- 6 Sep 2004 & 64 & 48\\
25 Jun 2005 -- 31 Aug 2005 & 67 & 86\\
24 Mar 2006 -- 22 May 2006 & 59 & 22 \\
\hline
\rule{0pt}{18pt}\it Summary & \it 806 & \it 1004 
\end{tabular}
\caption{The overview of the processed data. Note that on each suitable day, two 24-hours data series sampled by 12 hours were processed. }
\label{tab:series}
\end{table}

In summary, 2.82~TB of primary data were processed. The result of the whole processing consists of 1004 horizontal full-disc two-component-velocity maps in Sanson-Flamsteed projection with an effective spatial resolution of 60\arcsec{}. Each map has a size of 1520$\times$1520~px, which makes together 18.5~GB of disk space.

\subsection{Perspective removal}

\begin{figure}[!t]
\resizebox{\textwidth}{!}{\includegraphics{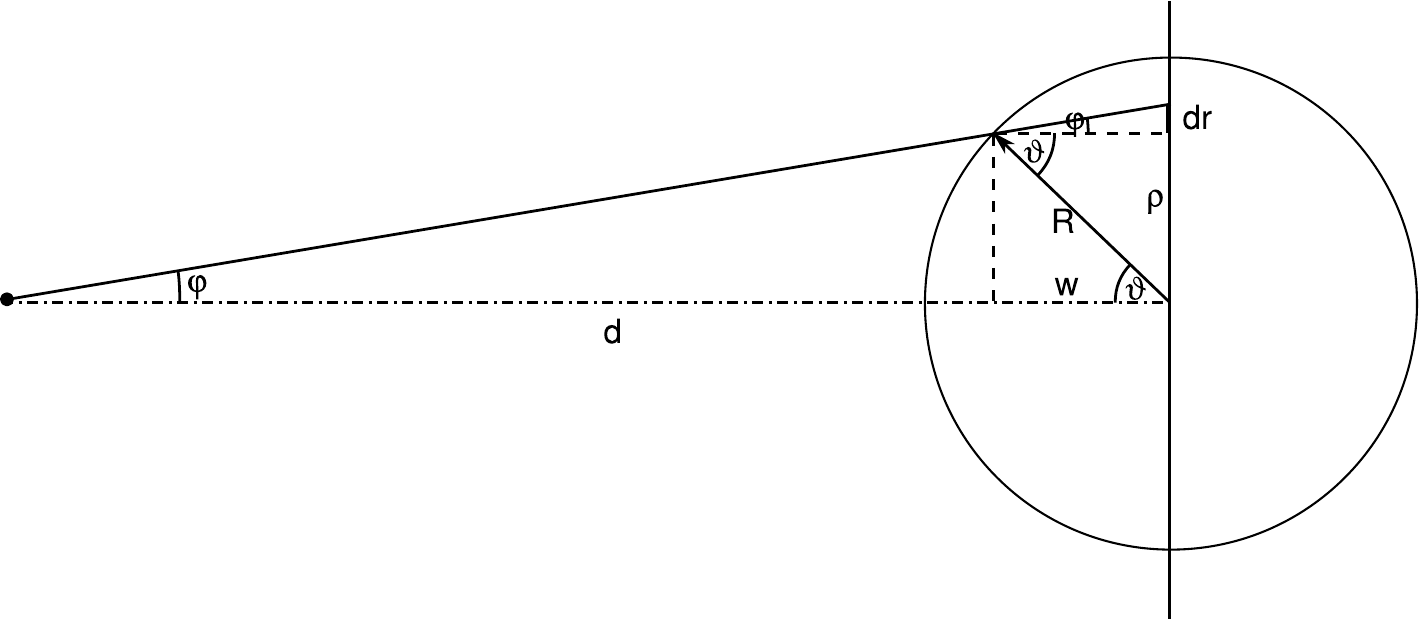}}
\caption{Perspective effect.}
\label{fig:perspective}
\end{figure}

The Sun is not a bright point and we observe it from the finite distance, therefore the positions are influenced by the perspective effect (Fig.~\ref{fig:perspective}). The point, which should be seen in the distance $\rho$ from the axis, is due to the perspective in projection to the disc seen in the distance $\rho+dr$. Therefore the positions in the projection to the disc should be corrected by $dr$ to get the correct positions in the heliographic coordinate system. The correction may be easily calculated from the geometry displayed in Fig.~\ref{fig:perspective}.

Two equations have to be solved:
\begin{equation}
\frac{dr+\rho}{d}=\tan\varphi=\frac{dr}{w} ,
\end{equation}
and
\begin{equation}
R^2=\rho^2+w^2 .
\end{equation}
As the result we get
\begin{equation}
dr=\frac{\rho\sqrt{R^2-\rho^2}}{d-\sqrt{R^2-\rho^2}}.
\end{equation}
We assume that $w$ is always positive, because only one hemisphere of the Sun is relevant for the observations from the Earth or L$_1$ point, where SoHO is located.

All the images have to be corrected to this effect, otherwise the displacement due to the perspective effect may reach the value of 1~px, which corresponds to 0.12\,$^\circ$ in heliographic coordinates. 

\subsection{Oscillations removal}

From all frames, the line-of-sight component of the Carrington rotation is subtracted and the effect of a perspective is corrected. The frames are transformed so that the heliographic latitude of the disc centre $b_0=0$ and the position angle of the solar rotation axis $P=0$. In the dopplergrams, the signal of $p$-modes of the solar oscillations is very well visible. The pattern of oscillations causes a lot of troubles during the application of the local correlation tracking method. To exclude this, the $p$-modes are suppressed using weighted average \citep[see][]{1988SoPh..117....1H}. The weights have a Gaussian form given by the formula (\ref{eq:average}). We sample averaged images in the interval of 15 minutes. The filter suppresses more than five hundred times the solar oscillations in the 2--4 mHz frequency band.

The frames with suppressed signal of solar oscillations are processed in the way described in detail in Section~\ref{sect:method}. Here I bring just a brief summary.

The processing of averaged frames consists of two main steps. In the first main step the mean zonal velocities are calculated and, on the basis of expansion to the Fay's formula $\omega=c_0+c_1 \sin^2 b + c_2\sin^4 b$, the differential rotation is removed. In the second main  step, the LCT algorithm with an enhanced sensitivity is applied. Finally, the differential rotation (obtained in the first step) is added to the vector velocity field obtained in the second main step. Both main steps can be divided into several sub-steps, which are mostly common.

\begin{enumerate}
\item The data series containing 96 averaged frames is ``derotated'' using the Carrington rotation rate in the first step and using the calculated differential rotation in the second step.

\item Derotated data are transformed into the Sanson-Flamsteed coordinate system to remove the geometrical distortion caused by the projection to the disc. The Sanson-Flamsteed (also known as sinusoidal) pseudo-cylindrical projection conserves the areas and therefore is suitable for the preparation
of the data used by LCT.

\item Remapped data undergo the $k$-$\omega$ filtering \citep[e.~g.][]{1989ApJ...336..475T} with the cut-off velocity 1\,500~\mps{} for suppression of the noise coming from the evolutionary changes of supergranules, of the numerical noise, and for the partial removal of the ``blind spot'' (an effect at the centre of the disc, where the supergranular structures are almost invisible in dopplergrams due to the prevailing horizontality of their internal velocity field).

\item Finally,  the LCT is applied: the lag between correlated frames is 4~hours, the correlation window has \emph{FWHM} 60\arcsec, the measure of correlation is the sum of absolute differences and the nine-point method for calculation of the subpixel value of displacement is used. The calculated velocity field is averaged over the period of one day.

\item The resulting velocity field is corrected using the formula (\ref{eq:calibration}) in Section~\ref{sect:synthetic_results}. The directions of the vectors before and after the correction are without change. Finally, $v_x$ component is corrected for the data-processing bias of $-15$~\mps{} determined in Section~\ref{sect:synthetic_results}.
\end{enumerate} 

\subsection{Visualization}

The processed data may be visualised in many ways. Some of them that were used during the analysis of the data are shown in this subsection. As an example the 24-hours averaged horizontal large-scale flow field centered at 00.15~UT of March 6th 2001 is taken. 

\begin{figure}[!h]
\resizebox{0.5\textwidth}{!}{\includegraphics{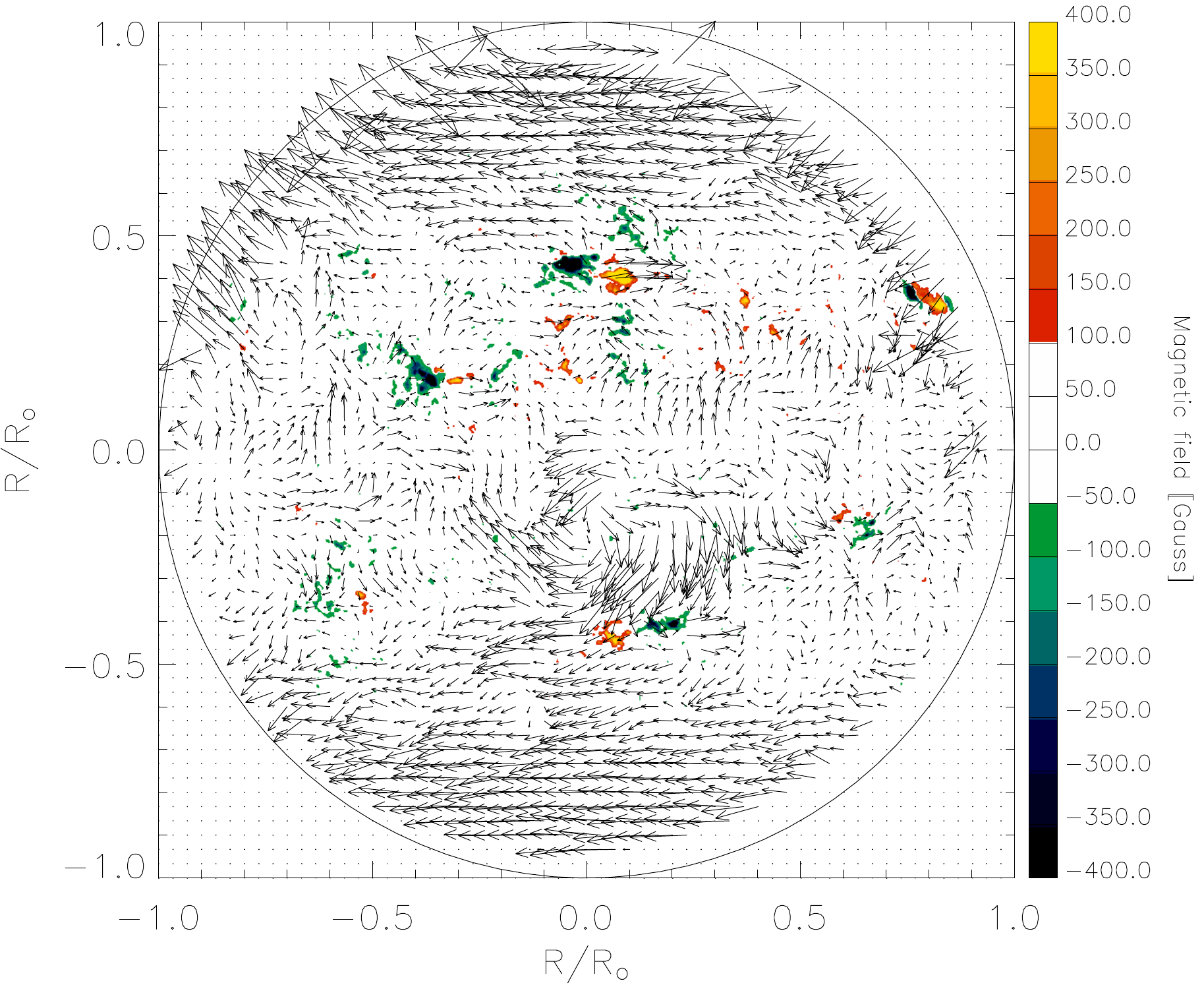}}
\resizebox{0.5\textwidth}{!}{\includegraphics{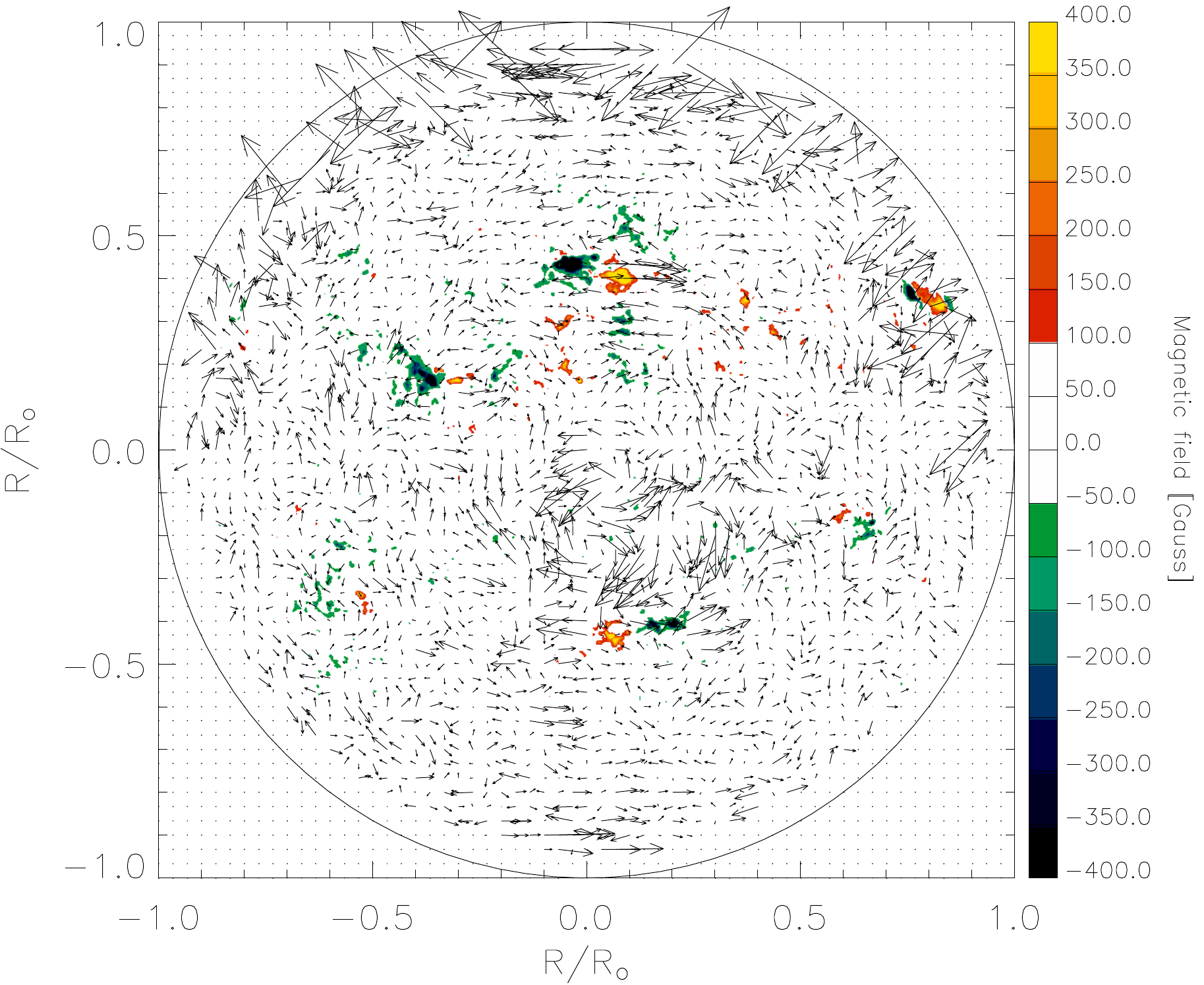}}
\caption{\emph{Left} -- Arrow-plot of the surface horizontal large-scale flow field calculated for 6th March 2001. \emph{Right} -- ``Surface weather'' at the same instant. The MDI magnetogram is superimposed.}
\label{fig:examples_arrows}
\end{figure}

\begin{figure}[!h]
\resizebox{0.5\textwidth}{!}{\includegraphics{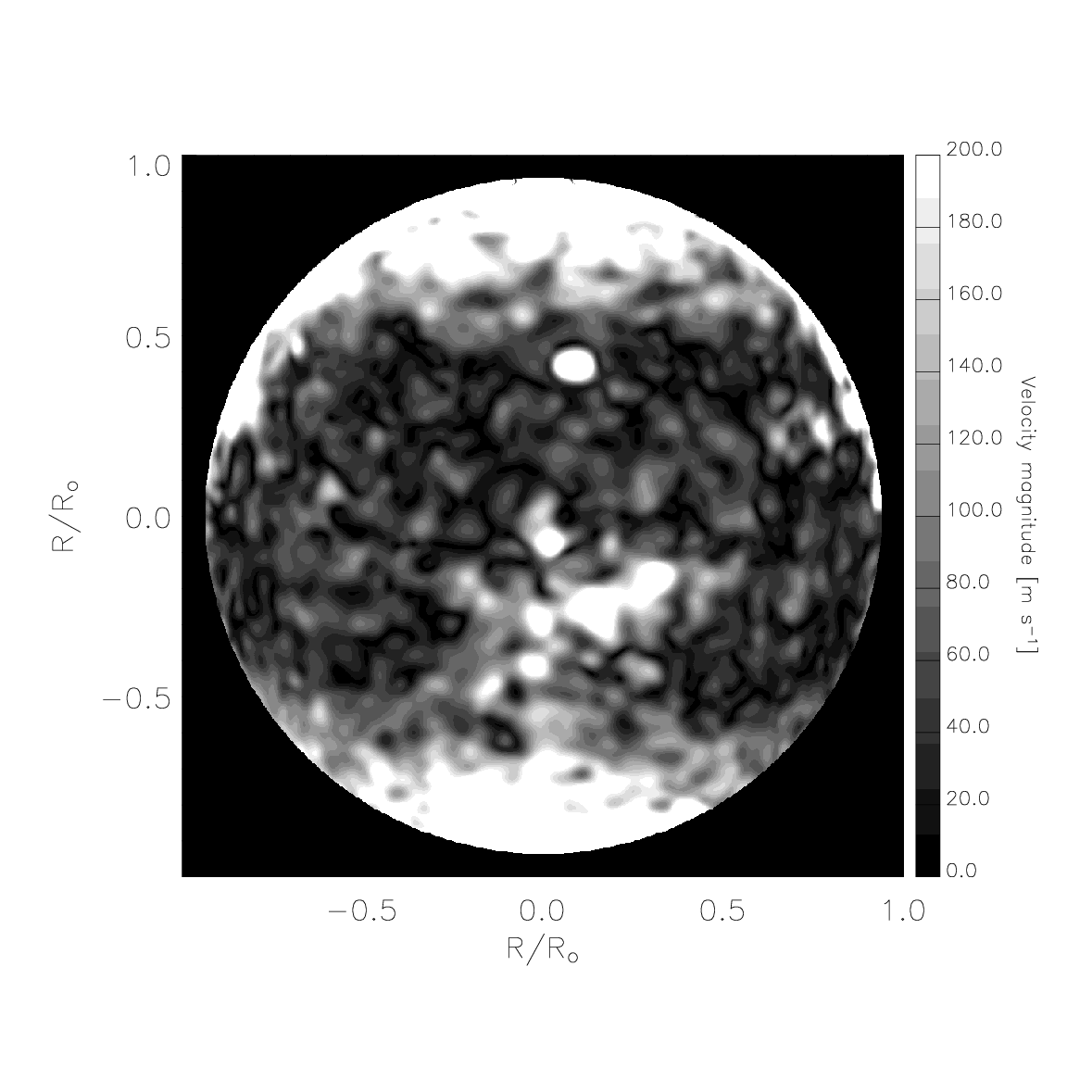}}
\caption{Magnitude of the flow field. Note the increased speed in the northern (top) hemisphere near the central meridian. This area corresponds to the preceding polarity of the complex sunspot group. It shows that the preceding polarity of young sunspot group is moving faster than the following one and stretches the bipolar magnetic field. }
\label{fig:examples_magnitude}
\end{figure}

The commonly used type of visualization is the vector arrow-plot (Fig.~\ref{fig:examples_arrows}). This type of visualization is not useful for the study of the details at the full disc, but provides an important overview of the behaviour of the detected flows. The same field with mean flows (differential rotation, meridional circulation) removed is in local helioseismology usually called as the ``sub-surface weather'' (SSW). Since motions of the supergranules are measured on the surface of the Sun, it is probably possible to call it a ``surface weather''. It provides an overview about the motions with respect to the very-large-scale background. An additional information comes from the map of absolute magnitudes of the velocities (Fig.~\ref{fig:examples_magnitude}).

The largest flows consist of differential rotation and meridional circulation. The differential rotation is calculated as the integral of the zonal component over the longitudes. To get a smooth curve, a parabolic fit in even powers of sinus of latitude -- Eq.~(\ref{eq:difrot}) -- is usually calculated. This fit is performed mostly from the historical reason. As it was already mentioned in Section~\ref{sect:introduction}, this type of fitting has a lot of disadvantages. For example it does not introduce the asymmetries that can be present between the two hemispheres. 

The meridional circulation is calculated as an integral of the meridional component of the velocity field over longitudes. It brings some information about the transfer of the angular momentum and the magnetic flux towards the solar poles. It is considered mostly poleward with one cell per hemisphere.

\begin{figure}[!h]
\resizebox{0.5\textwidth}{!}{\includegraphics{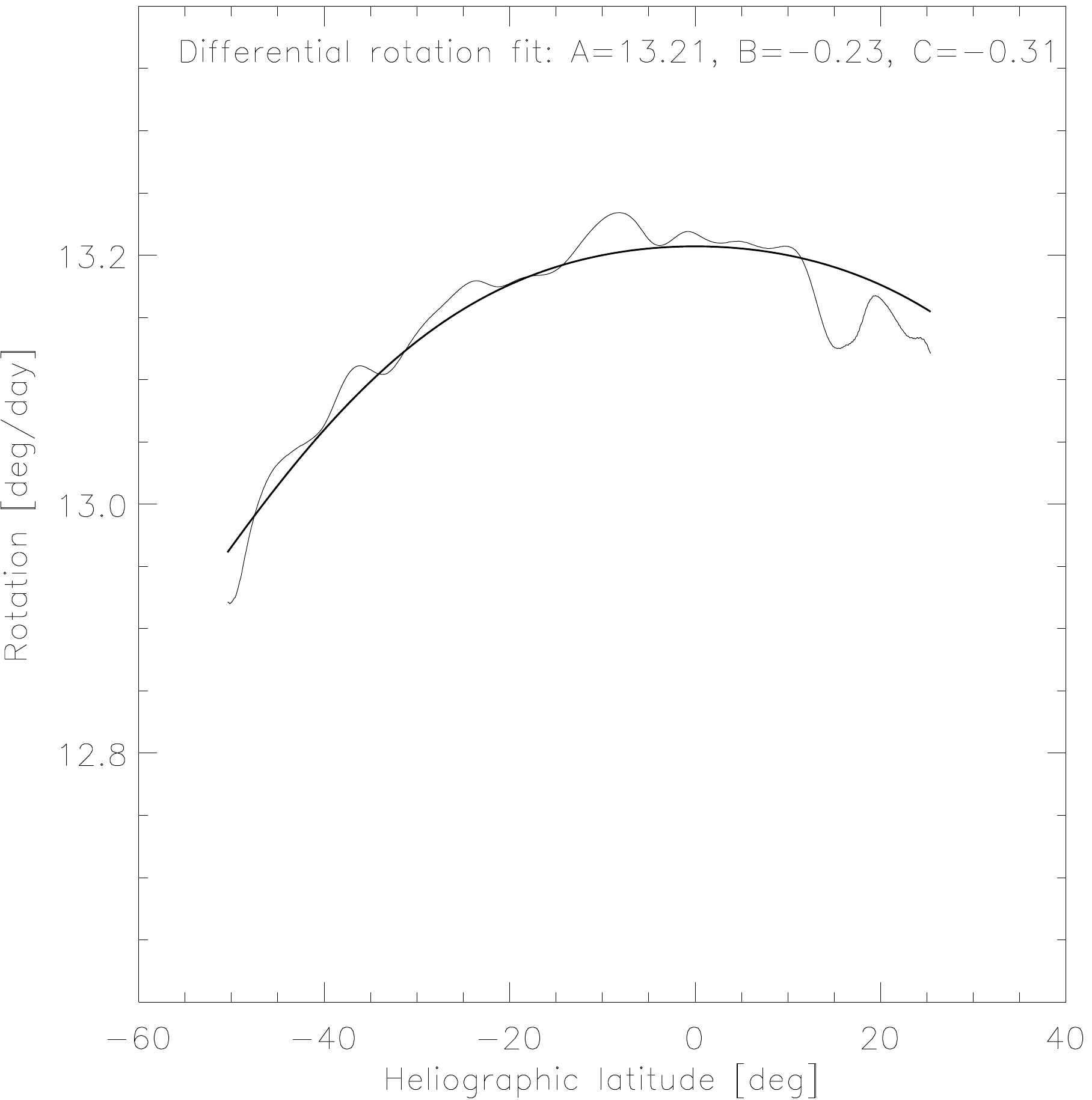}}
\resizebox{0.5\textwidth}{!}{\includegraphics{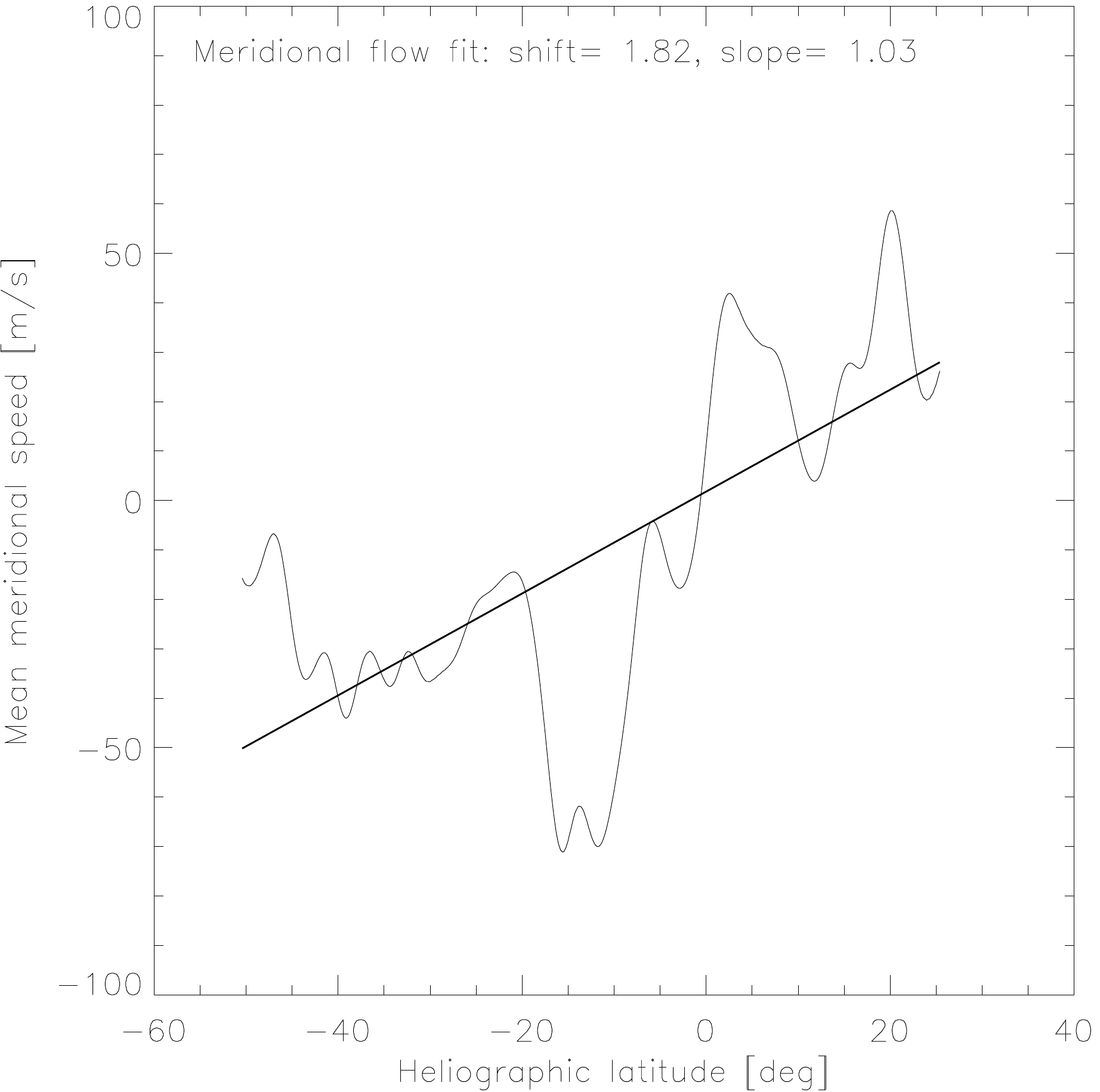}}
\caption{Mean componets of the surface flow field. \emph{Left} -- differential rotation and its parabolic fit, \emph{right} -- meridional circulation and its linear fit. }
\label{fig:examples_means}
\end{figure}

The mean flows calculated from the example result are displayed in Fig.~\ref{fig:examples_means}.

\begin{figure}[!h]
\resizebox{0.5\textwidth}{!}{\includegraphics{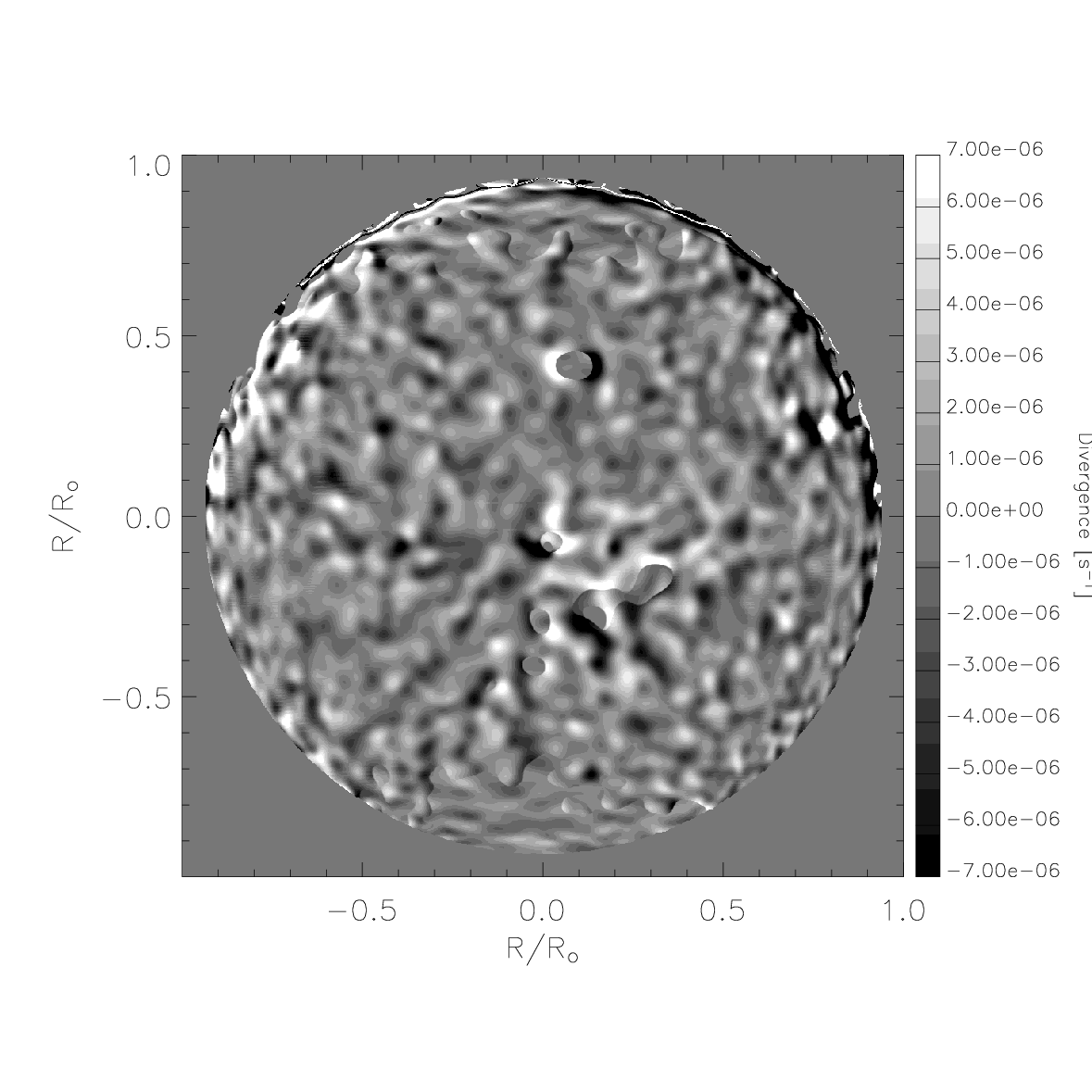}}
\resizebox{0.5\textwidth}{!}{\includegraphics{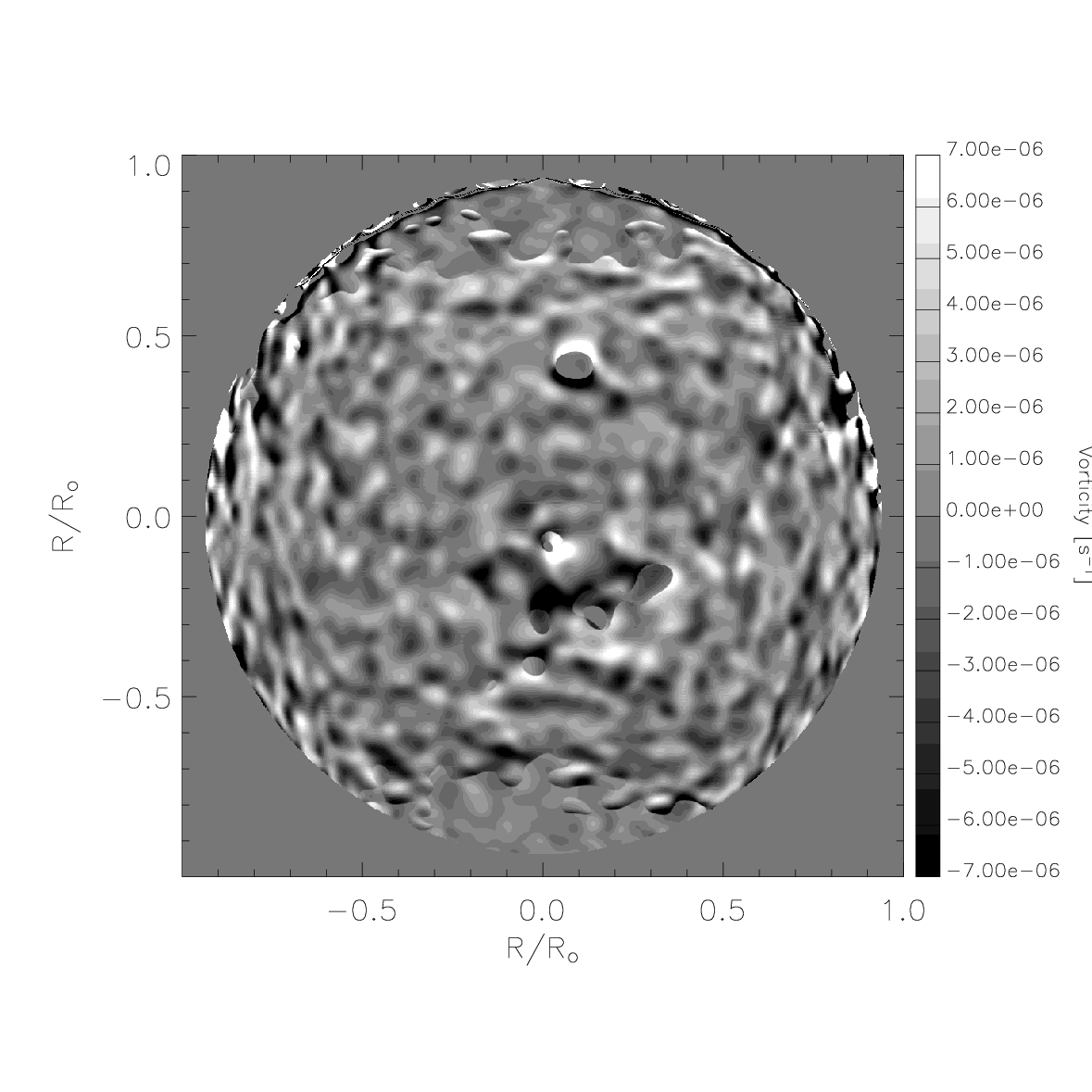}}
\caption{\emph{Left} -- horizontal divergence of the flow field, \emph{right} -- vorticity of the flow field.}
\label{fig:examples_divvort}
\end{figure}

The LCT technique basically provides only two components of the generally three-dimensional vector field. Nonetheless, it is possible to recover some information about the third (vertical) components also from the two-dimensional data. For this, the horizontal divergence $D_h$ (Fig.~\ref{fig:examples_divvort} left) is calculated.
\begin{equation}
D_h=-\left( \frac{\partial v_x}{\partial x}+\frac{\partial v_y}{\partial y} \right).
\end{equation}

Assuming the continuity equation in the form of $\nabla \cdot {\mathbf v} = 0$ we obtain:
\begin{equation}
\frac{\partial v_z}{\partial z} = D_h.
\end{equation}

Therefore in the regions, where $D_h$ is negative, we may expect a sink -- i.~e. a \emph{downflow} -- and vice versa.

An information about the twist comes from \emph{vorticity} (Fig.~\ref{fig:examples_divvort} right). Vorticity $V_z$ is defined as
\begin{equation}
V_z=\left( \nabla\times {\mathbf v} \right)_z.
\end{equation}
It denotes the twist in the flow field. If $V_z$ is negative, then the clockwise motion may be expected. Vortical structures are often present in areas of upflows and downflows. They are a consequence of the influence of the Coriollis force. 

\begin{figure}[!b]
\resizebox{0.5\textwidth}{!}{\includegraphics{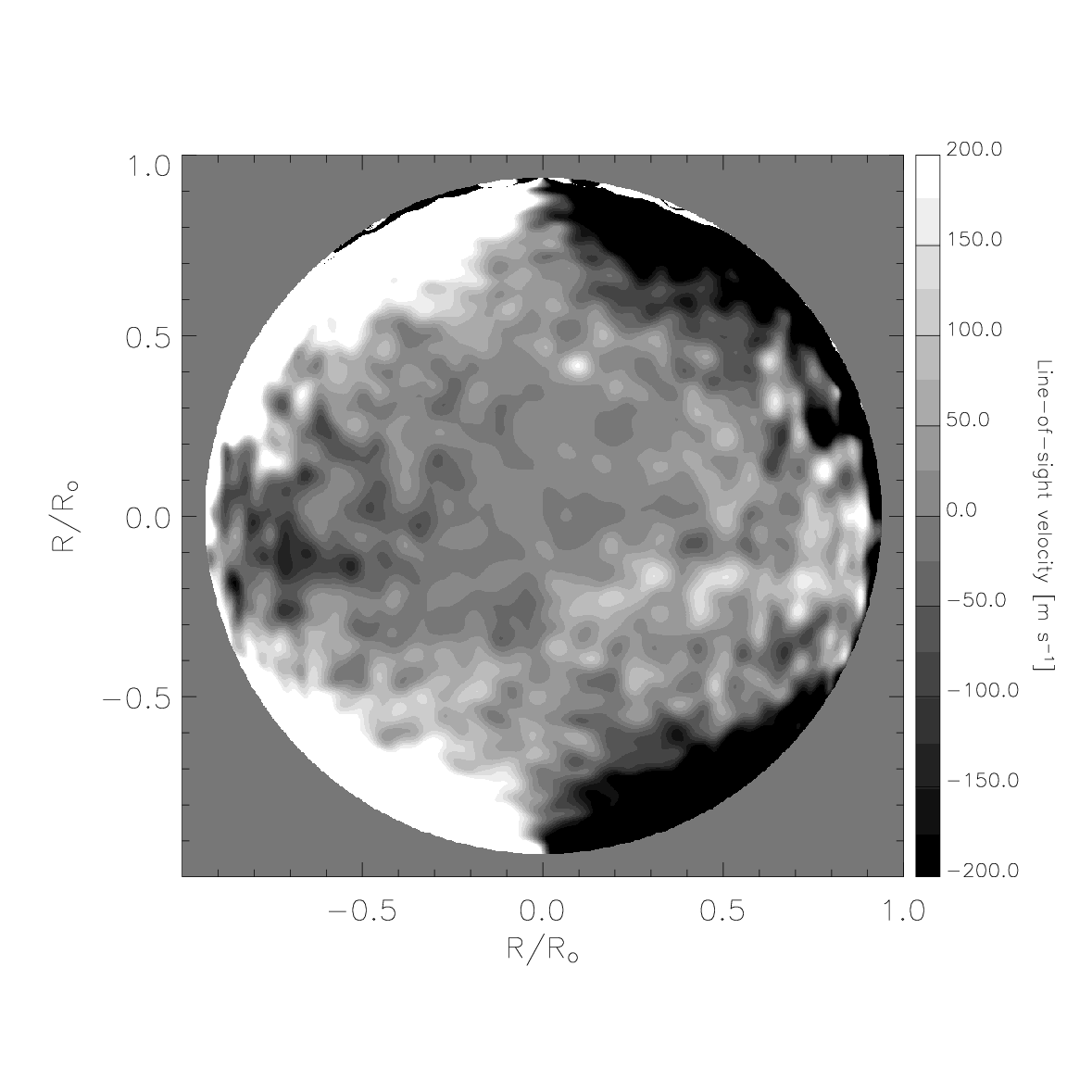}}
\resizebox{0.5\textwidth}{!}{\includegraphics{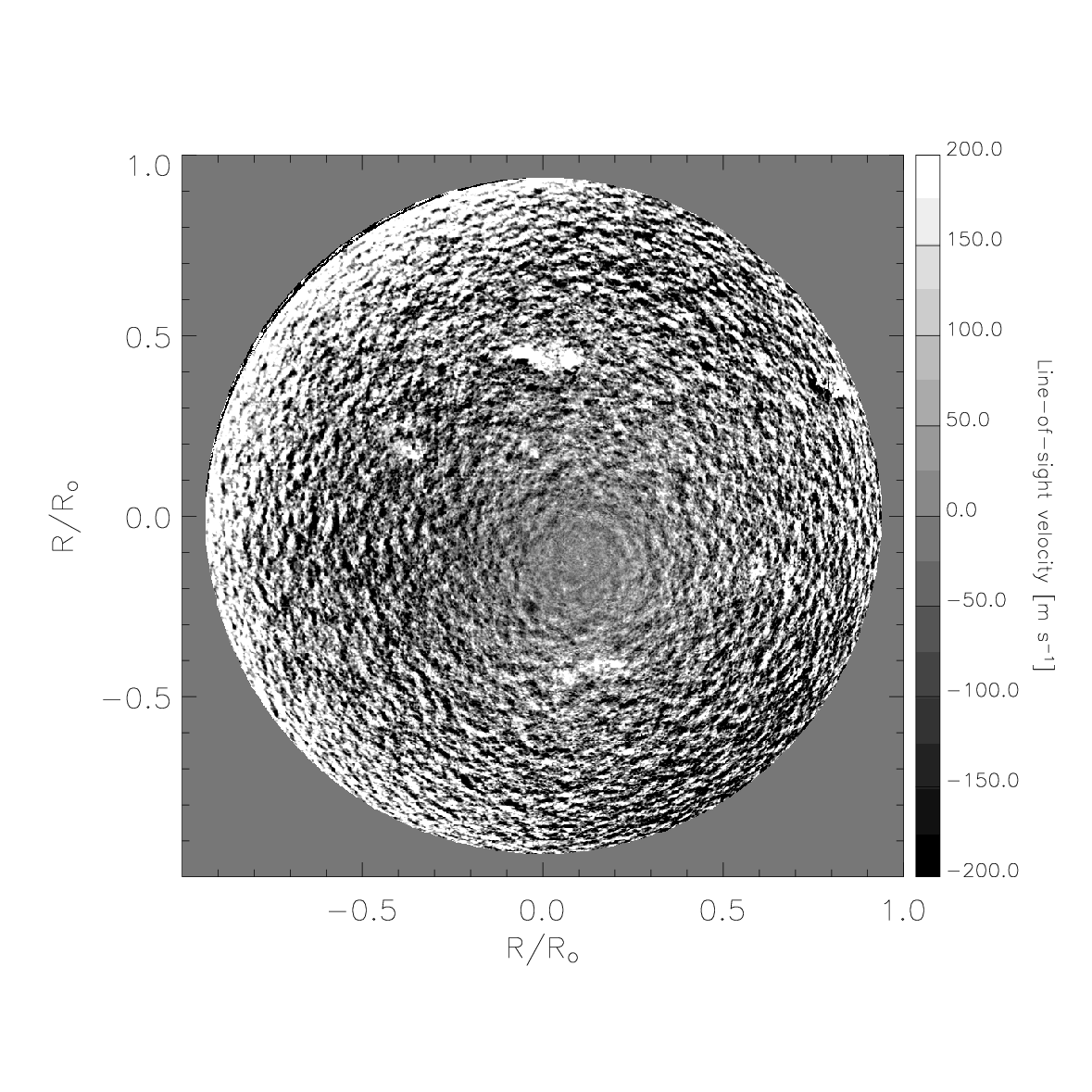}}
\caption{Maps of line-of-sight components. \emph{Left} -- artificial dopplergram constructed from the horizontal flow field with rigid rotation removed, \emph{right} -- original dopplergram with rigid rotation removed.}
\label{fig:examples_dopplergrams}
\end{figure}

The LCT technique was applied to the full-disc dopplergrams. From the resulting horizontal flow field (and neglecting the vertical component) it is possible to calculate the artificial dopplegram. 
\begin{equation}
v_d=v_h\sin\rho ,
\end{equation}
where $v_d$ is the line-of-sight velocity, $v_h$ is the horizontal velocity, and $\rho$ the heliocentric angle.

Comparing the artificial dopplergram to the measured one (Fig.~\ref{fig:examples_dopplergrams}) provides an information about the difference of the plasma speed (measured by the line-of-sight velocity in measured dopplergrams) with the speed derived from the motion of the supergranules. The difference (white and black areas near the disc edges in Fig.~\ref{fig:examples_dopplergrams}) may be caused by the wave-like motions or by the bias in the supergranules tracking procedure. This effect is known as the super-rotation of supergranules and was mentioned by several papers, first by \cite{1980SoPh...66..213D}, lately confirmed by e.~g.~\cite{1990ApJ...351..309S}, or recently by \cite{2000SoPh..193..333B}. The measured rotation of supergranules seems to be higher than the rotation speed of the plasma in them. Although this super-rotation was recently explained as the effect of projection in the dopplergrams \citep{2006ApJ...644..598H}, some uncertainties about this phenomenon still remain \citep[e.~g.][]{2007AA...466..691M}. When analysing the velocity powerspectra, some evidence about the traveling waves was found by \cite{2003Natur.421...43G}.

%% file: long.tex
\section{Long-term behaviour}
\label{sect:long}
\symbolfootnotetext[0]{\hspace*{-7mm} $\star$ This chapter was submitted to Astronomy \& Astrophysics as  {\v S}vanda,~M., Klva{\v n}a,~M., Sobotka,~M., and Bumba,~V., \emph{Large-scale horizontal flows in the solar photosphere. II. Long-term behaviour and magnetic activity response}.}

The subject of this chapter is a verification of the performance of the method described in Section~\ref{sect:method} on the real data and the investigation of long-term properties of the flows at largest scales obtained with this method. We shall also discuss the influence of magnetic fields on the measured zonal flow in the equatorial region. 

The data are processed as described in Section~\ref{sect:real}. In this chapter, the interest was focused on the properties of the mean zonal and meridional components. Therefore, from each two-component horizontal velocity field the mean zonal and meridional components depending only on heliographic latitude were calculated as the longitudinal average of the flow map, using 135 longitudinal degrees around the central meridian. As stated in Section~\ref{sect:method}, the  accuracy for each velocity vector is 15~\mps{} for velocities under 100~\mps{} and 25~\mps{} for velocities above 100~\mps. These inaccuracies have a character of a random error, therefore for the mean zonal and meridional components the accuracy is in the worst case 1~\mps.

The calculated surface flows may be biased by the projection effects, although the tests on the synthetic data did not show any signs of them. \cite{2006ApJ...644..598H} showed that the apparent superrotation of the structures tracked in dopplergrams reported by many studies can be explained as the projection effect. However, this bias would produce the systematic error, which should influence neither the periodic analysis, nor the relative motions of the active regions with respect to their surroundings.

\subsection{Results}

\subsubsection{Long-term properties}
\begin{figure}[!]
\centering
\resizebox{\textwidth}{!}{\includegraphics{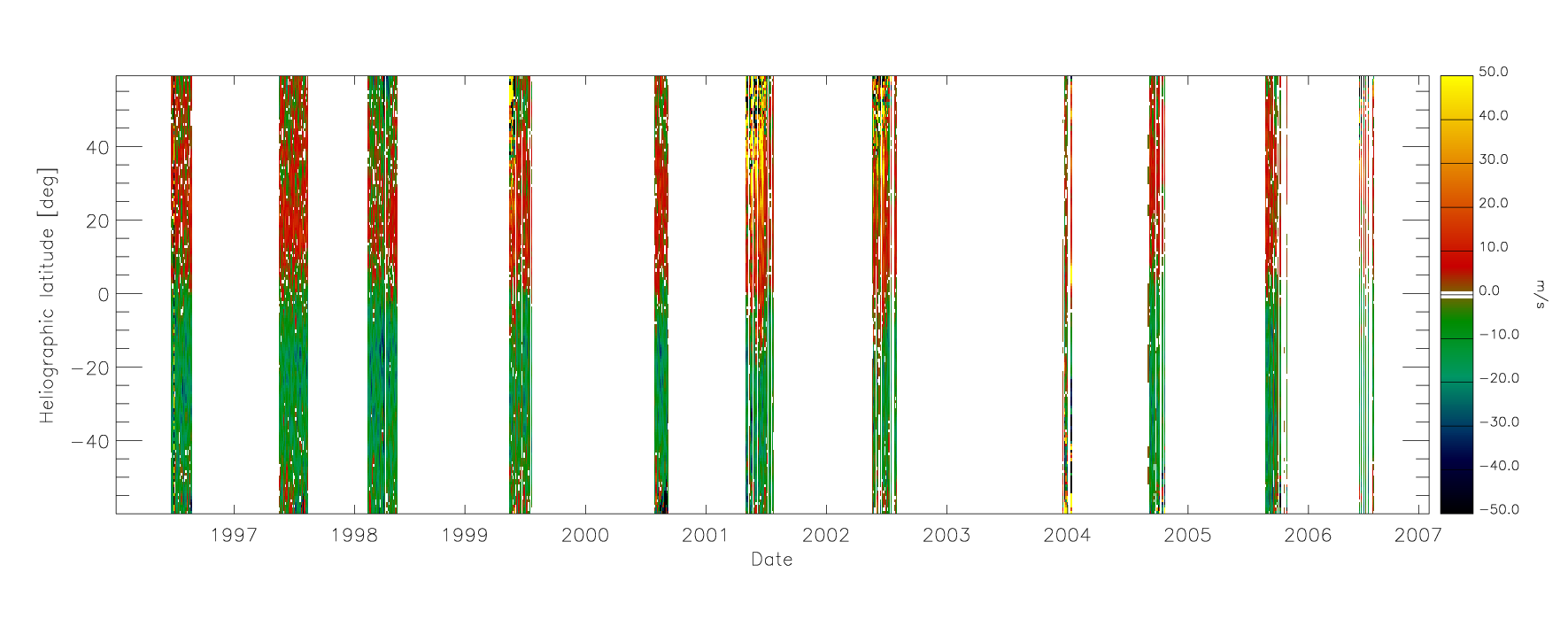}}
\caption{Mean meridional flow in time and heliographic latitude. It can be clearly seen that for almost all the processed measurements a simple model of one meridional cell per hemisphere would be sufficient. However, some local corruptions of this simple idea can be noticed on both hemispheres.}
\label{fig:meridional}
\end{figure}

For the study of long-term evolution of surface flows maps, the mean zonal and meridional components were calculated. The maps of mean meridional component in time and heliographic latitude are shown in Fig.~\ref{fig:meridional}. It can be clearly seen that on the northern hemisphere the flow towards the northern pole dominates while on the southern hemisphere the flow towards southern pole prevails. The ``zero line'', the boundary between the flow polarities, is not located exactly on the solar equator and seems to be shifted to the south in the period of increased solar activity (2001 and 2002). In agreement with \cite{1997AA...319..683M} and \cite{1998SoPh..183..263C} the meridional flow is found stronger in the periods of increased solar activity by about $~10$~\mps{} than in periods with lower magnetic activity.

\begin{figure}[!]
\centering
\resizebox{\textwidth}{!}{\includegraphics{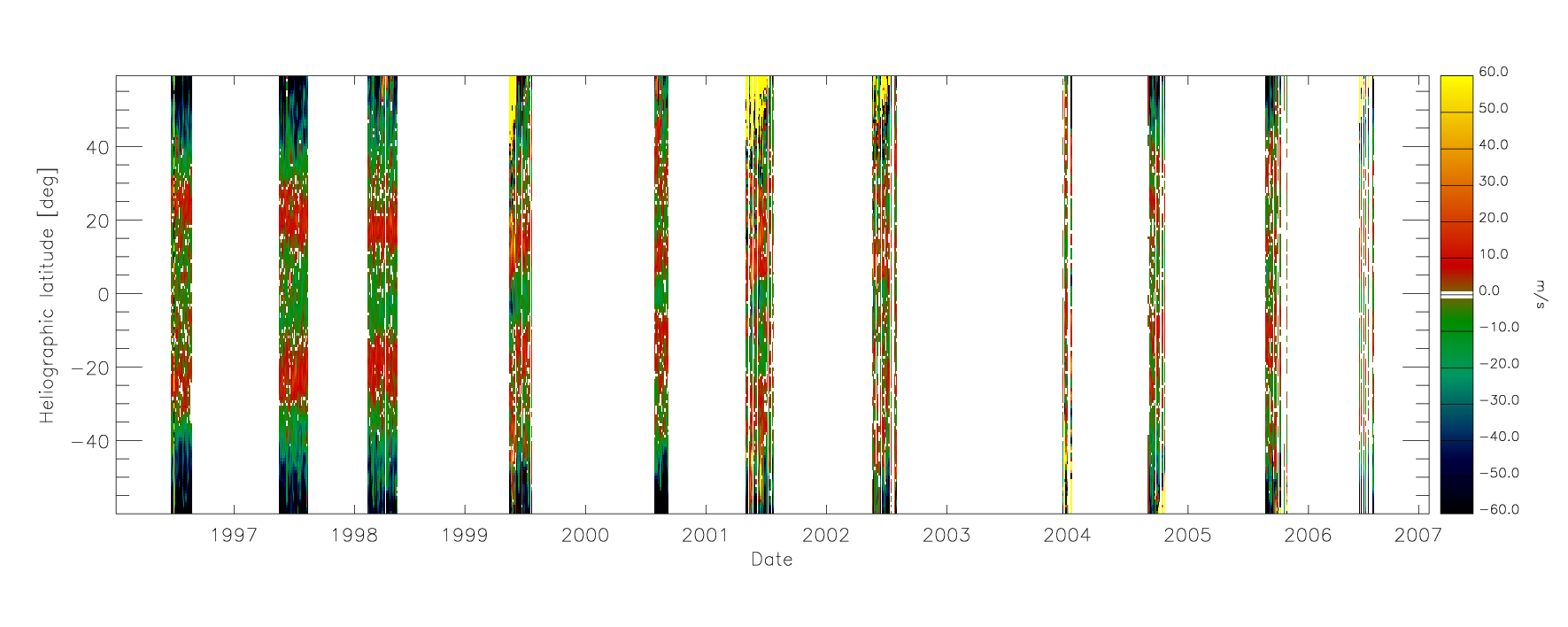}}
\caption{Torsional oscillations. The residua of the mean zonal flow with respect to its parabolic fit displayed in time and heliographic latitude. In the period of weak magnetic activity the pattern of belts propagating towards the equator is very clear. In the periods of stronger magnetic activity the flow field is influenced by local motions in active regions and therefore the pattern of torsional oscillations is not clearly seen.}
\label{fig:torsional}
\end{figure}

A similar map was made for the zonal component. The mean equatorial zonal velocity for all the data is 1900~\mps. For all the processed data the dependence on latitude is close to a parabolic shape, parameters of which change slowly in time. The residua of the zonal velocity with respect to its parabolic fit given by
\begin{equation}
v_b=a_0+a_1 b + a_2 b^2,
\end{equation}
where $b$ is the heliographic latitude and $v_b$ the mean zonal velocity in the given latitude, were calculated in order to see if the torsional oscillations can be detected in the measurements. As it is displayed in Fig.~\ref{fig:torsional}, the method clearly reveals torsional oscillations as an excess of the mean zonal velocity with respect to the zonal velocity in the neighbourhood. The behaviour of torsional oscillations is in agreement with their usual description -- the excess in magnitude is of the order of 10~\mps, they start at the beginning of the solar cycle in high latitudes and propagate towards the equator with the progression of the 11-year cycle. However, due to the used method, the visibility of torsional oscillations decrease with increasing solar activity. In the periods of strong activity both belts are not so clearly visible since the large-scale velocity field and its parabolic fit are strongly influenced by the presence of magnetic regions. However, the torsional oscillations belts still remain visible when the mean zonal component is symmetrised with respect to the solar equator. The meridional flow or torsional oscillations depending on time and latitude are not in the focus of this study, they are just used to check the ability and performance of our method.

\subsubsection{Periods in the mean components}
The mean zonal and meridional components in the equatorial area (averaged in the belt $b=-5\,^\circ - +5\,^\circ$) were analysed in order to examine the periods contained in the data. Since the data are not equidistant at all, a simple harmonic analysis cannot be used, so that the \emph{Stellingwerf method} \citep{1978ApJ...224..953S} was applied. It works on the principle of the phase dispersion minimization. The method sorts the data for every searched period into the phase diagram. Then the phase diagram is divided in a few (mostly ten) parts and for every part the mean dispersion is calculated. If the studied period is reasonable, the data-points group along the periodic curve and the dispersion in each part of the phase diagram is smaller than the dispersion of the whole data series. The normalized parameter $\theta \in \left(0,1\right>$ describes the quality of a given period. 

\begin{figure}[!t]
\centering
\resizebox{0.49\textwidth}{!}{\includegraphics{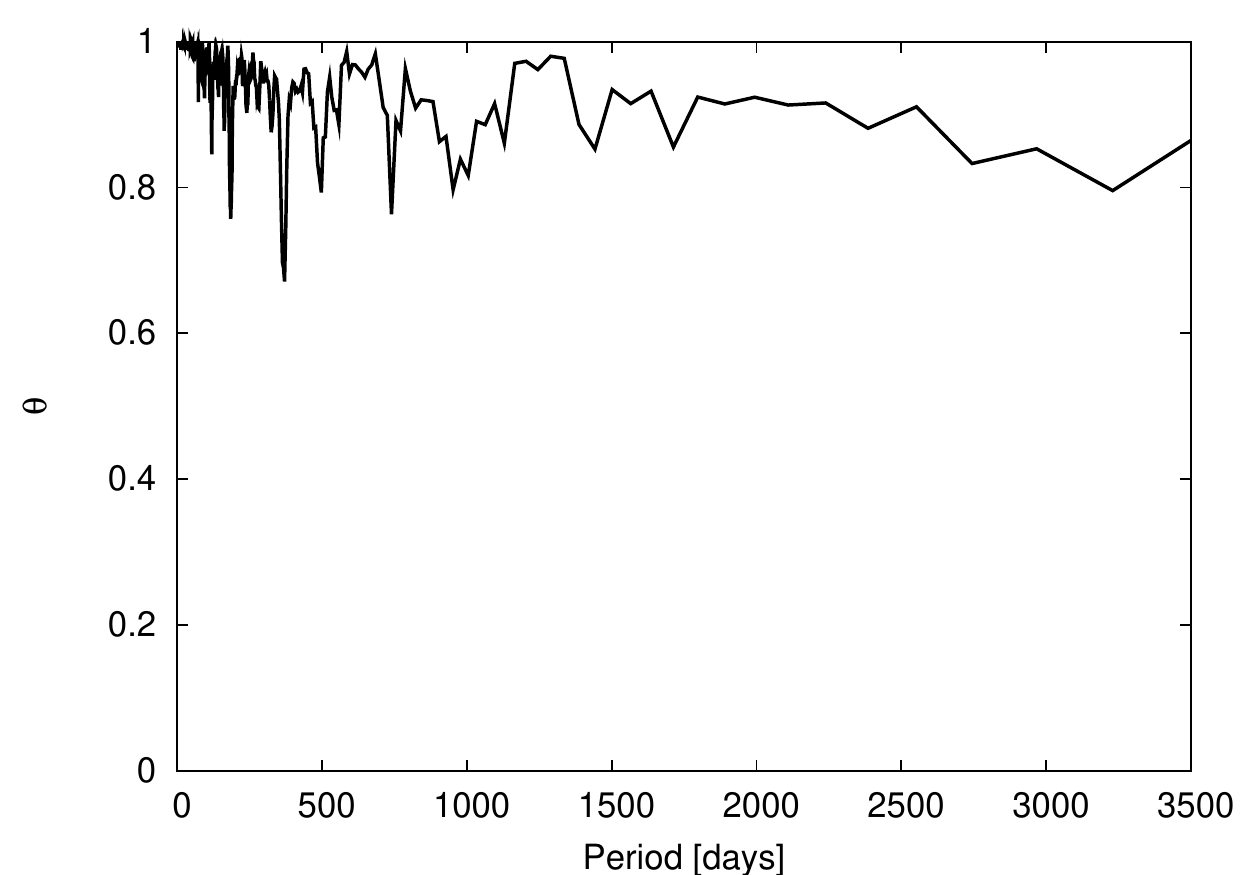}}
\resizebox{0.49\textwidth}{!}{\includegraphics{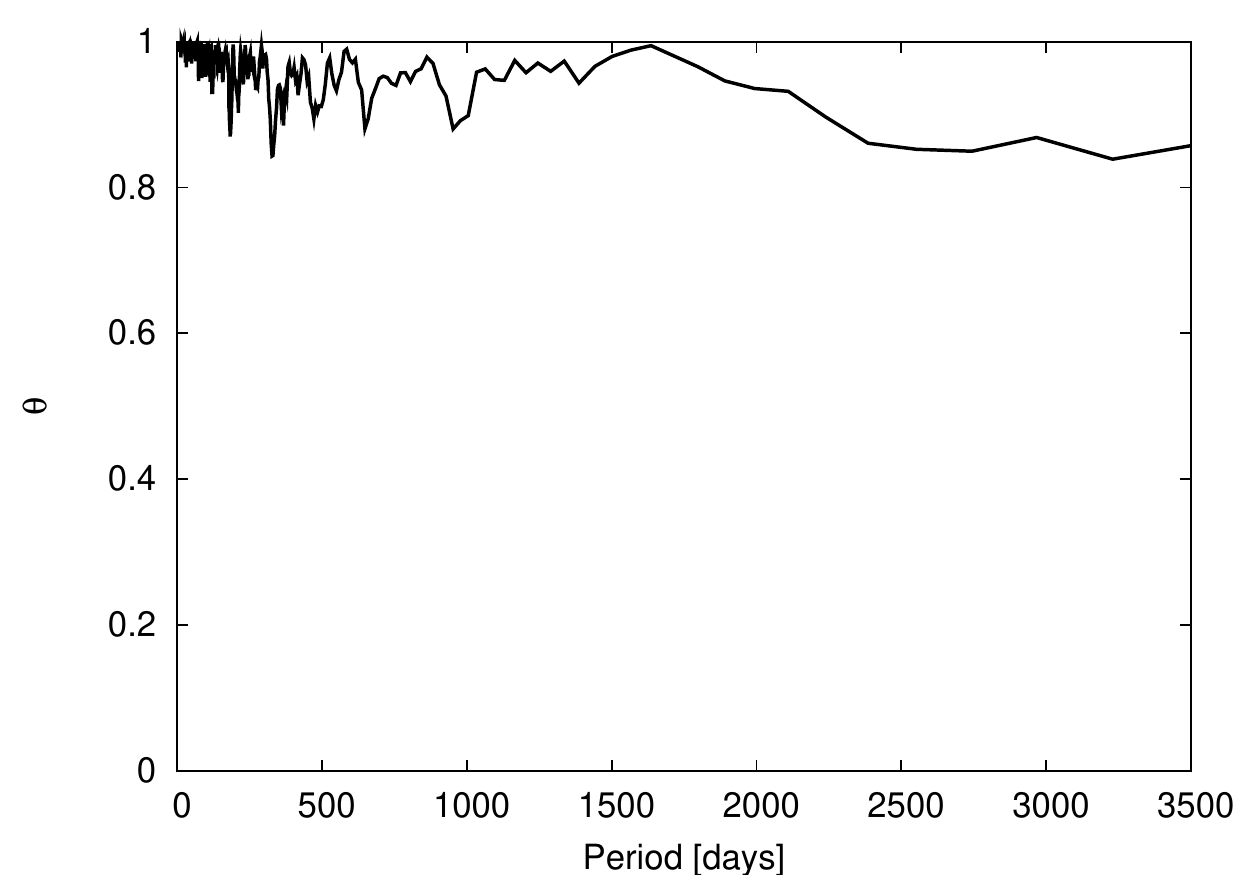}}\\
\resizebox{0.49\textwidth}{!}{\includegraphics{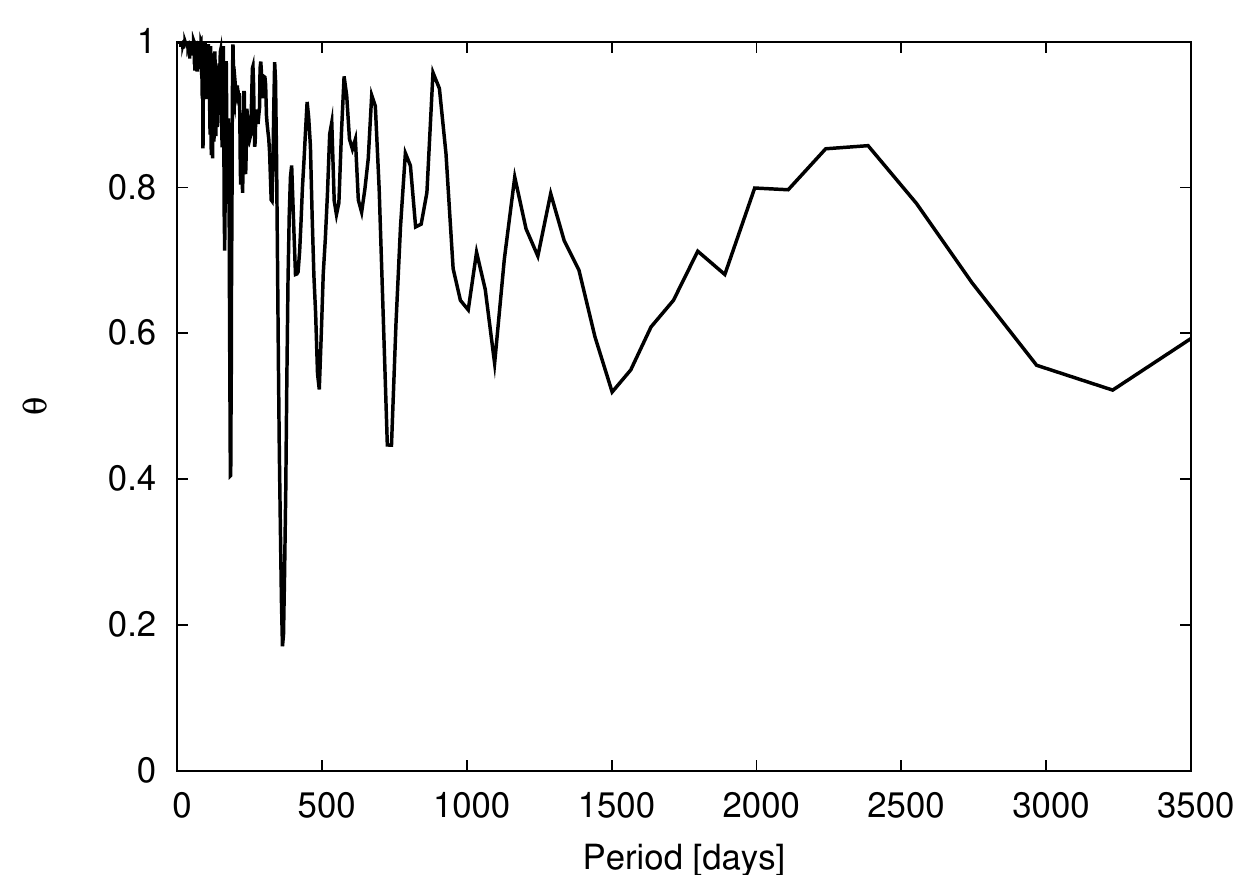}}
\resizebox{0.49\textwidth}{!}{\includegraphics{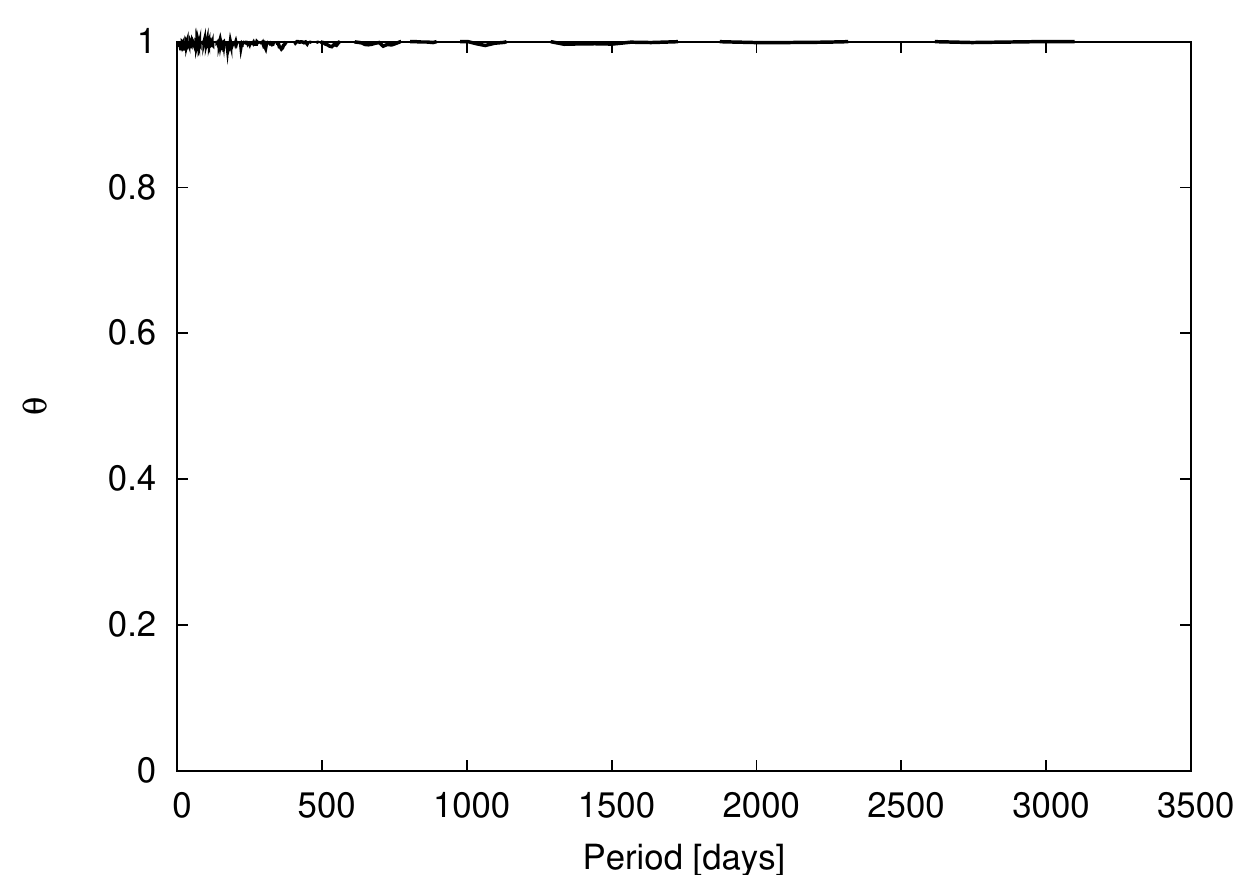}}\\
\caption{Periodograms determined using the Stellingwerf method. Parameter $\theta$ signifies the normalized phase variation. \emph{Upper left:} Periodogram of mean equatorial zonal velocity. \emph{Upper right:} Periodogram of mean equatorial meridional velocity. \emph{Bottom left:} Periodogram of sampled heliographic latitude of the centre of the solar disc. \emph{Bottom right:} Periodogram related to the sampling of data.}
\label{fig:periodograms}
\end{figure}

The influence of the calculated flow fields by the the position and orientation of the solar disc (position angle of the rotation axis $P$, heliographic latitude of the centre $b_0$) cannot be completely excluded, so the series of both parameters sampled on the same dates when the measurements of surface flows exist also passed the period analysis. Also the periods caused by sampling of the data using the same method were verified. Periodograms are displayed in Fig.~\ref{fig:periodograms}. 

Unfortunately, it has to be concluded that no significant period in the available data set was detected. As it can be seen in Fig.~\ref{fig:periodograms}, there exist non-convincing (the values of the parameter $\theta$ are quite high, which means that probably the periods are not significant) signs of periods detected in the real data, which are not present in
the control data set ($P$, $b_0$). Values of the suspicious periods are 657~days (1.80~years) in the meridional component and 1712~days (4.69~years) in the zonal component. Note that the 1.8-year period was also detected by \cite{2004AA...418L..17K}. It is claimed to be related to a possible Rossby wave $r$-mode signature in the photosphere with azimuthal order $m \sim 50$ reported by \cite{2000Natur.405..544K}, but lately disputed e.\,g. by \cite{2006SPD....37.3002W}. The period estimate for such an $r$-mode is close to 1.8 years. According to \cite{2005AA...438.1067K}, such a periodicity was observed in the total magnetic flux only on the southern hemisphere from 1997 to 2003. The coupling between the zonal flow and the meridional circulation could transfer the signal of the $r$-mode motion to the mean meridional component.

The detected suspicious periods may not be of solar origin. The sparse data set suffers from aliasing caused by a bad coverage of the studied interval. To confirm the periods, a far more homogeneous data set is needed. This may be a task for the ongoing space borne experiment Helioseismic Michelson Imager (HMI), which will be a successor of MDI. 

Detected periodicities are absent in Mt.~Wilson torsional oscillations time series, which is a far more homogeneous material than the one used in this study. \cite{1990ApJ...351..309S} also did not find any time variations in the study tracking the features in the low resolution dopplergrams covering homogeneously 20 years of Mt.~Wilson observations. All the arguments written above led to leave the detected periods as suspicious, as they cannot be confirmed from the current data set.

\subsection{Relation to the magnetic activity}
The coupling of equatorial zonal velocity (average equatorial solar rotation) and the solar activity in the near-equatorial area (belt of heliographic latitudes from $-10\,^\circ$ to $+10\,^\circ$) was also investigated. The average equatorial zonal velocity incorporates the average supergranular network rotation and also the movement of degenerated supergranules influenced by a local magnetic field with respect to their non-magnetic vicinity. Indexes of the solar activity were extracted from the daily reports made by \emph{Space Environment Center National Oceanic and Atmospheric Administration (SEC NOAA)}. Only the days when the measurements of horizontal flows exist were taken into account. As the index of the activity we have considered the total area of sunspots in the near-equatorial belt and also their type.

\begin{figure}
\resizebox{0.5\textwidth}{!}{\includegraphics{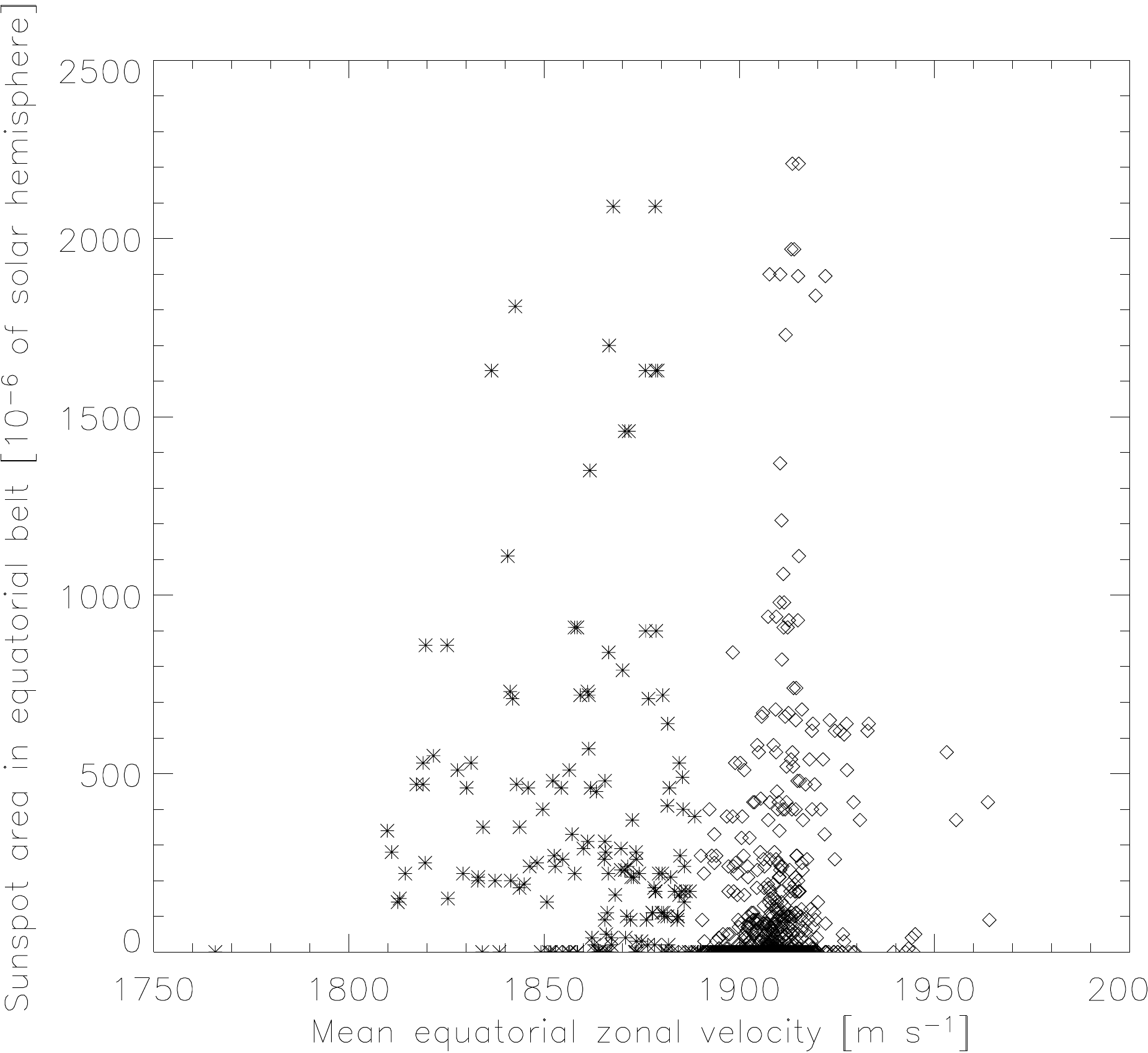}}
\caption{Mean zonal equatorial velocity versus the sunspot area in the near-equatorial belt. We decided to divide the data in two regimes along the velocity axis. Although the division is arbitrary, we believe that it is supported by the theory of the dynamical disconnection of sunspots from their roots.}
\label{fig:activity_velocity}
\end{figure}

\begin{figure}[!]
\resizebox{\textwidth}{!}{\includegraphics{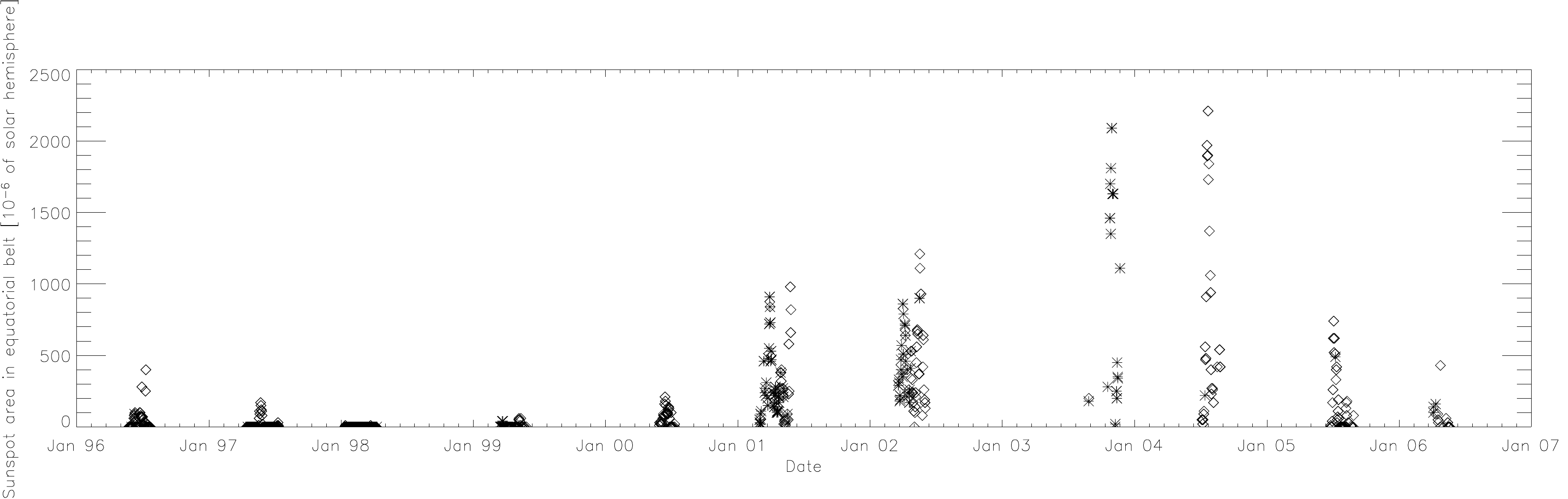}}
\caption{Sunspot area sampled in the same times when the measurements of the horizontal flows exist. Two regimes of the near-equatorial belt rotation are displayed. Diamonds denote the ``fast'' rotating equatorial belts, crosses the ``scattered'' group.}
\label{fig:velocity_regimes}
\end{figure}

First of all, the correlation coefficient $\rho$ between the mean equatorial zonal velocity and the sunspot area in the near-equatorial belt was computed with a value of $\rho=-0.17$. There is no significant linear relation between these two indices. The dependence of both quantities is plotted in Fig.~\ref{fig:activity_velocity}. There can be clearly found two different regimes, which are divided by the velocity of approximately 1890~\mps. In one regime (77~\% of the cases), the equatorial belt rotates about 60~\mps{} faster ($1910 \pm 9$~\mps; hereafter a ``fast group'') than the Carrington rotation, in the other one (23~\%) the rotation rate is scattered around the Carrington rate ($1860 \pm 20$~\mps; hereafter a ``scattered group''). The division in these two suggested groups using the speed criterion is arbitrary. If there exist only two groups, they certainly overlap and only a very detailed study could resolve their members. One may also see more than two groups in Fig.~\ref{fig:activity_velocity}. The arguments for the division in just two groups will follow.

For both regimes a typical sunspot area does not exist. The distribution of both regimes in time is displayed in Fig.~\ref{fig:velocity_regimes}. The data in the periods of larger solar activity (years 2001 and 2002, these are also the only years when the data cover two Carrington rotations continuously) show that both regimes alternate with a period of one Carrington rotation. 

The histogram of the mean zonal equatorial velocity has a similar, i.~e. bimodal, character like in Fig.~\ref{fig:activity_velocity} with a greater second peak, because such a histogram is constructed not only from belts containing magnetic activity but also from the belts where no magnetic activity was detected. The mean equatorial rotation for all the data is 1900~\mps, 1896~\mps{} for the equatorial rotation in the presence of sunspots, and 1904~\mps{} for days without sunspots in the equatorial region. 

Such bimodal velocity distribution is in disagreement with the results obtained by time-distance helioseismology by \cite{2004ApJ...607L.135Z}. In this work the authors found that the stronger the magnetic field the faster such a magnetic element rotates. They observed about 70~\mps{} faster rotation than the average for magnetic areas with magnetic fields stronger than 600~G. It might be possible that the size of the magnetic area and its magnetic field strength play different roles in the influencing of the plasma motions. However, \cite{2005AA...436.1075M} showed that the size of the magnetic area and the maximum magnetic field strength or the total magnetic flux correlate quite well.

\begin{figure}[!]
\resizebox{0.48\textwidth}{!}{\includegraphics{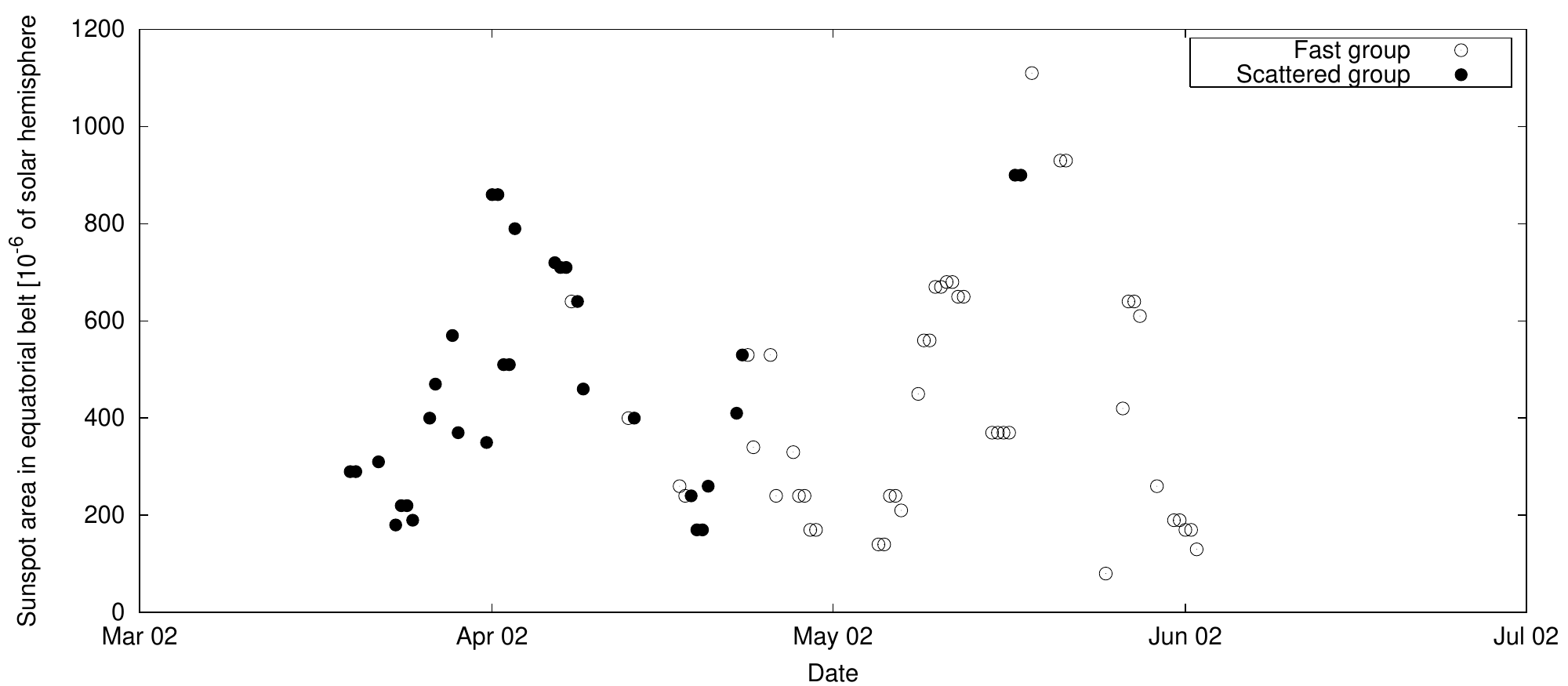}}
\resizebox{0.505\textwidth}{!}{\includegraphics{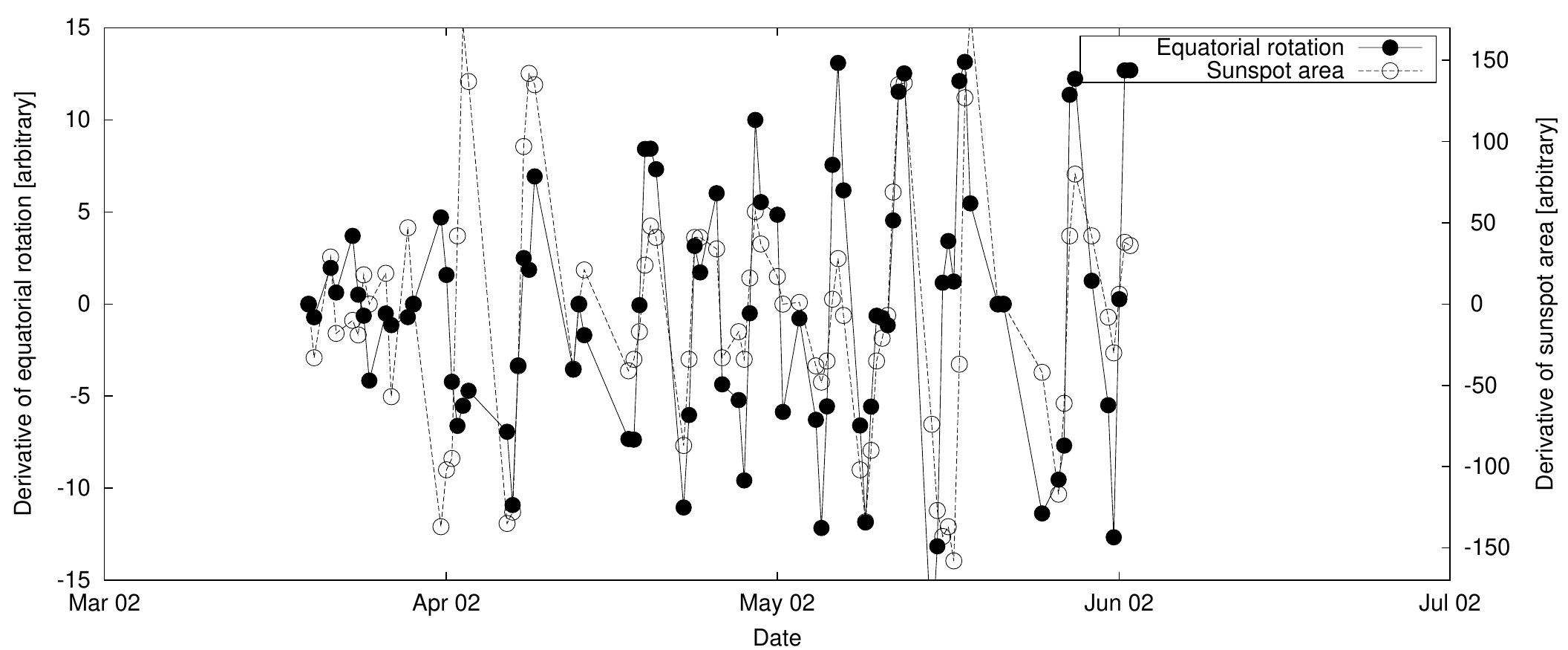}}
\caption{\emph{Left:} Distribution of two equatorial rotation modes in the year 2002. \emph{Right:} Derivatives of the mean zonal velocity (solid curve) and the sunspot area in the near-equatorial region (dashed curve) in 2002. Both quantities correlate with each other quite nicely.}
\label{fig:gradients2002}
\end{figure}

\begin{figure}[!]
\resizebox{0.48\textwidth}{!}{\includegraphics{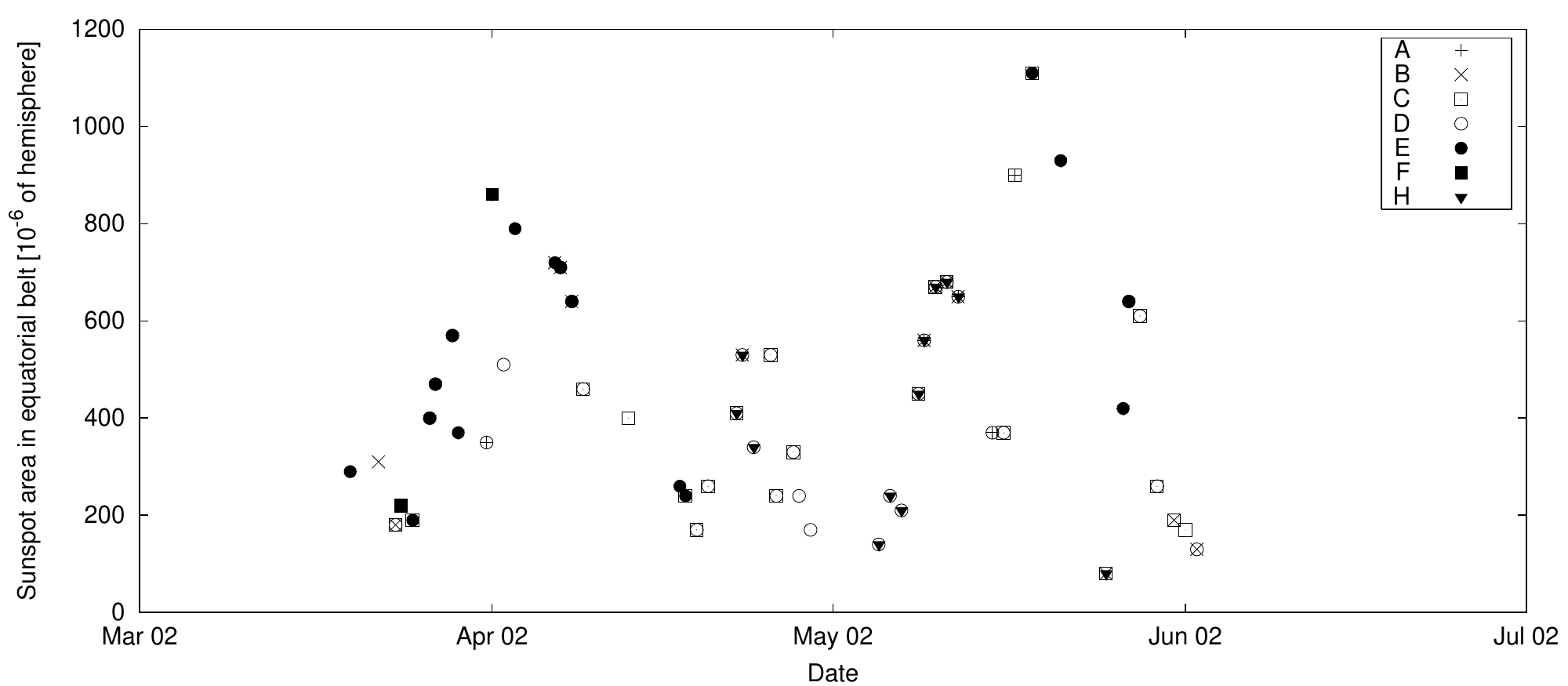}}
\resizebox{0.48\textwidth}{!}{\includegraphics{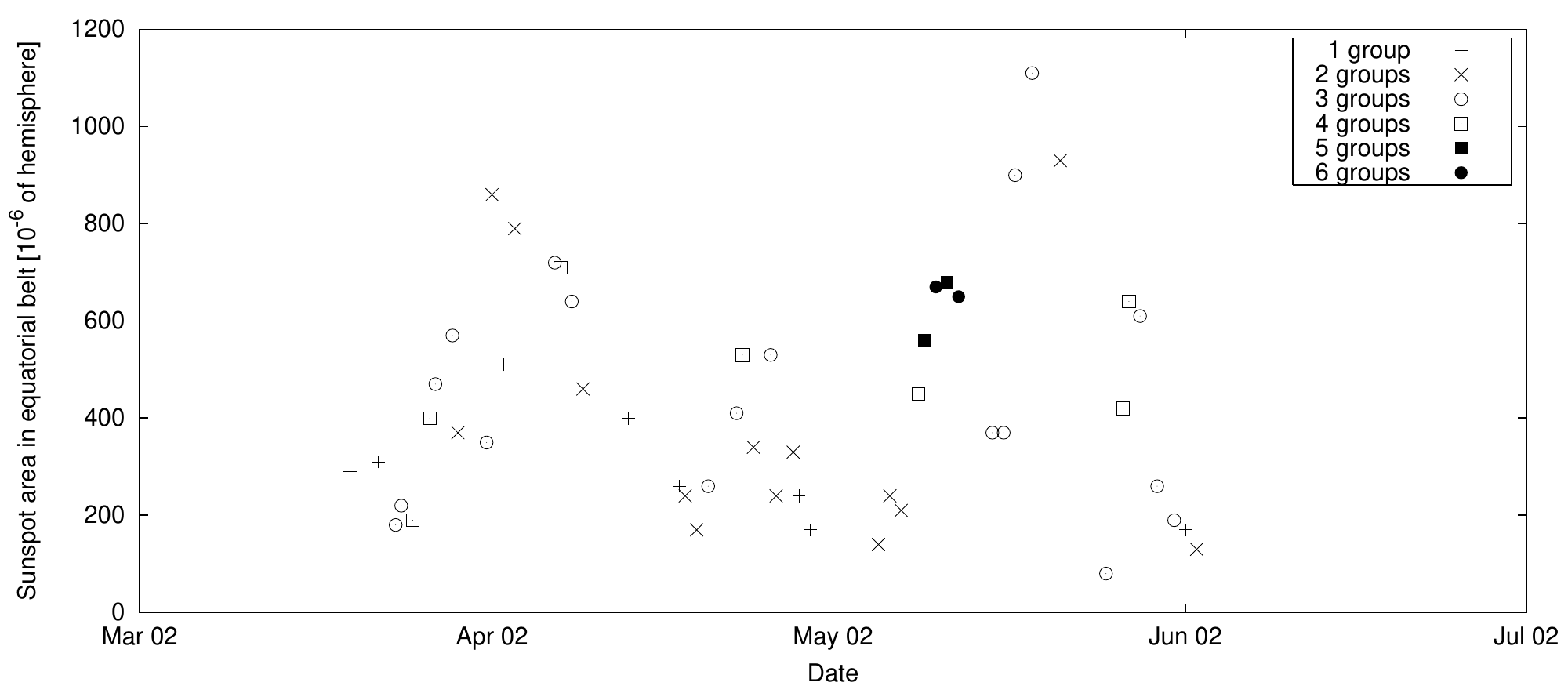}}
\caption{Active region morphological types (\emph{left}) and the number of active regions in the near-equatorial belt distributed as a function of time and sunspot area (\emph{right}).}
\label{fig:spottypes2002}
\end{figure}

Detailed studies of the sunspot drawings obtained from the Patrol Service of Ond\v{r}ejov Observatory and the Mt.~Wilson Observatory drawings archive revealed that in the ``fast'' group, the new or growing young active regions were present in the equatorial belt. On the contrary, in the ``scattered'' group the decaying or recurrent active regions prevailed in the equatorial area. The deceleration of the sunspot group with its evolution was noticed e.\,g. by \cite{2004SoPh..221..225R}. Moreover, the results suggest that the new and rapidly growing sunspots in the studied sample (March to May 2001 and April to June 2002) move with the same velocity. This behaviour could be explained by an emergence of the local magnetic field from a confined subphotospheric layer. According to the rough estimate \citep{2002SoPh..205..211C} the speed of $1910 \pm 9$~\mps{} corresponds to the layer of $0.946 \pm 0.008\ R_\odot$, where the angular velocity of rotation suddenly changes. During the evolution, the magnetic field is disrupted by the convective motions. An interesting behaviour is displayed by the alternation of the ``fast'' and ``scattered'' regimes (see Fig.~\ref{fig:gradients2002}) with the period of one Carrington rotation. It suggests that active regions in the equatorial region emerge in groups. We have to keep in mind that this study produces a very rough information due to the averaging of all effects in the equatorial belt. 

The observed behaviour could be a manifestation of the disconnection of magnetic field lines from the base of the surface shear during the evolution of the growing sunspot group. This behaviour was theoretically studied by \cite{2005AA...441..337S}. They suggested the dynamical disconnection of bipolar sunspot groups from their magnetic roots deep in the convection zone by upflow motions within three days after the emergence of the new sunspot group. The motion of sunspots changes during those three days from ``active'' to ``passive''. The active mode is displayed by motions reasonable faster with respect to the non-magnetic surroundings. The passive mode means mostly the deceleration of sunspot motions and influence of the sunspot motions only by the shallow surface plasma dynamics. The theory of the disconnection of sunspot groups from their magnetic root supports the division of the data set in two groups. 

\begin{figure}
\centering
\resizebox{0.49\textwidth}{!}{\includegraphics{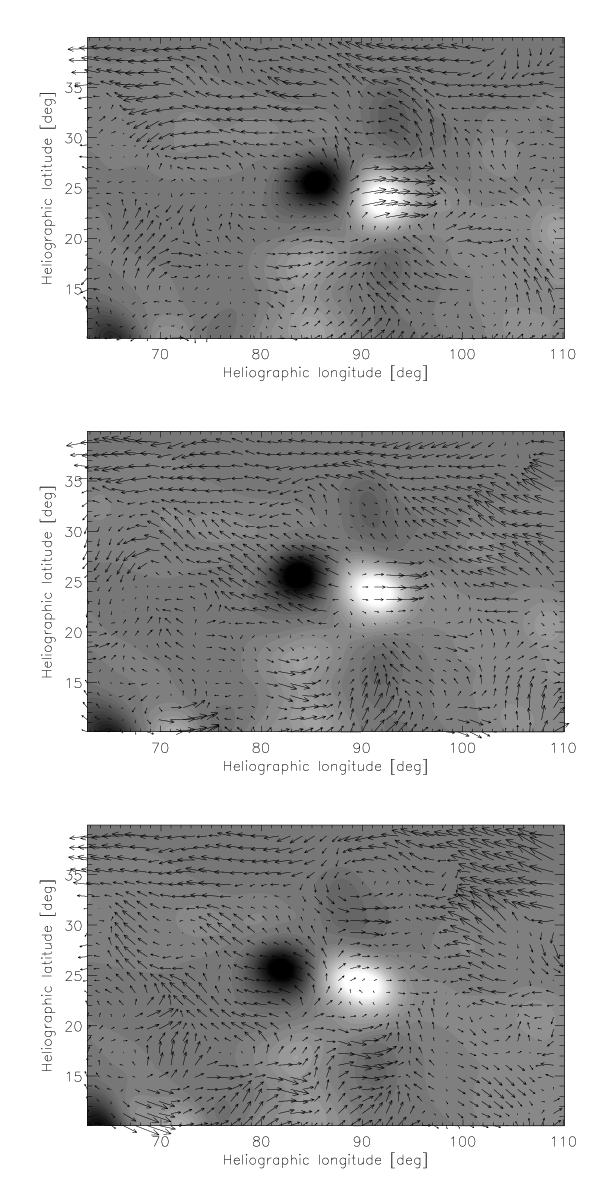}}
\caption{Case study: Evolution of the flows in and around the active region NOAA~9368 on March 6, March 7, and March 8, 2001. The leading polarity rotate significantly faster than the following one and the non-magnetic surroundings in the first day. The whole group slows down in the other two days. As the background image, the MDI magnetogram smoothed to the resolution of the measured flow field is used. The magnetic field intensities are displayed in the range of $-$1800 (black) to $+$1800 (white) Gauss in the linear scale.}
\label{fig:group_evolution}
\end{figure}

As an example, the active region NOAA~9368 (Fig.~\ref{fig:group_evolution}) was selected to show the behaviour of large-scale velocities in time. One see that the leading part of the active region rotates faster than the surroundings in the first day of observation and the whole group slows down in next two days. The inspection of details in the behaviour of selected active regions in the whole data set will be the subject of the ongoing studies.

The division of the equatorial belt into 10 sectors helped the investigation of the behaviour of different active region types. The dependence of the mean zonal velocity in the sector containing the studied active region on the morphological type of active region is shown in Table~\ref{tab:sunspot_types}. It can be clearly seen that, in average, more evolved active regions rotate slower than less evolved or young active regions, what is in agreement with e.~g. \cite{1986AA...155...87B}. No particular behaviour for various active region types was found (see Fig.~\ref{fig:spottypes2002}).
\begin{table}[!b]
\caption{Mean synodic rotation velocities of different active regions types and the average equatorial rotation of all data. The measured speed values may be systematically biased by the projection effect \cite[see][]{2006ApJ...644..598H}.}
\centering
\vspace*{3mm}
\begin{tabular}{cc}
\hline
\hline
Sunspot & Mean rotation \\
type & [\mps]\\
\hline
A & $1893 \pm 48$ \\
B & $1893 \pm 49$ \\
C & $1890 \pm 71$ \\
D & $1880 \pm 73$ \\
E & $1880 \pm 50$ \\
F & $1874 \pm 73$ \\
H & $1872 \pm 51$ \\
\hline
Average & $1900$ \\
\hline
\end{tabular}
\label{tab:sunspot_types}
\end{table}

We have also studied how the presence of the magnetic active areas will influence the average flow field. Since we found that a direct correlation is weak due to the existence of two different regimes, we decided to study the temporal change of both quantities. The aim is to study whether an emerging active region in the near-equatorial belt will influence the average equatorial rotation. We computed numerical derivatives of the total sunspot area in the near-equatorial belt and of the average zonal equatorial flow. We have found that the correlation coefficient between both data series is $\rho=0.36$ and is higher for the ``fast group'' ($\rho=0.41$) than for the ``scattered group'' ($\rho=0.24$). The correlation is higher in periods of increased magnetic activity in the equatorial belt. For example, for data in the year 2001 the correlation coefficient is $\rho_{2001}=0.58$ and for the year 2002 $\rho_{2002}=0.52$; see Fig.~\ref{fig:gradients2002}. In both particular cases, the correlation is higher for the ``fast regime'' ($\rho \sim 0.7$) than for the second group. 

It is important to know that LCT in our method measures basically the motions of supergranules influenced by the magnetic field, not by spots that are recorded in the used solar activity index. Therefore the results can be biased by the fact that the presence of magnetic field does not necessarily mean the presence of a sunspot in the photosphere. 

We think that despite an apparent disagreement, our apparently conflicting results can be valid and are in agreement with the results published earlier. Basically, as described by e.\,g. \cite{1990ApJ...357..271H} and explained in the model of \cite{2004SoPh..220..333B}, the solar rotation in the lower latitudes is slower in the presence of magnetic field. This is summarized in Table~\ref{tab:sunspot_types} and displayed in Fig.~\ref{fig:activity_velocity}. In most cases, ``spotty'' equatorial belts seem to rotate slower than the average for the whole data series. However, it is clear that emerging active regions cause in most cases the increase of the rotation rate. This is in agreement with a generally accepted statement found first by \cite{1970SoPh...12...23H} and \cite{1978ApJ...219L..55G}. The relation, obtained using a linear fit on our data set, can be described by the equation
\begin{equation}
\Delta v \sim 0.2\, \Delta A_{\rm sunspots}\ {\rm m\,s^{-1}},
\end{equation}
where $\Delta v$ is a change of the equatorial rotation speed with respect to the Carrington rotation and $\Delta A_{\rm sunspots}$ is a change of sunspot area in the equatorial belt (in 10$^{-6}$ of solar hemisphere). We estimate that strong local magnetic areas rotate few tens of \mps{} faster than the non-magnetic surroundings.

The results seem to be in contradiction to the previous study by \cite{1990ApJ...351..309S}, who tracked the features in the low-resolution dopplergrams measured at the Mt.~Wilson 46~m tower telescope for the period of 20~years. The authors interpret the detected flows as the velocity of the supergranular network, although the spatial resolution was lower than the size of individual supergranules. They found two regimes taking place in the rotation of the photosphere -- the quiet Sun and the active regions. The results showed that the regions occupied by the magnetic field displayed a slower rotation than the non-magnetic vicinity. \cite{1992ApJ...393..782T} used the high resolution data obtained at the Swedish Vacuum Solar Telescope on La Palma, Canary Islands, and found that the velocity field on the granular scale is different in the regions of the quiet Sun and in an active-region plage. Horizontal flow speeds measured by LCT were a factor of 2 slower in magnetic field regions than in the quiet Sun.

\subsection{Conclusions}
We have verified that the method developed and tested using the synthetic data is suitable for application to real data obtained by the MDI onboard SoHO and maybe also to the data that will be produced by its successor Helioseismic Michelson Imager (HMI) onboard the Solar Dynamic Observatory (SDO). HMI will have a greater resolution and will cover larger time span than two months each year. We verified that the long-term evolution of the horizontal velocity fields measured using our method is in agreement with generally accepted properties. 

During the periodic analysis of the equatorial area we found two suspicious periods in the real data, which are not present in the control data set containing the inclination of the solar axis towards the observer, the quantity that can bias systematically and periodically the results by a few \mps. The periods of 1.8~year and 4.7~years need to be confirmed using a more homogeneous data set.

We also found that the presence of the local magnetic field generally speeds-up the region occupied by the magnetic field. However, we cannot conclude that there exists a dependence of this behaviour for different types of sunspots. We can generally say that the more evolved types of active regions rotate slower than the young ones, however the variance of the typical rotation rate is much larger than the differences between the rates for each type. We have found that the distribution of active regions rotation is bimodal. The faster-rotating cases correspond to new and growing active regions. Their almost constant rotation speed suggests that they emerge from the base of the surface radial shear at $0.95\ R_\odot$. The decaying and recurrent regions rotate slower with a wider scatter in their velocities. This behaviour suggests that during the sunspot evolution, sunspots loose the connection to their magnetic roots. Both regimes alternate with a period of approximately one Carrington rotation in years 2001 and 2002, which suggests that new active regions emerge in groups and may have a linked evolution.

%% file: filament.tex
\section{Motions around the footpoints of the eruptive filament}
\label{sect:filament}
\symbolfootnotetext[0]{\hspace*{-7mm} $\star$ This chapter was done together with Thierry Roudier from Laboratoire d'Astrophysique de l'Observatoire Midi-Pyr\'en\'ees, Tarbes, France, and is a part of a paper submitted to Astronomy \& Astrophysics as  Roudier.,~Th., \v{S}vanda,~M., Meunier,~N., Keil,~S., Rieutord,~M., Malherbe,~J.~M., Rondi,~S., Molodij,~G., Bommier,~V., and Schmieder,~B., \emph{Large-scale horizontal flows in the solar photosphere. III. Effects on filament destabilization}.}

Dynamic processes on the Sun are linked to the evolution of the magnetic field as it is passing through the different layers from the convection zone to the solar atmosphere. In the photosphere,
magnetic fields are subject to diffusion due to supergranular flows and to the large-scale motions of
differential rotation and meridional circulation. The action of these surface motions on magnetic fields plays an important role in the formation of large-scale filaments \citep{2003SoPh..216..121M}. In particular, the magnetic fields that are transported across the solar surface can be sheared by dynamic surface motions, which result in shearing of the coronal field. This corresponds to the formation of coronal
flux ropes in models which can be compared with H$\alpha$ filament observations \citep{2006ApJ...642.1193M}.  Many theoretical models try to reproduce the basic structure and the stability of filaments by taking into account surface motions as quoted above. These models predict that magnetic flux ropes involved in solar filament formation may be stable for many days and then
suddenly become unstable resulting in a filament eruption. Observations show that twisting motions are
a very common characteristic of eruptive prominences \citep[see for example][]{2002SoPh..208..253P}. However, it is still unknown whether the magnetic flux ropes emerge already twisted or if it is only the photospheric motion that drives the twisting of the filament magnetic field. Therefore, the mechanisms that drive filament disappearance remain uncertain. The destabilization can come from the interior of the structure or by an outside flare.

In the paper by \cite{2007AA...467.1289R}, local horizontal photospheric flows were measured
at high spatial resolution (0.5\arcsec) in the vicinity of and beneath a filament before and during the filament's eruptive phases (the international JOP178 campaign). It was shown that the disappearance of the filament initiates in a filament gap. Both parasitic and normal magnetic polarities were continuously swept into the gap by the diverging supergranular flow. We also observed  the interaction of opposite polarities in the same region, which could be a candidate for initiating the destabilization of the filament by causing a reorganization of the magnetic field. Here we investigate the large-scale photospheric flows at moderate spatial resolution (2\arcsec) 
beneath and in the vicinity of the same  eruptive filament. 

\subsection{Observations}

During three consecutive days of the JOP 178 campaign, Oct 6, 7, and 8, 2004 (http://gaia. bagn.obs-mip.fr/jop178/index.html), we observed the evolution of a filament that was close to 
the central meridian. We also observed the photospheric flows directly below the filament and in its immediate area.  The filament extends from $-$5\degr{} to $-$30\degr{} in latitude. A filament eruption was observed on October 7, 2004 at 16:30~UT at   multiple wavelengths from ground and space instruments. The eruption produced a Coronal Mass Ejection (CME) at approximately 19:00~UT that was observed with LASCO-2/SOHO and two ribbon flares observed with SOHO/EIT.  MDI/SOHO longitudinal magnetic field and Doppler velocity were recorded  with a cadence of one minute during the 3 days.  The Air Force ISOON telescope located at the National Solar Observatory/Sacramento Peak provided a full-disc H$\alpha$ image every minute. The pixel sizes were respectively  1.96\arcsec{} for MDI magnetograms and dopplergrams and 1.077\arcsec{} for ISOON  H$\alpha$ images. 

\begin{figure}
\centering
\resizebox{0.9\textwidth}{!}{\includegraphics{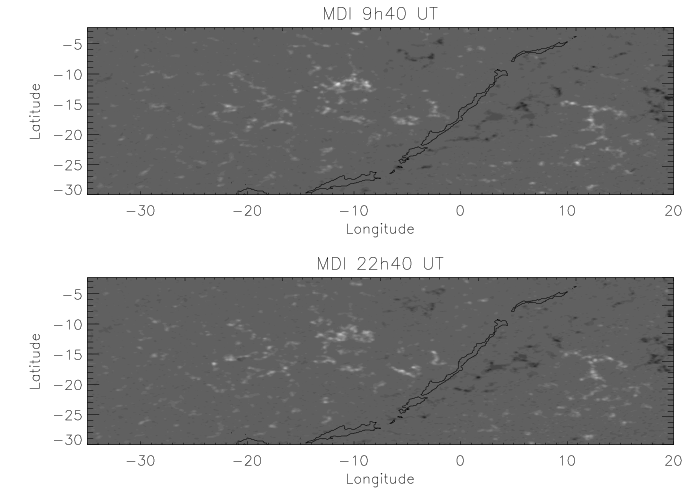}}
\caption{MDI longitudinal magnetic field data on October 7, 2004
at the beginning of the sequence at 9:40~UT and at its end at 22:40~UT. In order
to see the magnetic field evolution we overploted the shape of the 
filament at 13:30~UT in both figures.}
\label{fig:filmagneto}
\end{figure}

Our primary goal was to derive the horizontal flow field below and around the filament. Co-alignment between SOHO/MDI magnetograms and ISOON data was accomplished by adjusting the chromospheric network visible in H$\alpha$ (ISOON) and the amplitude of longitudinal MDI magnetograms to an accuracy of one pixel (1.96\arcsec). The general magnetic context before and after the filament eruption is shown Fig.~\ref{fig:filmagneto}.   

The local correlation tracking technique was applied to the set of full-disc MDI dopplergrams measured in three days around the time of filament eruption in high cadence. So the data in this campaign are suitable to be processed with the procedure described in Section~\ref{sect:real}. Due to the time series length and the required temporal resolution we had to adjust some parameters of data processing. The most significant difference is a sampling of averaged $p$-modes-free frames in 1-minute cadence instead of 15~minutes. The averaged frames were tracked using the Carrington rotation rate (with an angular velocity of 13.2 degrees per day), so that all the frames have the same heliographic longitude of the central meridian ($l_0=62.24\,^\circ$). The LCT method applied to full-disc dopplergrams is characterized by a Gaussian correlation window ($FHWM=60^{\prime\prime}$) and a time lag between correlated frames of 1~hour (basically 60 frames). In all cases, one half of the intervals were before the eruption and the second half after the eruption. All the pairs of correlated frames in the studied intervals were averaged to increase the signal-to-numerical-noise ratio.

\subsection{Photospheric flow pattern below and around the filament}

\begin{figure}
\centering\resizebox{0.5\textwidth}{!}{\includegraphics{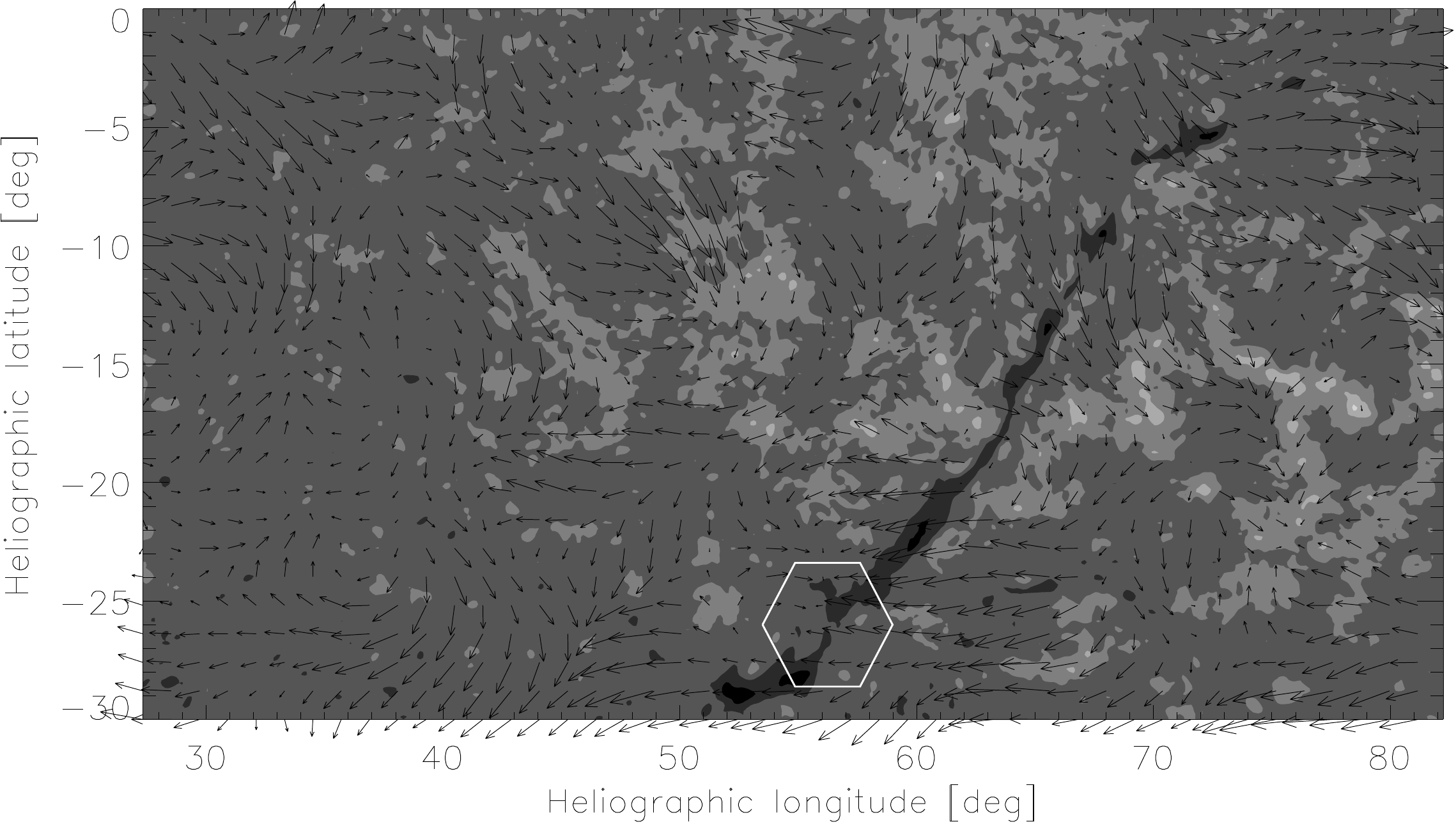}}\\
\caption[]{Horizontal photospheric flow field derived using LCT applied to full-disc dopplergrams. The filament observed by ISOON on October 7, 2004 is superimposed. The hexagon indicates the location where the filament eruption started.}
\label{fig:filvelo}
\end{figure}

In this subsection we describe the flows associated with the filament eruption, with particular emphasis on the filament evolution and the mean east--west (zonal) velocities properties.

The 13-hours averaged flow fields in the vicinity of the filament are shown in Fig.~\ref{fig:filvelo}. The measurements were compared with similar measurements based on tracking of individual magnetic features in MDI magnetograms. All the methods provide similar results. The large-scale flows are well structured and show both converging and diverging velocity patterns. We observe in particular a large-scale stream in the north--south direction parallel to the filament located about 10\degr{} to the east between $-$20\degr{} and $-$30\degr{} in latitude and 58\degr{} and 47\degr{} in longitude. This flow stream is clearly visible in Fig.~\ref{fig:filvelo}, and its dynamics can be seen at http://gaia.bagn.obs-mip.fr/jop178/oct7/mdi/7oct-mdi.htm. Around $-$20\degr{} in latitude the velocities of the differential rotation amplitude start to dominate. However, in both measurements we observe that the north--south large-scale stream on the eastern edge of the filament disturbs the regular differential rotation. The location of the north--south stream is close to the location where the filament eruption begins (longitude $l=56^\circ$, latitude $b=-26^\circ$ in Carrington coordinates). The amplitude of the southward motions is approximately  40~m\,s$^{-1}$, which is of the same order as the mean observed flows. Below the latitude of $-$20\degr{}, the combination of differential rotation and the north--south stream cause opposite polarities to move closer, which strongly increases the tension in the magnetic field very close to the starting point of the filament eruption.

\begin{figure}[!t]
\centering
\resizebox{0.6\textwidth}{!}{\includegraphics{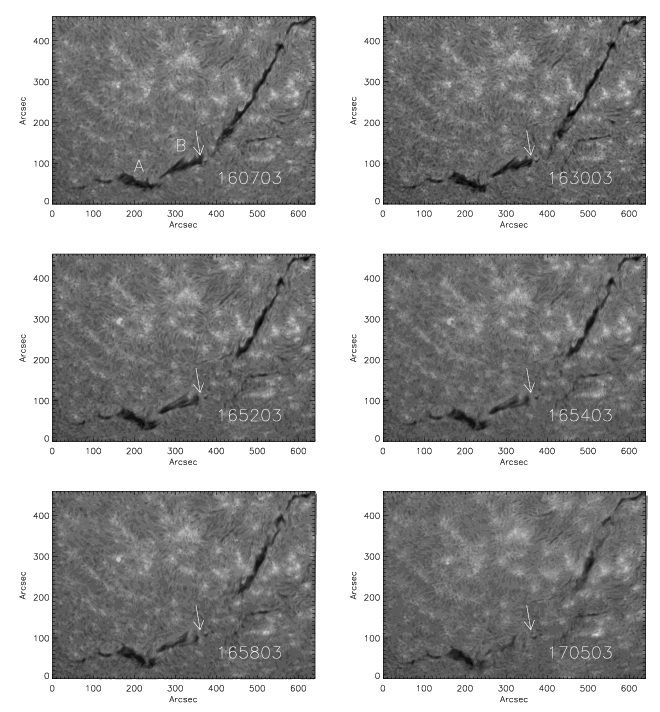}}
\caption[]{ Evolution of the filament during its eruption around 16:30~UT on October 7, 2004.
The arrow indicates a fixed position for all the subframes. A and B denote two parts of the filament.}
\label{fig:fileruption}
\end{figure}

The filament's evolution can be seen in  Fig.~\ref{fig:fileruption}. The north--south stream flow visible in  Fig.~\ref{fig:filwider} (left) crosses over the part of the filament labeled A. The arrow on Fig.~\ref{fig:fileruption} indicates the same fixed point (325\arcsec,167\arcsec) in all of the subframes. We observe a general southward motion of both the A and B segments of the filament. More precisely, we measure a tilt of these two filament segments at the point of their separation. Between 16:07~UT and 16:58~UT the long axis of segment A  rotates by an angle of 12\degr{} (clockwise) relatively to its western end, and the long axis of segment B of the filament rotates by an angle of 5.5\degr{} (clockwise) relatively to its western end.  These rotations are compatible with the surface flow shown in Fig.~\ref{fig:filwider} (left) and in particular the north--south stream flow.

We constructed the mean zonal velocity profile, i.~e. a profile of the differential rotation, to see if there is any correspondence to the presence of the filament. In this profile we found a strong secondary maximum visible at $-$23\degr{} of latitude. The secondary maximum indicates a decrease in the amplitude of the $v_x$ component because the flows in that region are oriented more in the north--south direction. That is partly due to the presence of the north--south stream described above and to the local organization of the flow. In particular, converging and diverging flows in this region seem to have a north--south orientation.

\begin{figure}[!t]
\centering
\resizebox{0.6\textwidth}{!}{\includegraphics{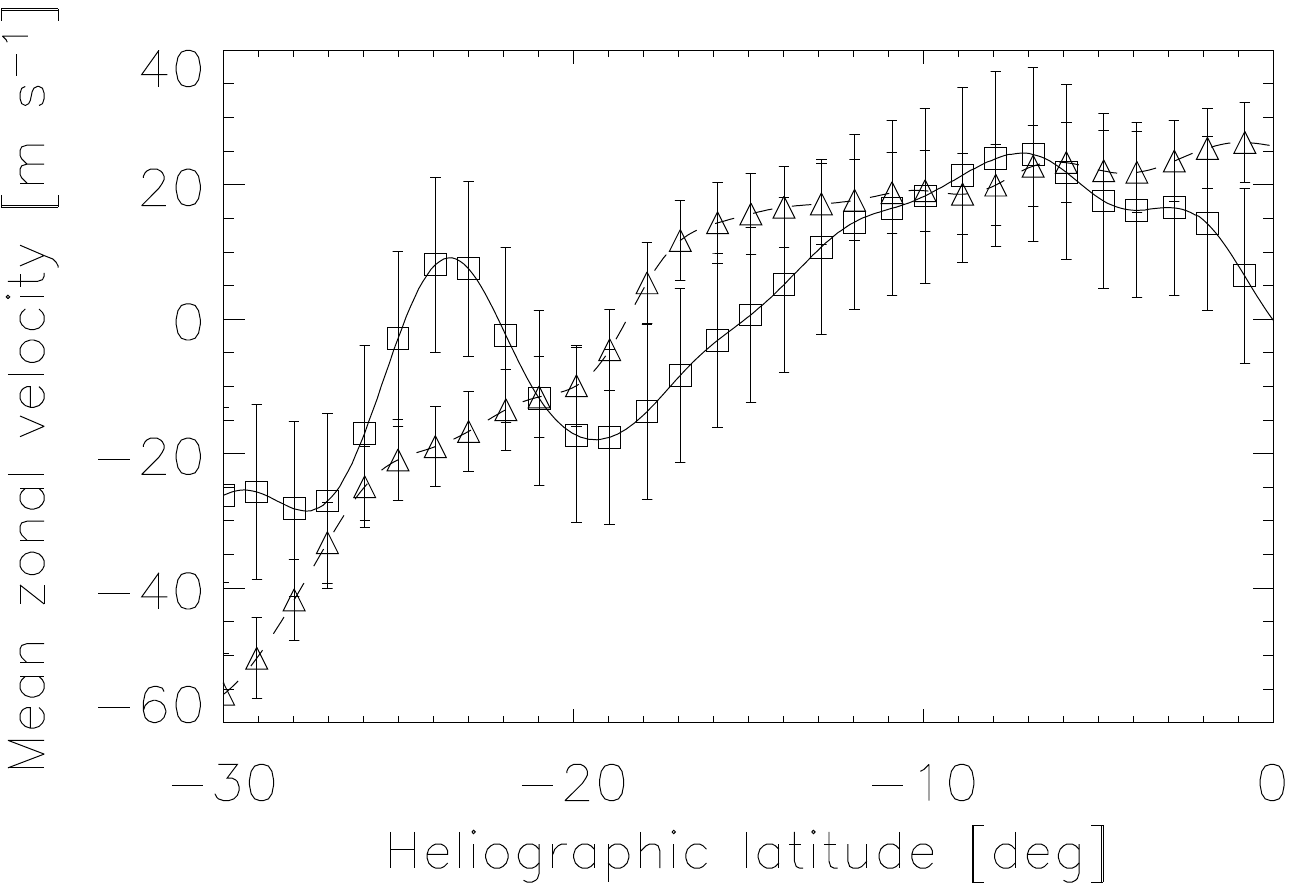}}
\caption{Profiles for 7 Oct in different longitudinal belts: solid line and squares for $-$25\degr{} to $-$17\degr{} degrees with respect to the centre of the disc (not Carrington coordinates), dashed line and triangles for 0\degr{} to 20\degr{}. }
\label{fig:filzonal}
\end{figure}

To distinguish the effects of the north--south stream from differential rotation, we computed the  mean zonal velocities from $-$25\degr{} to $-$17\degr{} in longitude with respect to the centre, where the north-south stream is present, and from 0\degr{} to 20\degr{} in longitude, where the differential rotation is not influenced by the current. The  mean zonal velocities in the longitudinal belt, where the  north--south stream is visible, exhibits clearly a secondary maximum (Fig.~\ref{fig:filzonal}) indicating that the solar rotation rate at this location is closer to that of the equator. As a consequence, the plasma in the north--south stream, which transports magnetic structures, rotates faster at about $-$23\degr{} latitude, than do the magnetic structure located in the belt of longitude between 0\degr{} to 20\degr. The combination of these different  surface motions (stream and differential rotation) tends to bring together fields with opposite polarities,  this in turn constrains the magnetic field lines.

We noted that the location of the starting-point of the filament eruption is around $-$26\degr{} in latitude, which is very close to that secondary maximum in the mean zonal velocity. Thus surface motions that bring together opposite polarities may play a role in triggering the filament eruption.

\subsection{Flow fields before and after the eruption}

\begin{figure}[!t]
\resizebox{0.5\textwidth}{!}{\includegraphics{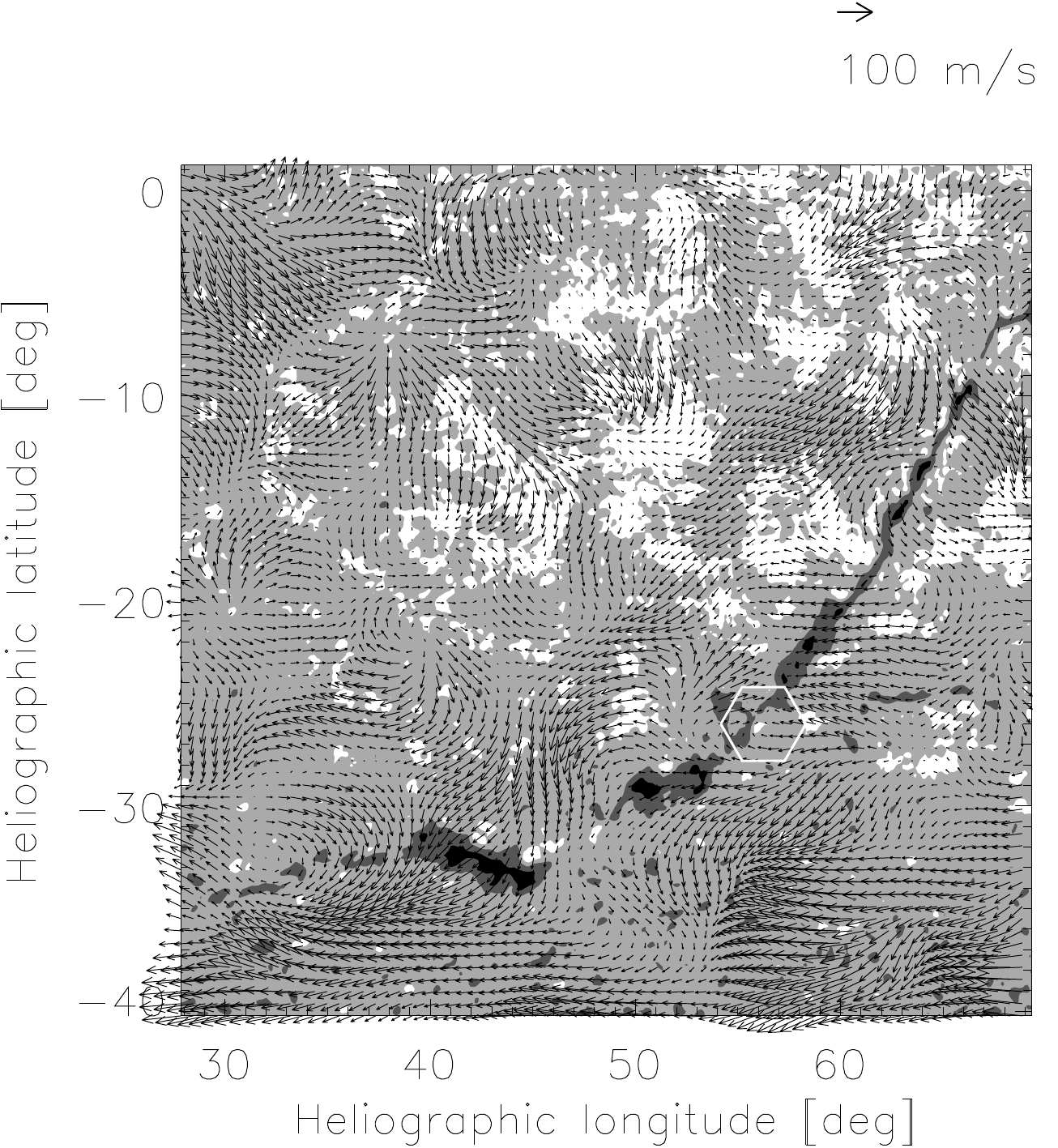}}
\resizebox{0.5\textwidth}{!}{\includegraphics{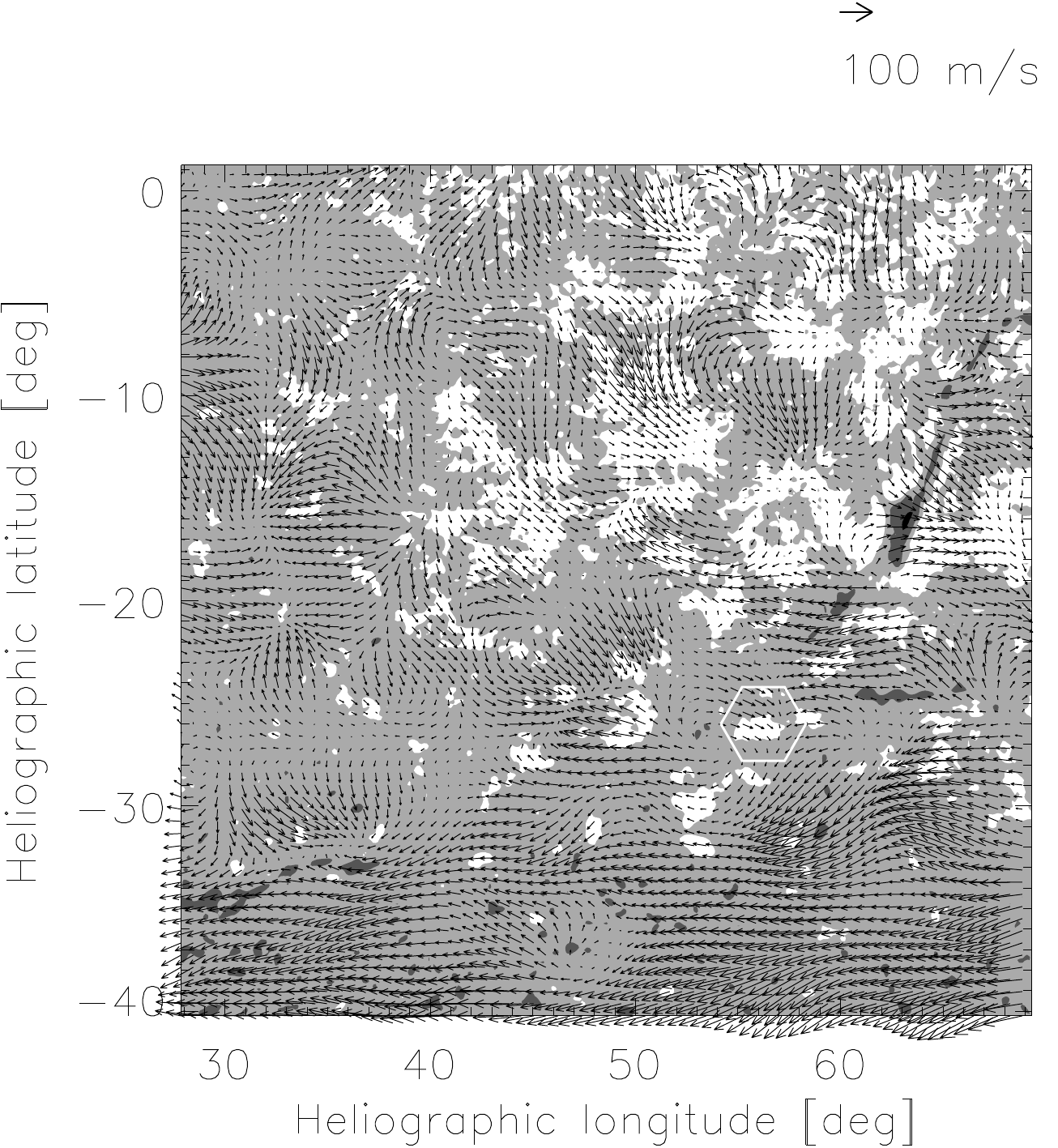}}\\
\resizebox{0.585\textwidth}{!}{\includegraphics{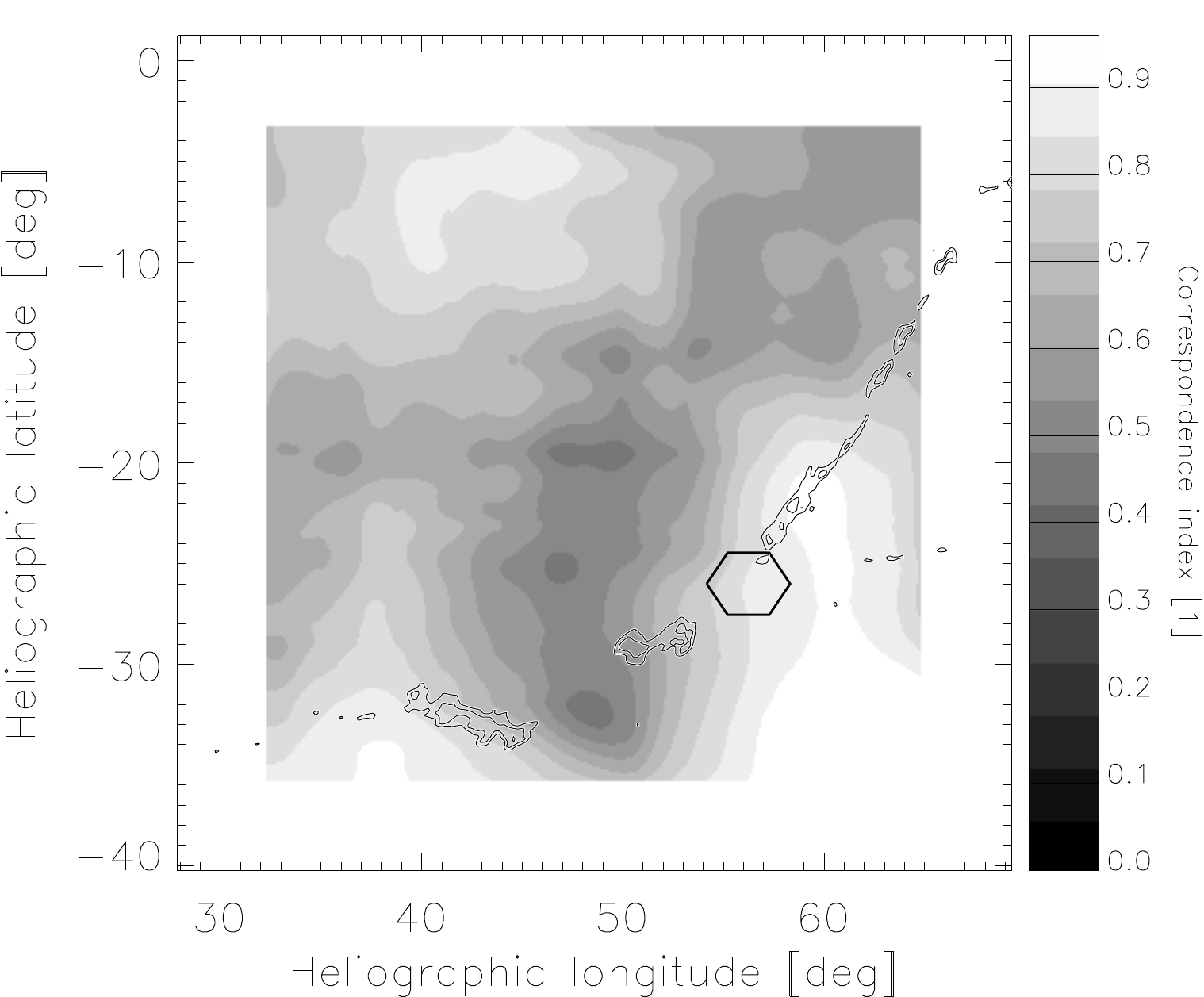}}\\
\caption[]{Horizontal motions measured before (\emph{left}) and after (\emph{right}) the eruption in the wide field of view. The filament observed by ISOON on October 7 2004 at 13:30~UT is superimposed. \emph{Bottom} -- The differences in the directions of both vector fields are evaluated using the weighted cosine defined by formula (\ref{eq:weightedcosine}) in the sliding window with size of 3.5 heliographics degrees. It is seen that the worst correspondence of both vector fields is in the area of the north--south current. 
It demonstrates that the flow field in the field-of-view remained more-or-less stable during the filament eruption except for the
close vicinity of the starting point.}
\label{fig:filwider}
\end{figure}

\begin{figure}[!t]
\resizebox{0.5\textwidth}{!}{\includegraphics{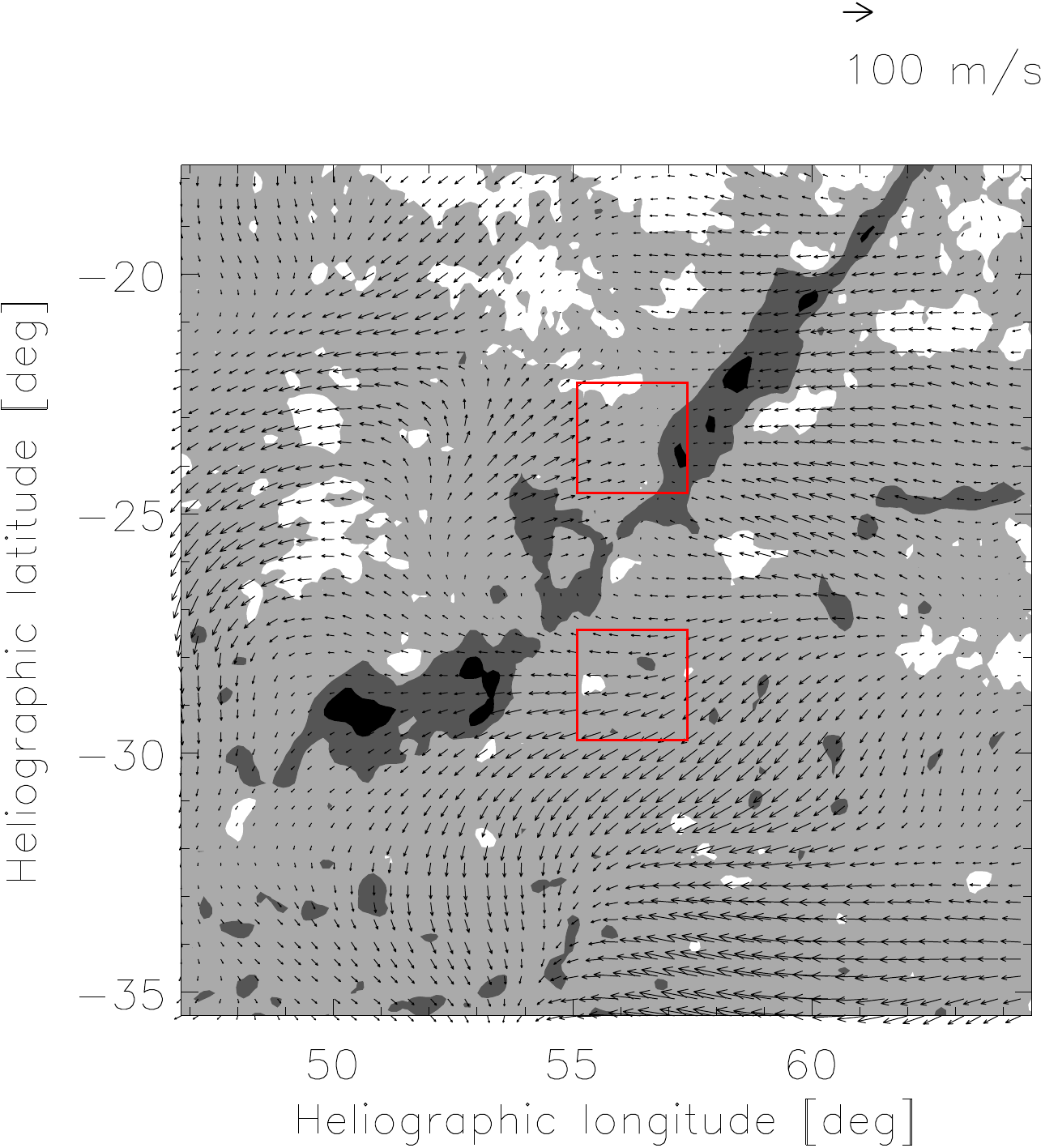}}
\resizebox{0.5\textwidth}{!}{\includegraphics{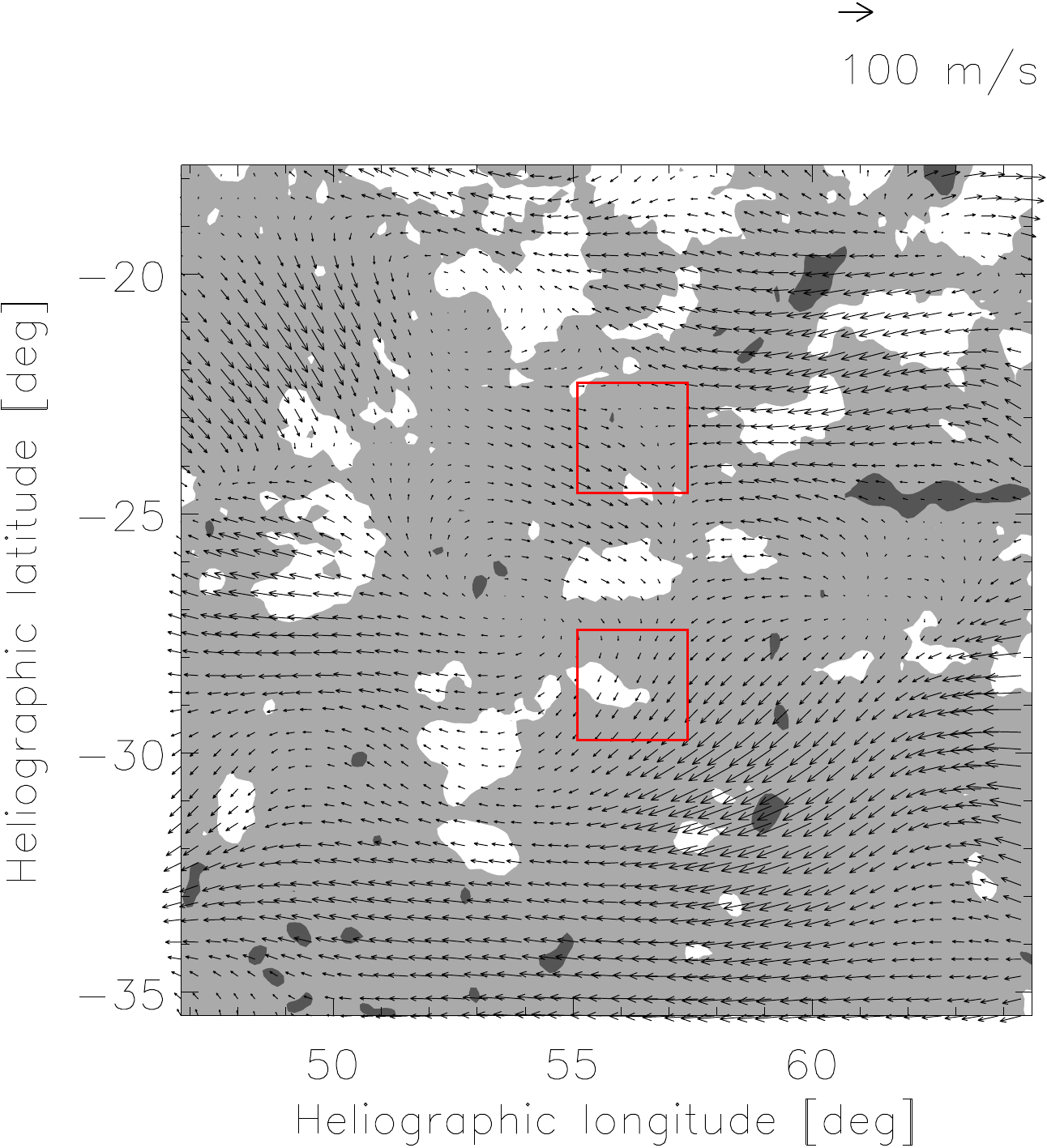}}\\
\caption[]{ The 3-hour average of the flow field in the close vicinity of the starting point ($l=56^\circ$, $b=-26^\circ$) 
before (\emph{left}) and after (\emph{right}) filament eruption. The filament observed by ISOON on October 7 2004  at 15.20~UT is superimposed. The red boxes denote areas used for the zonal shear calculation.}
\label{fig:fildetail}
\end{figure}
In this subsection we discuss the properties of the flow field just before and after the filament eruption, which occurred at about 16:30 UT. At the point where the filament eruption begins ($l=56^\circ$, $b=-26^\circ$ in Carrington coordinates), we detected a steepening of the gradient in the differential rotation curve. During the eruption, the gradient flattens out and a dip forms. Although the differential rotation curves describe the mean zonal velocities on the full disc, the change of its gradient signifies the change in the stretch influencing the magnetic field in the loop over the area under study. We can express the surface rotation  as an even power of $ \sin b$ : $\omega=A + B \sin^2 b+ C \sin^4 b$, where $A$ is the angular velocity rate of the equatorial rotation and $b$ is the heliographic latitude. From the data we find that the constant values (with their errors in parentheses) are before the eruption $A=13.375(0.010)$, $B=-1.46(0.10)$, $C=-1.42(0.20)$ and after the eruption $A=13.404(0.010)$, $B=-1.78(0.10)$, $C=-1.24(0.20)$. All the rates are synodical in deg\,day$^{-1}$. The full-disc profiles did not change significantly from before to  after the eruption. For example, for a latitude of $-$30\degr{} the zonal velocity has values of 12.92 (resp. 12.88)~deg\,day$^{-1}$ ($-$34~m\,s$^{-1}$, resp. $-$39~m\,s$^{-1}$ in the Carrington coordinate system), for a latitude of $-$20\degr{}  the values are 13.18 ($-$2~m\,s$^{-1}$) resp. 13.18 ($-$3~m\,s$^{-1}$)~deg\,day$^{-1}$. Although the parameters of the smooth fitted curve did not change too much, the local residual with respect to the smooth curve changed at the latitude where the filament eruption starts.

Fig.~\ref{fig:filwider} displays the horizontal flows before and after the eruption in a wide field of view measured using the LCT method on supergranular structures and averaging the resulting velocities over 3~hours. Before the eruption, we can clearly see the north--south stream parallel to and about  10\degr{} to the east of the filament. This stream disturbs differential rotation and brings plasma and magnetic structures to the south. Although differential rotation tends to spread the magnetic lines to the east, the observed north--south stream tends to shear the magnetic lines. After the eruption, only a northern segment of the filament is visible and the north--south stream has disappeared.

Fig.~\ref{fig:fildetail} shows the flow field in more detail, at the site where the eruption starts. The shear in the zonal component at the point where the eruption starts ($l=56^\circ$, $b=-26^\circ$ in Carrington coordinates) is clearly visible and prevails before and after the eruption, although the shape of the apparent vorticity has changed. This location corresponds to the area of upflows observed in the Meudon H$\alpha$ dopplergram \citep[Fig. 8 in ][]{2007AA...467.1289R}.

\begin{figure}[!t]
\centering
\resizebox{0.5\textwidth}{!}{\includegraphics{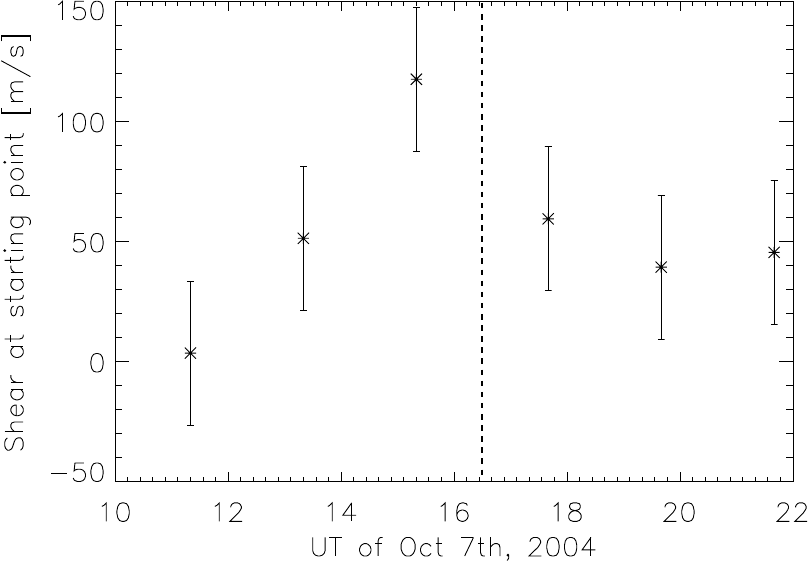}}
\caption[]{Temporal evolution of the velocity shear in zonal components. The eruption of filament took place at 16:30 UT.}
\label{fig:filshear}
\end{figure}

We measured the shear as a difference between the mean zonal component $v_x$  in the areas just to the north and just to the south of the starting point. We measured the average zonal flow over boxes with a side of 2.3\degr{} located 2.9\degr{} to the north and to the south of the point where the filament eruption appeared to start. The evolution in the shear velocity measured as the difference between the mean flow in the two boxes as a function of time can be seen in Fig.~\ref{fig:filshear}. Six 2-hours averages of the flow fields were used to determine this figure. One can see that the shear velocity is increasing before the eruption and decreasing after the eruption. One hour before the eruption, the shear reached the value of (120$\pm$15)~m\,s$^{-1}$ in a distance of 5.2\degr{} (62\,000~km in the photosphere). After the filament eruption, we observe a restoration of an ordinary differential rotation below $-$30\degr{} in latitude.

\subsection{Evolution on 6, 7 and 8 October 2004}

\begin{figure}[!t]
\resizebox{0.49\textwidth}{!}{\includegraphics{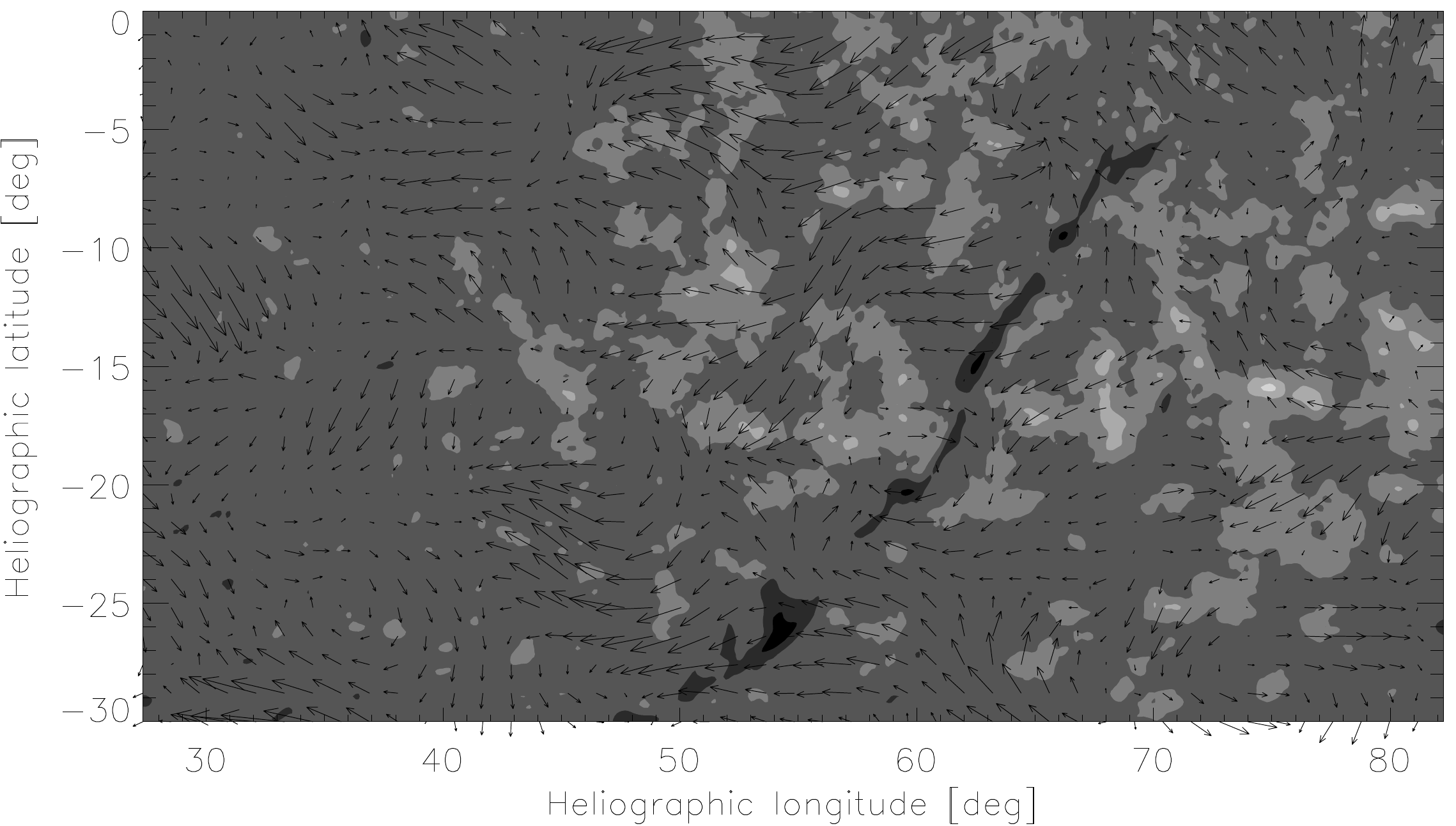}}
\resizebox{0.49\textwidth}{!}{\includegraphics{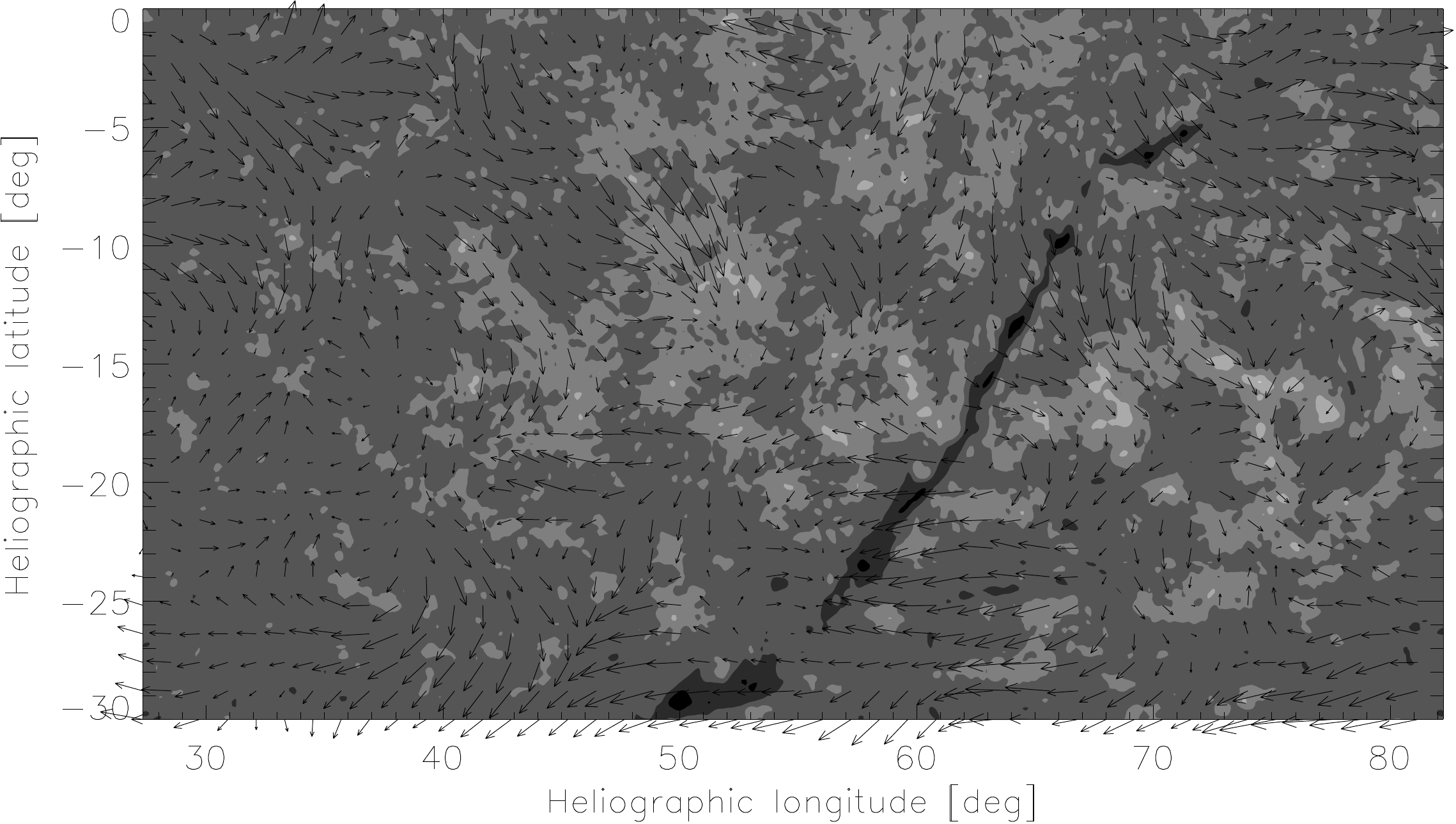}}\\
\resizebox{0.49\textwidth}{!}{\includegraphics{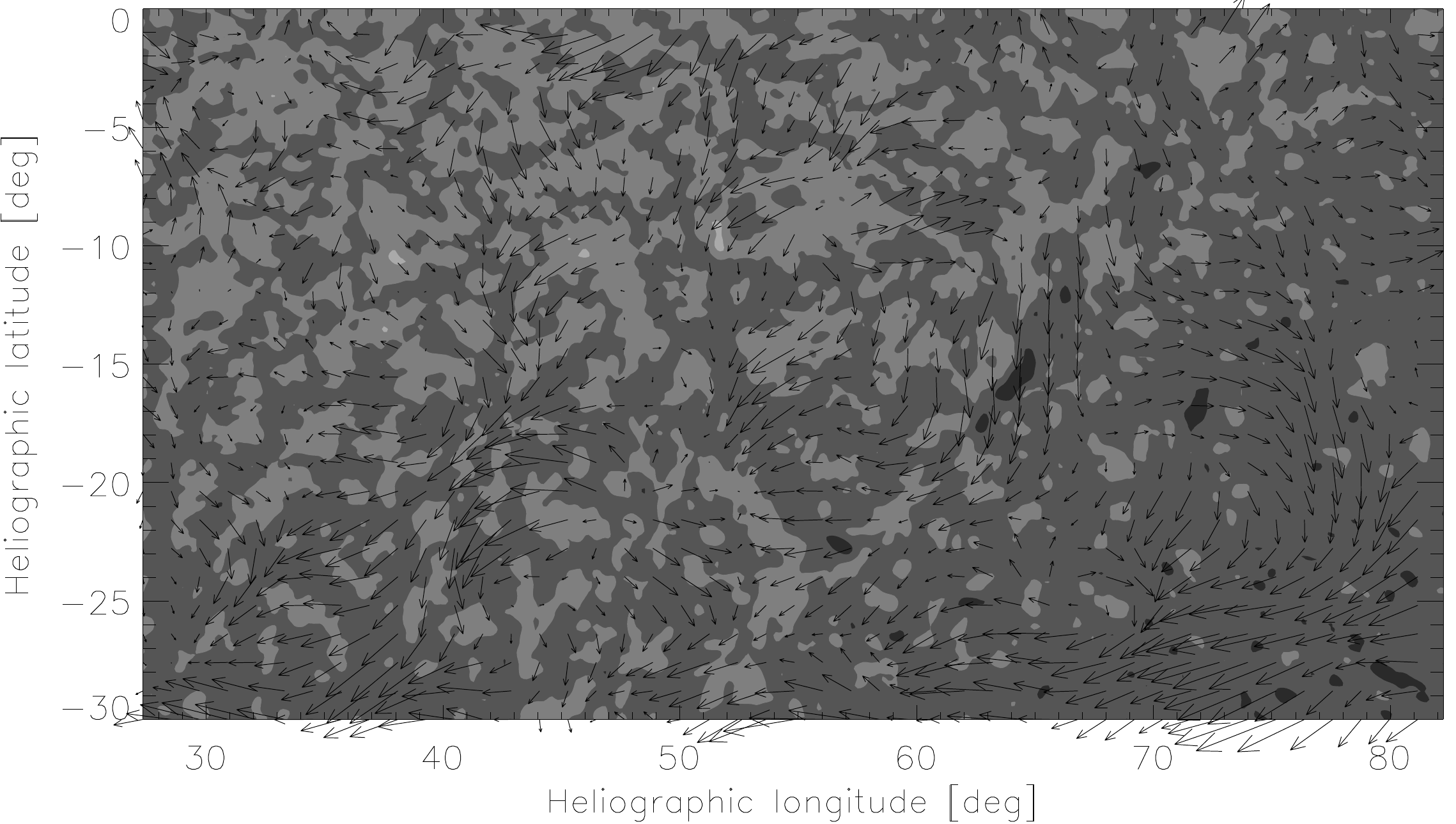}}
\resizebox{0.49\textwidth}{!}{\includegraphics{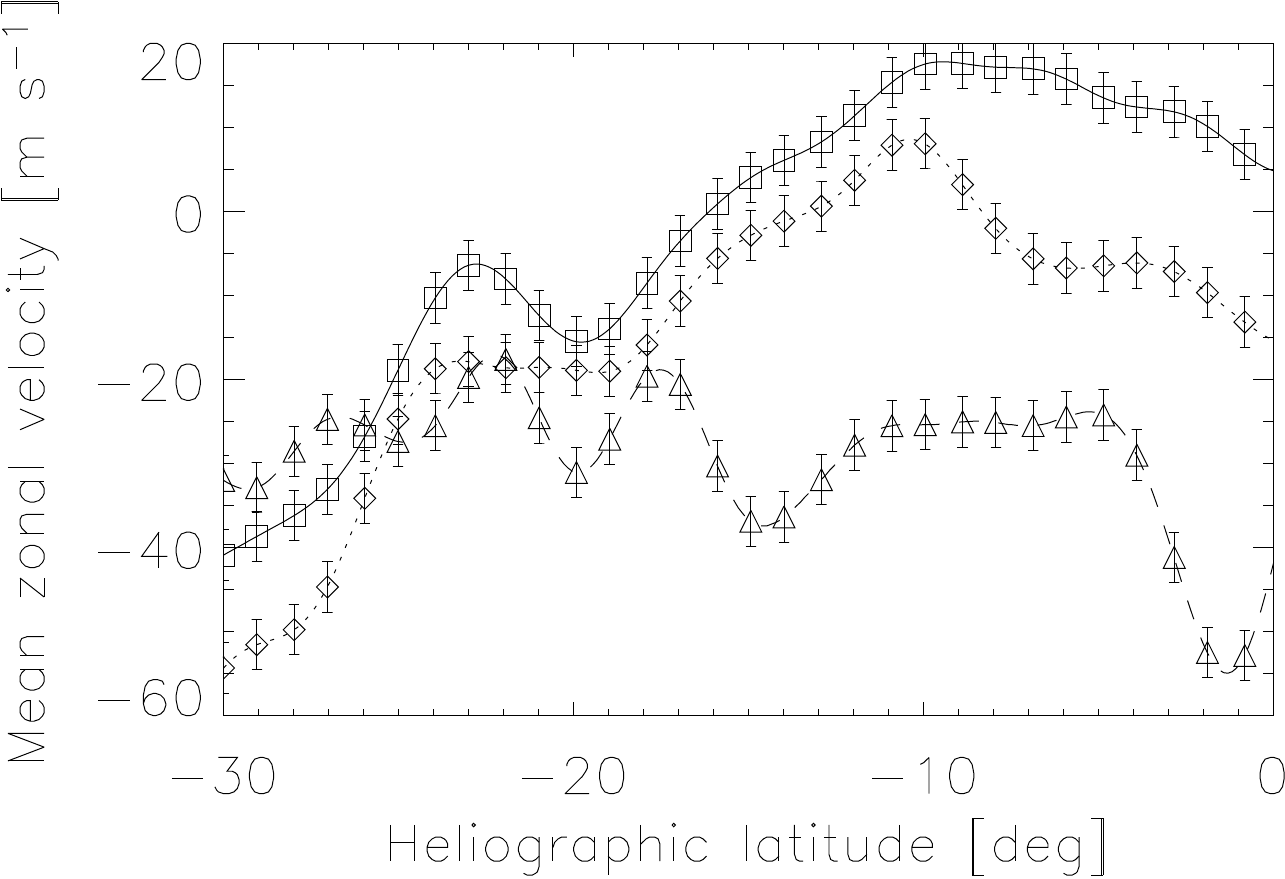}}
\caption[]{Horizontal flow evolution in three days 6 (\emph{upper left}), 7 (\emph{upper right}) and 8 (\emph{bottom left}) 
October 2004. \emph{Bottom right} -- Mean zonal velocities, computed
between $-$35\degr{} and $+$20\degr{} in longitude and $-$30\degr{} to 0\degr{} in latitude.
Dashed line (and triangles) represents mean zonal velocity for the LCT applied  on dopplergrams  
on 6 Oct 2004. Solid line (and squares) represents mean zonal velocity for the LCT applied 
on Doppler on 7 Oct 2004. Dash-dot line (and diamonds) represents mean zonal velocity for the 
LCT applied on Doppler on 8 Oct 2004.}
\label{fig:filevol}
\end{figure}In Fig.~\ref{fig:filevol} we compare the flow field in the filament area for the day of the eruption (October 7, 2004), with the flow fields on the preceding and following days.  We see that the topology of the flows changed in the filament area:  the daily evolution of the mean zonal profiles can be seen in Fig.~\ref{fig:filevol}  (bottom right). The dashed line with triangles is the mean zonal velocity for 6 October 2004. This profile is relatively flat probably due to the short time sequence as only 3 hours of data were available. The differential rotation profile for 7 October shows a secondary maximum around $-$23\degr{} in latitude, as discussed before. The 8 October profile exhibits the same trend, but with a smaller amplitude and an eastward velocity for latitudes greater than $-$10\degr{}. The secondary maximum appears strongly reduced indicating a restoration to a more regular differential rotation pattern in that zone.

One day before the eruption, the shear began to form at the site, where the filament eruption is triggered ($l=56^\circ$, $b=-26^\circ$ in Carrington coordinates). The north--south stream is also visible. Both phenomena may store free energy in the coronal magnetic field configuration. The topology of the flow and the stream is different a day after the filament eruption suggesting that after the disappearance of the southern part of the filament the conditions in the photosphere below the filament became more relaxed. This may suggest the mutual coupling of the photospheric flow and the configuration of the coronal magnetic field. To confirm this idea, high-cadence high-resolution images and magnetograms covering the eruption time would be needed.

\subsection {Discussion and conclusion}

Filaments and prominences are important complex structures of the solar atmosphere because they are linked to  CMEs, which can influence the Earth's atmosphere and near-space environment. Surface motions acting on pre-existing coronal fields play a critical role in the formation of filaments. They appear to reconfigure existing coronal fields, they can twist and stretch them and thereby deposit energy in the topology of the coronal magnetic field. Photospheric motions can also initiate coronal magnetic field disruption. Surface motions play an important role in formation Type~B filaments \citep{1974GAM....12.....T}, which are located between young and old dipoles and are long stable structure. This class of filaments requires surface motions to gradually reconfigure preexisting coronal fields \citep[e.~g.][]{2000ApJ...539..954D}.

In a paper by \cite{2007AA...467.1289R} authors removed all the large-scale flows in order to focus on smaller-scale flow, such as mesogranulation and supergranulation. In this study we have retained the large-scale flows in order to study their influence on the triggering of a filament eruption.

The eruption started around 16:30~UT at the latitude around $-$25\degr{} where the measurements of the horizontal flows based on dopplergram tracking show a modification of the slope in the differential rotation of the plasma. It seems to be a consequence of the presence of a north--south stream along the filament's position, which is easily measured by tracing structures in the dopplergrams.

The observed north--south stream has an amplitude of 30--40~m\,s$^{-1}$. In the sequence of H$\alpha$ images that record 
the filament's evolution, the part of the filament, which is in the north--south stream, is rotated in a direction compatible with the flow direction of the stream. This behavior suggests that the foot-points of the filament are carried by the surface flows. The influence of the stream is strengthened by differential rotation. We should keep in mind that the filament extends from $-$5\degr{} to $-$30\degr{} in latitude and that the northern part of the filament is subjected to a faster rotation than the southern part. The stream, with a contribution of the differential rotation, causes the stretching of the coronal magnetic field in the filament and therefore contributes to destabilisation of the filament. The topology of the north--south stream changed after the filament eruption and the filament almost disappeared.

We have measured the increase of the zonal shear at the site where the filaments eruption begins before the eruption. After the eruption, the shear suddenly decreased. This result suggests that the shear in the zonal component of the flow field is the most important component of the surface flow affecting the stability of the coronal magnetic field, and can lead to its eruption,  which in turn can drive active phenomena such as ribbon flares and CMEs. This evolution of the shear in the flow field is probably related to the re-orientation by 70\degr~ (or 110\degr) of the transverse field after the eruption seen in the daily vector magnetograms obtained with THEMIS \citep{2007AA...467.1289R}. 

All of the features observed in the topology of the horizontal velocity fields at the starting-point site could contribute to destabilise the filament resulting in its eruption. We propose that the stability and evolution of filaments are influenced by  surface flows that carry the footpoints of the filament.

Filaments or prominences are important complex structures of the solar atmosphere. Several mechanisms are probably involved in the creation of filament eruptions: the action of surface motions to create or increase the helicity of the flux rope (\citeauthor{2005ASSL..320...57V}, \citeyear{2005ASSL..320...57V} or \citeauthor{2005AA...433..683R}, \citeyear{2005AA...433..683R}), reconnecting field lines in the corona \citep{2006ApJ...641..577M}, the chirality evolution of the barb \citep{2005ApJ...630L.101S}, etc. The coronal magnetic field is generally thought to be anchored in the photosphere and the flux  transport on the solar surface \citep{1989Sci...245..712W} is the natural mechanism to explain the evolution of filament. Recent models of large scale corona \citep{2006ApJ...641..577M} consider the action of the large-scale surface motions such as differential rotation, meridional flow and supergranular surface diffusion. Our present observation indicates that large-scale surface flows  are structured (not uniform) in the areas of divergence or stream flows, which should be taken into account in the numerical simulations. 
  
A better understanding of the mechanisms which lead to filament eruptions requires simultaneous multi-wavelength and multi-spatial resolution observations (both high-resolu\-tion of the filament and low-resolution of the full Sun) over a wide range of latitudes. Indeed, previous works showed that different phenomena are observed at high resolution such as magnetic 
reconnection close to the starting location of the filament eruption \citep{2007AA...467.1289R}. In this study, observing a
larger area at lower resolution it was shown that at the same location where the filament first begins to erupt. We have found that a steep gradient in differential rotation, a north-south stream, and a shear in the zonal component were present.

%% file: meridional.tex
\section{Meridional magnetic flux transport}
\label{sect:meridional}
\symbolfootnotetext[0]{\hspace*{-7mm} $\star$ This chapter was submitted to Astrophysical Journal Letters as \v{S}vanda,~M., Kosovichev,~A.~G., Zhao,~J., \emph{Speed of meridional flows and magnetic flux transport on the Sun}.}

\subsection{Introduction}

The largest-scale velocity fields on the Sun consist of the differential
rotation and meridional circulation. Both, the differential rotation and meridional circulation, are key
ingredients of the solar dynamo. The differential rotation plays an
important role in generating and strengthening toroidal magnetic
field inside the Sun, while the meridional flow
transports the magnetic flux towards the solar poles resulting in
cyclic polar field reversals \citep[for a recent review, 
see][]{Brandenburg2005}.

The meridional flux transport seems to be an essential agent 
influencing the length, strength and other properties of the solar
magnetic cycles. Generally, the slower the meridional flow, the
longer the next magnetic cycle can be expected. Dynamo models showed
that the turn-round time of the meridional cell is between 17 and 21
years, and that the global dynamo may have some kind of memory lasting
longer than one cycle \citep{2006GeoRL..3305102D}.

The speed of the meridional flow and its variation with the solar cycle measured by local helioseismology 
in the subsurface layers of the Sun are used 
as an input in the recent flux-transport models \citep{2006ApJ...649..498D}. In local helioseismology measurements
(e.g. \citeauthor{2004ApJ...603..776Z} 
\citeyear{2004ApJ...603..776Z}, \citeauthor{2006ApJ...638..576G} \citeyear{2006ApJ...638..576G}), the meridional flow is
derived from a general subsurface flow field by averaging the
north-south component of plasma velocity over longitude for a Carrington
rotation period. The studies reveal that the mean meridional flow
varies with the solar activity cycle. These variations may
significantly affect solar-cycle predictions based on the solar dynamo models, which assume that the
magnetic flux is transported with the mean meridional flow speed
\citep{2006ApJ...649..498D}.

Our goal is to verify this assumption and to investigate the
relationship between the subsurface meridional flows and the flux
transport. In this study, we show that the mean meridional flows
derived from the time-distance helioseismology subsurface flow maps 
are affected 
by strong local flows around active regions in the activity belt. However,
these local flows have much less significant effect on the magnetic flux
transport.

\subsection{Method of measurements}

The magnetic field data were obtained from Kitt Peak synoptic maps of the longitudinal
magnetic field. The magnetic butterfly diagram is continuously
constructed from synoptic magnetic maps measured at National Solar
Observatory by averaging the magnetic flux in longitude at each
latitude position for each solar rotation since 1976.

\begin{figure}[!t]
\centering
\resizebox{0.8\textwidth}{!}{\includegraphics{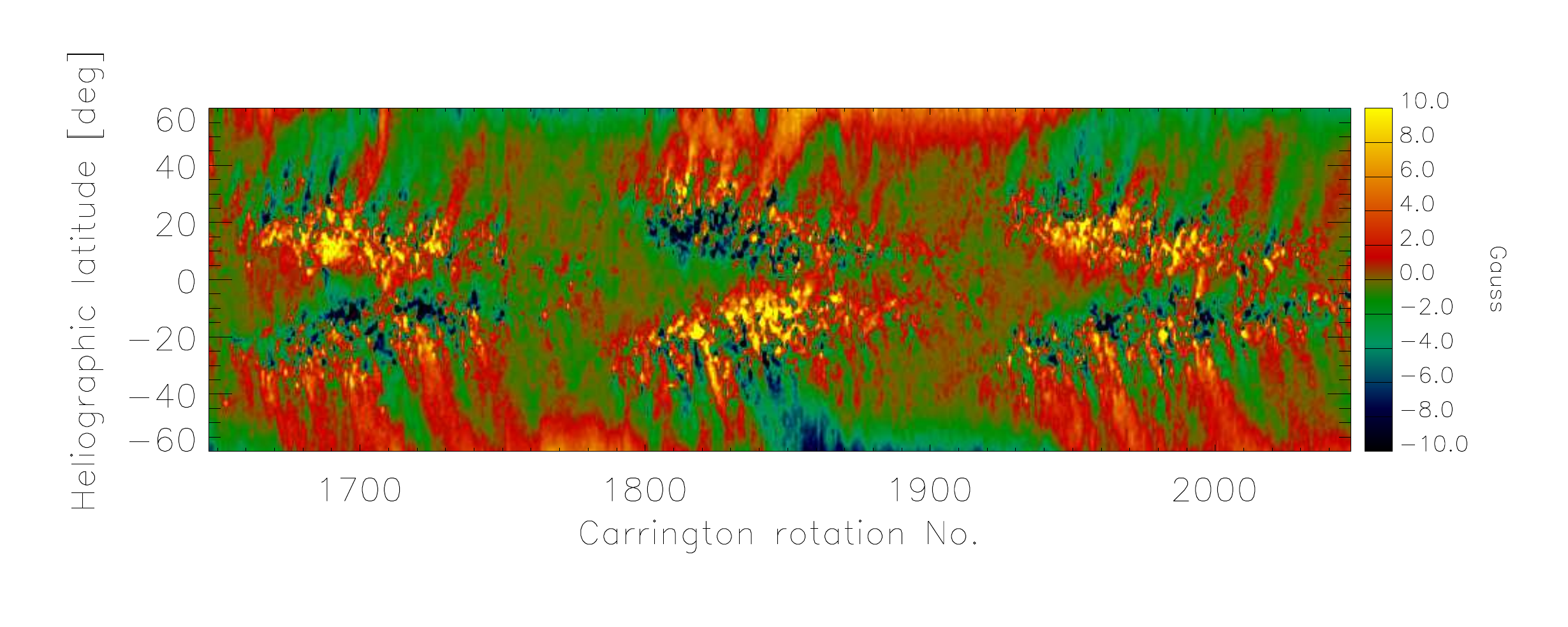}}\\
\resizebox{0.8\textwidth}{!}{\includegraphics{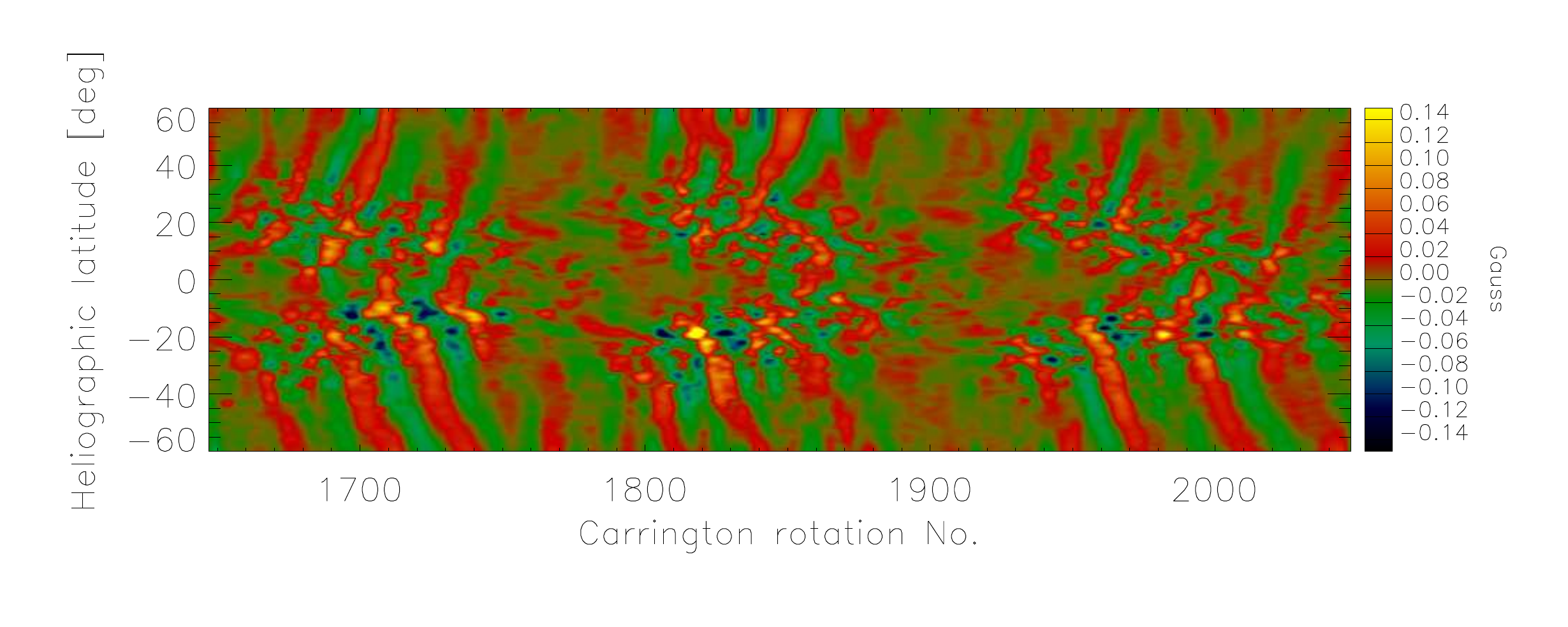}}\\
\resizebox{0.8\textwidth}{!}{\includegraphics{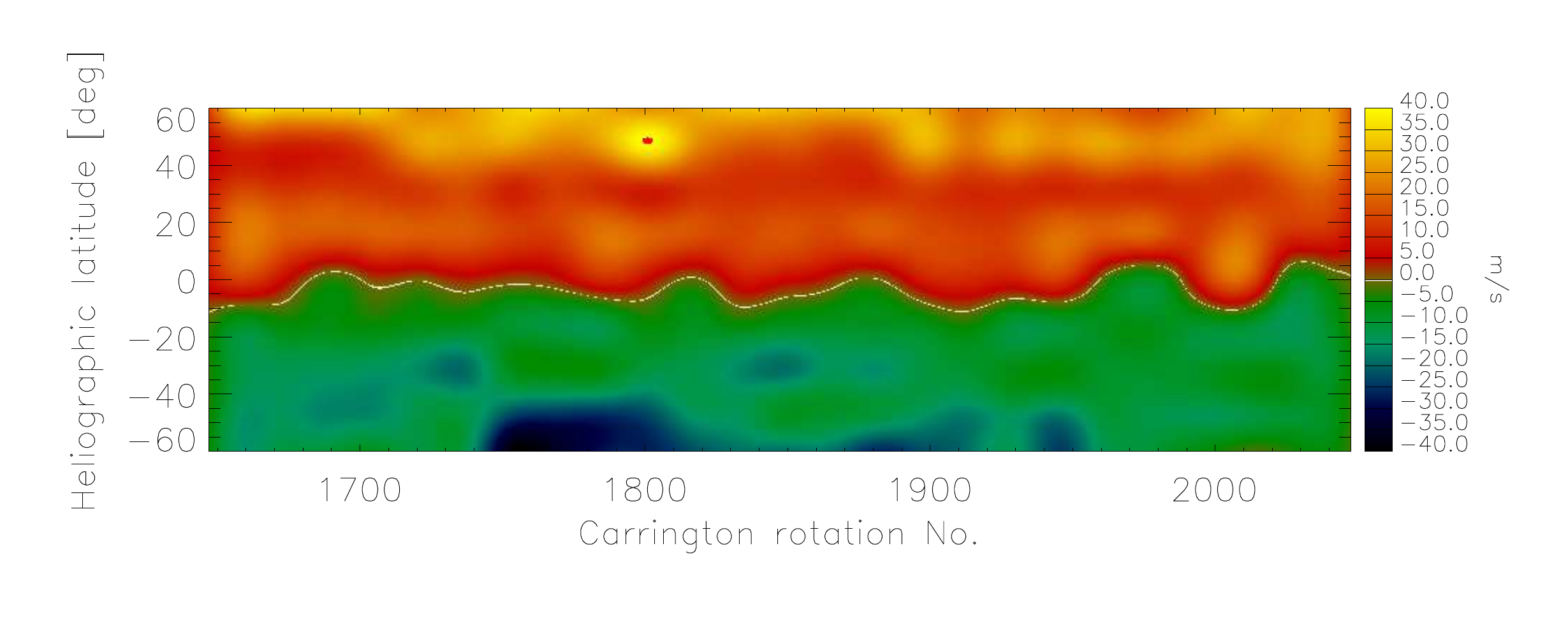}}\\
\caption{\emph{Upper panel} -- The magnetic butterfly diagram. \emph{Middle panel} -- The filtered magnetic butterfly diagram showing enhancements of the flux transport elongated structures. \emph{Bottom panel} -- The measured meridional flux transportation speed (in the south--north direction)  for Carrington rotations 1900--2048.}
\label{fig:butterflymaps}
\end{figure}

At mid-latitudes of the magnetic butterfly diagram (Fig.~\ref{fig:butterflymaps} upper panel), between the active region zone and polar regions,
we clearly see elongated structures corresponding to the
poleward magnetic flux transport. The aim of our method is to measure the slopes of these structures
and to derive the speed of the meridional magnetic flux transport.

In addition to the large-scale structures, the original diagram 
contains small-scale relatively short-living local
magnetic field structures, which appear as a `noise' in the diagram. 
To improve the signal-to-noise ratio for
the magnetic flux structures we applied a frequency band-pass filter for the
frequencies between $1.06\times 10^{-8}$~s$^{-1}$ (period of
1093~days) and $3.17\times 10^{-7}$~s$^{-1}$ (period of 36.5 days).
The filtering procedure is performed separately for each individual
latitudinal cut on the diagram. We tried also other methods of enhancement of structures
and found that they all provide comparable results.

\begin{figure}[!t]
\centering
\resizebox{0.8\textwidth}{!}{\rotatebox{270}{\includegraphics{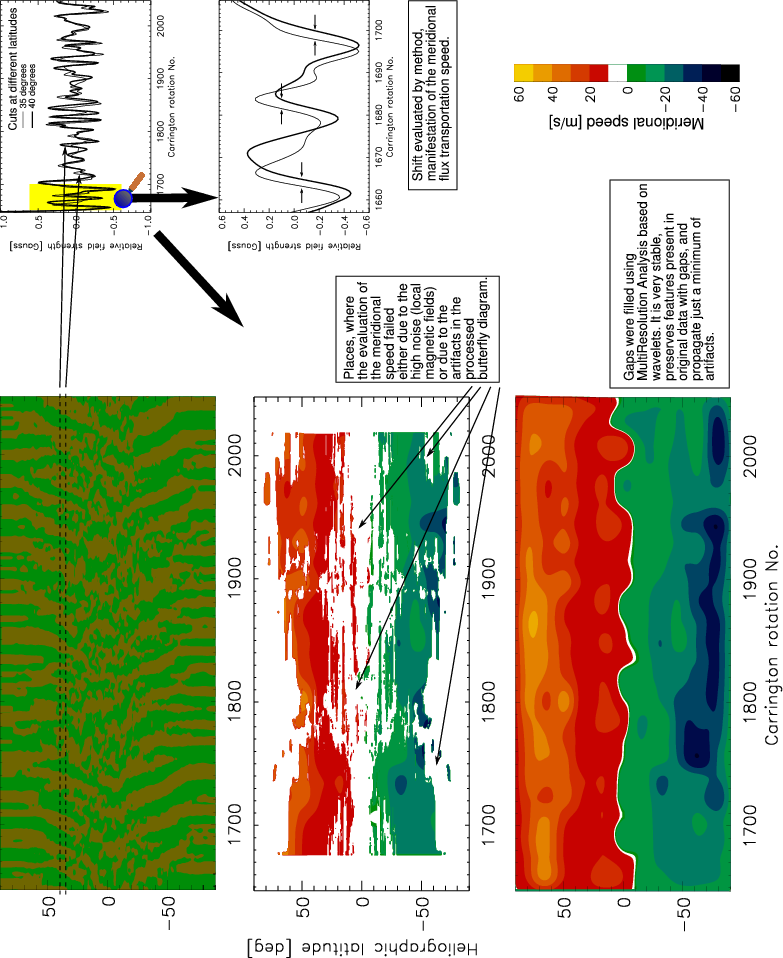}}}
\caption{Cartoon describing the method of measurement of the meridional flux transport speed from the magnetic butterfly diagram.}
\label{fig:butterfly_tracking}
\end{figure}

The difference between  the original butterfly diagram and the filtered
one can be seen in Fig.~\ref{fig:butterflymaps} (upper and middle panel). The flux-transport elongated structures
are more easily visible after the filtering, and therefore more
suitable for analysis.

The meridional flux transport is measured on the basis of 
cross-correlation of two latitude rows in the map. We assume that the
flux-transport structures in two different rows on the same
hemisphere are similar in shape, but their positions are different due to 
the meridional transport. We cross-correlate pairs of rows separated
by heliographic latitude $\Delta b$ in a sliding 
window with the size of 55 Carrington rotations. The edges of
the correlation window are apodized by a smooth function to avoid
the boxcar effects. The extremal position is calculated as a maximum
of the parabolic fit of the set of correlation coefficients of
correlated windows in five discrete displacements. If the
distribution of the correlation coefficients does not have a maximum,
or if the normalized quality of the fit is too low (under 0.8), the meridional velocity in this
pixel is not evaluated. We have chosen $\Delta b=5^\circ$ as the
best tradeoff between the spatial resolution and precision.

To make the procedure more robust, we average the calculated
meridional velocity for five consecutive frames separated by
0.5\degr{} and centered at $\Delta b$ from the studied row. If any
of the speeds in averaged five rows is far out of the expected range
($-$60 to $+$60 m\,s$^{-1}$), then it is not used in the averaging.
From the fit, the accuracy of the measured flow speed is evaluated
and the maximum value of the set of five independent measurements at
different rows is taken.

The same procedure is done with the processed map rotated by
180\,$^\circ$ to avoid any possible preferences in the direction
determination, and both results are averaged. The measured errors
were taken as the maximum value of both independent measurements.

For the ongoing analysis, only the speeds that were measured with
the error lower than 3~m\,s$^{-1}$ were taken into account. This criterion and some
failures of the slope measurement introduce gaps in the data, which
we need to fill. For this purpose we need to determine the best
continuous differentiable field that approximates the data. The
determination of such a field can be done in various ways, but we
wish to avoid possible artifacts. For filling the gaps we used the  
MultiResolution Analysis. It is based on wavelet analysis and we have chosen the
Daubechies wavelet due to its compact support. This property is
important since it minimizes edge effects. Moreover, using
these wavelets also preserves the location of zero-crossing and
maxima of the signal during the analysis. Daubechies are claimed to
be very stable in the noisy environments. For details see
\cite{rieutord07}.

\begin{figure}[!t]
\centering
\resizebox{0.9\textwidth}{!}{\includegraphics{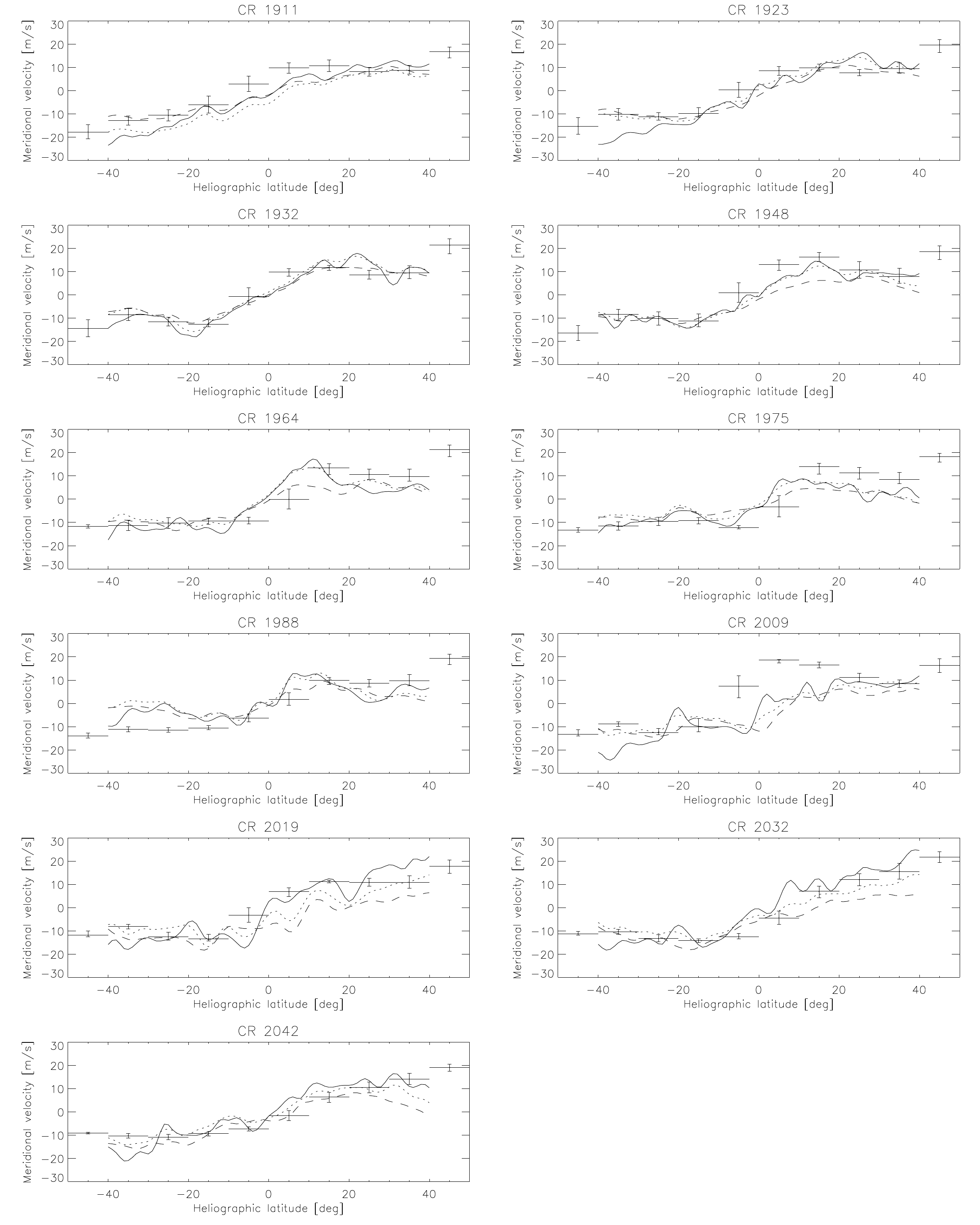}} 
\caption{The longitudinally averaged meridional flow
speed measured for a set of Carrington rotations by time-distance helioseismology. Solid line plots 
the time-distance mean meridional flow at 3--4.5~Mm depth, dotted at 6--9~Mm, and 
dashed at 9--12~Mm. The dots with error-bars represent 
the 10-degree-bin-averaged values of the flux transport speed
derived from the magnetic butterfly diagram (Fig~\ref{fig:butterflymaps}).}
\label{fig:butterfly_rotations}
\end{figure}

For the comparison between the meridional flow obtained from
time-distance helioseismology \citep{2004ApJ...603..776Z} and the
magnetic flux transport from our method, we have calculated the
averaged values of both quantities in bins of 10 heliographic degrees.
For the period of 1996--2006, only eleven Carrington rotations have been evaluated 
(one per year) by time-distance helioseismology using the full-disc \emph{Dynamics
data} from the MDI instrument on SoHO spacecraft. These data are available
only for 60 days per year. We compared the measurements of the magnetic flux 
transport and the meridional flows in those
particular  non-consecutive Carrington rotations. The plots are
displayed in Fig.~\ref{fig:butterfly_rotations}.

\subsection{Results}

The correlation between the magnetic flux and flow speed is high (correlation coefficients are in 
range of 0.7--0.9). We have to keep in mind that while the time-distance meridional circulation profiles
represent the behaviour of the plasma during particular Carrington rotations, the magnetic butterfly diagram tracking
profiles represent the flux transport smoothed over 10 Carrington rotations. Therefore, the agreement
cannot be perfect in principle. To make our results more accurate, the continuous helioseismic data are needed. 

The speeds of the meridional flux transport in the near-equatorial region are less reliable, since the elongated structures in the magnetic butterfly diagram extend from the activity belts toward poles. In the equatorial region, significant part of the measurements were excluded from the analysis due to their large measured error. The gaps were filled using MultiResolution Analysis from well-measured points. Although the results seem reasonable here, their lower reliability has to be kept in mind.

In the minimum of activity (such as CR 1911, 1923, or 2032), the profile of the meridional flux transport speed is very consistent with the mean longitudinally averaged profile of the meridional flow from helioseismology. The best agreement if found for the depth 9--12~Mm. This suggests that the flux transport may be influenced by flows in the deeper layers. 

With increasing magnetic activity in the photosphere of the Sun, the gradient of the mean meridional circulation profile derived from time-distance helioseismology becomes steeper. This is consistent with the results obtained
by numerical simulations by \cite{2004SoPh..220..333B}. The simulations show that with increasing magnetic activity, the 
Maxwell stresses oppose the Reynolds stresses, causing an acceleration of the meridional circulation and deceleration of 
the rotation in low latitudes. Our measurements show that the variations of the slope of the mean meridional flux transport speed in latitute are lower with the progression of the solar cycle. 

\begin{figure}[!t]
\resizebox{!}{3.7cm}{\includegraphics{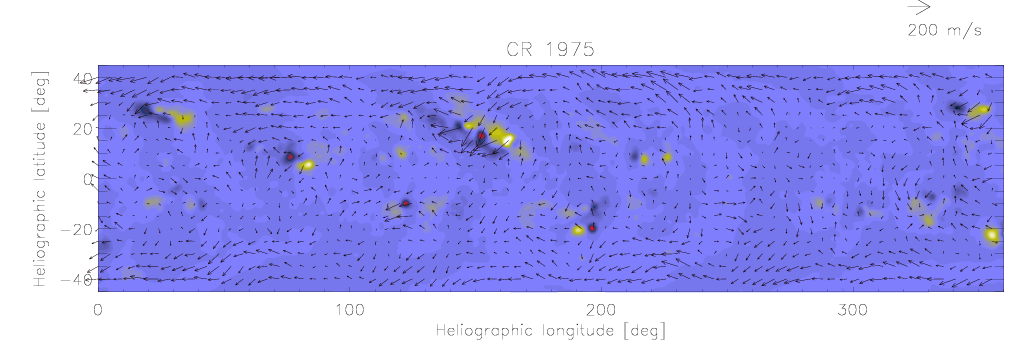}}
\resizebox{!}{3.5cm}{\includegraphics{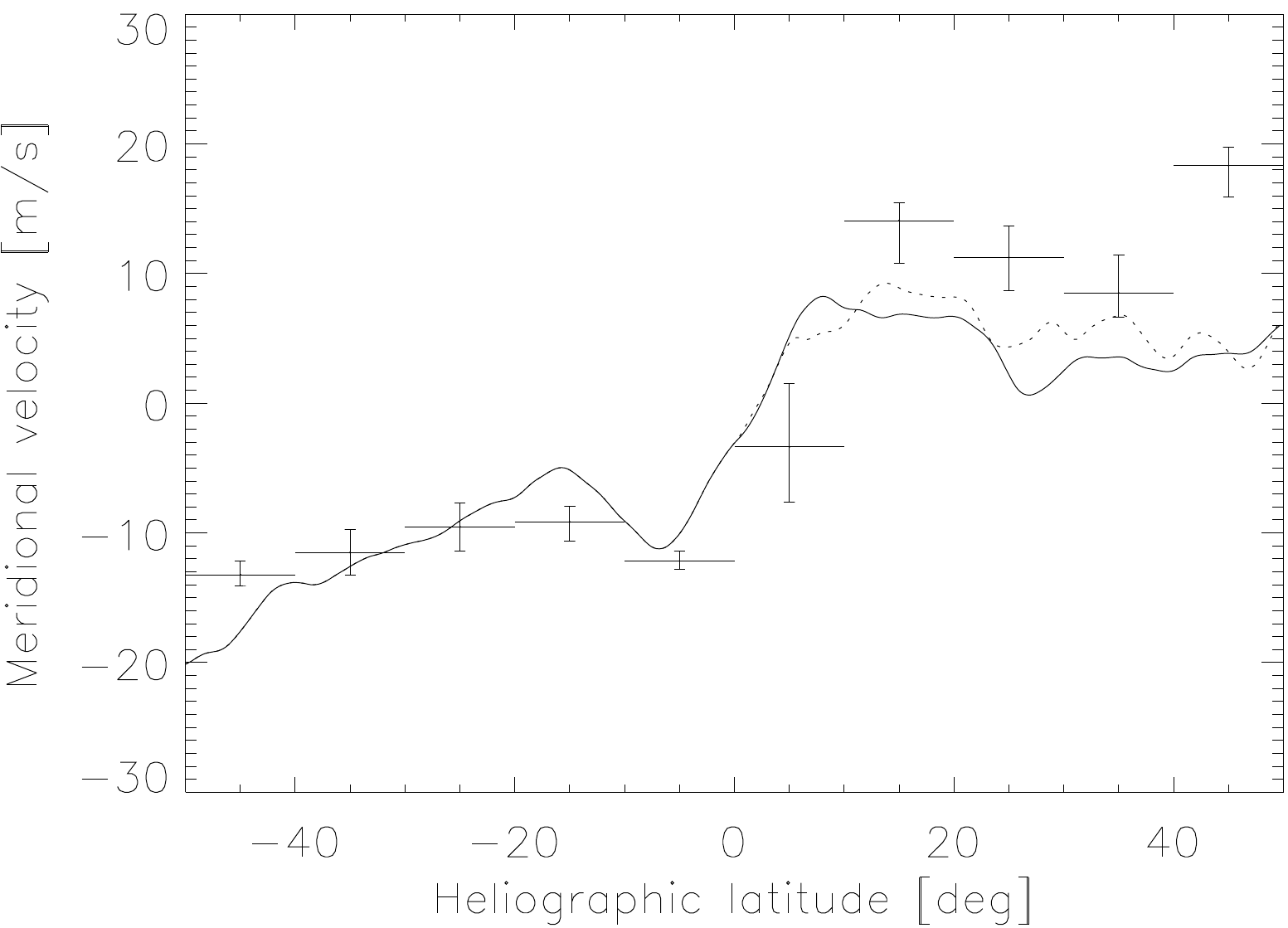}}
\caption{\emph{Left} -- A large-scale flow map at depth 3--4.5~Mm for Carrington rotation No. 1975 (April 2001) and the corresponding MDI magnetogram in the grey-scale background. Large-scale flows towards the equator in the magnetic regions are visible around the large active region. \emph{Right} --  The longitudinally averaged meridional circulation profile for the same Carrington rotation. The southern hemisphere depicts almost no magnetic activity, so the meridional circulation profile obtained by averaging the time-distance flow map (solid line) almost fits the magnetic flux transport profile (points with error-bars) there, while on the northern hemisphere they differ. After masking the magnetic regions on the northern hemisphere, the recalculated profile (dotted line) tends to fit the butterfly tracking one also on the northern hemisphere.} 
\label{fig:butterfly_casestudy}
\end{figure}

When the large-active regions emerge in the activity belt, the flow towards equator is formed on the equatorial side of the magnetic regions (see example of the subsurface flow map in Fig.~\ref{fig:butterfly_casestudy} left). This equatorward flow acts as a counter-cell of the meridional flow (present at the same longitudes as the corresponding magnetic region) and causes a descrease of mean meridional flow amplitude in the activity belt. This behaviour is noticed in all studied cases recorded during eleven non-consecutive solar rotations, for which the \emph{Dynamics data} useful for helioseismic inversion exist. Therefore the formation of the apparent counter-cell seems to be a common property of all large active regions in depths 3--12~Mm. 

Flows in this counter-cell do not influence the magnetic flux transport, which can be demonstrated when the magnetic region is excluded from the synoptic map (Fig.~\ref{fig:butterfly_casestudy} right). The calculated meridional circulation profile is then closer to the profile of the meridional flux transport speed derived from the magnetic butterfly diagram.

\subsection{Conclusions}

We have compared the measurements of meridional speed derived by two
different techniques: by time-distance local helioseismology and by
measuring the flux transport speed using the magnetic butterfly
diagram. We have found that the results agree quite well in
general, but they differ in the regions occupied by local
magnetic fields. The detail flow maps from helioseismology show that this is due to the presence of meridional counter-cells at the
equatorward side of magnetic regions, which influence the time-distance derived meridional
flow profile, but does not influence the magnetic flux transport. 

\begin{figure}[!t]
\centering
\resizebox{0.6\textwidth}{!}{\includegraphics{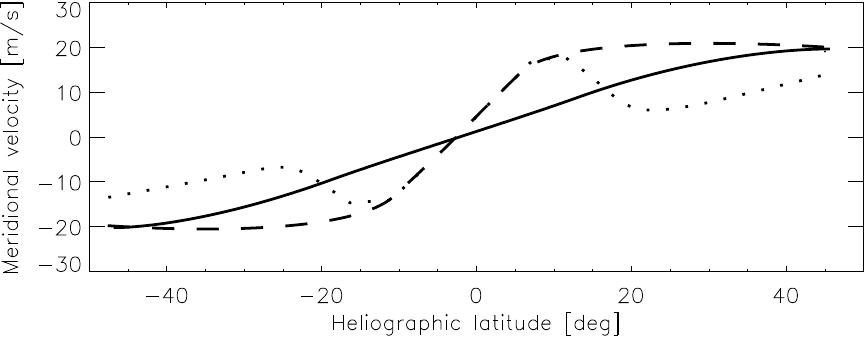}}
\caption{Cartoon of the mean meridional flow derived as the average of the north--south component of helioseismology flow maps. Flow in the periods without significant magnetic activity (solid line), in periods of increased magnetic activity (dashed line) and when the counter-cell in the active regions are present (dotted line). Measured meridional flux transport speed remains more-or-less similar to the solid curve.} 
\label{fig:merflow_model}
\end{figure}

We have studied eleven non-consecutive Carrington rotations covering one solar cycle. The effect of the 
local flows around active regions and especially on their equatorial side is noticed in all the studied 
cases (Fig.~\ref{fig:merflow_model}). Therefore, this behaviour seems to be a common property of the subsurface dynamics around active regions
located in the activity belt. However, we have to keep in mind that
both datasets are not directly comparable, since the time-distance flow maps represent the behaviour of
flows during one Carrington rotation, while the butterfly diagram tracking procedure provide results 
averaged over few Carrington rotations. The more homogenous data for local helioseismology are needed 
to study this effect in more detail. 

The results show that the speed of the magnetic flux transport towards the solar poles may 
significantly deviate from the longitudinally averaged meridional flow speed derived from 
local helioseismology measurements, which are affected by local circulation flows around
active regions in the activity belt. Therefore using the mean meridional flow profile from
helioseismology may bias the results of the dynamo models.

%% file: conclusions.tex
\section{Conclusions and perspectives}

During my Ph.\,D. studies I have developed the method suitable for the measurements of horizontal velocity fields in the solar photosphere, which is based on the local correlation tracking (LCT) method. The method consists of several separated steps covering the data processing procedure, including noise removal and coordinate transformations. The values of free parameters were adjusted using the synthetic data with known properties. They come from the SISOID numerical simulation, developed especially for this purpouse. From the comparison between the model velocity field and the calculated one the calibration curve was determined. At this point I have confirmed the fact that the LCT method underrepresents the magnitudes of the velocity field due to the spatial averaging during the calculation.

I have also compared the results of the proposed method with the results of time-distance local helioseismology, the method, which is considered the most powerfull in the topic of velocity fields measurements. It is shown that both methods reasonably match, which is encouraging for both methods.

The long-term behaviour of the surface flow field was studied. The generally accepted long-term properties were confirmed by the proposed method. The periodic analysis of the mean velocity components in the equatorial region did not show any conclusive results. We have detected few suspicious periods that might be related to the global Rossby waves pattern, but we cannot confirm this idea from the present material.

When a slightly modified version of the method was applied to the data describing the temporal evolution of solar photosphere under the filament in the time before and after its eruption, it was found that the topology of the surface flows have changed significantly. All the measured changes cause a stretch in the place, where the filament eruption started. The measurements support the theory of influence of the coronal magnetic field topology by the photospheric motions.

The study of the meridional flux transport measured from the magnetic butterfly diagram showed that the flux transport speed may significantly differ from the meridional speed of the plasma measured by the time-distance local helioseismology. The results suggest that in the dynamo models, the mean latitudinally averaged meridional flow speed coming from helioseismology cannot be used directly as the quantity describing the meridional flux transport. The longitudinal structure of the meridional flows seems to be very important.

\subsection{Perspectives}

The material obtained during the work related with this thesis provide a great perspective in the future investigations. In the future studies we would like to use it particularly for the investigation of the flows in active regions and of the coupling between the magnetic and velocity fields in magnetic areas. Especially, we will work on possible confirmation of the phenomenon of the dynamical disconnection of the sunspots from their magnetic roots, for which we believe that the current dataset is suitable.
 
By now, we have processed a three-day series of the great active region NOAA~9393, the largest active region of the current solar cycle, which are waiting for the detailed study, in which we would like to incorporate also the three-dimensional magnetic field reconstruction. 

The filament eruption, which was studied in the Section~\ref{sect:filament}, is a perfect material for the non-potential modelling of the coronal magnetic field and its evolution in time. We expect that it will provide a lot of information about the particular coupling of the magnetic field in corona and the horizontal photospheric large-scale flows.

The data coming from the magnetic butterfly diagram tracking analysis are suitable as an input for the flux transport dynamo models. We have started to cooperate with Dr.~Mausumi Dikpati, an author of the well-behaving flux tranport dynamo model, which is able to reproduce the properties of last 12 solar cycles. We expect that the input of the real measured meridional flux tranport speed, which is essential, instead of that measured by local helioseimology will improve the results and the ability of the magnetic activity forecast. The longitudinal structure of the meridional flux transport speed and meridional flows need also a  careful study.

%% file: appendix_a.tex
%
\section{Appendix -- Flows on the stellar surfaces?}
\label{sect:stellar}

\symbolfootnotetext[0]{\hspace*{-7mm} $\star$ This work was done together with Zsolt K\H{o}v\'ari from Konkoly Observatory, Hungary, and Klaus~G.~Strassmeier, Astrophysikalische Institut Potsdam, Germany, with contribution of Michi Weber, Katalin Ol\'ah, Kristi\'an Vida, and Janus Bartus.}

\noindent Large-scale flows play and important role in the dynamics of the surface and sub-surface magnetic field. Their observation and measurements in the photosphere of the Sun bring information that can help theorists to model the processes involved in the solar dynamo. Assuming that the same physical principles like in our Sun take place also in the interior of other stars, it would be interesting also to obtain some information about the behaviour of the plasma in the photosphere of other stars.
 
Recently, the progress in the Doppler imaging method (\citeauthor{1987ApJ...321..496V}, \citeyear{1987ApJ...321..496V} or \citeauthor{1989AA...208..179R}, \citeyear{1989AA...208..179R}) allows the acquisition of time series of the images mapping the evolution and motions of surface features on the stellar surfaces \citep[e.~g.][]{2004AA...417.1047K}. Although the resolution of surface maps is low and they contain lots of noise, they can be attributed to the presence of solar-like active regions caused by the presence of the magnetic field. The large-scale low resolution magnetic maps were used for the measurement of the surface flow on the Sun by the LCT method by \cite{2001SoPh..198..253A}. Following this similarity, we may try to use the LCT method to track the surface features in the series of the Doppler imaging maps to derive the surface flow structure in the photospheres of other stars. 

\subsection{LQ Hydrae}

As the pilot target the solar-like rapid rotator LQ~Hydrae, for which a time-series of consecutive surface Doppler maps exists
\citep{2004AA...417.1047K}, was chosen. LQ~Hya can be regarded as a very young proxy of our Sun. Astrophysical properties for LQ~Hya are adopted from \cite{2004AA...417.1047K} and references therein. Relevant parameters for the present study are summarized in Table~\ref{tab:lqhyaparameters}.

Spot variability on LQ\,Hya has been studied since the discovery of its light variability. \cite{1997AAS..125...11S} and \cite{1999AAS..140...29S} detected rapid spot evolution on time-scales from weeks to months and estimated a possible spot
cycle length of 6--7 years. With recent updates of the photometric data base to a total length of 21 years, \cite{2004AA...417.1047K}
suggested an even shorter cycle length of 3.7 years, in agreement with the fundamental-mode oscillation period of 3.2 years predicted by \cite{2000MNRAS.318.1171K} from a distributed-dynamo model applied to LQ~Hya. The first Doppler image of LQ~Hya was presented by \cite{1993AA...268..671S} and afterwards the star has become a favored target of different Doppler imaging campaigns until nowadays \citep[the early history of Doppler imaging studies for LQ~Hya was summarized by][]{2002AN....323..309S}. The most recent Doppler imaging results were published in \cite{2004AA...417.1047K}, where a time-series analysis was performed. From the cross-correlation of many consecutive Doppler images a weak solar-type surface differential rotation was recovered.

\begin{table}
\begin{center}
\begin{tabular}{ll}
\hline
\hline
Parameter & Value \\
\hline
Classification & K2V\\
Distance (Hipparcos) & $18.35 \pm 0.35$~pc \\
Luminosity, $L$ & $0.270 \pm 0.009\ L_\odot$\\
$\log g$ & $4.0 \pm 0.5$ \\
$T_{\rm eff}$ & $5070 \pm 100$~K\\
$v \sin i$ & $28.0 \pm 1.0$~k\mps \\
Inclination, $i$ & $65\,^\circ \pm 10\,^\circ$\\
Rotation period, $P_{\rm rot}$ & $1.60066 \pm 0.00013$~days\\
Radius, $R$ & $0.97 \pm 0.07\ R_\odot$\\
Age & $\sim 100$~Myr\\
Mass & $\sim 0.8\ M_\odot$\\
\hline
\end{tabular}
\end{center}
\caption{Astrophysical data for LQ Hydrae. Adopted from
\cite{2004AA...417.1047K} and  \cite{1997AA...323L..49P}.}
\label{tab:lqhyaparameters}
\end{table}

For the time-series Doppler images, \cite{2004AA...417.1047K} used 52 spectra taken during a 57 nights long observing run at NSO in
November$-$December 1996. The data covered 35 consecutive rotations of LQ~Hya and were used to recover a series of 28 overlapping Doppler
images with the inversion code \emph{TempMap} (\citeauthor{2000AAS..147..151R}, \citeyear{2000AAS..147..151R} and  \citeauthor{2002AN....323..220R}, \citeyear{2002AN....323..220R}). The Fe\,{\sc i}-6430\AA\ line is more reliable mapping line
for the cool and relatively small $v\sin i$-star LQ~Hya and resulted in maps with higher resolution compared to Ca\,{\sc
i}-6439\AA . Therefore, the iron maps for the present pilot study were adopted. All maps show spot activity preferably at low
latitudes, between $-20^\circ$ and $+50^\circ$, with a concentration in a band around the star centered at $+30^\circ$, and with only occasional evidence for a higher-latitude spot extension. Further information on the data, its reduction and preparations, and on the limits of the Doppler images of LQ~Hya, can be found in the original paper by \cite{2004AA...417.1047K}.

The Doppler surface maps of LQ~Hya were tracked with a fixed rotation period of 1.60066 days (cf. Table~\ref{tab:lqhyaparameters}). Each of the 28 maps was transformed to the Sanson-Flamsteed coordinate system. The advantage of using sinusoidal projection is the conservation of areas. Only then maps are suitable for the application of LCT for the optimization of its best free parameters. Usually it is a trade-off between the sensitivity, i.~e. the spatial resolution, and the signal-to-noise ratio -- or equivalent -- in the maps. In a parameter study, the trade-offs were determined that set the size of the correlation window $FWHM$ and the time lag $\Delta t$ between correlated images and found the best values with $FWHM=10^\circ$ and $\Delta t=1$~frame. These tests showed that the results are not very sensitive to the selection of $FWHM$ but they are sensitive to the selection of $\Delta t$, which means that the difference structures are probably driven by rapid spot evolution not resolved in phase-averaged Doppler images. The mean time lag
between consecutive images is 1.660~days, which equals to 1.037~rotations. Such consecutive images differ by only a single spectrum and differences between them are hardly representative. However, the trends of evolution are preserved in the entire time series.

\begin{figure}[!t]
\resizebox{!}{6.5cm}{\includegraphics{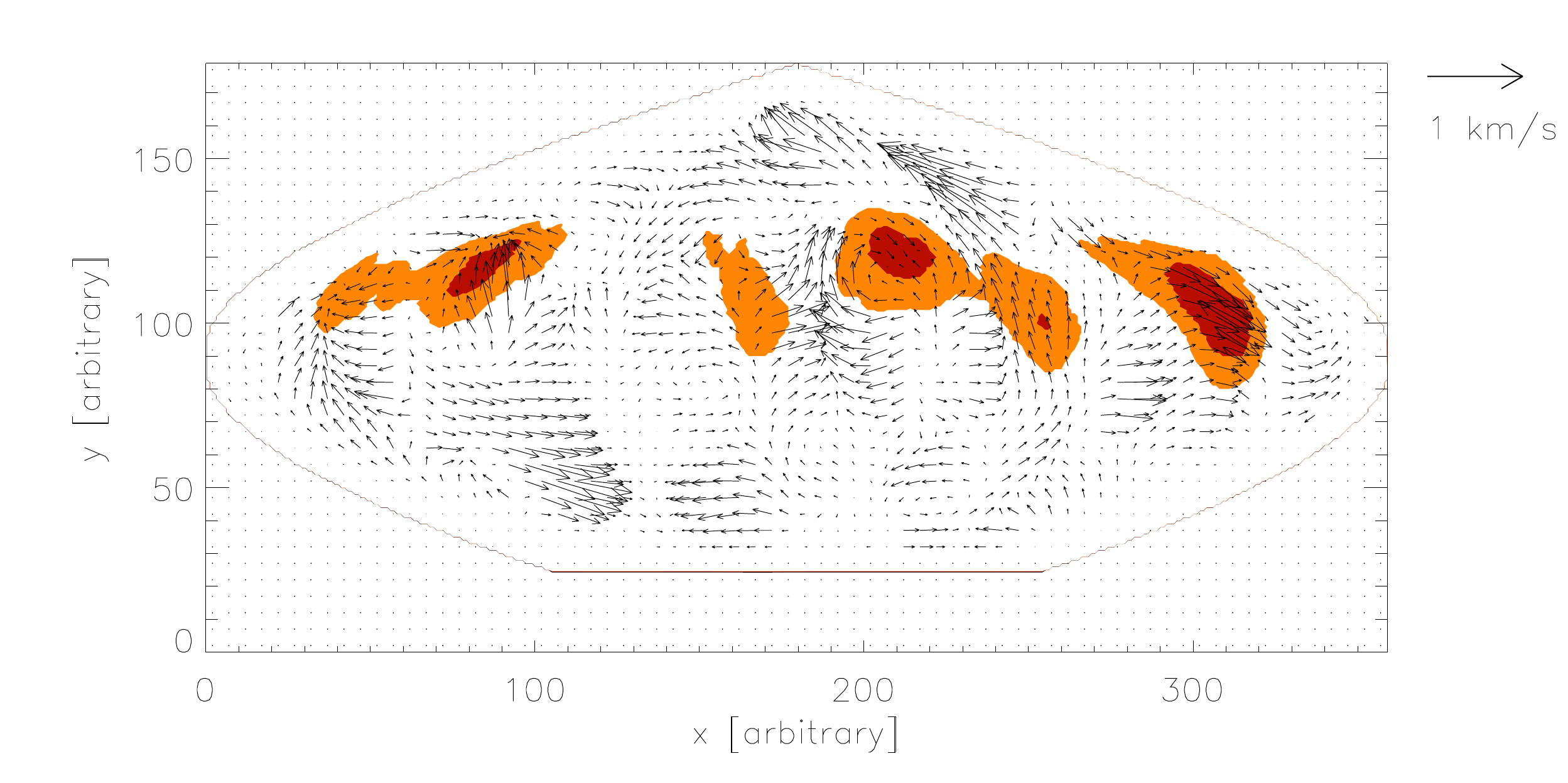}}
\resizebox{!}{6.5cm}{\includegraphics{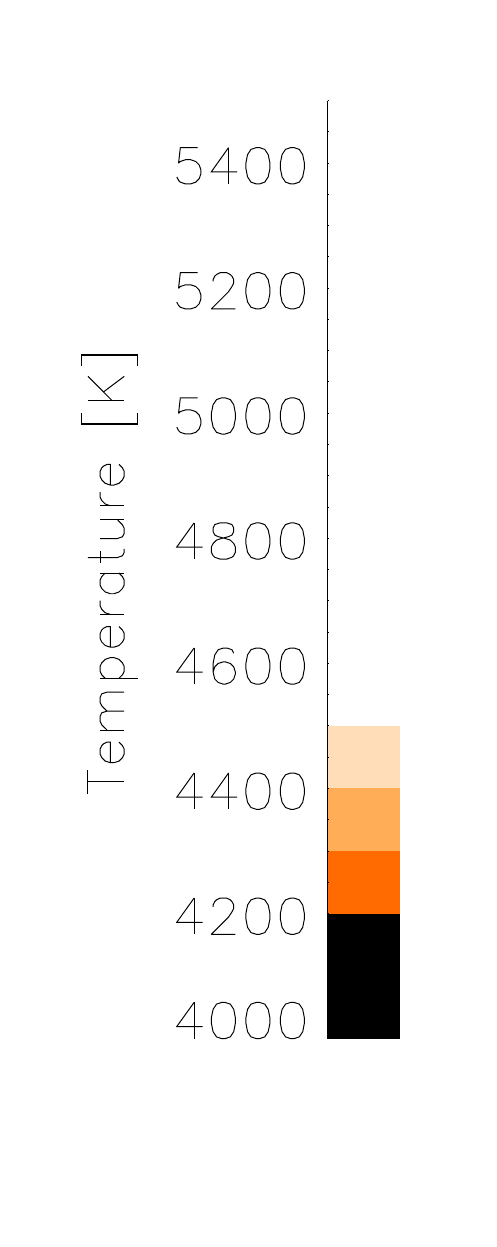}}\\
\resizebox{!}{6.5cm}{\includegraphics{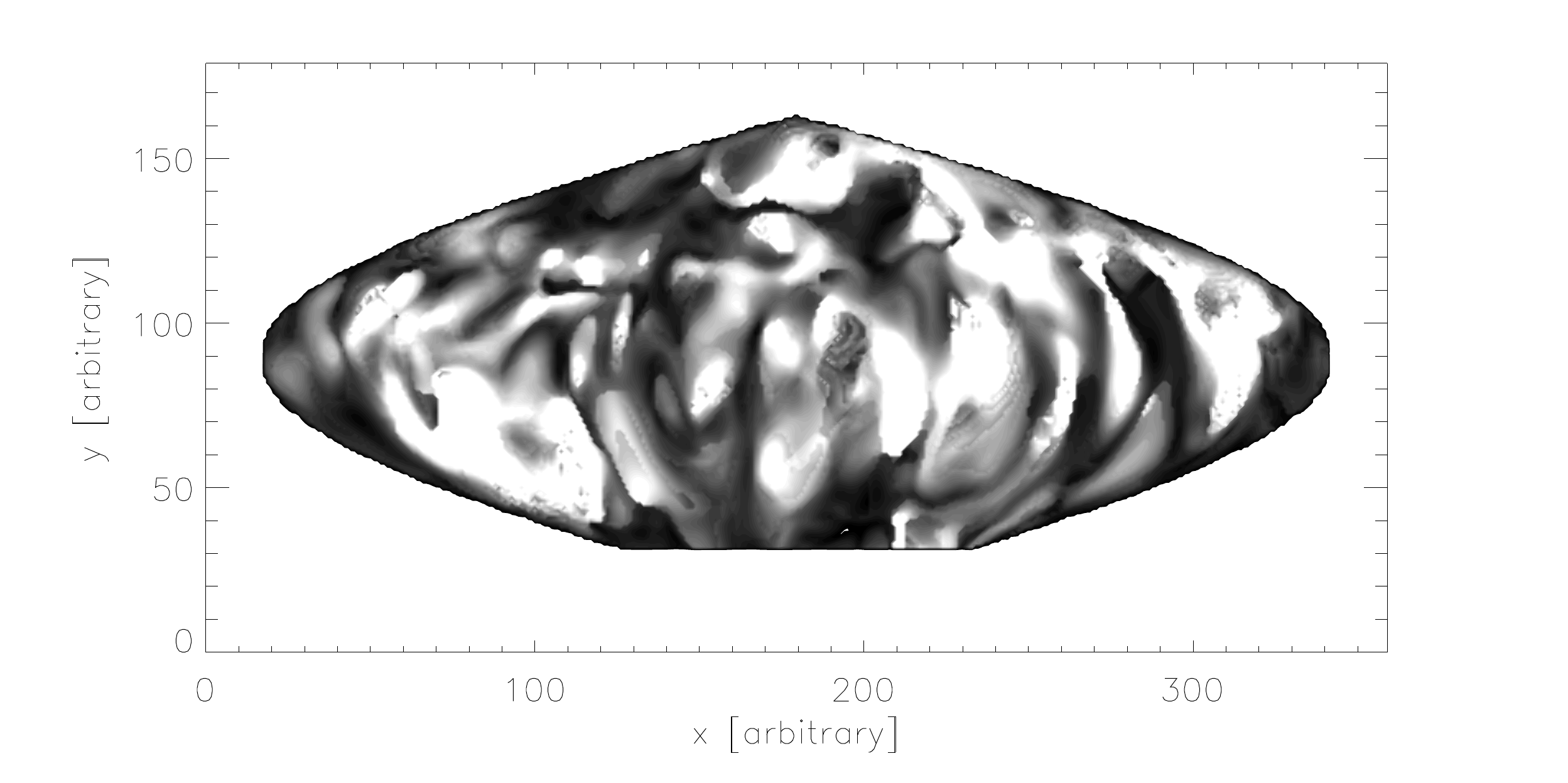}}
\resizebox{!}{6.5cm}{\includegraphics{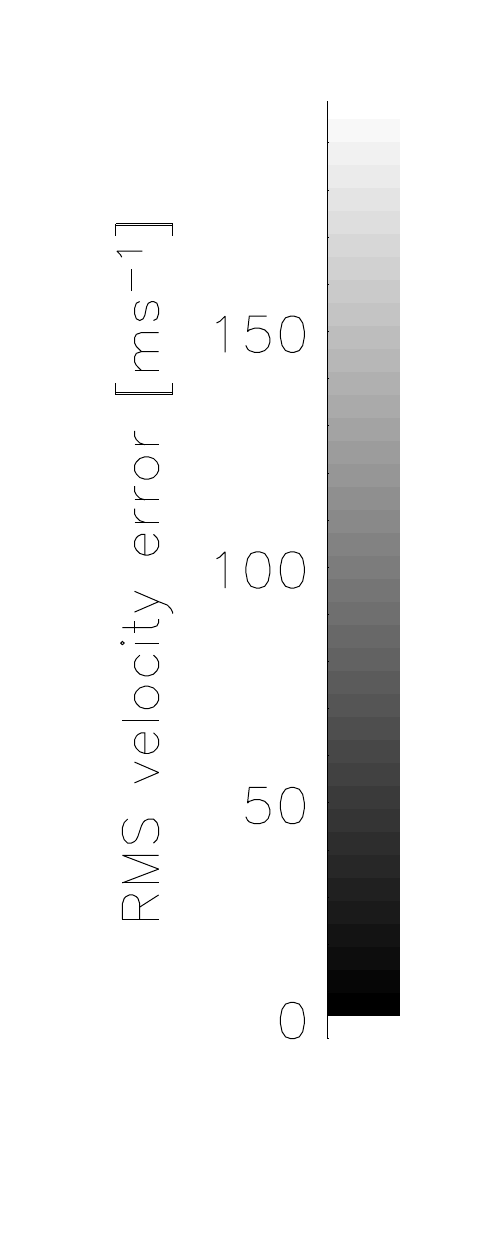}}
\caption{\emph{Top}: Flows in the photosphere of the rapidly rotating K-dwarf LQ~Hya (arrows). The length of the arrows is
scaled by the actual radius of the stellar disc in solar units which is adopted to be $0.97 R_\odot$. The background image is the segmented average temperature map used in the series of 28 Doppler maps to recover the flow field. We have truncated its temperature plotting range to emphasize the positions of its spots. There appear to be two distinct flow structures; a general poleward flow and a
network of convergent flows around spots. \emph{Bottom}: The root-mean-square deviations of individual flow vectors.}
\label{fig:lqhyaflows}
\end{figure}

\subsubsection{Flow field}

For the entire series of 28 maps, 27 correlation pairs were calculated and averaged with equal weight. The resulting difference network is then interpreted as the surface flow field of LQ~Hya and is shown as a vector field in Fig.~\ref{fig:lqhyaflows}.
The average Doppler image of the series of 28 frames is used as the reference frame and is plotted as a background. The
calculated root-mean-square error of individual vectors is plotted in the bottom panel of the same figure. The vectors in
Fig.~\ref{fig:lqhyaflows} indicate that the large-scale flow network is possibly related to the spatially resolved surface structures,
which is expected from solar analogy. The amplitude of the velocity vectors of typically several hundreds of \mps{} also seems
reasonable. Flows of the order of 100~\mps{} are observed in the photosphere of the Sun. 

Fig.~\ref{fig:lqhyaflows} is indicative of the existence of convergent flows towards the starspot regions. Again, this is consistent with the situation in the solar photosphere, where convergent flows towards active regions are seen in the results obtained by local
helioseismology \citep[e.~g.][]{2004ApJ...603..776Z}. However, our usage of indirect Doppler maps of low temporal resolution incorporates that these convergent flows may not be uniquely interpretable as velocity fields and could also be due to the waxing or waning of spots. Because of the low spatial resolution of indirect stellar surface maps one cannot distinguish between a single large spot or a group of smaller spots. Consequently, we could have recorded a systematic migration pattern of small spots towards the center of an activity herd, e.~g. in a giant convection cell. Nevertheless, it presents the first evidence and stimulates further study.

\begin{figure}
\resizebox{0.5\textwidth}{!}{\includegraphics{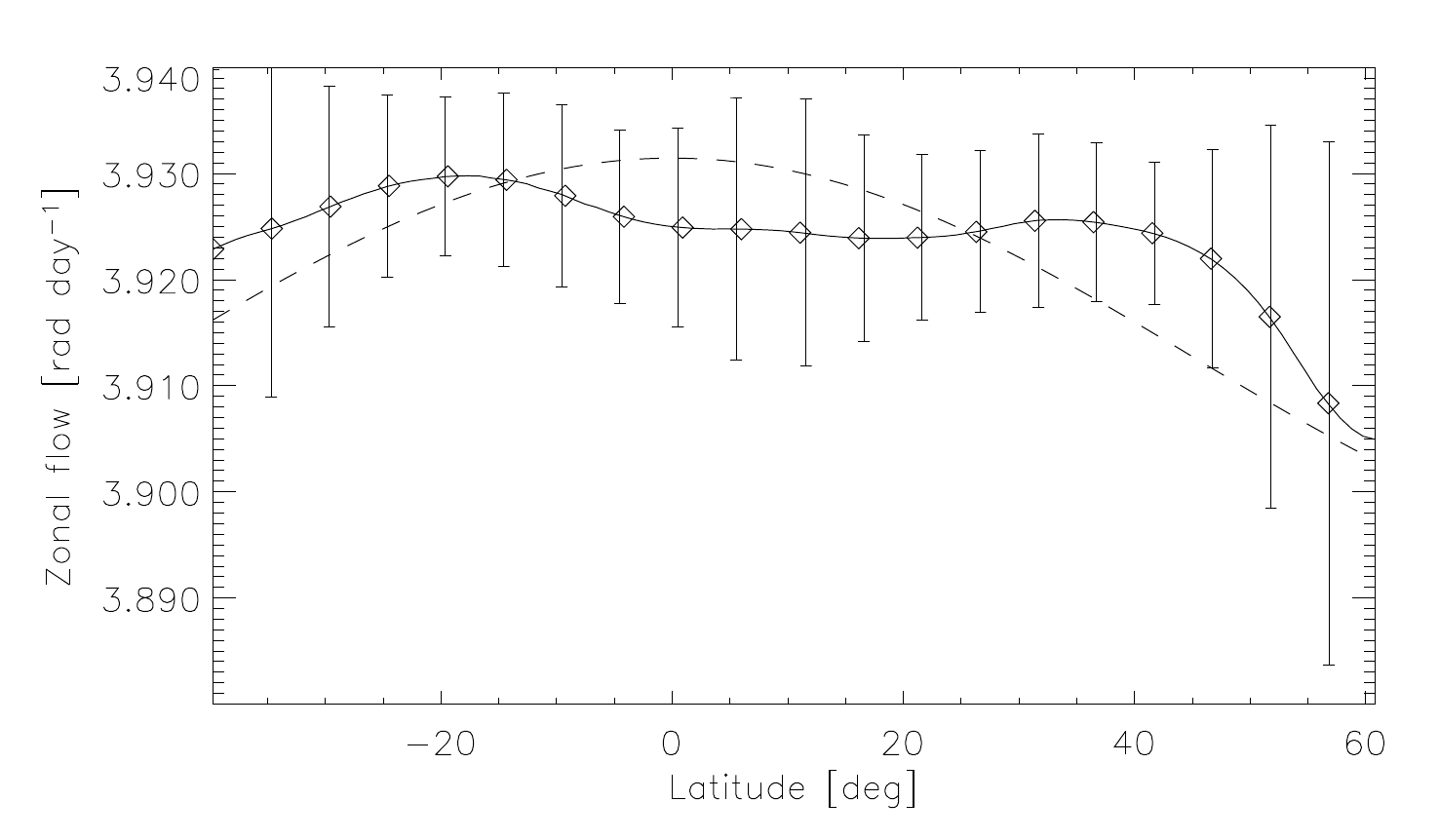}}
\resizebox{0.5\textwidth}{!}{\includegraphics{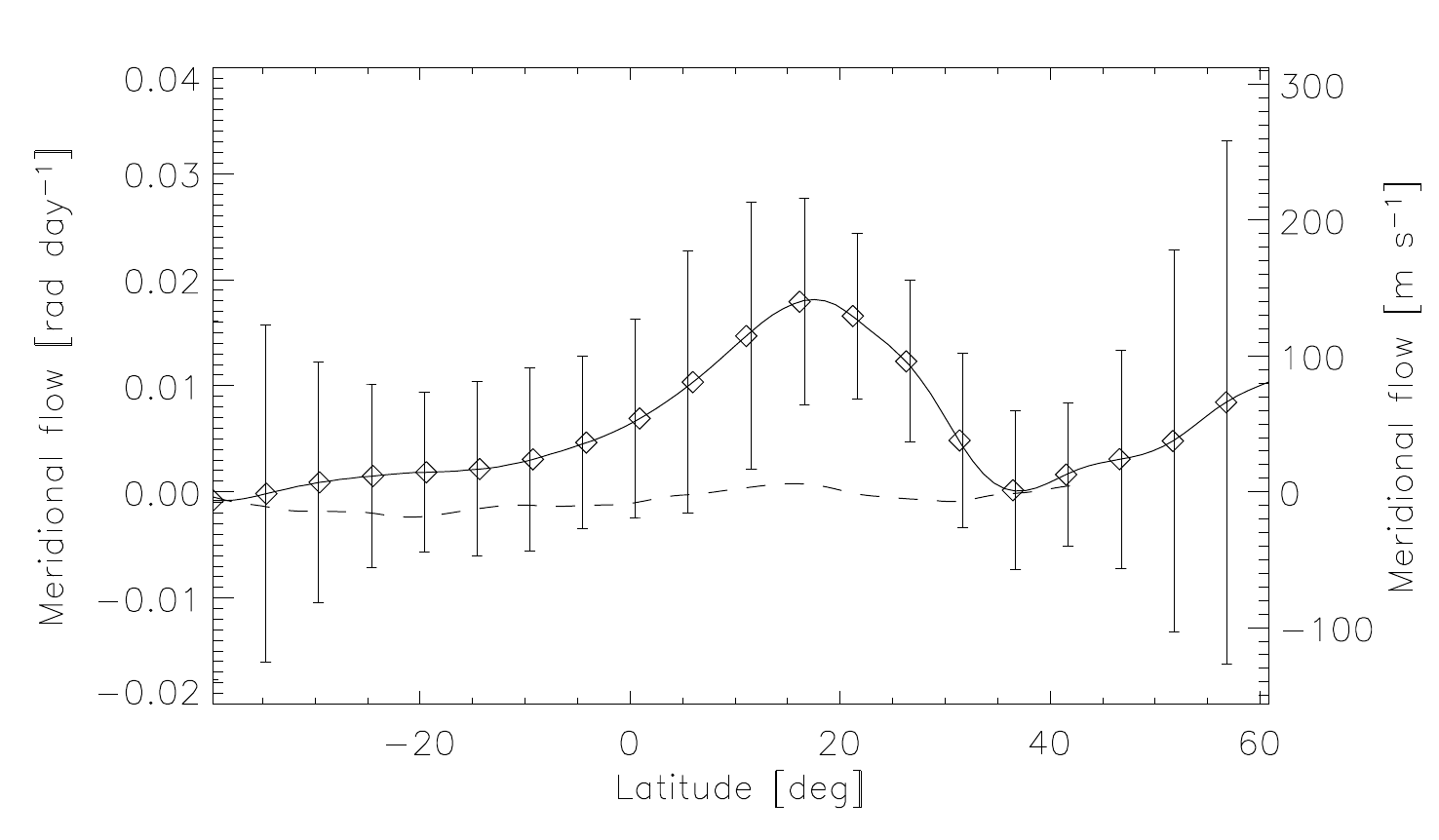}}
\caption{\emph{Left}: Mean azimuthal component (full line) of the flow field from Fig.~\ref{fig:lqhyaflows}. The profile shows the signature of a solar-like differential rotation with higher angular velocity at the equator with respect to the poles. The best fit $\sin^2\varphi$ differential-rotation law is overplotted as a dashed line. \emph{Right}: Mean meridional component of the flow field in Fig.~\ref{fig:lqhyaflows}. A northward flow prevailed on almost the entire stellar disc during the imaging period in 1996. As a
comparison, the solar meridional flow is over-plotted as a dashed line.} 
\label{fig:lqhyamean}
\end{figure}

\subsubsection{Mean flows}

The average azimuthal velocity profile is displayed in Fig.~\ref{fig:lqhyamean} in the left panel. The curve can be fitted by a solar-like
differential-rotation law of 
\begin{equation}
\Omega (\varphi) = \Omega_{\rm eq}-\Delta\Omega \sin^2 \varphi,
\end{equation}
where $\Omega (\varphi)$ is the angular velocity at a given latitude $\varphi$, $\Omega_{\rm eq}$ is the equatorial rotation and $\Delta\Omega$ is the rotational shear between the equator and the poles. The best fit is obtained with $\Omega_{\rm eq}=3.928
\pm 0.001\ \rm rad\ day^{-1}$ and $\Delta\Omega=0.019 \pm 0.002\ \rm rad\ day^{-1}$, in accordance with a lap time of $\sim 330$~days, i.~e. the time the equator needs to lap the pole by one full rotation. As expected, the equatorial rotation rate is in
perfect agreement with the phasing period of the Doppler images. The surface shear is consistent with that found by \emph{ACCORD} 
(Average Cross-CORrelation of contiguous Doppler images) technique in the study of \cite{2004AA...417.1047K}:  $\Delta\Omega=0.022 \pm 0.008\ \rm rad\ day^{-1}$. After all, the two results from the two different methods confirm the principal reliability for both methods.

The average meridional flow profile is displayed in Fig.~\ref{fig:lqhyamean} in the right panel. A poleward flow pattern seems to
prevail for almost the entire stellar surface. The velocity peak near a latitude of +20\degr\ in Fig.~\ref{fig:lqhyamean} right is due
to the poleward components of the converging flows around the many low-latitude spots. The underlying global meridional component may
likely just be seen above latitudes of, say, +40\degr{}. Fig.~\ref{fig:lqhyamean} left indicates a peak velocity of 0.01~rad/day (80~\mps{}) at a latitude of above 60\degr{}.

Rapid poleward migration of spots was first recognized by \cite{1999ApJS..121..547V} on the sub-giant V711~Tau (HR~1099), and later verified through time-series Doppler imaging by \cite{2000AA...354..537S}. In both studies it was associated with the anti-solar differential rotation, i.~e. the poles rotating faster than the equator. This was also seen on another giant, HD~31993 \citep{2003AA...408.1103S}, and the issue was discussed recently by \cite{2006IAUJD...8E..30W} suggesting that the stronger the
anti-solar differential rotation, the faster the meridional flow towards the poles. On the Sun the meridional-flow pattern is
generally poleward and symmetric to the equator on both hemispheres. At that point, it has to be noted that Doppler reconstructions
are more reliable near the visible polar hemisphere and become less reliable approaching the invisible part of the star.
Therefore, detecting a solar-like north-south symmetry from Doppler images is beyond reason. On the other hand, variability of
the differential rotation of active stars, including LQ~Hya, is found on a timescale of few years by \cite{2006IAUJD...8E..73P}.

Possibly connected, meridional flows on LQ~Hya may also exhibit temporal fluctuations, just like on the Sun where the meridional
flow is also observed to change its direction \citep[e.~g.][]{1988SoPh..117..291U}. Therefore, the findings should be seen as a snapshot of an always changing pattern. More data in a further study will hopefully show more details on the general properties of the meridional flow of LQ~Hya.

\subsubsection{Estimation of noise and sensitivity to free parameters}

Since the flow map in Fig.~\ref{fig:lqhyaflows} is the average of 27 correlated pairs of frames, it is important to know the
sensitivity to the selection of frames. The flow map from even- and odd-numbered frames was separately calculated, i.~e. with time
steps $\Delta t$ of two frames. The flow maps from these two data sets are very similar, their median root-mean-square error for the
whole map is only 10~\mps.

In the data set, two neighboring Doppler maps are not independent, they always differ by one spectrum (and overlap with usually seven
spectra). As another test of reliability, the flow map using just the fully independent Doppler maps was calculated, i.~e. running the calculations with $\Delta t=8$~frames. In detail the flow map obtained this way is less reliable and more noisy. Although the surface features changed during the eight rotations covered, the median root-mean-square of difference of this map and the map calculated with $\Delta t=1$ is still approximately only 50~\mps.

\subsubsection{Discussion}

This is a very preliminary study, which shows that the application of LCT on the Doppler imaging data might work. It is clear from the root-mean-square errors displayed in Fig.~\ref{fig:lqhyaflows} in the bottom part that the results are quite noisy. At this moment, the calculated flow field cannot be considered as reliable. There are a few points, at which the technique could be significantly improved.

\begin{itemize}
\item The Doppler images used in the pilot study are reconstructed on the basis of only one spectral line. It would be more reliable to use during the inversions as many spectral lines as possible. The new version of \emph{TempMap} code, which is in development, should enable this and therefore produce more reliable and less noisy temperature maps.
\item The used LCT code is the same as used for solar studies. Its use does not completely satisfy the assumption, under which it is considered reliable. For example the size of the correlation window is not larger than the tracers -- starspots -- but it is two or three times smaller. Although it has been tested on solar data that when using a sufficient averaging, also smaller \emph{FWHM}s can be used, it is not clear enough, how this noise influences the calculated flow maps based on Doppler temperature maps.
\item The code does not take into account that temperature maps cover the whole stellar surface, not just the visible hemisphere. Therefore the whole code should be overwritten for this purpose.
\end{itemize}

\noindent It is shown that an application of LCT technique on Doppler imaging maps provides mean flows comparable to those measured by different techniques. Therefore it can be concluded that the results are not completely arbitrary, however they may be arbitrary in details. This topic of research would be very important if it would be able to produce reliable surface flow maps that could be used as an observational input to the numerical simulations of global dynamo action. 

\subsection{UZ Librae and $\mathbf\sigma$ Geminorum}

\symbolfootnotetext[0]{\hspace*{-7mm} $\star$ The results from this section were presented at 5th Potsdam Thinkshop in June 2007 as parts of contributions \emph{K. Vida, Zs.~K\"{o}v\'ari, M.~\v{S}vanda, K.~Ol\'ah, J.~Bartus and K.~G.~Strassmeier: Differential rotation and surface flow pattern on UZ Librae} and \emph{Zs.~K\H{o}v\'ari, J.~Bartus, M.~\v{S}vanda, K.~Vida, K.~G.~Strassmeier, K.~Ol\'ah and E.~Forg\'acs-Dajka: Surface velocity network with anti-solar differential rotation on the active K-giant $\sigma$~Geminorum}.} 

\begin{figure}[!t]
\resizebox{!}{6.5cm}{\includegraphics{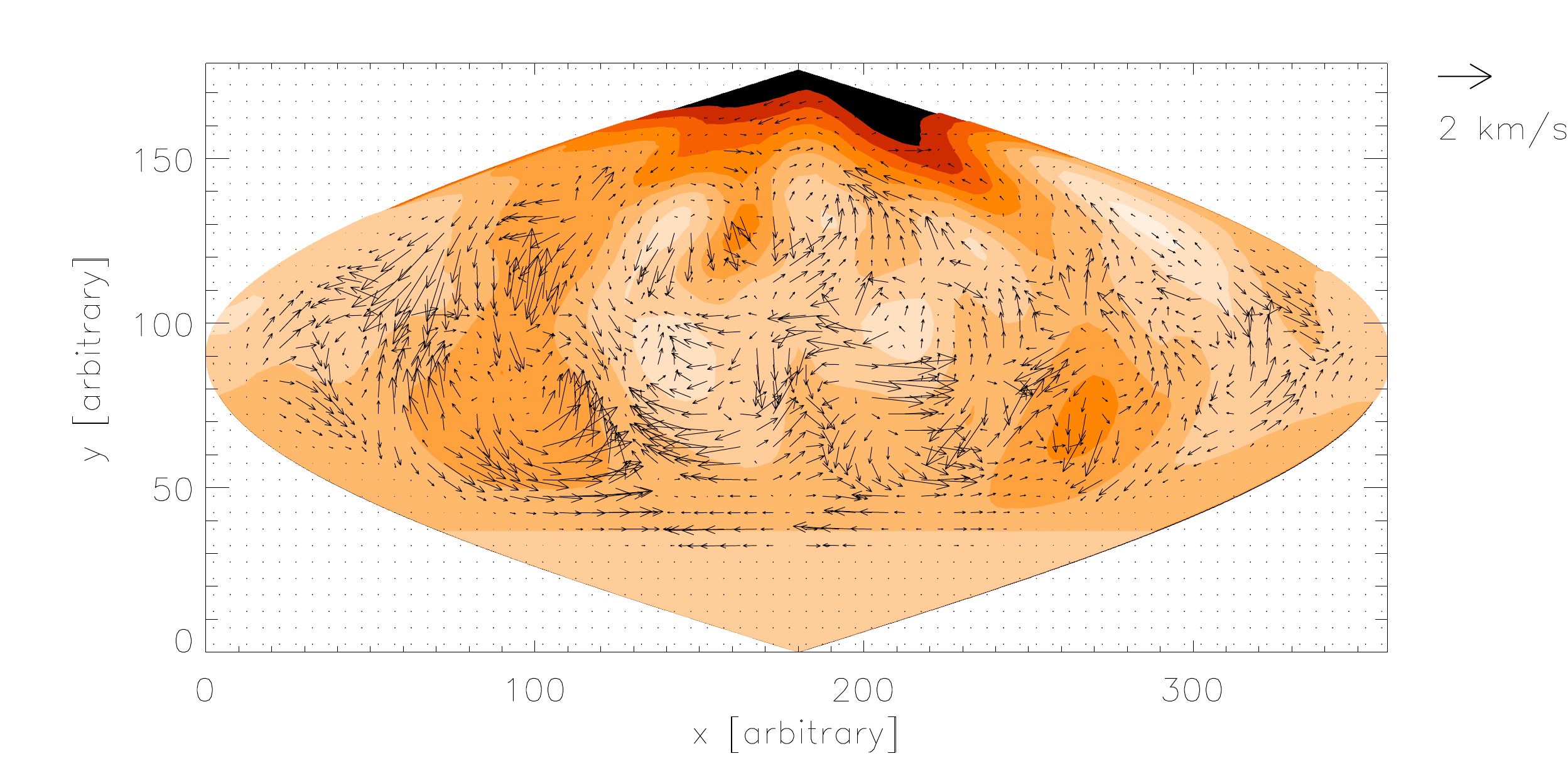}}
\resizebox{!}{6.5cm}{\includegraphics{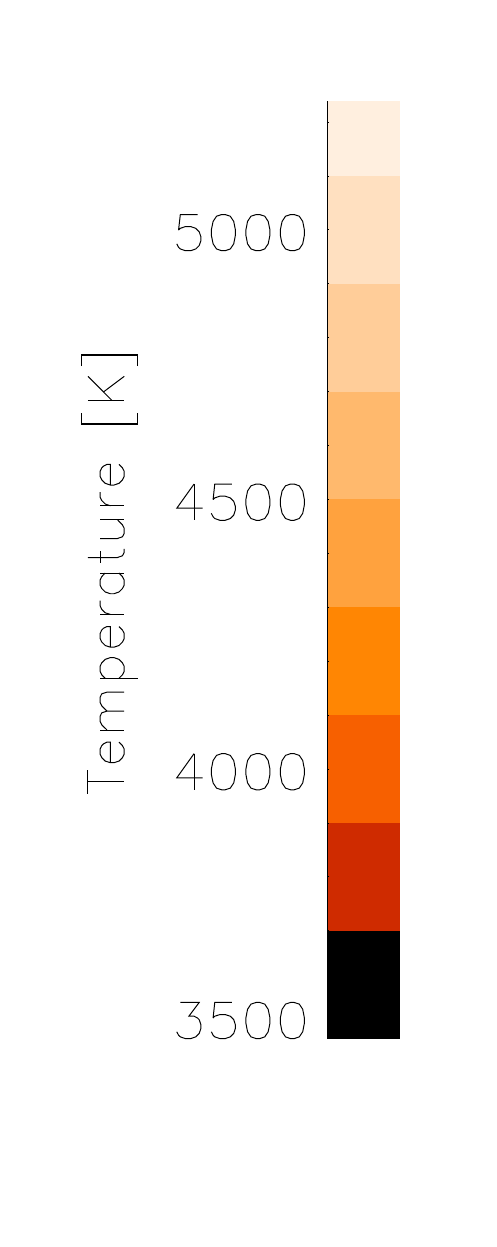}}\\
\caption{Surface flow pattern on UZ Librae, K0III star.}
\label{fig:uzlibflows}
\end{figure}

\begin{figure}
\resizebox{0.5\textwidth}{!}{\includegraphics{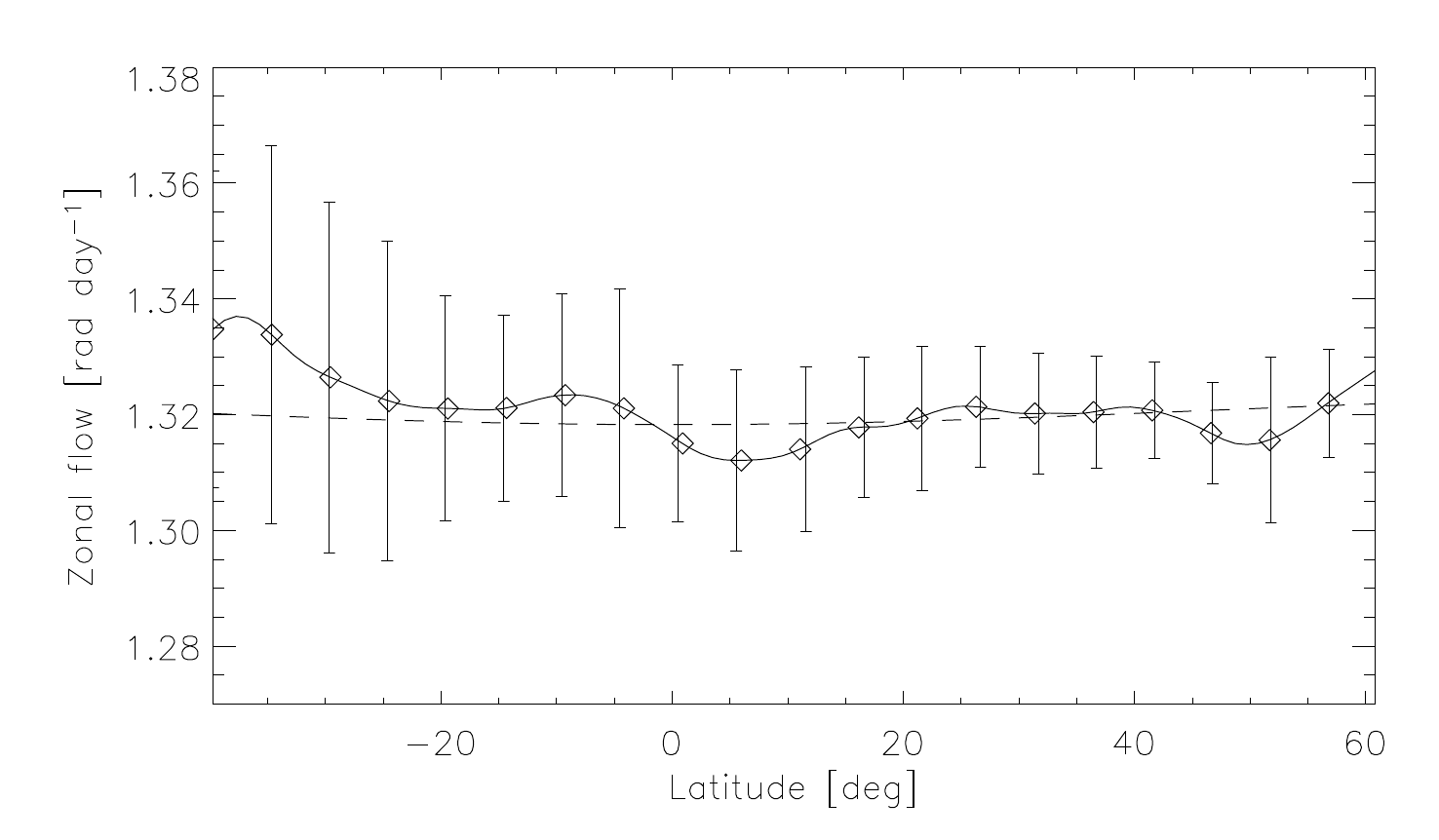}}
\resizebox{0.5\textwidth}{!}{\includegraphics{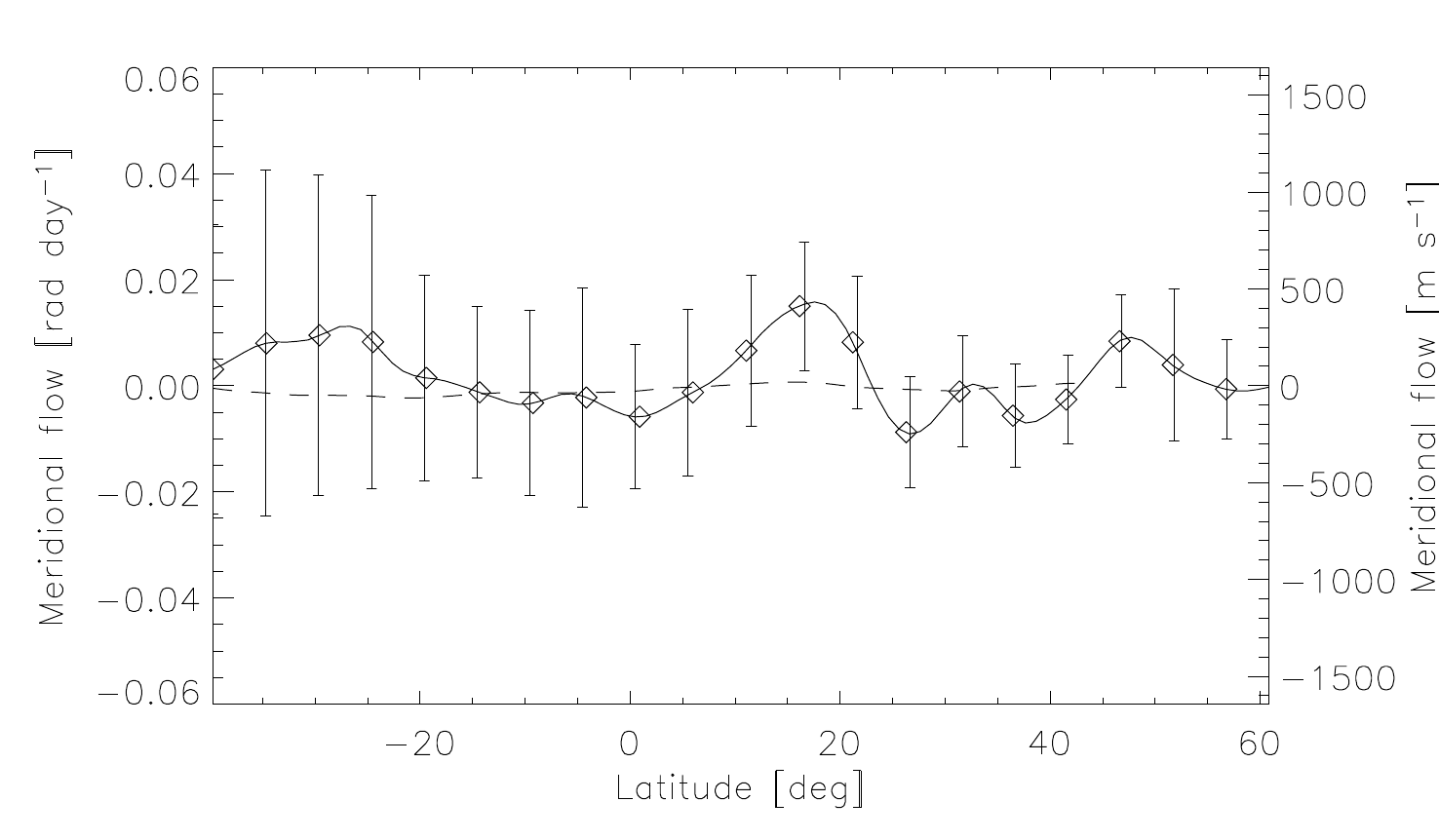}}
\caption{Differential rotation (\emph{left}) and its parabolic fit and meridional circulation of UZ~Lib and the solar meridional circulation (\emph{right}).} 
\label{fig:uzlibmean}
\end{figure}

UZ~Lib is a K0III RS~CVn-type close binary variable, 3.4-times bigger than the Sun. The star is in detail described in the catalogue of chromospherically active binary stars by \cite{1993AAS..100..173S}. The period of rotation of the primary is the same as the orbital period of the system ($P_{\rm rot}=P_{\rm orb}=4.768241$~days) with $v\sin i = 67\pm 1$~km\,s$^{-1}$, $i=50\,^\circ \pm 10\,^\circ$, and photospheric temperature $T_{\rm eff}=4800$~K. In 1998, 11 consecutive Doppler images were obtained and now used for the recovering of the surface flows pattern. Each 6th frame was reconstructed from the independent set of spectral profiles in Fe\,{\sc i}-6393~\AA, Ca\,{\sc i}-6439~\AA, and Fe\,{\sc i}-6411~\AA{} lines; separately reconstructed images in each line were simply combined to get a better signal-to-noise ratio. The resulting flow map is displayed in Fig.~\ref{fig:uzlibflows} and a calculated differential rotation profile and meridional flow in Fig.~\ref{fig:uzlibmean}. For the meridional flow there are no convincing results, the error-bars are much larger than the mean value in each sampling point. The differential rotation is inconclusively anti-solar with parameters of $\Omega_{\rm eq}=1.318 \pm 0.002\ \rm rad\ day^{-1}$ and $\Delta\Omega=-0.005 \pm 0.006\ \rm rad\ day^{-1}$. A stronger anti-solar rotation is found by ACCORD analysis of the same data set. 

For the calculation, just 5 maps were averaged, so that the signal-to-noise ratio is much lower than in the case of LQ~Hya. Thus, the calculated maps are even less reliable, which depict itself in large errors of the differential rotation and meridional flow measurements. 

The star $\sigma$~Gem is a K1III RS~CVn-type long-period binary 12.3-times bigger than the Sun. As UZ~Lib, $\sigma$~Gem is also locked ($P_{\rm rot}=P_{\rm orb}=19.604$~days) with $v\sin i = 27\pm 1$~km\,s$^{-1}$, $i=60\,^\circ \pm 15\,^\circ$, and photospheric temperature $T_{\rm eff}=4630$~K. Together 6 consecutive temperature maps were taken in 1997--1998 covering 3.6 rotations of the primary component \citep{2001AA...373..199K}. Each second map was obtained using independent set of line profiles in spectral lines Ca\,{\sc i}-6439~\AA{} and Fe\,{\sc i}-6430~\AA{} and combined together to decrease statistical uncertainties of individual Doppler reconstructions. 

\begin{figure}[!t]
\resizebox{!}{6.5cm}{\includegraphics{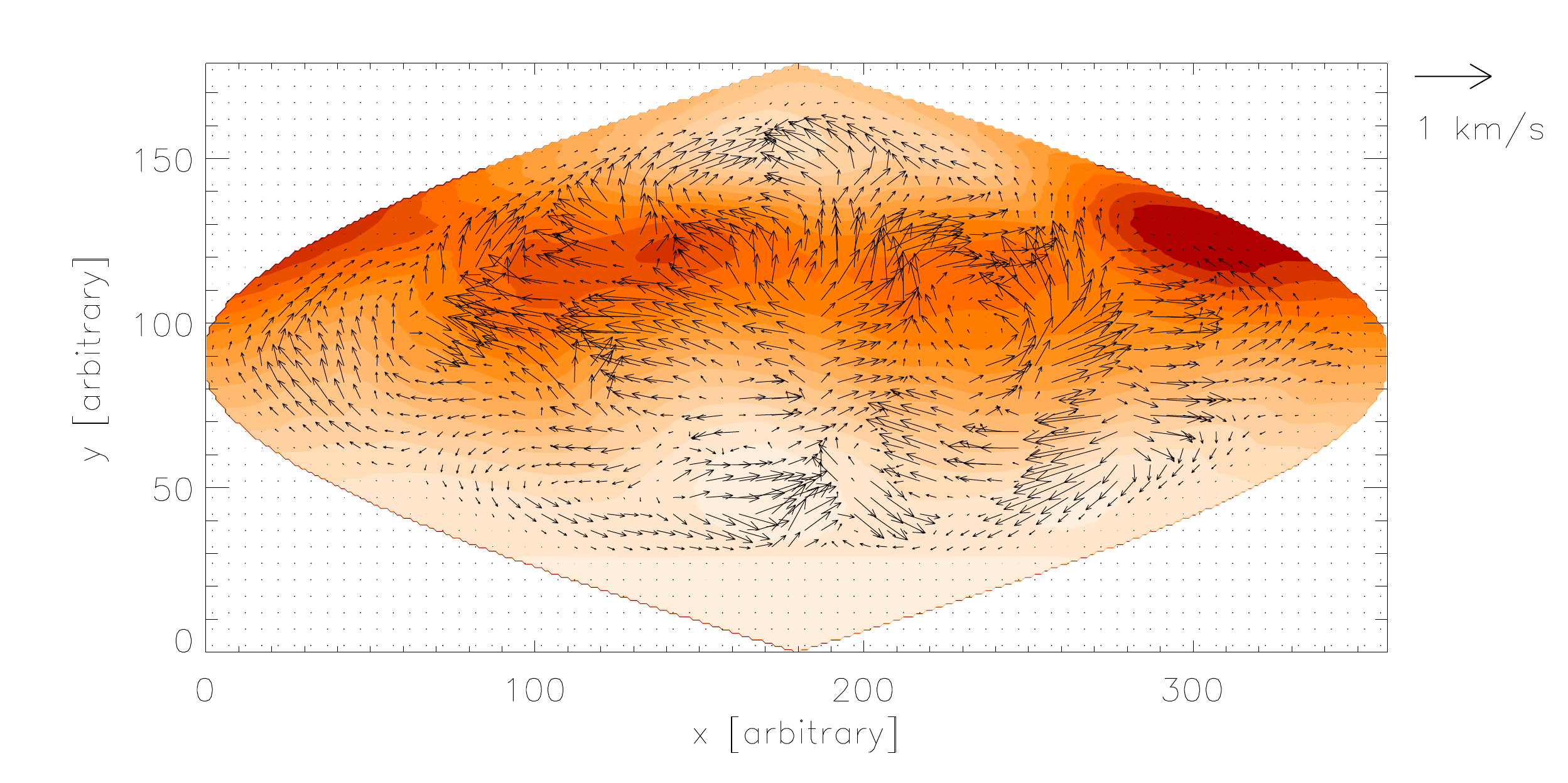}}
\resizebox{!}{6.5cm}{\includegraphics{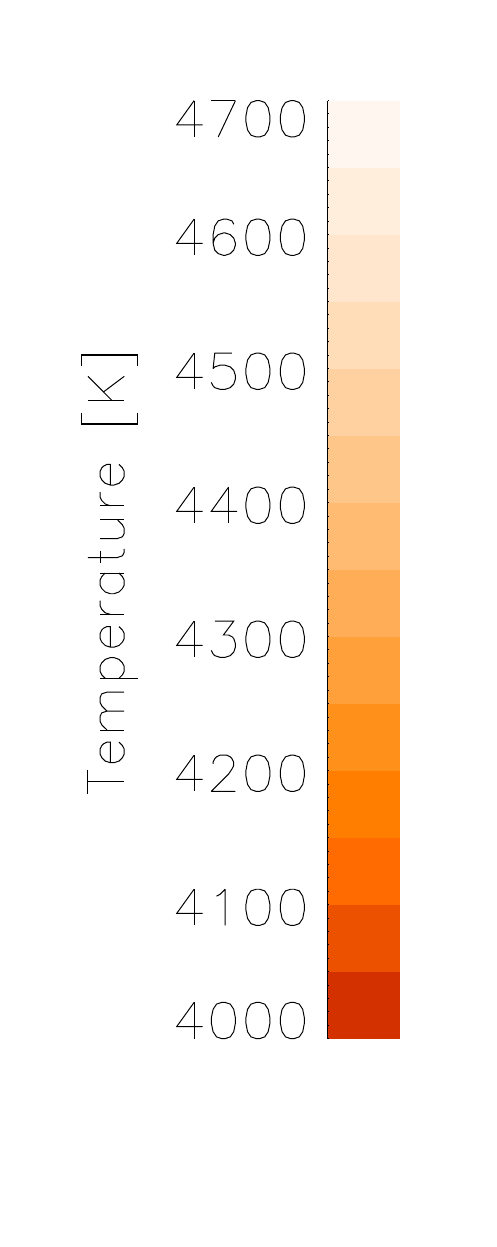}}\\
\caption{Surface flow pattern on $\sigma$~Geminorum, K1III star.}
\label{fig:sgemflows}
\end{figure}

\begin{figure}
\resizebox{0.5\textwidth}{!}{\includegraphics{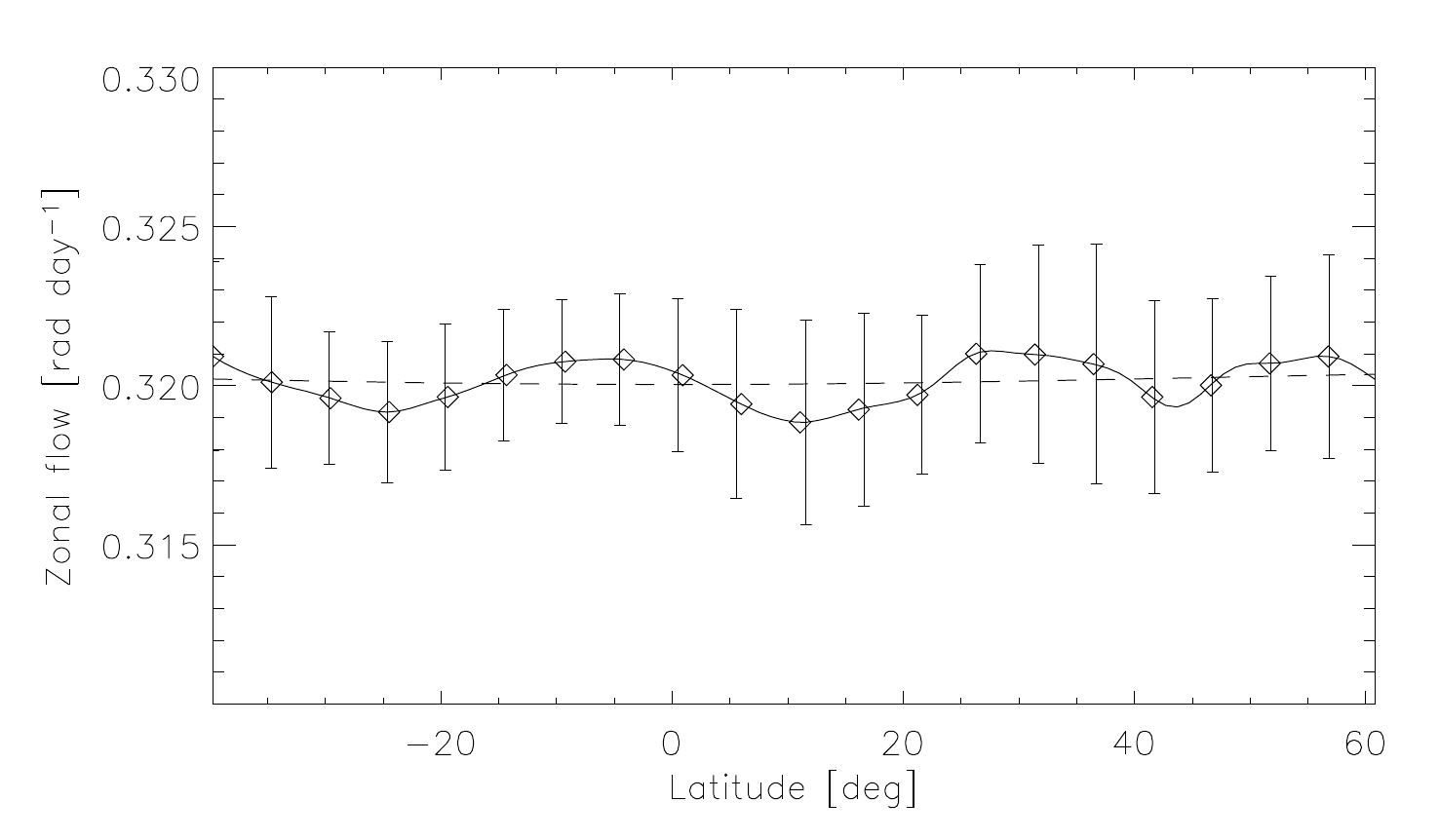}}
\resizebox{0.5\textwidth}{!}{\includegraphics{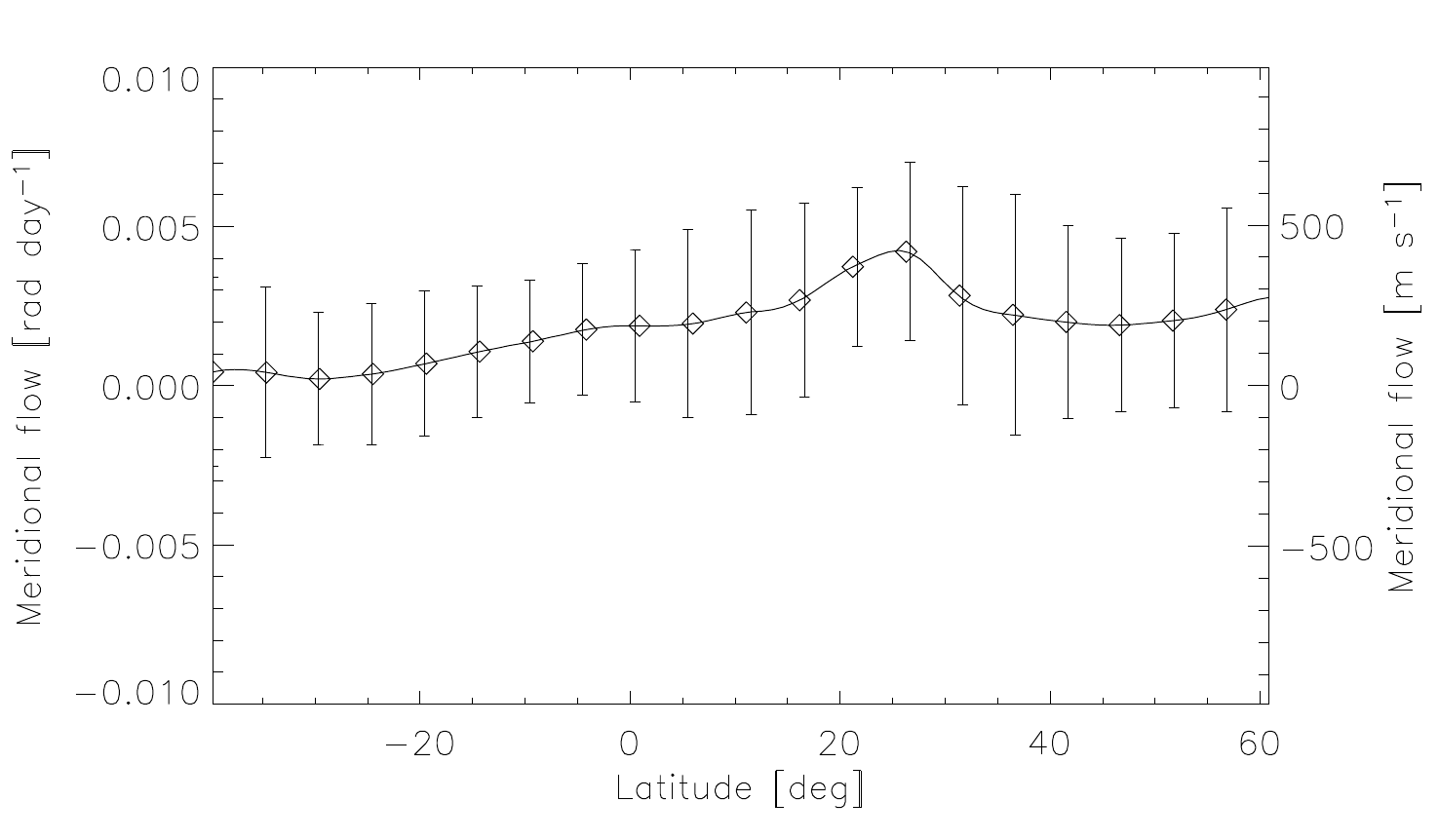}}
\caption{Differential rotation (\emph{left}) and its parabolic fit and meridional circulation of $\sigma$~Gem (\emph{right}).} 
\label{fig:sgemmean}
\end{figure}

Four independent pairs of frames were used for the LCT analysis and averaged to decrease the numerical noise in individual LCT maps. The flow map is displayed in Fig.~\ref{fig:sgemflows}. The differential rotation (Fig.~\ref{fig:sgemmean} left) shows anti-solar rotation with parameters of $\Omega_{\rm eq}=13.83 \pm 0.01\ \rm rad\ day^{-1}$ and $\Delta\Omega=-0.04 \pm 0.03\ \rm rad\ day^{-1}$. Almost ten times stronger anti-solar differential rotation is found using ACCORD analysis. The meridional flow (Fig.~\ref{fig:sgemmean} right) shows a poleward migration of the spot with an average speed of $220\pm 10$~m\,s$^{-1}$. 

\subsection{Conclusions}
The results in the field of stellar surface flows investigation are in the state of individual pilot studies. There is a lot of issues, which has to be solved before the method will provide enough reliable results. With improving quality of the Doppler imaging procedures, for a lot of stars the series of evolving surface features may be expected in a short time period. The robotic telescopes such as STELLA (STELLar Activity; e.~g. \citeauthor{2001AGM....18.P232S}, \citeyear{2001AGM....18.P232S}, \citeauthor{2004AN....325..598W}, \citeyear{2004AN....325..598W}, or \citeauthor{2006PADEU..17..101B}, \citeyear{2006PADEU..17..101B}) will provide large amount of data suitable for stellar surface flow analysis. By that time it would be great to have a powerfull tool for the stellar spots tracking and recovering the signal of differential rotation and meridional circulation. Recently, some theoretical dynamo models of main sequence stars including the differential rotation appeared (e.~g.~\citeauthor{2006MNRAS.373..819L}, \citeyear{2006MNRAS.373..819L} or \citeauthor{2003ApJ...599.1449C}, \citeyear{2003ApJ...599.1449C}) and their results would be easily comparable to the measurements. 

Although the results obtained with the ``standard'' LCT code are not reliable, they are encouraging to develop the new code suitable for the tracking of features in the Doppler imaging temperature maps. Synthetic data would also provide important information about the reliability of the combination of Doppler imaging reconstruction and LCT application.

This topic of astrophysics will be continued in next years in cooperation with a group of Prof.~Strassmeier in Potsdam.

%% file: mypubs.tex
\newcounter{cislo}
\newcommand{\publikace}[4]
{\addtocounter{cislo}{1} \noindent\sloppypar \makebox[1cm]{\thecislo
.}\hangindent 1.0cm \hangafter 1 #1: #3, \emph{#2}, #4
\\}

\section*{\Large List of publications and conference contributions related to this thesis}
\addcontentsline{toc}{section}{List of publications and conference contributions related to this thesis}

%

\publikace{{\bf \v{S}vanda, M.}; Sobotka, M.; Klva\v{n}a, M.}{Experiences with the use of the local correlation tracking method when studying large-scale velocity
fields}{2005}{WDS'05 Proceedings of Contributed Papers: Part III -- Physics (ed. J. \v{S}afr\'ankov\'a), Prague, Matfyzpress, p.~457--462}

\publikace{{\bf \v{S}vanda, M.}; Klva\v{n}a, M.; Sobotka, M.}{Mapping of large-scale photospheric velocity fields}{2005}{in Proceedings of the 11th European Solar Physics Meeting ``The Dynamic Sun: Challenges for Theory and Observations'' (ESA SP-600). 11--16 September 2005, Leuven, Belgium. Editors: D. Danesy, S. Poedts, A. De Groof and J. Andries. Published on CDROM, p.~71.1}

\publikace{{\bf \v{S}vanda, M.}; Klva\v{n}a, M.; Sobotka, M.}{Large-scale horizontal flows in the solar photosphere -- I. Method and tests on synthetic data}{2006}{Astronomy \& Astrophysics, {\bf 458}, p.~301--306}

\publikace{{\bf \v{S}vanda, M.}; Zhao, J.; Kosovichev, A. G.}{Comparison of Large-Scale Flows on the Sun Measured by Time-distance Helioseismology and Local Correlation Tracking}{2007}{Solar Physics, {\bf 241(1)}, p.~27--37}

\publikace{K\H{o}v\'ari, Zs.; Bartus, J.; Strassmeier, K. G.; Vida, K.; {\bf \v{S}vanda, M.}; Ol\'ah, K.}{Anti-solar differential rotation on the active K-giant $\sigma$\,Geminorum}{accepted 23.\,6.\,2007}{Astronomy \& Astrophysics}

\publikace{{\bf \v{S}vanda, M.}; Klva\v{n}a, M.; Sobotka, M.; Bumba, V.}{Large-scale horizontal flows in the solar photosphere -- II. Long-term behaviour and magnetic activity response}{submitted 25.\,4.\,2007, 1st review sent 19.\,6.\,2007}{Astronomy \& Astrophysics}

\publikace{Roudier, Th.; {\bf \v{S}vanda, M.}; Meunier, N.; Keil, S.; Rieutord, M.; Malherbe, J. M.;, Rondi, S.; Molodij, G.; Bommier, V.; Schmieder, B.}{Large-scale horizontal flows in the solar photosphere -- III. Effects on filament destabilization}{submitted 30.\,5.\,2007, 1st review sent 14.\,9.\,2007}{Astronomy \& Astrophysics}

\publikace{Vida, K.; K\"{o}v\'ari, Zs.; {\bf \v{S}vanda, M.}; Ol\'ah, K.; Bartus, J.; Strassmeier, K.~G.}{Differential rotation and surface flow pattern on UZ Librae}{2007}{poster, 5th Potsdam Thinkshop -- Meridional flow, differential rotation, solar and stellar activity}

\publikace{K\H{o}v\'ari, Zs.; Bartus, J.; {\bf \v{S}vanda, M.}; Vida, K;  Strassmeier, K.~G.; Ol\'ah, K.; Forg\'acs-Dajka, E.}{Surface velocity network with anti-solar differential rotation on the active K-giant $\sigma$~Geminorum}{2007}{poster, 5th Potsdam Thinkshop -- Meridional flow, differential rotation, solar and stellar activity}

\publikace{{\bf \v{S}vanda, M.}; Kosovichev, A.~G.; Zhao, J.}{Measurement of the meridional flux transport in the solar photosphere}{2007}{poster, 5th Potsdam Thinkshop -- Meridional flow, differential rotation, solar and stellar activity}

\publikace{{\bf \v{S}vanda, M.}; Kosovichev, A.~G.; Zhao, J.}{Speed of meridional flows and magnetic flux transport on the Sun}{submitted 2.\,9.\,2007}{Astrophysical Journal Letters}